\documentclass[twocolumn,traditabstract,longauth]{aa}

\usepackage{graphicx}
\usepackage[breaklinks, colorlinks, citecolor=blue]{hyperref}
\usepackage{epsfig}     % for figure inclusion
\usepackage{amsmath}
\usepackage{txfonts}
\usepackage{natbib}
\usepackage{color}
\usepackage{url}

\def\setsymbol#1#2{\expandafter\def\csname #1\endcsname{#2}}
\def\getsymbol#1{\csname #1\endcsname}

\def\Planck{{\it Planck\/}}

\def\allearlypapers{\nocite{planck2011-1.1, planck2011-1.3, planck2011-1.4, planck2011-1.5, planck2011-1.6, planck2011-1.7, planck2011-1.10, planck2011-1.10sup, planck2011-5.1a, planck2011-5.1b, planck2011-5.2a, planck2011-5.2b, planck2011-5.2c, planck2011-6.1, planck2011-6.2, planck2011-6.3a, planck2011-6.4a, planck2011-6.4b, planck2011-6.6, planck2011-7.0, planck2011-7.2, planck2011-7.3, planck2011-7.7a, planck2011-7.7b, planck2011-7.12, planck2011-7.13}}

\newbox\tablebox    \newdimen\tablewidth
\def\leaderfil{\leaders\hbox to 5pt{\hss.\hss}\hfil}
\def\endPlancktable{\tablewidth=\columnwidth 
    $$\hss\copy\tablebox\hss$$
    \vskip-\lastskip\vskip -2pt}
\def\endPlancktablewide{\tablewidth=\textwidth 
    $$\hss\copy\tablebox\hss$$
    \vskip-\lastskip\vskip -2pt}
\def\tablenote#1 #2\par{\begingroup \parindent=0.8em
    \abovedisplayshortskip=0pt\belowdisplayshortskip=0pt
    \noindent
    $$\hss\vbox{\hsize\tablewidth \hangindent=\parindent \hangafter=1 \noindent
    \hbox to \parindent{\sup{\rm #1}\hss}\strut#2\strut\par}\hss$$
    \endgroup}
\def\doubleline{\vskip 3pt\hrule \vskip 1.5pt \hrule \vskip 5pt}

\def\L2{\ifmmode L_2\else $L_2$\fi}

\def\DeltaT{\ifmmode \Delta T\else $\Delta T$\fi}
\def\deltat{\ifmmode \Delta t\else $\Delta t$\fi}
\def\fknee{\ifmmode f_{\rm knee}\else $f_{\rm knee}$\fi}
\def\Fmax{\ifmmode F_{\rm max}\else $F_{\rm max}$\fi}
\def\solar{\ifmmode{\rm M}_{\mathord\odot}\else${\rm M}_{\mathord\odot}$\fi}

\def\inv{\ifmmode^{-1}\else$^{-1}$\fi}
\def\mo{\ifmmode^{-1}\else$^{-1}$\fi}
\def\sup#1{\ifmmode ^{\rm #1}\else $^{\rm #1}$\fi}
\def\expo#1{\ifmmode \times 10^{#1}\else $\times 10^{#1}$\fi}
\def\,{\thinspace}
\def\lsim{\mathrel{\raise .4ex\hbox{\rlap{$<$}\lower 1.2ex\hbox{$\sim$}}}}
\def\gsim{\mathrel{\raise .4ex\hbox{\rlap{$>$}\lower 1.2ex\hbox{$\sim$}}}}

\def\simprop{\mathrel{\raise .4ex\hbox{\rlap{$\propto$}\lower 1.2ex\hbox{$\sim$}}}}
\def\deg{\ifmmode^\circ\else$^\circ$\fi}
\def\pdeg{\ifmmode $\setbox0=\hbox{$^{\circ}$}\rlap{\hskip.11\wd0 .}$^{\circ}
          \else \setbox0=\hbox{$^{\circ}$}\rlap{\hskip.11\wd0 .}$^{\circ}$\fi}
\def\arcs{\ifmmode {^{\scriptstyle\prime\prime}}
          \else $^{\scriptstyle\prime\prime}$\fi}
\def\arcm{\ifmmode {^{\scriptstyle\prime}}
          \else $^{\scriptstyle\prime}$\fi}
\newdimen\sa  \newdimen\sb
\def\parcs{\sa=.07em \sb=.03em
     \ifmmode \hbox{\rlap{.}}^{\scriptstyle\prime\kern -\sb\prime}\hbox{\kern -\sa}
     \else \rlap{.}$^{\scriptstyle\prime\kern -\sb\prime}$\kern -\sa\fi}
\def\parcm{\sa=.08em \sb=.03em
     \ifmmode \hbox{\rlap{.}\kern\sa}^{\scriptstyle\prime}\hbox{\kern-\sb}
     \else \rlap{.}\kern\sa$^{\scriptstyle\prime}$\kern-\sb\fi}
\def\ra[#1 #2 #3.#4]{#1\sup{h}#2\sup{m}#3\sup{s}\llap.#4}
\def\dec[#1 #2 #3.#4]{#1\deg#2\arcm#3\arcs\llap.#4}
\def\deco[#1 #2 #3]{#1\deg#2\arcm#3\arcs}
\def\rra[#1 #2]{#1\sup{h}#2\sup{m}}

\def\dots{\relax\ifmmode \ldots\else $\ldots$\fi}
\def\WHzsr{\ifmmode $W\,Hz\mo\,sr\mo$\else W\,Hz\mo\,sr\mo\fi}
\def\mHz{\ifmmode $\,mHz$\else \,mHz\fi}
\def\GHz{\ifmmode $\,GHz$\else \,GHz\fi}
\def\mKs{\ifmmode $\,mK\,s$^{1/2}\else \,mK\,s$^{1/2}$\fi}
\def\muKs{\ifmmode \,\mu$K\,s$^{1/2}\else \,$\mu$K\,s$^{1/2}$\fi}
\def\muKRJs{\ifmmode \,\mu$K$_{\rm RJ}$\,s$^{1/2}\else \,$\mu$K$_{\rm RJ}$\,s$^{1/2}$\fi}
\def\muKHz{\ifmmode \,\mu$K\,Hz$^{-1/2}\else \,$\mu$K\,Hz$^{-1/2}$\fi}
\def\MJysr{\ifmmode \,$MJy\,sr\mo$\else \,MJy\,sr\mo\fi}
\def\MJysrmK{\ifmmode \,$MJy\,sr\mo$\,mK$_{\rm CMB}\mo\else \,MJy\,sr\mo\,mK$_{\rm CMB}\mo$\fi}
\def\microns{\ifmmode \,\mu$m$\else \,$\mu$m\fi}

\def\muK{\ifmmode \,\mu$K$\else \,$\mu$\hbox{K}\fi}
\def\microK{\ifmmode \,\mu$K$\else \,$\mu$\hbox{K}\fi}
\def\muW{\ifmmode \,\mu$W$\else \,$\mu$\hbox{W}\fi}
\def\kms{\ifmmode $\,km\,s$^{-1}\else \,km\,s$^{-1}$\fi}
\def\kmsMpc{\ifmmode $\,\kms\,Mpc\mo$\else \,\kms\,Mpc\mo\fi}
\setsymbol{LFI:center:frequency:70GHz:units}{70.3\,GHz}
\setsymbol{LFI:center:frequency:44GHz:units}{44.1\,GHz}
\setsymbol{LFI:center:frequency:30GHz:units}{28.5\,GHz}

\setsymbol{LFI:center:frequency:70GHz}{70.3}
\setsymbol{LFI:center:frequency:44GHz}{44.1}
\setsymbol{LFI:center:frequency:30GHz}{28.5}

\setsymbol{LFI:center:frequency:LFI18:Rad:M:units}{71.7\GHz}
\setsymbol{LFI:center:frequency:LFI19:Rad:M:units}{67.5\GHz}
\setsymbol{LFI:center:frequency:LFI20:Rad:M:units}{69.2\GHz}
\setsymbol{LFI:center:frequency:LFI21:Rad:M:units}{70.4\GHz}
\setsymbol{LFI:center:frequency:LFI22:Rad:M:units}{71.5\GHz}
\setsymbol{LFI:center:frequency:LFI23:Rad:M:units}{70.8\GHz}
\setsymbol{LFI:center:frequency:LFI24:Rad:M:units}{44.4\GHz}
\setsymbol{LFI:center:frequency:LFI25:Rad:M:units}{44.0\GHz}
\setsymbol{LFI:center:frequency:LFI26:Rad:M:units}{43.9\GHz}
\setsymbol{LFI:center:frequency:LFI27:Rad:M:units}{28.3\GHz}
\setsymbol{LFI:center:frequency:LFI28:Rad:M:units}{28.8\GHz}
\setsymbol{LFI:center:frequency:LFI18:Rad:S:units}{70.1\GHz}
\setsymbol{LFI:center:frequency:LFI19:Rad:S:units}{69.6\GHz}
\setsymbol{LFI:center:frequency:LFI20:Rad:S:units}{69.5\GHz}
\setsymbol{LFI:center:frequency:LFI21:Rad:S:units}{69.5\GHz}
\setsymbol{LFI:center:frequency:LFI22:Rad:S:units}{72.8\GHz}
\setsymbol{LFI:center:frequency:LFI23:Rad:S:units}{71.3\GHz}
\setsymbol{LFI:center:frequency:LFI24:Rad:S:units}{44.1\GHz}
\setsymbol{LFI:center:frequency:LFI25:Rad:S:units}{44.1\GHz}
\setsymbol{LFI:center:frequency:LFI26:Rad:S:units}{44.1\GHz}
\setsymbol{LFI:center:frequency:LFI27:Rad:S:units}{28.5\GHz}
\setsymbol{LFI:center:frequency:LFI28:Rad:S:units}{28.2\GHz}

\setsymbol{LFI:center:frequency:LFI18:Rad:M}{71.7}
\setsymbol{LFI:center:frequency:LFI19:Rad:M}{67.5}
\setsymbol{LFI:center:frequency:LFI20:Rad:M}{69.2}
\setsymbol{LFI:center:frequency:LFI21:Rad:M}{70.4}
\setsymbol{LFI:center:frequency:LFI22:Rad:M}{71.5}
\setsymbol{LFI:center:frequency:LFI23:Rad:M}{70.8}
\setsymbol{LFI:center:frequency:LFI24:Rad:M}{44.4}
\setsymbol{LFI:center:frequency:LFI25:Rad:M}{44.0}
\setsymbol{LFI:center:frequency:LFI26:Rad:M}{43.9}
\setsymbol{LFI:center:frequency:LFI27:Rad:M}{28.3}
\setsymbol{LFI:center:frequency:LFI28:Rad:M}{28.8}
\setsymbol{LFI:center:frequency:LFI18:Rad:S}{70.1}
\setsymbol{LFI:center:frequency:LFI19:Rad:S}{69.6}
\setsymbol{LFI:center:frequency:LFI20:Rad:S}{69.5}
\setsymbol{LFI:center:frequency:LFI21:Rad:S}{69.5}
\setsymbol{LFI:center:frequency:LFI22:Rad:S}{72.8}
\setsymbol{LFI:center:frequency:LFI23:Rad:S}{71.3}
\setsymbol{LFI:center:frequency:LFI24:Rad:S}{44.1}
\setsymbol{LFI:center:frequency:LFI25:Rad:S}{44.1}
\setsymbol{LFI:center:frequency:LFI26:Rad:S}{44.1}
\setsymbol{LFI:center:frequency:LFI27:Rad:S}{28.5}
\setsymbol{LFI:center:frequency:LFI28:Rad:S}{28.2}

\setsymbol{LFI:white:noise:sensitivity:70GHz:units}{152.6\muKs}
\setsymbol{LFI:white:noise:sensitivity:44GHz:units}{173.1\muKs}
\setsymbol{LFI:white:noise:sensitivity:30GHz:units}{146.8\muKs}

\setsymbol{LFI:white:noise:sensitivity:70GHz}{152.6}
\setsymbol{LFI:white:noise:sensitivity:44GHz}{173.1}
\setsymbol{LFI:white:noise:sensitivity:30GHz}{146.8}

\setsymbol{LFI:white:noise:sensitivity:LFI18:Rad:M:units}{512.0\muKs}
\setsymbol{LFI:white:noise:sensitivity:LFI19:Rad:M:units}{581.4\muKs}
\setsymbol{LFI:white:noise:sensitivity:LFI20:Rad:M:units}{590.8\muKs}
\setsymbol{LFI:white:noise:sensitivity:LFI21:Rad:M:units}{455.2\muKs}
\setsymbol{LFI:white:noise:sensitivity:LFI22:Rad:M:units}{492.0\muKs}
\setsymbol{LFI:white:noise:sensitivity:LFI23:Rad:M:units}{507.7\muKs}
\setsymbol{LFI:white:noise:sensitivity:LFI24:Rad:M:units}{462.2\muKs}
\setsymbol{LFI:white:noise:sensitivity:LFI25:Rad:M:units}{413.6\muKs}
\setsymbol{LFI:white:noise:sensitivity:LFI26:Rad:M:units}{478.6\muKs}
\setsymbol{LFI:white:noise:sensitivity:LFI27:Rad:M:units}{277.7\muKs}
\setsymbol{LFI:white:noise:sensitivity:LFI28:Rad:M:units}{312.3\muKs}
\setsymbol{LFI:white:noise:sensitivity:LFI18:Rad:S:units}{465.7\muKs}
\setsymbol{LFI:white:noise:sensitivity:LFI19:Rad:S:units}{555.6\muKs}
\setsymbol{LFI:white:noise:sensitivity:LFI20:Rad:S:units}{623.2\muKs}
\setsymbol{LFI:white:noise:sensitivity:LFI21:Rad:S:units}{564.1\muKs}
\setsymbol{LFI:white:noise:sensitivity:LFI22:Rad:S:units}{534.4\muKs}
\setsymbol{LFI:white:noise:sensitivity:LFI23:Rad:S:units}{542.4\muKs}
\setsymbol{LFI:white:noise:sensitivity:LFI24:Rad:S:units}{399.2\muKs}
\setsymbol{LFI:white:noise:sensitivity:LFI25:Rad:S:units}{392.6\muKs}
\setsymbol{LFI:white:noise:sensitivity:LFI26:Rad:S:units}{418.6\muKs}
\setsymbol{LFI:white:noise:sensitivity:LFI27:Rad:S:units}{302.9\muKs}
\setsymbol{LFI:white:noise:sensitivity:LFI28:Rad:S:units}{285.3\muKs}

\setsymbol{LFI:white:noise:sensitivity:LFI18:Rad:M}{512.0}
\setsymbol{LFI:white:noise:sensitivity:LFI19:Rad:M}{581.4}
\setsymbol{LFI:white:noise:sensitivity:LFI20:Rad:M}{590.8}
\setsymbol{LFI:white:noise:sensitivity:LFI21:Rad:M}{455.2}
\setsymbol{LFI:white:noise:sensitivity:LFI22:Rad:M}{492.0}
\setsymbol{LFI:white:noise:sensitivity:LFI23:Rad:M}{507.7}
\setsymbol{LFI:white:noise:sensitivity:LFI24:Rad:M}{462.2}
\setsymbol{LFI:white:noise:sensitivity:LFI25:Rad:M}{413.6}
\setsymbol{LFI:white:noise:sensitivity:LFI26:Rad:M}{478.6}
\setsymbol{LFI:white:noise:sensitivity:LFI27:Rad:M}{277.7}
\setsymbol{LFI:white:noise:sensitivity:LFI28:Rad:M}{312.3}
\setsymbol{LFI:white:noise:sensitivity:LFI18:Rad:S}{465.7}
\setsymbol{LFI:white:noise:sensitivity:LFI19:Rad:S}{555.6}
\setsymbol{LFI:white:noise:sensitivity:LFI20:Rad:S}{623.2}
\setsymbol{LFI:white:noise:sensitivity:LFI21:Rad:S}{564.1}
\setsymbol{LFI:white:noise:sensitivity:LFI22:Rad:S}{534.4}
\setsymbol{LFI:white:noise:sensitivity:LFI23:Rad:S}{542.4}
\setsymbol{LFI:white:noise:sensitivity:LFI24:Rad:S}{399.2}
\setsymbol{LFI:white:noise:sensitivity:LFI25:Rad:S}{392.6}
\setsymbol{LFI:white:noise:sensitivity:LFI26:Rad:S}{418.6}
\setsymbol{LFI:white:noise:sensitivity:LFI27:Rad:S}{302.9}
\setsymbol{LFI:white:noise:sensitivity:LFI28:Rad:S}{285.3}

\setsymbol{LFI:knee:frequency:70GHz:units}{29.5\mHz}
\setsymbol{LFI:knee:frequency:44GHz:units}{56.2\mHz}
\setsymbol{LFI:knee:frequency:30GHz:units}{113.7\mHz}

\setsymbol{LFI:knee:frequency:70GHz}{29.5}
\setsymbol{LFI:knee:frequency:44GHz}{56.2}
\setsymbol{LFI:knee:frequency:30GHz}{113.7}

\setsymbol{LFI:knee:frequency:LFI18:Rad:M:units}{16.3\mHz}
\setsymbol{LFI:knee:frequency:LFI19:Rad:M:units}{15.1\mHz}
\setsymbol{LFI:knee:frequency:LFI20:Rad:M:units}{18.7\mHz}
\setsymbol{LFI:knee:frequency:LFI21:Rad:M:units}{37.2\mHz}
\setsymbol{LFI:knee:frequency:LFI22:Rad:M:units}{12.7\mHz}
\setsymbol{LFI:knee:frequency:LFI23:Rad:M:units}{34.6\mHz}
\setsymbol{LFI:knee:frequency:LFI24:Rad:M:units}{46.2\mHz}
\setsymbol{LFI:knee:frequency:LFI25:Rad:M:units}{24.9\mHz}
\setsymbol{LFI:knee:frequency:LFI26:Rad:M:units}{67.6\mHz}
\setsymbol{LFI:knee:frequency:LFI27:Rad:M:units}{187.4\mHz}
\setsymbol{LFI:knee:frequency:LFI28:Rad:M:units}{122.2\mHz}
\setsymbol{LFI:knee:frequency:LFI18:Rad:S:units}{17.7\mHz}
\setsymbol{LFI:knee:frequency:LFI19:Rad:S:units}{22.0\mHz}
\setsymbol{LFI:knee:frequency:LFI20:Rad:S:units}{8.7\mHz}
\setsymbol{LFI:knee:frequency:LFI21:Rad:S:units}{25.9\mHz}
\setsymbol{LFI:knee:frequency:LFI22:Rad:S:units}{15.8\mHz}
\setsymbol{LFI:knee:frequency:LFI23:Rad:S:units}{129.8\mHz}
\setsymbol{LFI:knee:frequency:LFI24:Rad:S:units}{100.9\mHz}
\setsymbol{LFI:knee:frequency:LFI25:Rad:S:units}{38.9\mHz}
\setsymbol{LFI:knee:frequency:LFI26:Rad:S:units}{58.9\mHz}
\setsymbol{LFI:knee:frequency:LFI27:Rad:S:units}{104.4\mHz}
\setsymbol{LFI:knee:frequency:LFI28:Rad:S:units}{40.7\mHz}

\setsymbol{LFI:knee:frequency:LFI18:Rad:M}{16.3}
\setsymbol{LFI:knee:frequency:LFI19:Rad:M}{15.1}
\setsymbol{LFI:knee:frequency:LFI20:Rad:M}{18.7}
\setsymbol{LFI:knee:frequency:LFI21:Rad:M}{37.2}
\setsymbol{LFI:knee:frequency:LFI22:Rad:M}{12.7}
\setsymbol{LFI:knee:frequency:LFI23:Rad:M}{34.6}
\setsymbol{LFI:knee:frequency:LFI24:Rad:M}{46.2}
\setsymbol{LFI:knee:frequency:LFI25:Rad:M}{24.9}
\setsymbol{LFI:knee:frequency:LFI26:Rad:M}{67.6}
\setsymbol{LFI:knee:frequency:LFI27:Rad:M}{187.4}
\setsymbol{LFI:knee:frequency:LFI28:Rad:M}{122.2}
\setsymbol{LFI:knee:frequency:LFI18:Rad:S}{17.7}
\setsymbol{LFI:knee:frequency:LFI19:Rad:S}{22.0}
\setsymbol{LFI:knee:frequency:LFI20:Rad:S}{8.7}
\setsymbol{LFI:knee:frequency:LFI21:Rad:S}{25.9}
\setsymbol{LFI:knee:frequency:LFI22:Rad:S}{15.8}
\setsymbol{LFI:knee:frequency:LFI23:Rad:S}{129.8}
\setsymbol{LFI:knee:frequency:LFI24:Rad:S}{100.9}
\setsymbol{LFI:knee:frequency:LFI25:Rad:S}{38.9}
\setsymbol{LFI:knee:frequency:LFI26:Rad:S}{58.9}
\setsymbol{LFI:knee:frequency:LFI27:Rad:S}{104.4}
\setsymbol{LFI:knee:frequency:LFI28:Rad:S}{40.7}

\setsymbol{LFI:slope:70GHz:units}{$-1.03$\mHz}
\setsymbol{LFI:slope:44GHz:units}{$-0.89$\mHz}
\setsymbol{LFI:slope:30GHz:units}{$-0.87$\mHz}

\setsymbol{LFI:slope:70GHz}{$-1.03$}
\setsymbol{LFI:slope:44GHz}{$-0.89$}
\setsymbol{LFI:slope:30GHz}{$-0.87$}

\setsymbol{LFI:slope:LFI18:Rad:M:units}{$-1.04$\mHz}
\setsymbol{LFI:slope:LFI19:Rad:M:units}{$-1.09$\mHz}
\setsymbol{LFI:slope:LFI20:Rad:M:units}{$-0.69$\mHz}
\setsymbol{LFI:slope:LFI21:Rad:M:units}{$-1.56$\mHz}
\setsymbol{LFI:slope:LFI22:Rad:M:units}{$-1.01$\mHz}
\setsymbol{LFI:slope:LFI23:Rad:M:units}{$-0.96$\mHz}
\setsymbol{LFI:slope:LFI24:Rad:M:units}{$-0.83$\mHz}
\setsymbol{LFI:slope:LFI25:Rad:M:units}{$-0.91$\mHz}
\setsymbol{LFI:slope:LFI26:Rad:M:units}{$-0.95$\mHz}
\setsymbol{LFI:slope:LFI27:Rad:M:units}{$-0.87$\mHz}
\setsymbol{LFI:slope:LFI28:Rad:M:units}{$-0.88$\mHz}
\setsymbol{LFI:slope:LFI18:Rad:S:units}{$-1.15$\mHz}
\setsymbol{LFI:slope:LFI19:Rad:S:units}{$-1.00$\mHz}
\setsymbol{LFI:slope:LFI20:Rad:S:units}{$-0.95$\mHz}
\setsymbol{LFI:slope:LFI21:Rad:S:units}{$-0.92$\mHz}
\setsymbol{LFI:slope:LFI22:Rad:S:units}{$-1.01$\mHz}
\setsymbol{LFI:slope:LFI23:Rad:S:units}{$-0.95$\mHz}
\setsymbol{LFI:slope:LFI24:Rad:S:units}{$-0.73$\mHz}
\setsymbol{LFI:slope:LFI25:Rad:S:units}{$-1.16$\mHz}
\setsymbol{LFI:slope:LFI26:Rad:S:units}{$-0.79$\mHz}
\setsymbol{LFI:slope:LFI27:Rad:S:units}{$-0.82$\mHz}
\setsymbol{LFI:slope:LFI28:Rad:S:units}{$-0.91$\mHz}

\setsymbol{LFI:slope:LFI18:Rad:M}{$-1.04$}
\setsymbol{LFI:slope:LFI19:Rad:M}{$-1.09$}
\setsymbol{LFI:slope:LFI20:Rad:M}{$-0.69$}
\setsymbol{LFI:slope:LFI21:Rad:M}{$-1.56$}
\setsymbol{LFI:slope:LFI22:Rad:M}{$-1.01$}
\setsymbol{LFI:slope:LFI23:Rad:M}{$-0.96$}
\setsymbol{LFI:slope:LFI24:Rad:M}{$-0.83$}
\setsymbol{LFI:slope:LFI25:Rad:M}{$-0.91$}
\setsymbol{LFI:slope:LFI26:Rad:M}{$-0.95$}
\setsymbol{LFI:slope:LFI27:Rad:M}{$-0.87$}
\setsymbol{LFI:slope:LFI28:Rad:M}{$-0.88$}
\setsymbol{LFI:slope:LFI18:Rad:S}{$-1.15$}
\setsymbol{LFI:slope:LFI19:Rad:S}{$-1.00$}
\setsymbol{LFI:slope:LFI20:Rad:S}{$-0.95$}
\setsymbol{LFI:slope:LFI21:Rad:S}{$-0.92$}
\setsymbol{LFI:slope:LFI22:Rad:S}{$-1.01$}
\setsymbol{LFI:slope:LFI23:Rad:S}{$-0.95$}
\setsymbol{LFI:slope:LFI24:Rad:S}{$-0.73$}
\setsymbol{LFI:slope:LFI25:Rad:S}{$-1.16$}
\setsymbol{LFI:slope:LFI26:Rad:S}{$-0.79$}
\setsymbol{LFI:slope:LFI27:Rad:S}{$-0.82$}
\setsymbol{LFI:slope:LFI28:Rad:S}{$-0.91$}

\setsymbol{LFI:FWHM:70GHz:units}{13\parcm01}
\setsymbol{LFI:FWHM:44GHz:units}{27\parcm92}
\setsymbol{LFI:FWHM:30GHz:units}{32\parcm65}

\setsymbol{LFI:FWHM:70GHz}{13.01}
\setsymbol{LFI:FWHM:44GHz}{27.92}
\setsymbol{LFI:FWHM:30GHz}{32.65}

\setsymbol{LFI:FWHM:LFI18:units}{13\parcm39}
\setsymbol{LFI:FWHM:LFI19:units}{13\parcm01}
\setsymbol{LFI:FWHM:LFI20:units}{12\parcm75}
\setsymbol{LFI:FWHM:LFI21:units}{12\parcm74}
\setsymbol{LFI:FWHM:LFI22:units}{12\parcm87}
\setsymbol{LFI:FWHM:LFI23:units}{13\parcm27}
\setsymbol{LFI:FWHM:LFI24:units}{22\parcm98}
\setsymbol{LFI:FWHM:LFI25:units}{30\parcm46}
\setsymbol{LFI:FWHM:LFI26:units}{30\parcm31}
\setsymbol{LFI:FWHM:LFI27:units}{32\parcm65}
\setsymbol{LFI:FWHM:LFI28:units}{32\parcm66}

\setsymbol{LFI:FWHM:LFI18}{13.39}
\setsymbol{LFI:FWHM:LFI19}{13.01}
\setsymbol{LFI:FWHM:LFI20}{12.75}
\setsymbol{LFI:FWHM:LFI21}{12.74}
\setsymbol{LFI:FWHM:LFI22}{12.87}
\setsymbol{LFI:FWHM:LFI23}{13.27}
\setsymbol{LFI:FWHM:LFI24}{22.98}
\setsymbol{LFI:FWHM:LFI25}{30.46}
\setsymbol{LFI:FWHM:LFI26}{30.31}
\setsymbol{LFI:FWHM:LFI27}{32.65}
\setsymbol{LFI:FWHM:LFI28}{32.66}

\setsymbol{LFI:FWHM:uncertainty:LFI18:units}{0.170\arcm}
\setsymbol{LFI:FWHM:uncertainty:LFI19:units}{0.174\arcm}
\setsymbol{LFI:FWHM:uncertainty:LFI20:units}{0.170\arcm}
\setsymbol{LFI:FWHM:uncertainty:LFI21:units}{0.156\arcm}
\setsymbol{LFI:FWHM:uncertainty:LFI22:units}{0.164\arcm}
\setsymbol{LFI:FWHM:uncertainty:LFI23:units}{0.171\arcm}
\setsymbol{LFI:FWHM:uncertainty:LFI24:units}{0.652\arcm}
\setsymbol{LFI:FWHM:uncertainty:LFI25:units}{1.075\arcm}
\setsymbol{LFI:FWHM:uncertainty:LFI26:units}{1.131\arcm}
\setsymbol{LFI:FWHM:uncertainty:LFI27:units}{1.266\arcm}
\setsymbol{LFI:FWHM:uncertainty:LFI28:units}{1.287\arcm}

\setsymbol{LFI:FWHM:uncertainty:LFI18}{0.170}
\setsymbol{LFI:FWHM:uncertainty:LFI19}{0.174}
\setsymbol{LFI:FWHM:uncertainty:LFI20}{0.170}
\setsymbol{LFI:FWHM:uncertainty:LFI21}{0.156}
\setsymbol{LFI:FWHM:uncertainty:LFI22}{0.164}
\setsymbol{LFI:FWHM:uncertainty:LFI23}{0.171}
\setsymbol{LFI:FWHM:uncertainty:LFI24}{0.652}
\setsymbol{LFI:FWHM:uncertainty:LFI25}{1.075}
\setsymbol{LFI:FWHM:uncertainty:LFI26}{1.131}
\setsymbol{LFI:FWHM:uncertainty:LFI27}{1.266}
\setsymbol{LFI:FWHM:uncertainty:LFI28}{1.287}

\setsymbol{HFI:center:frequency:100GHz:units}{100\,GHz}
\setsymbol{HFI:center:frequency:143GHz:units}{143\,GHz}
\setsymbol{HFI:center:frequency:217GHz:units}{217\,GHz}
\setsymbol{HFI:center:frequency:353GHz:units}{353\,GHz}
\setsymbol{HFI:center:frequency:545GHz:units}{545\,GHz}
\setsymbol{HFI:center:frequency:857GHz:units}{857\,GHz}

\setsymbol{HFI:center:frequency:100GHz}{100}
\setsymbol{HFI:center:frequency:143GHz}{143}
\setsymbol{HFI:center:frequency:217GHz}{217}
\setsymbol{HFI:center:frequency:353GHz}{353}
\setsymbol{HFI:center:frequency:545GHz}{545}
\setsymbol{HFI:center:frequency:857GHz}{857}

\setsymbol{HFI:Ndetectors:100GHz}{8}
\setsymbol{HFI:Ndetectors:143GHz}{11}
\setsymbol{HFI:Ndetectors:217GHz}{12}
\setsymbol{HFI:Ndetectors:353GHz}{12}
\setsymbol{HFI:Ndetectors:545GHz}{3}
\setsymbol{HFI:Ndetectors:857GHz}{4}

\setsymbol{HFI:FWHM:Maps:100GHz:units}{9\parcm88}
\setsymbol{HFI:FWHM:Maps:143GHz:units}{7\parcm18}
\setsymbol{HFI:FWHM:Maps:217GHz:units}{4\parcm87}
\setsymbol{HFI:FWHM:Maps:353GHz:units}{4\parcm65}
\setsymbol{HFI:FWHM:Maps:545GHz:units}{4\parcm72}
\setsymbol{HFI:FWHM:Maps:857GHz:units}{4\parcm39}
\setsymbol{HFI:FWHM:Maps:100GHz}{9.88}
\setsymbol{HFI:FWHM:Maps:143GHz}{7.18}
\setsymbol{HFI:FWHM:Maps:217GHz}{4.87}
\setsymbol{HFI:FWHM:Maps:353GHz}{4.65}
\setsymbol{HFI:FWHM:Maps:545GHz}{4.72}
\setsymbol{HFI:FWHM:Maps:857GHz}{4.39}

\setsymbol{HFI:beam:ellipticity:Maps:100GHz}{1.15}
\setsymbol{HFI:beam:ellipticity:Maps:143GHz}{1.01}
\setsymbol{HFI:beam:ellipticity:Maps:217GHz}{1.06}
\setsymbol{HFI:beam:ellipticity:Maps:353GHz}{1.05}
\setsymbol{HFI:beam:ellipticity:Maps:545GHz}{1.14}
\setsymbol{HFI:beam:ellipticity:Maps:857GHz}{1.19}

\setsymbol{HFI:FWHM:Mars:100GHz:units}{9\parcm37}
\setsymbol{HFI:FWHM:Mars:143GHz:units}{7\parcm04}
\setsymbol{HFI:FWHM:Mars:217GHz:units}{4\parcm68}
\setsymbol{HFI:FWHM:Mars:353GHz:units}{4\parcm43}
\setsymbol{HFI:FWHM:Mars:545GHz:units}{3\parcm80}
\setsymbol{HFI:FWHM:Mars:857GHz:units}{3\parcm67}

\setsymbol{HFI:FWHM:Mars:100GHz}{9.37}
\setsymbol{HFI:FWHM:Mars:143GHz}{7.04}
\setsymbol{HFI:FWHM:Mars:217GHz}{4.68}
\setsymbol{HFI:FWHM:Mars:353GHz}{4.43}
\setsymbol{HFI:FWHM:Mars:545GHz}{3.80}
\setsymbol{HFI:FWHM:Mars:857GHz}{3.67}

\setsymbol{HFI:beam:ellipticity:Mars:100GHz}{1.18}
\setsymbol{HFI:beam:ellipticity:Mars:143GHz}{1.03}
\setsymbol{HFI:beam:ellipticity:Mars:217GHz}{1.14}
\setsymbol{HFI:beam:ellipticity:Mars:353GHz}{1.09}
\setsymbol{HFI:beam:ellipticity:Mars:545GHz}{1.25}
\setsymbol{HFI:beam:ellipticity:Mars:857GHz}{1.03}

\setsymbol{HFI:CMB:relative:calibration:100GHz}{$\lsim 1\%$}
\setsymbol{HFI:CMB:relative:calibration:143GHz}{$\lsim 1\%$}
\setsymbol{HFI:CMB:relative:calibration:217GHz}{$\lsim 1\%$}
\setsymbol{HFI:CMB:relative:calibration:353GHz}{$\lsim 1\%$}
\setsymbol{HFI:CMB:relative:calibration:545GHz}{}
\setsymbol{HFI:CMB:relative:calibration:857GHz}{}

\setsymbol{HFI:CMB:absolute:calibration:100GHz}{$\lsim 2\%$}
\setsymbol{HFI:CMB:absolute:calibration:143GHz}{$\lsim 2\%$}
\setsymbol{HFI:CMB:absolute:calibration:217GHz}{$\lsim 2\%$}
\setsymbol{HFI:CMB:absolute:calibration:353GHz}{$\lsim 2\%$}
\setsymbol{HFI:CMB:absolute:calibration:545GHz}{}
\setsymbol{HFI:CMB:absolute:calibration:857GHz}{}

\setsymbol{HFI:FIRAS:gain:calibration:accuracy:statistical:100GHz}{}
\setsymbol{HFI:FIRAS:gain:calibration:accuracy:statistical:143GHz}{}
\setsymbol{HFI:FIRAS:gain:calibration:accuracy:statistical:217GHz}{}
\setsymbol{HFI:FIRAS:gain:calibration:accuracy:statistical:353GHz}{2.5\%}
\setsymbol{HFI:FIRAS:gain:calibration:accuracy:statistical:545GHz}{1\%}
\setsymbol{HFI:FIRAS:gain:calibration:accuracy:statistical:857GHz}{0.5\%}

\setsymbol{HFI:FIRAS:gain:calibration:accuracy:systematic:100GHz}{}
\setsymbol{HFI:FIRAS:gain:calibration:accuracy:systematic:143GHz}{}
\setsymbol{HFI:FIRAS:gain:calibration:accuracy:systematic:217GHz}{}
\setsymbol{HFI:FIRAS:gain:calibration:accuracy:systematic:353GHz}{}
\setsymbol{HFI:FIRAS:gain:calibration:accuracy:systematic:545GHz}{7\%}
\setsymbol{HFI:FIRAS:gain:calibration:accuracy:systematic:857GHz}{7\%}

\setsymbol{HFI:FIRAS:zero:point:accuracy:100GHz:units}{0.8\MJysr}
\setsymbol{HFI:FIRAS:zero:point:accuracy:143GHz:units}{}
\setsymbol{HFI:FIRAS:zero:point:accuracy:217GHz:units}{}
\setsymbol{HFI:FIRAS:zero:point:accuracy:353GHz:units}{1.4\MJysr}
\setsymbol{HFI:FIRAS:zero:point:accuracy:545GHz:units}{2.2\MJysr}
\setsymbol{HFI:FIRAS:zero:point:accuracy:857GHz:units}{1.7\MJysr}

\setsymbol{HFI:FIRAS:zero:point:accuracy:100GHz}{0.8}
\setsymbol{HFI:FIRAS:zero:point:accuracy:143GHz}{}
\setsymbol{HFI:FIRAS:zero:point:accuracy:217GHz}{}
\setsymbol{HFI:FIRAS:zero:point:accuracy:353GHz}{1.4}
\setsymbol{HFI:FIRAS:zero:point:accuracy:545GHz}{2.2}
\setsymbol{HFI:FIRAS:zero:point:accuracy:857GHz}{1.7}

\setsymbol{HFI:unit:conversion:100GHz:units}{0.2415\MJysrmK}
\setsymbol{HFI:unit:conversion:143GHz:units}{0.3694\MJysrmK}
\setsymbol{HFI:unit:conversion:217GHz:units}{0.4811\MJysrmK}
\setsymbol{HFI:unit:conversion:353GHz:units}{0.2883\MJysrmK}
\setsymbol{HFI:unit:conversion:545GHz:units}{0.05826\MJysrmK}
\setsymbol{HFI:unit:conversion:857GHz:units}{0.002238\MJysrmK}

\setsymbol{HFI:unit:conversion:100GHz}{0.2415}
\setsymbol{HFI:unit:conversion:143GHz}{0.3694}
\setsymbol{HFI:unit:conversion:217GHz}{0.4811}
\setsymbol{HFI:unit:conversion:353GHz}{0.2883}
\setsymbol{HFI:unit:conversion:545GHz}{0.05826}
\setsymbol{HFI:unit:conversion:857GHz}{0.002238}

\setsymbol{HFI:colour:correction:alpha=-2:V1.01:100GHz}{0.9893}
\setsymbol{HFI:colour:correction:alpha=-2:V1.01:143GHz}{0.9759}
\setsymbol{HFI:colour:correction:alpha=-2:V1.01:217GHz}{1.0007}
\setsymbol{HFI:colour:correction:alpha=-2:V1.01:353GHz}{1.0028}
\setsymbol{HFI:colour:correction:alpha=-2:V1.01:545GHz}{1.0019}
\setsymbol{HFI:colour:correction:alpha=-2:V1.01:857GHz}{0.9889}

\setsymbol{HFI:colour:correction:alpha=0:V1.01:100GHz}{1.0008}
\setsymbol{HFI:colour:correction:alpha=0:V1.01:143GHz}{1.0148}
\setsymbol{HFI:colour:correction:alpha=0:V1.01:217GHz}{0.9909}
\setsymbol{HFI:colour:correction:alpha=0:V1.01:353GHz}{0.9888}
\setsymbol{HFI:colour:correction:alpha=0:V1.01:545GHz}{0.9878}
\setsymbol{HFI:colour:correction:alpha=0:V1.01:857GHz}{1.0014}

\def\cc{\ifmmode{\,{\rm cm}^{-3}}\else{$\,{\rm cm}^{-3}$}\fi}
\def\cq{\ifmmode{\,{\rm cm}^{-2}}\else{$\,{\rm cm}^{-2}$}\fi}
\def\mic{\ifmmode{\,\mu{\rm m}}\else{$\mu$m}\fi}
\def\eccs{\ifmmode{\,{\rm erg}\,{\rm cm}^{-3} {\rm s}^{-1}}\else{$\,{\rm
erg}\,{\rm cm}^{-3} {\rm s}^{-1}$}\fi}
\def\ecqs{\ifmmode{\,{\rm erg}\,{\rm cm}^{-2}\,{\rm s}^{-1}\,{\rm
sr}^{-1}}\else{$\,{\rm erg}\,{\rm cm}^{-2}\,{\rm s}^{-1}\,{\rm sr}^{-1}$}\fi}
\def\deg{\ifmmode{^{\circ}}\else{$^{\circ}$}\fi} 
\def\pc{\ifmmode{\,{\rm pc}}\else{$\,{\rm pc}$}\fi} 
\def\kms{\ifmmode{\,{\rm km}\,{\rm s}^{-1}}\else{km s$^{-1}$}\fi} 
\def\mlin{\ifmmode{\,{\rm M}_\odot\,{\rm pc}^{-1}}
 \else{M$_\odot$\,pc$^{-1}$}\fi}
\def\msol{\ifmmode{\,{\rm M_\odot}}\else{M$_\odot$}\fi}  
\def\lsol{\ifmmode{\,{\rm L_\odot}}\else{L$_\odot$}\fi}  
\def\kmspc{\ifmmode{\,{\rm km}\,{\rm s}^{-1}\,{\rm pc}^{-1}}\else{km
s$^{-1}$ pc$^{-1}$}\fi} 
\def\MJysr{\ifmmode{\,{\rm MJy\,sr}^{-1}}\else{$\,{\rm MJy\,sr}^{-1}$}\fi} 
\def\Kkms{\ifmmode{\,{\rm K\,km\,s}^{-1}}\else{$\,{\rm K\,km\,s}^{-1}$}\fi} 
\def\twCO{\ifmmode{\rm ^{12}CO}\else{$\rm^{12}CO$}\fi} 
\def\thCO{\ifmmode{\rm ^{13}CO}\else{$\rm^{13}CO$}\fi} 
\def\CeiO{\ifmmode{\rm C^{18}O}\else{$\rm C^{18}O$}\fi} 
\def \Cp{\ifmmode{\rm C^+}\else{$\rm C^+$}\fi} 
\def \CHp{\ifmmode{\rm CH^+}\else{$\rm CH^+$}\fi}
\def \thCHp{\ifmmode{\rm ^{13}CH^+}\else{$\rm ^{13}CH^+$}\fi}
\def \CHtp{\ifmmode{\rm CH_2^+}\else{$\rm CH_2^+$}\fi} 
\def\CHthp{\ifmmode{\rm CH_3^+}\else{$\rm CH_3^+$}\fi} 
\def \HCOp{\ifmmode{\rm HCO^+}\else{$\rm HCO^+$}\fi} 
\def \HtOp{\ifmmode{\rm H_3O^+}\else{$\rm H_3O^+$}\fi} 
\def \HCfiN{\ifmmode{\rm HC_5N}\else{$\rm HC_5N$}\fi} 
\def\wat{\ifmmode{\rm H_2O}\else{$\rm H_2O$}\fi} 
\def \oxy{\ifmmode{\rm O_2}\else{$\rm O_2$}\fi} 
\def \HH{\ifmmode{\rm H_2}\else{$\rm H_2$}\fi}
\def \Jone{\ifmmode{\rm {(J=1--0)}}\else{{(J=1--0)}}\fi} 
\def\Jtwo{\ifmmode{\rm {(J=2--1)}}\else{{(J=2--1)}}\fi} 
\def\Jfou{\ifmmode{\rm {J=4--3}}\else{{J=4--3}}\fi} 
\def \Jthr{\ifmmode{\rm {J=3--2}}\else{{J=3--2}}\fi} 
\def \Ta{\ifmmode{\rm T_A}\else{$\rm T_A$}\fi} 
\def \Tas{\ifmmode{\rm T_A^*}\else{$\rm T_A^*$}\fi} 
\def \Tmb{\ifmmode{\rm T_{mb}}\else{$\rm T_{mb}$}\fi} 
\def \Tr{\ifmmode{\rm T_r}\else{$\rm T_r$}\fi} 
\def \Trs{\ifmmode{\rm T_r^*}\else{$\rm T_r^*$}\fi}

\begin{document}
%This preliminary author list corresponds to \title{Author list for Proj. Ref. 7.7: The mm/submm properties of a sample of Galactic cold cores}
%Prepared by R. Leonardi (rleonardi@sciops.esa.int), ESAC/ESA, on 05JAN2011
%This version is frozen on 5 Jan 2011 at 9:30 CET, any further changes will be delayed to after submission
%\subtitle{There are 203 co-authors in this list}
\author{\small
Planck Collaboration:
P.~A.~R.~Ade\inst{68}
\and
N.~Aghanim\inst{44}
\and
M.~Arnaud\inst{55}
\and
M.~Ashdown\inst{53, 74}
\and
J.~Aumont\inst{44}
\and
C.~Baccigalupi\inst{66}
\and
A.~Balbi\inst{26}
\and
A.~J.~Banday\inst{72, 6, 60}
\and
R.~B.~Barreiro\inst{50}
\and
J.~G.~Bartlett\inst{3, 51}
% ...
}
\institute{\small
Aalto University Mets\"{a}hovi Radio Observatory, Mets\"{a}hovintie 114, FIN-02540 Kylm\"{a}l\"{a}, Finland\\
\and
Agenzia Spaziale Italiana Science Data Center, c/o ESRIN, via Galileo Galilei, Frascati, Italy\\
\and
Astroparticule et Cosmologie, CNRS (UMR7164), Universit\'{e} Denis Diderot Paris 7, B\^{a}timent Condorcet, 10 rue A. Domon et L\'{e}onie Duquet, Paris, France\\
\and
Atacama Large Millimeter/submillimeter Array, ALMA Santiago Central Offices Alonso de Cordova 3107, Vitacura, Casilla 763 0355, Santiago, Chile\\
\and
CITA, University of Toronto, 60 St. George St., Toronto, ON M5S 3H8, Canada\\
\and
CNRS, IRAP, 9 Av. colonel Roche, BP 44346, F-31028 Toulouse cedex 4, France\\
\and
California Institute of Technology, Pasadena, California, U.S.A.\\
\and
DAMTP, Centre for Mathematical Sciences, Wilberforce Road, Cambridge CB3 0WA, U.K.\\
\and
DSM/Irfu/SPP, CEA-Saclay, F-91191 Gif-sur-Yvette Cedex, France\\
\and
DTU Space, National Space Institute, Juliane Mariesvej 30, Copenhagen, Denmark\\
\and
Departamento de F\'{\i}sica, Universidad de Oviedo, Avda. Calvo Sotelo s/n, Oviedo, Spain\\
\and
Department of Astronomy and Astrophysics, University of Toronto, 50 Saint George Street, Toronto, Ontario, Canada\\
\and
Department of Earth Sciences and Astronomy, University of Tokyo, Komaba 3-8-1, Meguro, Tokyo, 153-8902, Japan\\
\and
Department of Physics \& Astronomy, University of British Columbia, 6224 Agricultural Road, Vancouver, British Columbia, Canada\\
\and
Department of Physics, Gustaf H\"{a}llstr\"{o}min katu 2a, University of Helsinki, Helsinki, Finland\\
\and
Department of Physics, Princeton University, Princeton, New Jersey, U.S.A.\\
\and
Department of Physics, Purdue University, 525 Northwestern Avenue, West Lafayette, Indiana, U.S.A.\\
\and
Department of Physics, University of California, Berkeley, California, U.S.A.\\
\and
Department of Physics, University of California, One Shields Avenue, Davis, California, U.S.A.\\
\and
Department of Physics, University of California, Santa Barbara, California, U.S.A.\\
\and
Department of Physics, University of Illinois at Urbana-Champaign, 1110 West Green Street, Urbana, Illinois, U.S.A.\\
\and
Dipartimento di Fisica G. Galilei, Universit\`{a} degli Studi di Padova, via Marzolo 8, 35131 Padova, Italy\\
\and
Dipartimento di Fisica, Universit\`{a} La Sapienza, P. le A. Moro 2, Roma, Italy\\
\and
Dipartimento di Fisica, Universit\`{a} degli Studi di Milano, Via Celoria, 16, Milano, Italy\\
\and
Dipartimento di Fisica, Universit\`{a} degli Studi di Trieste, via A. Valerio 2, Trieste, Italy\\
\and
Dipartimento di Fisica, Universit\`{a} di Roma Tor Vergata, Via della Ricerca Scientifica, 1, Roma, Italy\\
\and
Discovery Center, Niels Bohr Institute, Blegdamsvej 17, Copenhagen, Denmark\\
\and
Dpto. Astrof\'{i}sica, Universidad de La Laguna (ULL), E-38206 La Laguna, Tenerife, Spain\\
\and
E\"{o}tv\"{o}s Lor\'{a}nd University, Department of Astronomy, P\'{a}zm\'{a}ny P\'{e}ter s\'{e}t\'{a}ny 1/A, 1117 Budapest, Hungary\\
\and
European Southern Observatory, ESO Vitacura, Alonso de Cordova 3107, Vitacura, Casilla 19001, Santiago, Chile\\
\and
European Space Agency, ESAC, Planck Science Office, Camino bajo del Castillo, s/n, Urbanizaci\'{o}n Villafranca del Castillo, Villanueva de la Ca\~{n}ada, Madrid, Spain\\
\and
European Space Agency, ESTEC, Keplerlaan 1, 2201 AZ Noordwijk, The Netherlands\\
\and
Helsinki Institute of Physics, Gustaf H\"{a}llstr\"{o}min katu 2, University of Helsinki, Helsinki, Finland\\
\and
INAF - Osservatorio Astrofisico di Catania, Via S. Sofia 78, Catania, Italy\\
\and
INAF - Osservatorio Astronomico di Padova, Vicolo dell'Osservatorio 5, Padova, Italy\\
\and
INAF - Osservatorio Astronomico di Roma, via di Frascati 33, Monte Porzio Catone, Italy\\
\and
INAF - Osservatorio Astronomico di Trieste, Via G.B. Tiepolo 11, Trieste, Italy\\
\and
INAF/IASF Bologna, Via Gobetti 101, Bologna, Italy\\
\and
INAF/IASF Milano, Via E. Bassini 15, Milano, Italy\\
\and
INRIA, Laboratoire de Recherche en Informatique, Universit\'{e} Paris-Sud 11, B\^{a}timent 490, 91405 Orsay Cedex, France\\
\and
IPAG: Institut de Plan\'{e}tologie et d'Astrophysique de Grenoble, Universit\'{e} Joseph Fourier, Grenoble 1 / CNRS-INSU, UMR 5274, Grenoble, F-38041, France\\
\and
Imperial College London, Astrophysics group, Blackett Laboratory, Prince Consort Road, London, SW7 2AZ, U.K.\\
\and
Infrared Processing and Analysis Center, California Institute of Technology, Pasadena, CA 91125, U.S.A.\\
\and
Institut d'Astrophysique Spatiale, CNRS (UMR8617) Universit\'{e} Paris-Sud 11, B\^{a}timent 121, Orsay, France\\
\and
Institut d'Astrophysique de Paris, CNRS UMR7095, Universit\'{e} Pierre \& Marie Curie, 98 bis boulevard Arago, Paris, France\\
\and
Institute of Astronomy and Astrophysics, Academia Sinica, Taipei, Taiwan\\
\and
Institute of Space and Astronautical Science, Japan Aerospace Exploration Agency, 3-1-1 Yoshinodai, Chuo-ku, Sagamihara, Kanagawa, 252-5210, Japan\\
\and
Institute of Theoretical Astrophysics, University of Oslo, Blindern, Oslo, Norway\\
\and
Instituto de Astrof\'{\i}sica de Canarias, C/V\'{\i}a L\'{a}ctea s/n, La Laguna, Tenerife, Spain\\
\and
Instituto de F\'{\i}sica de Cantabria (CSIC-Universidad de Cantabria), Avda. de los Castros s/n, Santander, Spain\\
\and
Jet Propulsion Laboratory, California Institute of Technology, 4800 Oak Grove Drive, Pasadena, California, U.S.A.\\
\and
Jodrell Bank Centre for Astrophysics, Alan Turing Building, School of Physics and Astronomy, The University of Manchester, Oxford Road, Manchester, M13 9PL, U.K.\\
\and
Kavli Institute for Cosmology Cambridge, Madingley Road, Cambridge, CB3 0HA, U.K.\\
\and
LERMA, CNRS, Observatoire de Paris, 61 Avenue de l'Observatoire, Paris, France\\
\and
Laboratoire AIM, IRFU/Service d'Astrophysique - CEA/DSM - CNRS - Universit\'{e} Paris Diderot, B\^{a}t. 709, CEA-Saclay, F-91191 Gif-sur-Yvette Cedex, France\\
\and
Laboratoire Traitement et Communication de l'Information, CNRS (UMR 5141) and T\'{e}l\'{e}com ParisTech, 46 rue Barrault F-75634 Paris Cedex 13, France\\
\and
Laboratoire de Physique Subatomique et de Cosmologie, CNRS, Universit\'{e} Joseph Fourier Grenoble I, 53 rue des Martyrs, Grenoble, France\\
\and
Laboratoire de l'Acc\'{e}l\'{e}rateur Lin\'{e}aire, Universit\'{e} Paris-Sud 11, CNRS/IN2P3, Orsay, France\\
\and
Lawrence Berkeley National Laboratory, Berkeley, California, U.S.A.\\
\and
Max-Planck-Institut f\"{u}r Astrophysik, Karl-Schwarzschild-Str. 1, 85741 Garching, Germany\\
\and
MilliLab, VTT Technical Research Centre of Finland, Tietotie 3, Espoo, Finland\\
\and
National University of Ireland, Department of Experimental Physics, Maynooth, Co. Kildare, Ireland\\
\and
Niels Bohr Institute, Blegdamsvej 17, Copenhagen, Denmark\\
\and
Observational Cosmology, Mail Stop 367-17, California Institute of Technology, Pasadena, CA, 91125, U.S.A.\\
\and
Optical Science Laboratory, University College London, Gower Street, London, U.K.\\
\and
SISSA, Astrophysics Sector, via Bonomea 265, 34136, Trieste, Italy\\
\and
SUPA, Institute for Astronomy, University of Edinburgh, Royal Observatory, Blackford Hill, Edinburgh EH9 3HJ, U.K.\\
\and
School of Physics and Astronomy, Cardiff University, Queens Buildings, The Parade, Cardiff, CF24 3AA, U.K.\\
\and
Space Sciences Laboratory, University of California, Berkeley, California, U.S.A.\\
\and
Spitzer Science Center, 1200 E. California Blvd., Pasadena, California, U.S.A.\\
\and
Stanford University, Dept of Physics, Varian Physics Bldg, 382 Via Pueblo Mall, Stanford, California, U.S.A.\\
\and
Universit\'{e} de Toulouse, UPS-OMP, IRAP, F-31028 Toulouse cedex 4, France\\
\and
Universities Space Research Association, Stratospheric Observatory for Infrared Astronomy, MS 211-3, Moffett Field, CA 94035, U.S.A.\\
\and
University of Cambridge, Cavendish Laboratory, Astrophysics group, J J Thomson Avenue, Cambridge, U.K.\\
% ...
}

  \title{\textit{Planck} Early Results: The submillimetre properties of a sample of
   Galactic cold clumps}

\date{ }
\authorrunning{The \Planck\ collaboration}

\abstract
%CONTEXT
{
%}
%AIMS
%{
We perform a detailed investigation of sources from
the Cold Cores Catalogue of \Planck\ Objects (C3PO).  Our goal is to probe the reliability of the detections,
validate the separation between warm and cold dust emission components, 
provide the first glimpse at the nature, internal morphology and physical
characterictics of the \Planck-detected sources.
%}
%METHOD
%{
We focus on a sub-sample of ten sources from the C3PO list, selected to sample
different environments, from high latitude cirrus to nearby (150\,pc) and
remote (2\,kpc) molecular complexes.  We present \Planck\ surface brightness
maps and derive the dust temperature, emissivity spectral index, and column
densities of the fields.  With the help of higher resolution {\it Herschel\/}
and {\it AKARI\/} continuum observations and molecular line data, we
investigate the morphology of the sources and the properties of the
substructures at scales below the \Planck\ beam size.  
%}
%RESULTS
%{
The cold clumps detected by \Planck\ are found to be located on large-scale 
filamentary (or cometary) structures that extend up to 20\,pc in the remote
sources.  The thickness of these filaments ranges 
between 0.3 and 3\,pc, for column densities 
$N_{\HH} \sim 0.1$ to $1.6 \times 10^{22}$\cq,
and with linear mass density covering a broad range, between 
15 and 400\,\mlin. 
The dust temperatures are low (between 10 and 15\,K) and the \Planck\
cold clumps correspond to local minima of the line-of-sight averaged 
dust temperature in these fields.  
These low temperatures are confirmed when {\it AKARI\/} and {\it Herschel\/}
data are added to the spectral energy distributions.
{\it Herschel\/} data reveal a wealth of substructure within the 
\Planck\ cold clumps. 
In all cases (except two sources harbouring young stellar objects), 
the substructures are found to be colder, with temperatures as low as 7\,K.
Molecular line observations provide gas column densities which are consistent
with those inferred from the dust. The
linewidths are all supra-thermal, providing large virial linear mass densities 
in the range 10 to 300\,\mlin, comparable within factors of a few, to the
gas linear mass densities.
%}
%CONCLUSIONS
%{
The analysis of this small set of cold clumps 
already probes a broad variety of structures in the C3PO sample, 
probably associated with different evolutionary stages, from cold and
starless clumps, to young protostellar objects still embedded 
in their cold surrounding cloud.
Because of the all-sky coverage and its 
sensitivity, \Planck\ is able to detect and locate
the coldest spots in massive elongated structures that may be the
long-searched for progenitors of stellar clusters. }


\maketitle

\section{Introduction}\label{sec:intro}

The main difficulty in understanding star formation lies in the vast
range of scales involved in the process, including not only 
the gravitationally unstable densest and coldest structures, but also their 
large-scale environment.
The characteristics of star formation, described by the stellar mass
distribution, the formation efficiency, the evolutionary timescales,
and by the modes of formation (clustered or isolated,
 spontaneous or triggered) are linked  to the properties of the cold cores
and those of their Galactic surroundings. 
Theory predicts
that the initial mass function (IMF) is largely
determined during the pre-stellar core fragmentation phase, but the latter 
depends on the properties of the pre-stellar cores, particularly
the nature of their support against self-gravity, and their density
and temperature distributions.  
Understanding star formation therefore also requires an understanding
of the formation and evolution of dense cores.  Turbulence, magnetic
fields, and gravity all contribute at sub-parsec scales as well as Galactic
scales \citep[see reviews of][]{Bergin2007,McKee2007,WardThompson2007,
Crutcher2009}.

So far, 
ground-based \citep[e.g.,][]{Motte1998,Johnstone2000} and {\it Herschel\/}
\citep[e.g.,][]{Andre2010,DiFrancesco2010,Konyves2010}  
observations of nearby star-forming regions have
revealed a core mass spectrum with a slope similar to that of the IMF.
This intriguing result calls for broader investigations, in particular towards 
the inner Galaxy, where the Galactic environment differs from that of the Solar 
neighbourhood, and the star formation efficiency is anticipated to be different. 

The physical properties of cold cores are still poorly known, 
possibly because of the short duration of this phase, but mainly
because of observational difficulties. The thermal emission of cold dust, 
intrinsically weak and blended with that of warmer components, 
must be sought for in the submillimetre range.  A combination of
continuum and molecular line studies is also needed to develop a
global view of the clouds, from large scales (tens of parsecs) down to
the scale of dense cores.
During the past decade, the development of sensitive continuum and
heterodyne detectors at millimetre and submillimetre wavelengths has 
significantly increased our knowledge of the properties of the cold
dark clouds \citep[see Section~1 in][ hereafter Paper~I]{planck2011-7.7b}.
The \Planck\footnote{\Planck\ (http://www.esa.int/Planck) is a project of the
European Space Agency (ESA) with instruments provided by two scientific
consortia funded by ESA member states (in particular the lead countries France
and Italy), with contributions from NASA (USA) and telescope reflectors
provided by a collaboration between ESA and a scientific consortium led and
funded by Denmark.}
and {\it Herschel\/} satellites now provide a unique
opportunity for studying Galactic dust emission and in particular its
dense and cold component. The first {\it Herschel\/} surveys have already
provided new insights into cold cores, although
limited to specific fields  \citep{Andre2010, Bontemps2010,
Konyves2010, Menshchikov2010, Molinari2010, Motte2010, Peretto2010,
Schneider2010, Stutz2010, WardThompson2010, Zavagno2010}. 

The \Planck\ satellite \citep{tauber2010a, planck2011-1.1} provides
complementarity to {\it Herschel\/}
by carrying out an all-sky survey that is well suited
for the systematic detection of cold cores. 
In Paper~I, we present the
first statistical results of the cores from this survey. Combining \Planck\ and
{\it IRAS\/} $100\,\mu$m data, we have built a preliminary catalogue of
10{,}783 cold cores (the Cold Core Catalogue of \Planck\ Objects, C3PO). 
A sub-sample of the most reliable detections is delivered
as part of the \Planck\ Early Release Compact Source Catalogue
\citep[ERCSC, see][]{planck2011-1.10}, i.e. the Early Cold Cores (ECC) catalogue.
The cores from C3PO cover a wide range in properties, with: temperature, from 7\,K to
16\,K, with a peak around 13\,K; density from $30\,{\rm cm}^{-3}$
to $10^{5}\,{\rm cm}^{-3}$, with an average value of
$2\times10^{3}\,{\rm cm}^{-3}$; mass from $0.3$ to
$2.5 \times 10^{4} M_{\odot}$; and size
ranging from 0.2 to 18\,pc. The sources are
found to be extended, and more importantly are
elongated, with a distribution of aspect ratios extending up to 4 
and peaking at twice the beam size of the \Planck-HFI
instrument. As discussed in Paper~I, these properties better match the
definition of ``clumps'' rather than ``cores''
\citep[see][]{Williams2000,Bergin2007}. Clumps may contain dense cores,
which are likely the precursors of individual or multiple stars. 
Thus, the main population seen with
\Planck\ does not correspond to single pre-stellar cores, but ensembles 
of  cold substructures. A detailed study of
the \Planck\ cold objects requires observations at higher
resolution. This is the main objective of the {\it Herschel\/} Open Time
Key Program ``Galactic Cold Cores'' that is dedicated to following up of
a sample of \Planck\ cores.

In this paper we present the first detailed analysis of ten
representative sources from the C3PO catalogue. The targets have been
selected to cover different types of clump, spanning a wide range of mass,
temperature, and density, and located in different environments, ranging from
high latitude cirrus to nearby and distant molecular complexes. We
combine the \Planck\ data with available ancillary data (in
particular IRIS, 2MASS and {\it AKARI\/}) and dedicated follow-up observations
with {\it Herschel\/} and ground-based radio telescopes. Using these
ancillary data, we seek to confirm the \Planck\ detections and to
demonstrate the reliability of the method which used only \Planck\ and
{\it IRAS\/} data to determine the source parameters that enter
the C3PO catalogue (Paper~I) and ECC catalogue \citep{planck2011-1.10,
  planck2011-1.10sup}. 
The higher resolution continuum and line data make it possible
to examine the internal structure of the 
\Planck\ sources. They provide the first hints about the process of
internal fragmentation and the physical state of the
compact cores at scales below the size of the \Planck\
beam. These properties are essential for the interpretation of the
full C3PO catalogue and will be a major topic for the projects carrying out
follow-up studies of C3PO and ECC catalogue sources.

After describing the observational data set and the source sample 
selection (Section~\ref{sect:data}), we explain the
analysis methods used to derive physical properties for the cores
(Section~\ref{sect:analysis_methods}). The main results are
presented in Section~\ref{sect:results}. By fitting spectral enery
distributions (SEDs) we derive temperatures,
emissivity spectral indices, and column densities for the cores and
the surrounding fields, and estimate the linear mass densities, masses and 
bolometric luminosities
of the cores (Section~\ref{sect:ResWithPlanck}). The small-scale
structures within the \Planck\ clumps are studied with the help of
{\it Herschel\/} and {\it AKARI\/} data
(Section~\ref{sect:substruct_Herschel}), and the gas
properties of a few cores are derived from molecular line data
(Section~\ref{sect:gas_properties}). 
Following a discussion of the physical characteristics of the \Planck\ cores,
we present our summary and perspectives for the future
in Section~\ref{sect:conclusion}.

\section{Observations}\label{sect:data}

\subsection{The sources}

We have selected from the Cold Core Catalogue of \Planck\ Objects (C3PO)
ten sources with high reliability, signal-to-noise ratio ${\rm S/N}\ge 8$,
and low colour
temperature, $T \le 14\,$K (see Paper~I). An initial Monte Carlo
sampling of the full C3PO catalogue was performed to prepare a
candidate list that covered the full range in Galactic position,
temperature, flux, and column density. Further selection was made by
examining the \Planck\ data and ancillary information from
IRIS \citep{Miville2005}, 2MASS extinction \citep{Skrutskie2006},
and CO line data from \cite{Dame2001} and NANTEN surveys
\citep[e.g.,][]{Fukui1999}, as well as {\it IRAS\/}
and {\it AKARI\/} point
source catalogues and by cross-checking the sources with the SIMBAD
database. The selected sources represent different large-scale
morphologies and environments, including filaments, isolated and clustered
structures and high-latitude cirrus clouds. One key criterion was the
knowledge of the source distance, derived either with an
extinction method \citep{Marshall2006}, by association
with a known molecular cloud complex, or through a kinematic distance
estimate (see paper I).  Nine out of the ten sources have already been
observed with
{\it Herschel\/} as part of the Open Time Key Program ``Galactic cold cores.''
In particular, the sample includes the three targets observed during
the {\it Herschel\/} Science Demonstration Phase
\citep[SDP, see][]{Juvela2010, Juvela2011}.
Half of the sources in the sample are in the \Planck\ ECC. This is mainly due to 
the criteria used to select the most reliable detections from the C3PO full catalogue (see Paper I): 
with $T < 14\,$K for the colour temperature corresponding to the aperture photometry SED, and a high signal to noise ratio 
for the source detection ($S/N > 15$).

The targets are listed in Table~\ref{table:source_selection} and 
presented in detail in the Appendix.  The ten
sources that we focus on will hereafter be referred to by the labels
S1 through S10.

\begin{table*}
\caption{Source selection.}
\label{table:source_selection}
\nointerlineskip
%\centering
\begin{tiny}
\setbox\tablebox=\vbox{
\newdimen\digitwidth
\setbox0=\hbox{\rm 0}
\digitwidth=\wd0
\catcode`*=\active
\def*{\kern\digitwidth}
\newdimen\signwidth
\setbox0=\hbox{+}
\signwidth=\wd0
\catcode`!=\active
\def!{\kern\signwidth}
\halign{#\hfil\tabskip=1.0em&
\hfil#&
\hfil#\hfil&
\hfil#\hfil&
\hfil#\hfil&
\hfil#\hfil&
\hfil#\hfil&
\hfil#\hfil&
\hfil#\hfil&
\hfil#\hfil&
\hfil#\hfil&
\hfil#\hfil&
\hfil#\hfil\tabskip=0pt\cr
\noalign{\doubleline}
Source name & & $l$ & $b$ & S/N & $S_{857}$ & $S_{545}$ & $S_{353}$ &
 $\theta_{\rm Min}$ & $\theta_{\rm Maj}$ & $\left\langle\theta\right\rangle$
 & $d$ & complex \cr
& & (deg) & (deg) & & (Jy) & (Jy) & (Jy) & (arcmin) & (arcmin) & (arcmin) &
 (pc) & \cr
\noalign{\vskip 4pt\hrule\vskip 6pt}
\Planck$-$G126.6+24.5 &\bf{S1} & 126.62 &  !24.55  &  44.0  &  *62.0  &
 *22.5 &   *5.2 &    4.9  &  *7.6  &   6.1 & *150   &    Polaris flare \cr
\Planck$-$G20.7+20.0 &\bf{S2} & 316.53   & !20.68 &   20.0  &  *35.6  &
 *12.7  &  *2.9  &   7.2   & *9.1  &   8.1  & *550    &           Cometary Globule \cr
\Planck$-$G131.7+9.7 &\bf{S3} & 131.74 &   !*9.70   & 38.2  &  185.8  &
 *77.5 &   19.7  &   6.3  &  *8.0  &   7.1 &  *200   &    Cepheus flare \cr
\Planck$-$G215.2$-$16.4 &\bf{S4}  & 215.44 & $-$16.38 & 24.0 & *85.9  &
 *34.9  &  *9.2  &   5.0 &   *7.4  &   6.1  & *450   &              Orion \cr
\Planck$-$G276.9+1.7 &\bf{S5}  & 276.87   &  !*1.73   & *8.5 & *88.4  &
 *34.4  &  *9.5  &   4.6 &   18.8  &   9.3 &  2000     &           Vela \cr
\Planck$-$G176.2$-$2.1 &\bf{S6}  & 176.18  &  *$-$2.11 & 25.8 & 116.8 &
 *51.0  &  13.2  &   5.0  &  11.5  &   7.6   & 2000   &      Perseus Arm \cr
\Planck$-$G161.6$-$9.2 &\bf{S7}  & 161.56  &  *$-$9.29 & 15.0 & 196.5 &
 *78.1  &  20.9  &   5.5  &  14.9  &   9.1  &  *350    &          PerOB2 \cr
\Planck$-$G109.8+2.7 &\bf{S8}  & 109.79   &  !*2.71 & 16.1  &  272.4  &
 122.8  &  33.3  &   5.1  &  *9.3  &   6.9  &  *800    &          Cephee \cr
\Planck$-$G107.2+5.5 &\bf{S9} & 107.17  &  !*5.45  &  23.6  &  490.2  &
 185.5  &  49.5  &   4.4  &  *8.3  &   6.1  &  *800      &      Cep$-$OB3b \cr
\Planck$-$G300.9$-$9.0 &\bf{S10}  & 300.86 &  *$-$8.96 & 82.7 & 267.3 &
 *98.8  &  24.3 &    4.2  &  15.2  &   8.0   & *225   &          Musca \cr
\noalign{\vskip 3pt\hrule\vskip 4pt}
} }
\endPlancktablewide
\end{tiny}
\end{table*}

\subsection{Planck data} 

\Planck\ \citep{tauber2010a, planck2011-1.1} is the third generation space 
mission to measure the anisotropy of the cosmic microwave background (CMB).  
It observes the sky in nine frequency bands covering 30--857\,GHz with high 
sensitivity and angular resolution from 31$\arcmin$ to 5$\arcmin$.
The Low Frequency Instrument
\citep[LFI;][]{Mandolesi2010, Bersanelli2010, planck2011-1.4}
covers the 30, 44, and 70\,GHz bands with amplifiers cooled to 20\,\hbox{K}.
The High Frequency Instrument \citep[HFI;][]{Lamarre2010, planck2011-1.5}
covers the 100, 143, 217, 353, 545, and 857\,GHz bands with bolometers cooled
to 0.1\,\hbox{K}.  Polarisation is
measured in all but the highest two bands \citep{Leahy2010, Rosset2010}.  
A combination of radiative cooling and three mechanical coolers produces the 
temperatures needed for the detectors and optics \citep{planck2011-1.3}.  Two 
data processing centres (DPCs) check and calibrate the data and make maps 
of the sky \citep{planck2011-1.7, planck2011-1.6}.  \Planck's sensitivity,
angular resolution, and frequency coverage make it a powerful instrument for
Galactic and extragalactic astrophysics as well as cosmology.
Early astrophysics results are given in \Planck\ Collaboration, 2011h--z.

\begin{figure*}
\center
\includegraphics[width=12cm]{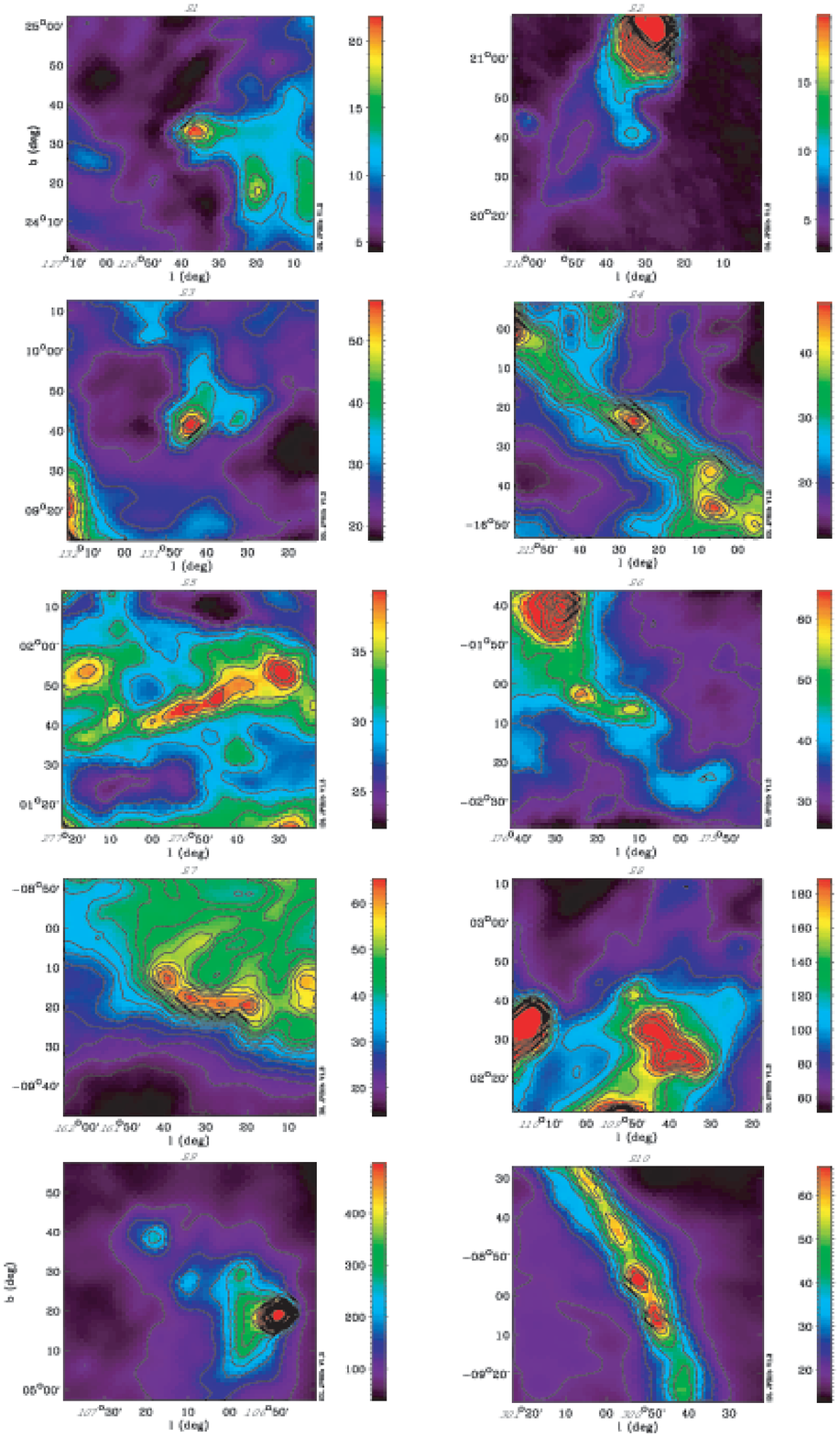}
\caption{
\Planck-HFI brightness emission maps at 857\,GHz. The colour scale is
in ${\rm MJy}\,{\rm sr}^{-1}$. Note the large dynamic range of the \Planck\
cold clumps from $10\,{\rm MJy}\,{\rm sr}^{-1}$ (S2) to
$250\,{\rm MJy}\,{\rm sr}^{-1}$ (S9). 
}
\label{fig:Planck_maps}
\end{figure*}

We use data from the \Planck-HFI bands at 857\,GHz, 545\,GHz and 353\,GHz
that cover the main peak of the cold dust emission. By restricting
ourselves to these highest frequencies, we can perform the analysis at
the best angular resolution provided by \Planck, i.e., ${\sim}\,
4.5\arcmin$ full width at half-maximum (FWHM).  

Fig.~\ref{fig:Planck_maps} displays the 857\,GHz surface brightness
maps for each $1^\circ \times 1^\circ$ field. The cores are
located at the centres of the maps and show dust emission that ranges
from 10 to 220\,${\rm MJy}\,{\rm sr}^{-1}$. They are embedded in extended
structures whose shape varies from large filaments (S4, S5, S6, and
S7) to more isolated and apparently compact morphologies. Most cores
are extended and elongated compared to the \Planck\ beam. 
As discussed in Paper I, the ellipticity and extension of the cores are
not biased by the local beam shape. As an illustration, Fig.~\ref{fig_FEBeCoP} shows the comparison 
between the local Point Spread Function provided by the FEBeCoP tool \citep{Mitra2010} 
with the elliptical Gaussian fit of the sub-sample detections. 
Typical sizes are from 0.2 to 11\,pc.  Some cores (S2, S8, and S9) are located
near bright, warmer regions, where the intensities are higher by up to
a factor of 10.  In a few cases, the environment exhibits 
sharp edges, at large scale (see in particular S1, S3 and S7).

The full set of HFI maps ($1^\circ \times 1^\circ$ fields from 857
to 100\,GHz) for each source are shown in Figs.~\ref{fig:allnu_sources1}
\ref{fig:allnu_sources2} and \ref{fig:allnu_sources3}. These maps have been 
derived from the HEALPix sky maps \citep{Gorski2005}.
With the exception of source S2, the
cold source at the centre of the maps is usually visible down to 143\,GHz.
At 100\,GHz the sources become difficult to detect because of the
falling intensity of the dust spectrum. Only the Musca filament is
clearly visible even at 100\,GHz, although the surface brightness of the
source is low (${\sim}\,0.15\,{\rm MJy}\,{\rm sr}^{-1}$).

Comparison with the $100\,\mu$m maps from IRIS \citep{Miville2005}
confirms the conclusion of previous submillimetre surveys (using PRONAOS,
ARCHEOPS, BLAST, and ground-based telescopes) that the cold dust emission
is not traced by the $100\,\mu$m data, but must be studied using longer
wavelengths.  The IRIS maps are dominated by the warmer and more
extended structure around the cores. This is the basis of the source
detection method CoCoCoDeT (Cold Core Colour Detection Tool), 
that was described in detail in \cite{Montier2010} and in Paper~I
(Section~2.4). It uses as a template the spectrum of the warm emission
component estimated from the IRIS $100\,\mu$m map.
Subtraction of the warm component results in residual maps of
the cold emission component from which source fluxes are
derived. The detection process is applied independently to the 857,
545, and 353\,GHz maps, after smoothing HFI and IRIS data to the same
angular resolution of $4.5^\prime$. The main steps of the method and the
core extraction process are illustrated in the Appendix (in
Fig.~\ref{fig:fig_detect}) for S1. The residual $100\,\mu$m
signal that remains when a background model is subtracted from the
IRIS map (see Paper~I) is still needed to constrain the temperature of
the cold core SEDs, which typically peak in the range 200--$300\,\mu$m.
The analysis of Section~\ref{sect:results} will also test the
accuracy of this procedure.  The coordinates, distances, fluxes densities, and
sizes of the selected \Planck\ cores are given in
Table~\ref{table:source_selection}. The values listed there are taken
directly from the C3PO catalogue (see Paper~I and
Section~\ref{sect:photometry_SEDs}).

\subsection{{\it Herschel\/} observations}

The {\it Herschel\/} photometric observations were carried out with the PACS
and SPIRE instruments \citep{Pilbratt2010, Poglitsch2010, Griffin2010}.
Three fields, S8, S9, S10 (corresponding to
the source names PCC249, PCC288, and PCC550) were observed in November
and December 2009 as part of the {\it Herschel\/} SDP.
The other fields were observed between July and September 2010.
Most observations were performed separately with PACS (100 and
$160\,\mu$m) and with SPIRE (250, 350 and $500\,\mu$m)
in scan mapping mode. Because of the larger field size, S5 and
S9 were observed in parallel mode using both instruments
simultaneously.  The observations employed two orthogonal scanning
directions, except for the PACS observations of PCC288 where three
scanning directions were used. The data were reduced with {\it Herschel\/}
Interactive Processing Environment (HIPE), using the official pipeline with the addition of
specialised reduction routines to take advantage of the orthogonal
scans for deglitching PACS data and to remove SPIRE scan baselines.
The PACS maps were created using the MADmap algorithm \citep{Cantalupo2010}.
The SPIRE maps are the product of direct projection onto the sky and
averaging of the time ordered data, with a baseline correction.

As for most bolometer observations without an absolute calibrator, the
zero level (or offset) of the PACS and SPIRE data are
arbitrary. We therefore compared the {\it Herschel\/} and PACS data with the
predictions of a model constrained by the \Planck\ and {\it IRAS\/} data. The
model uses the all-sky dust temperature maps described in
\cite{planck2011-7.0} to infer the average radiation field intensity
for each pixel at the common resolution level for the \Planck\ and {\it IRAS\/}
data. The DUSTEM model \citep{Compiegne2010} with the above
value for the radiation field intensity was used to predict the
expected brightness in the {\it Herschel}-SPIRE and PACS bands, using the
nearest available \Planck\ or {\it IRAS\/} band for normalisation and taking
into account the appropriate colour correction in the {\it Herschel\/}
filters. The predicted brightness was correlated with the observed maps
smoothed to the \Planck\ and {\it IRAS\/} resolution over the region observed
with {\it Herschel\/} and the gain and offsets were derived from this
correlation. We have used these gain and offset inter-calibration values in
order to convert the maps from ${\rm Jy}\,{\rm beam}^{-1}$ and
${\rm Jy}\,{\rm pix}^{-1}$ (SPIRE and PACS,
respectively) into brightness units (${\rm MJy}\,{\rm sr}^{-1}$).

\subsection{{\it AKARI\/} observations}

The {\it AKARI\/} satellite \citep{Murakami2007} has conducted
all-sky surveys at infrared wavelengths centred at 9$\,\mu$m, 18$\,\mu$m,
65$\,\mu$m, 90$\,\mu$m, 140$\,\mu$m, and 160$\,\mu$m. We use the observations
made by the FIS instrument in the wide far-infrared bands of 90$\,\mu$m
and 140$\,\mu$m. The accuracy of the calibration is currently estimated
to be 26\% at 90$\,\mu$m and 33\% at 140$\,\mu$m, and the beam sizes of these
two bands are ${\sim}\,39\arcsec$ and 58$\arcsec$, respectively.
For details of the {\it AKARI\/} far-infrared all
sky survey, see \cite{Doi2009}.

\subsection{Molecular line data}

We have carried out ``fast'' observations of different CO isotopic lines in
some 60 \Planck\ cold core candidate fields for the {\it Herschel\/} follow-up
programme, five of which are included in the present sample. 

The Onsala 20-m telescope was used for $^{12}$CO $J$=1$\rightarrow$0,
$^{13}$CO $J$=1$\rightarrow$0, and C$^{18}$CO $J$=1$\rightarrow$0
observations in December 2009
and April 2010. An area of a few arcmin in diameter was mapped
around the position of the \Planck\ source S3 and the sources in
the SDP fields S8 and S9. The Onsala beam size is approximately
${\sim}\,33\arcsec$
and the typical rms noise in the $^{13}$CO and C$^{18}$O spectra was
below 0.1\,K per a channel of 0.07\,km\,s$^{-1}$.

The APEX observations of field S10 were made in July 2010.  The size of the
$^{13}$CO $J$=2$\rightarrow$1 map is ${\sim}\,5\arcmin$ and the typical rms
noise is below 0.2\,K. The APEX beam size at 220\,GHz is
${\sim}\,28\arcsec$.

The IRAM-30m observations were performed in July and October 2010. 
The EMIR receivers E090
(3\,mm) and E230 (1\,mm) were used in parallel with the high resolution correlator
(VESPA). The most important parameters for the different settings are given in 
the Appendix in Table~\ref{30m}. 
Maps of 3\arcmin$\times$3\arcmin\ were performed 
using on-the-fly (OTF) mode combined with frequency switching. 
A summary of the observations and the main results obtained are
presented in Section~\ref{sect:gas_properties}.

\section{Methods}  \label{sect:analysis_methods}

\subsection{Photometry and SEDs}\label{sect:photometry_SEDs}

\subsubsection{\Planck\ and {\it IRAS\/} photometry}
\label{sect:Planck_photometry}

The method for estimating the photometry for the C3PO catalogue
has been described in detail in \cite{Montier2010} 
and Paper~I (Section~2.4).  We recall here only the main
steps of the detection and flux extraction process:
\begin{enumerate}
\item For each pixel, and for each frequency, the warm background colour
 ($C_{\rm bkg}$) is estimated as the median value of 
 the ratio of the \Planck\ to 100$\,\mu$m emission maps ($I_{\nu}/I_{100}$)
 within a 15$\arcmin$ radius disc;
\item For each \Planck\ frequency, the contribution of the warm component is
 obtained by extrapolation from $100\,\mu$m through
 $I_{\nu}^{\rm w} = C_{\rm bkg} \times I_{100}$;
\item The cold residual map is computed by subtracting the warm component
 from the \Planck\ map;
\item The cold source detection is performed using a thresholding method
 applied on the cold residual map, with the criterion ${\rm S/N}>4$;
\item The source shape is estimated by fitting a 2D elliptical Gaussian
 to the colour map $I_{857}/I_{100}$;
\item The flux density at $100\,\mu$m is derived by fitting an elliptical
 Gaussian plus a polynomial surface for the background;
\item The warm template at $100\,\mu$m is corrected by removing the 
 source contribution that was estimated in the previous step;
\item Aperture photometry at 857, 545, and 353 GHz is performed on the
 cold residual maps, with the aperture determined by the source shape from
 step~5.
\end{enumerate}

The photometric uncertainties associated with this method
have been estimated with a Monte Carlo analysis (see Paper~I, Section~2.5);
they are $40\%$ for {\it IRAS\/} $100\,\mu$m and $8\%$ in the \Planck\ bands.
The additional calibration uncertainties to be taken into account are
13.5\% and 7\%, respectively, for IRIS \citep{Miville2005} and
the HFI bands 857, 545, and 353\,GHz \citep{planck2011-1.7}.
As described in Paper~I, the elliptical Gaussian fit 
performed in step~5 leads to an estimate of the minor and major axis lengths
($\sigma_{\rm Min}$ and $\sigma_{\rm Maj}$, respectively), related to the 
FWHM values, i.e., $\theta_{\rm Min}$ and $\theta_{\rm Maj}$ of a Gaussian by
$\sigma = \theta  / \sqrt{8 \, \ln(2)}$.
The fluxes and FWHM values are given for each source in
Table~\ref{table:source_selection}. 

The 353\,GHz band includes some contribution from the CO $J$=3$\rightarrow$2
molecular line. The magnitude of this effect has been estimated in
\cite{planck2011-1.7}
using data from $^{12}$CO $J$=1$\rightarrow$0 surveys, together with
an estimated average
line ratio of $J$=3$\rightarrow$2 to $J$=1$\rightarrow$0, and
knowledge of the 353\,GHz band spectral response. The derived correction
factor is $171\,\mu$K in thermodynamic temperature for a CO $J$=1$\rightarrow$0
line area of 1\,K\,km\,s$^{-1}$. The correction is small, but the exact
effect is hard to estimate for our sources because of the lack of high
spatial resolution CO data and because the CO excitation towards
cold cores may differ significantly from the average values assumed in
deriving the above factor. We therefore present results without
making a correction for the CO contribution, but comment on its
possible influence later.

\subsubsection{Photometry with ancillary data} 

When using {\it Herschel\/} and {\it AKARI\/} data, the fluxes are estimated at
a different resolution and with direct aperture photometry. The maps are
convolved to a resolution of 37$\arcsec$ (for {\it Herschel\/} and {\it AKARI\/}
90\,$\mu$m), 58$\arcsec$ (for {\it Herschel\/} and {\it AKARI\/} 90\,$\mu$m and
140\,$\mu$m), or 4.5$\arcmin$ (when including \Planck\ and IRIS data). In
each case the aperture has a radius equal to the FWHM of the smoothed
data and the background is subtracted using the 30$\%$ quantile value within a
reference annulus that extends from $1.3\times{\rm FWHM}$ to
$1.8\times{\rm FWHM}$. 
In order not to be affected by possible emission from
transiently heated grains, the fits employ only data at wavelengths
longer than 100$\,\mu$m. 
The statistical uncertainty of the flux values is derived from the
surface brightness fluctuations in the reference annulus. As above,
the calibration uncertainty is taken to be 7\% for the \Planck\ channels
and 13.5\% for IRIS 100\,$\mu$m. For {\it Herschel\/} we assume a 15\%
calibration uncertainty for PACS and 12\% for SPIRE.

\subsection{Estimation of temperatures, spectral indices, and optical depths} \label{sect:T_beta_Nh}

Since the dust thermal emission is optically thin in the submillimetre range,  
the source SEDs can be modelled as a modified blackbody of the form
\begin{equation}
S_{\nu} = \tau_{\nu} B_{\nu}\left( T_{\rm c} \right) \Omega_{\rm cl}
\label{Eq:SED1}
\end{equation}
where $S_{\nu}$ is the flux density integrated over the clump solid angle
$\Omega_{\rm cl}$ (common to all frequencies),
$\tau_\nu= N_{\HH} \mu m_{\rm H}\kappa_{\nu_0} (\nu/\nu_0)^\beta$
is the dust optical depth, and $B_{\nu}$ is the Planck function at
the dust colour temperature, $T_{\rm c}$.  Here
$N_{\HH}$ is the \HH\ gas column density, $\mu=2.33$ the mean mass per
particle, $m_{\rm H}$ the mass of the proton, $\kappa_{\nu_0}$ the mass
absorption coefficient at frequency $\nu_0$
and $\beta$ the dust emissivity spectral index. 
The SED may be rewritten as
\begin{equation}
S_{\nu} =  A B_{\nu}\left( T_{\rm c} \right) \nu^{\,\beta}
\label{Eq:SED}
\end{equation}
to separate the three quantities (assumed independent) which can be derived
from the fit: amplitude $A \propto N_{\HH} \kappa_{\nu_0}$; spectral index
$\beta$; and effective temperature $T_{\rm c}$.
The $\chi^2$ minimisation search operates in the ($A$, $T_{\rm c}$, $\beta$)
space and takes into account the colour
corrections for the \Planck\ bands described in \cite{planck2011-1.5}. 
In the fitting procedure, we consider only the flux error bars associated
with the photometry method, i.e., $40\%$ for IRIS and $8\%$ for HFI bands.  
In a second step, we compute the final error bars on $T_{\rm c}$ and
$\beta$ by adding the contribution from calibration uncertainties;
these have to be considered separately because the HFI calibration
errors are not independent of eachother. In Paper~I the size of this effect was
estimated using a grid of $T_{\rm c}$ and $\beta$ values, and was found to be
small ($\le 2.5\%$ for $T_{\rm c}$ and $\le 2\%$ for $\beta$)
compared to the uncertainties of the flux extraction method.
The final uncertainties in
$T_{\rm c}$ and $\beta$ are obtained as a quadratic sum of the two
contributions. They do not include the correlations between $T_{\rm c}$ and $\beta$
inherent in the fitting procedure itself. 

Applying the same fitting method pixel by pixel to the IRIS $100\,\mu$m
and HFI 857, 545, and 353\,GHz surface brightness images ($1^\circ
\times 1^\circ$), we calculate maps of the dust colour temperature
and the emissivity index. Here the flux error bars are the
quadratic sum of the noise map and the calibration uncertainties.
The averaged column densities of the clumps are then 
derived from the observed flux at $\nu_0= 857$ GHz and the 
dust colour temperature inferred from the fit:
\begin{equation}
N_{\rm{H}_2} =  \frac{S_{\nu_0}}{  \Omega_{\rm cl}\, \mu m_{\rm H}
 \kappa_{\nu_0} \times B_{\nu_0}(T_{\rm c})}
\end{equation}
with $\Omega_{\rm cl} = \pi \sigma_{\rm Maj} \sigma_{\rm Min}$.

The value of $\kappa_{\nu}$ is a main source of uncertainty. Large
variations exist between dust models, depending on properties of the dust
\citep[see][]{Beckwith1990, Henning1995}:
composition (with or without ice mantles); structure
(compact or fluffy aggregates); and size.  Both models
and observations show that $\kappa_{\nu}$ increases from the diffuse
medium to dense and cold regions by a factor of 3--4 
\citep{Ossenkopf1994, Kruegel1994, Stepnik2003, Juvela2011,planck2011-7.13}.
For this study, we adopt the dust absorption coefficient 
of \cite{Beckwith1990} inferred for high-density environments
\citep{Preibisch1993, Henning1995, Motte1998}:
\begin{equation}
\kappa_{\nu} =  0.1{\rm cm}^{2}{\rm g}^{-1}\,(\nu/1000\,{\rm GHz})^{\beta},
\label{eq:beckwith}
\end{equation}
with a fixed emissivity index of $\beta=2$.  The 
reference frequency $\nu_{0} = 857\,$GHz chosen to estimate
$N_{\rm{H}_2}$ is such that the extrapolation from 1000\,GHz
with $\beta=2$ (instead of the fitted spectral index)
introduces an uncertainty on $N_{\rm{H}_2}$ of less than 30\%. 
At a distance $d$, the mass of the clump is simply
$M=N_{\rm{H}_2}^{cl} \Omega_{\rm cl}\, \mu m_{\rm H} d^2$ or
\begin{equation}
M =  \frac{S_{\nu_0} d^2}{  \kappa_{\nu_0} B_{\nu_0}(T_{\rm c})}.
\end{equation}

\section{Results}  \label{sect:results}

\subsection{Physical parameters derived from \Planck\
and IRIS data } \label{sect:ResWithPlanck}

\begin{figure*}
\centering
\includegraphics[width=12cm]{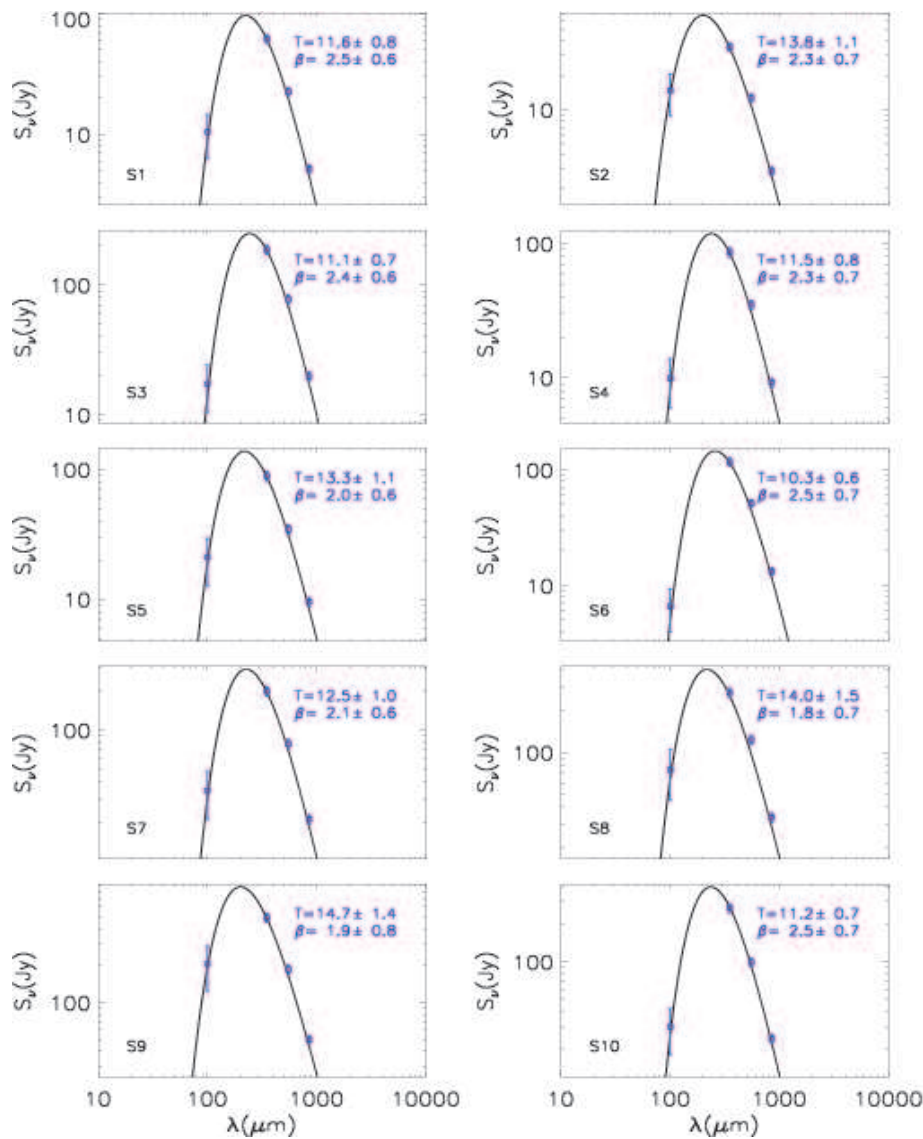}
\caption{
SEDs and fit parameters ($T$, $\beta$) obtained combining \Planck\ HFI and
IRIS flux densities integrated over the clump in the cold residual maps. 
}
\label{fig:Planck_all_seds}
\end{figure*}

The cold clump SEDs determined using the \Planck\ HFI and IRIS flux densities
are shown in Fig.~\ref{fig:Planck_all_seds}.  
As discussed in Section~\ref{sect:Planck_photometry}, 
these SEDs are those of the cold dust residual. 
The values obtained for the dust temperature and emissivity spectral
index are shown in the frames of the figure.  Using the method described in
Section~\ref{sect:T_beta_Nh}, we infer the gas column density,
bolometric luminosity, and mass of each clump (Table~\ref{table:Phys_param}).
All these parameters ($T$, $\beta$, $N_{\rm{H}_{2}}$) are values
which are {\em averaged} over the line-of-sight and the extent of the
cold clumps, which reach the parsec-scale. Denser and/or colder structure is
to be expected on smaller scales.
These averaged temperatures vary between 10.3 and 14.7\,K, 
and the dust emissivity spectral indices range
between 1.8 and 2.5.

The column densities are distributed around a mean of
$7.5\times10^{21}\,{\rm cm}^{-2}$, with a few of the largest values being
above ${\sim}\,10^{22}$\,cm$^{-2}$.  These relatively moderate column densities
also correspond to averaged values over the extent of the clumps.  The clumps
are likely to be heterogeneous, with denser substructures that will be
studied with higher angular resolution data in
Section~\ref{sect:substruct_Herschel}.  The masses of the sample cover a
large range from 3.5 to 1800 M$_{\odot}$.  The highest-mass object, S6,
is also the coldest source of the sample, and it is quite extended
(${\sim}\,2.9 \times 6.7\,$pc); these properties make this source an
interesting candidate for being a high mass star-forming precursor in
its early stages.
More generally, the sizes of the source sample are rather large, in
the range 0.2--11 parsecs, which confirms that they should be better
classified as ``clumps'' rather than ``cores,'' according to the
terminology used for nearby molecular clouds
\citep[e.g.,][]{Williams2000, Motte2007}.  Scales typically considered for
``dense cores'' and ``clumps'' are ${\sim}\,0.1\,$pc and ${\sim}\,1\,$pc,
respectively.

As noted in Paper~I, most of the cold clumps found with the CoCoCoDet
procedure are elongated, with aspect ratios significantly larger than
unity.  The maps of Fig. 1 show that these clumps are not isolated,
but are, in many cases, substructures of long filaments, not always
straight.  Without direct estimates of the density it is difficult to
assess that they are actual filaments, denser than their
environment. However, their brightness contrast above the background
is a few, in the case of S4, S5, S6 and S10, suggesting actual density
enhancements given their small thickness.
Their length, estimated from their angular size in the maps of Fig.~1 
and their distance, ranges from about 5\,pc for S10 to 18\,pc for S5.

Because of this, we choose to focus not on
the clump mass, that depends on the size of the clump derived from the 
detection procedure, but on the mass per unit 
length of the filament at the position of the cold clump.  This mass per
unit length (or linear mass density) 
is given by $m = 16.7 \mlin\ N_{21} \theta_{\rm Min}$, where
$N_{21}$ is the clump column density expressed in units of $10^{21}$\,\cq\ 
and $\theta_{\rm Min}$, the half-power thickness of the clump identified in
the \Planck\ maps, is taken as the width of the filament. The linear mass
densities are given in Table~\ref{table:Phys_param} for all clumps,
including those with an aspect ratio only slightly larger than unity.
However, they should be considered only as an approximate
guide to the linear mass density at the location of the clump.

The linear mass densities vary by a factor ${\sim}\,30$ and compare well with
those inferred from molecular lines. The range of values (15 to 400\,\mlin)
is characteristic of regions of massive or intermediate-mass star formation,
such as Orion and $\rho$ Oph \citep{HilyBlant2004}.
The linear mass densities of the most tenuous filaments
\citep[$<1$ \mlin,][]{Falgarone2001}
are not found in the present sub-sample of \Planck\ cold clumps. 

Many of the cold clumps detected by \Planck\ therefore appear
as cold substructures within larger scale filamentary structures
that have length up to $\sim$ 20\,pc in the \Planck\ maps and with parsec-scale thickness (S5, S6, S7).
The thinnest filaments found
are only a few parsec long and with thickness of a few 0.1\,pc (S10). 

At this stage of the analysis, it is  difficult to assess the
exact nature of the sources, i.e., discriminating between protostellar
objects and starless clumps. We can however use the temperature and
bolometric luminosity estimates as first indicators of the evolutionary stage
of the sources.  Such an approach was proposed by \cite{Netterfield2009}
and \cite{Roy2010} in their analysis of the cold
cores detected with the balloon-borne experiment, BLAST, in the
Cygnus-X and Vela surveys.  Among the values obtained
for the ratio $L/M$ in our sample (ranging from 0.13 to 0.91 \lsol/\msol),
it is interesting to note that 
the three highest (with $L/M>0.6 $\lsol/\msol) are associated with $T>14\,$K dust
sources S2, S8 and S9). As we will see in
Section~\ref{sect:substruct_Herschel}, the \Planck\
clumps in these fields harbour warm dust sources, and two of them are
associated with young stellar objects.

\begin{table*}
\begingroup
\caption{Clump physical parameters derived from \Planck\ SEDs.}
\label{table:Phys_param}
\vspace{-12pt}
\setbox\tablebox=\vbox{
\newdimen\digitwidth
\setbox0=\hbox{\rm 0}
\digitwidth=\wd0
\catcode`*=\active
\def*{\kern\digitwidth}
\newdimen\signwidth
\setbox0=\hbox{+}
\signwidth=\wd0
\catcode`!=\active
\def!{\kern\signwidth}
\halign{#\hfil\tabskip=1.0em&
\hfil#\hfil&
\hfil#\hfil&
\hfil#\hfil&
\hfil#\hfil&
\hfil#\hfil&
\hfil#\hfil&
\hfil#\hfil&
\hfil#\hfil&
\hfil#\hfil\tabskip=0pt\cr
\noalign{\doubleline}
Source & $T_{\rm c}$ & $\beta$ & $\theta_{\rm Min}$ & $\theta_{\rm Maj}$ &
 $N_{\rm{H}_{2}}$ & $m$ & $M$ & $L$  &  $L/M$ \cr
& (K) & & (pc) & (pc) & (cm$^{-2}$) & (\mlin) &
 (M$_{\odot}$) & (L$_{\odot}$) & (L$_{\odot}$/M$_{\odot}$) \cr
\noalign{\vskip 3pt\hrule\vskip 4pt}
S1  &  $11.6\pm0.8$ & $2.5\pm0.6$ & 0.2 & *0.3 & $4.4\times10^{21}$ & *15 &
 ***3.5  & **1.0  &  0.28 \cr
S2  &  $13.8\pm1.1$ & $2.3\pm0.7$ & 1.2 & *1.5 & $8.0\times10^{20}$ & *16 &
 **15.0  & *11.0  &  0.72 \cr
S3  &  $11.1\pm0.7$ & $2.4\pm0.6$ & 0.4 & *0.5 & $1.1\times10^{22}$ & *77 &
 **22.0  & **4.3  &  0.19 \cr
S4  &  $11.5\pm0.8$ & $2.3\pm0.7$ & 0.7 & *1.0 & $6.3\times10^{21}$ & *74 &
 **44.0  & *11.0  &  0.24 \cr
S5  &  $13.3\pm1.1$ & $2.0\pm0.6$ & 2.7 & 11.0 & $1.7\times10^{21}$ & *77 &
 *520.0  & 270.0  &  0.52 \cr
S6 &   $10.3\pm0.6$ & $2.5\pm0.7$ & 2.9 & *6.7 & $8.4\times10^{21}$ & 400 &
 1800.0  &  230.0 &  0.13 \cr
S7 &   $12.5\pm1.0$ & $2.1\pm0.6$ & 0.6 & *1.5 & $4.9\times10^{21}$ & *49 &
 **45.0  & *17.0  &  0.37 \cr
S8 &   $14.0\pm1.5$ & $1.8\pm0.7$ & 1.2 & *2.2 & $8.1\times10^{21}$ & 160 &
 *210.0  &  140.0 &  0.65 \cr
S9 &   $14.7\pm1.4$ & $1.9\pm0.8$ & 1.0 & *1.9 & $1.6\times10^{22}$ & 270 &
 *330.0  &  300.0 &  0.91 \cr
S10 &  $11.2\pm0.7$ & $2.5\pm0.7$ & 0.3 & *1.0 & $1.3\times10^{22}$ & *64 &
 **40.0   & **8.4 &  0.21 \cr
\noalign{\vskip 3pt\hrule\vskip 4pt}
} }
\endPlancktablewide
\tablenote a Minor and major axis FWHM values of the Gaussian ellipse, converted to
length units.\par
\tablenote b Linear mass density.\par
\endgroup
\end{table*}

%DS- I reduced the number of sig figs in some of the columns

\subsection{Colour temperature and spectral index maps} \label{sect:Tmaps}

\begin{figure}
\includegraphics[width=8.5cm]{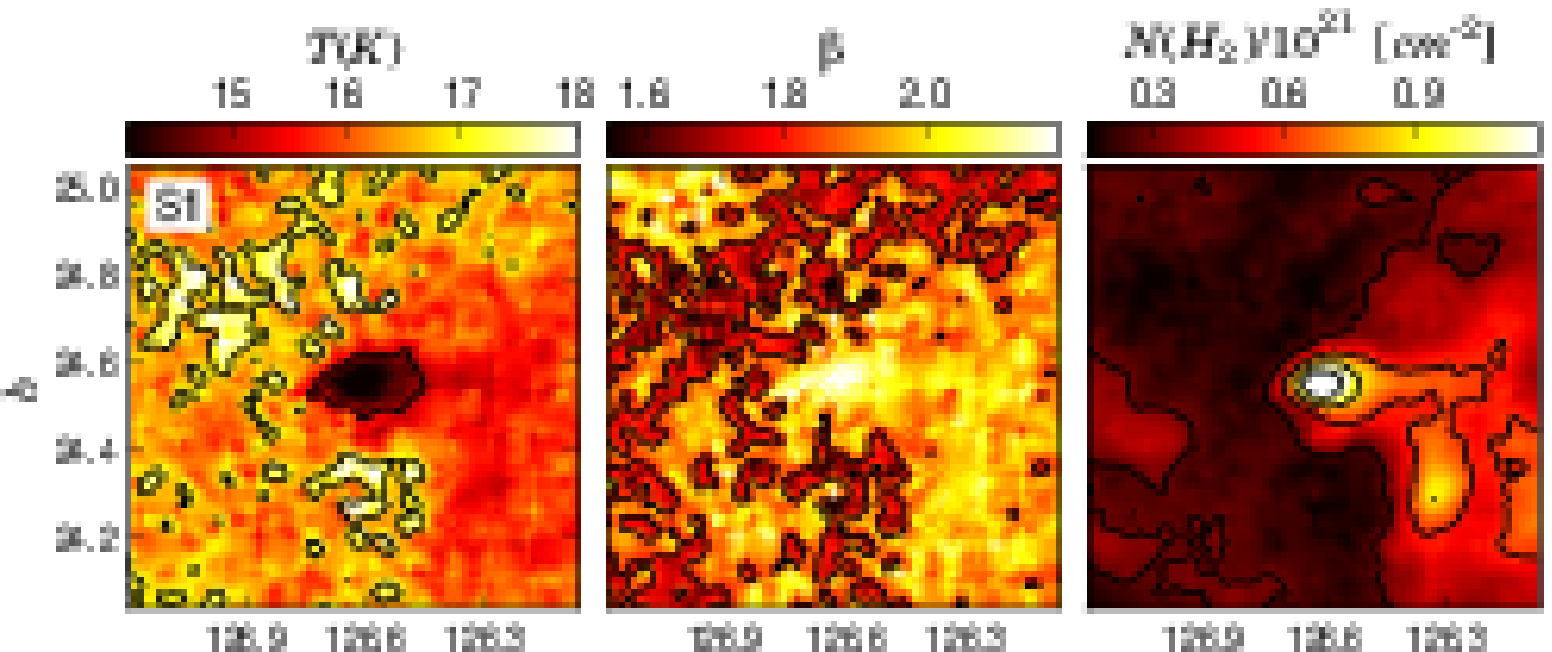}
\includegraphics[width=8.5cm]{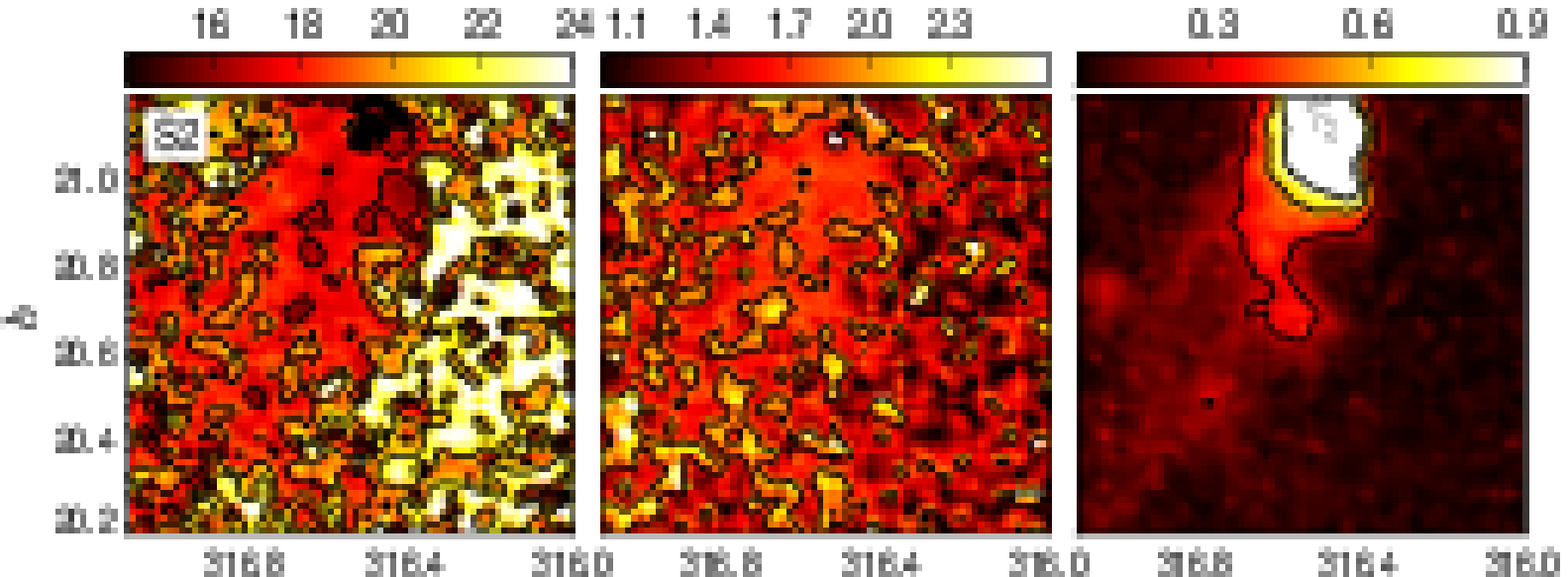}
\includegraphics[width=8.5cm]{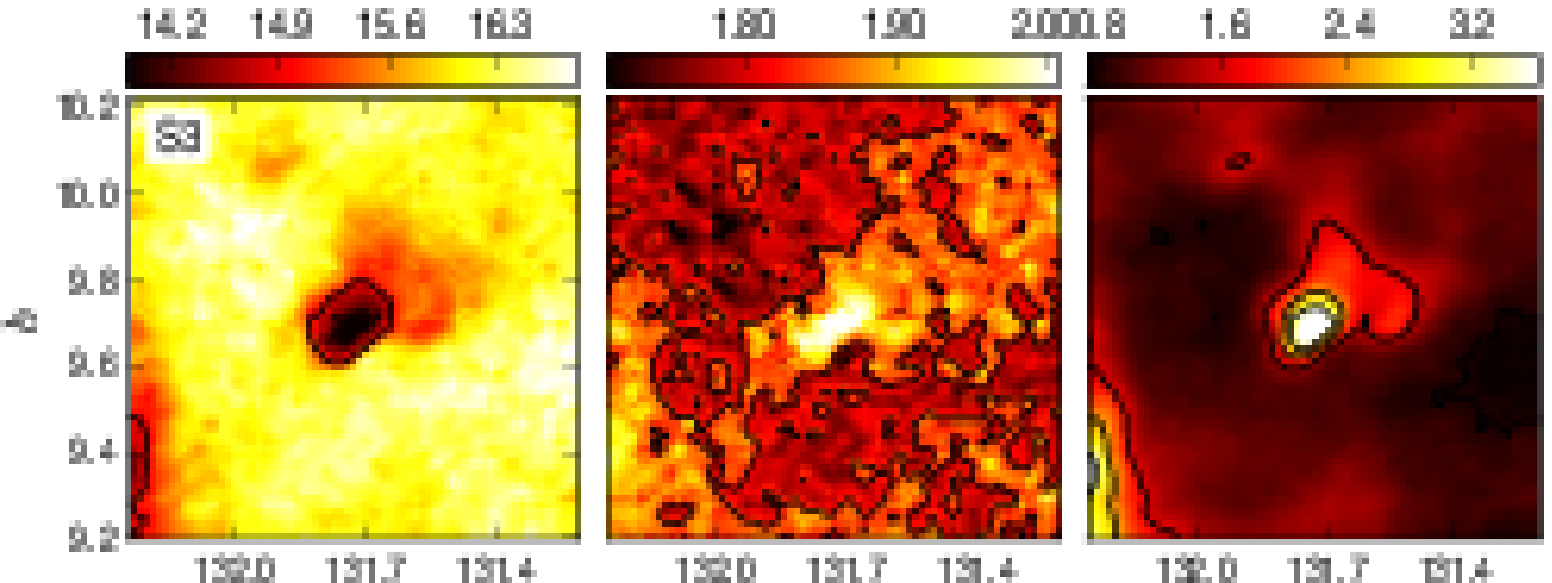}
\includegraphics[width=8.5cm]{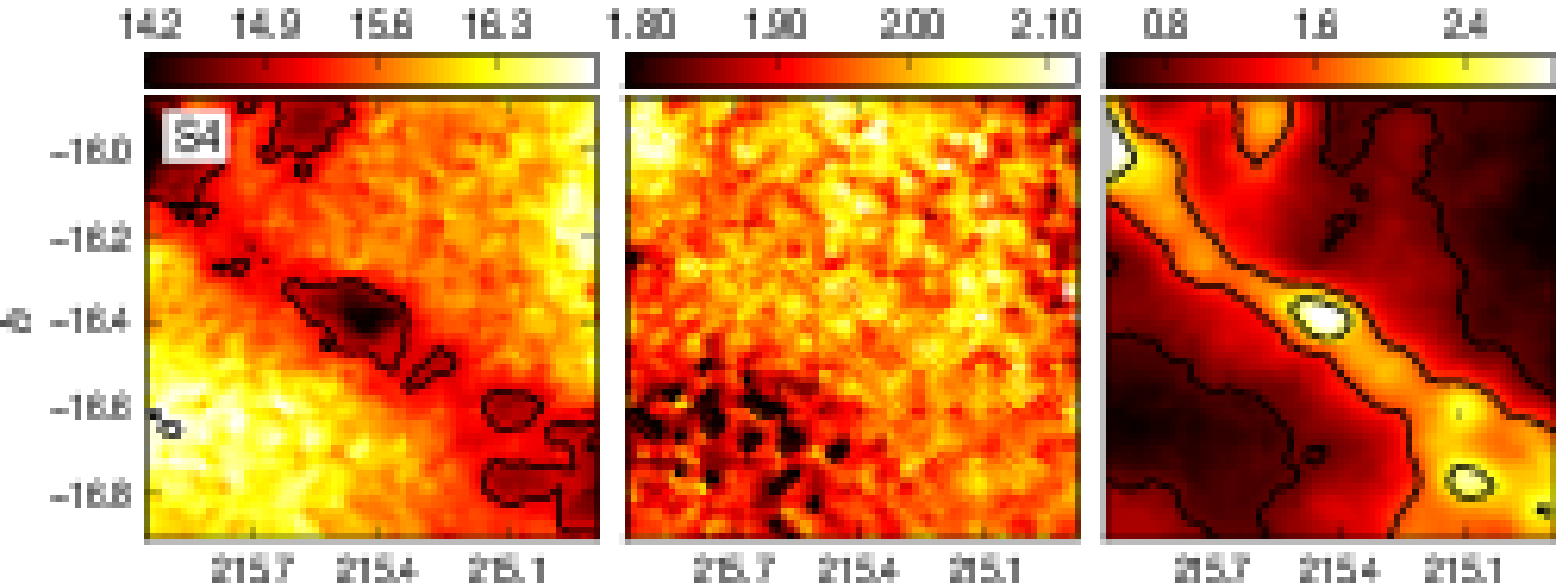}
\includegraphics[width=8.5cm]{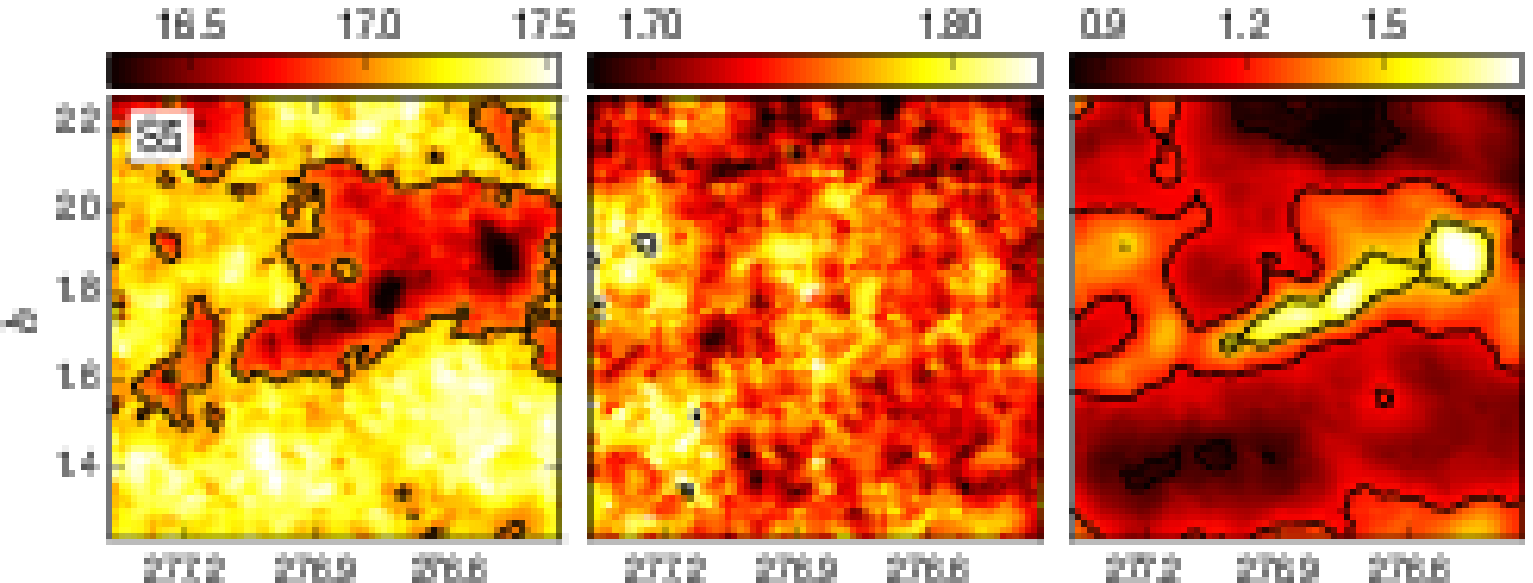}
\includegraphics[width=8.5cm]{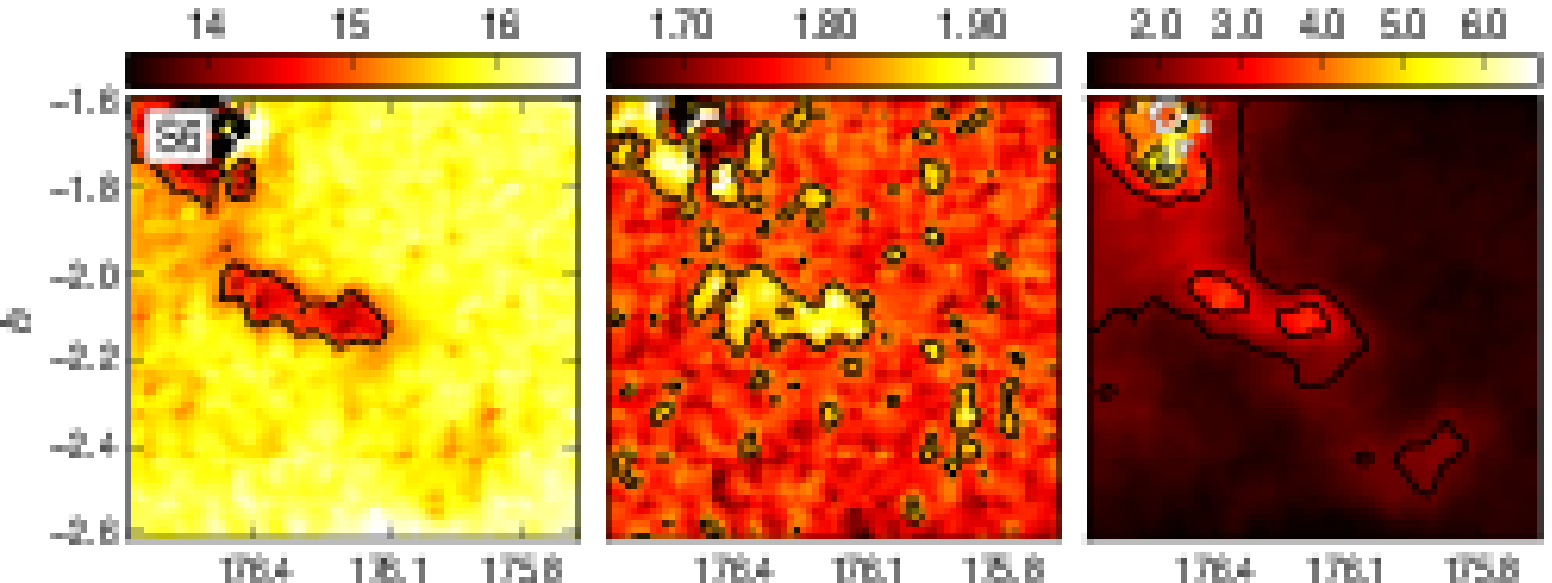}
\includegraphics[width=8.5cm]{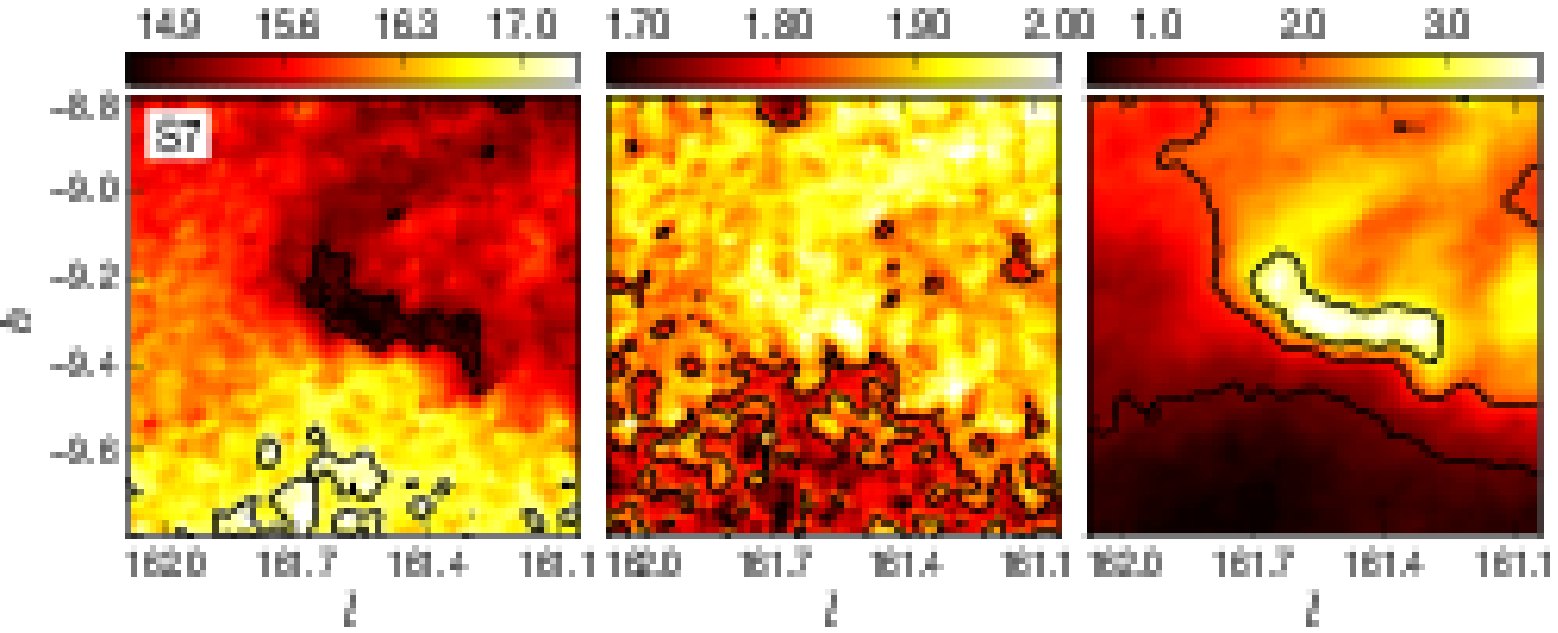}
\caption[]{
Maps of dust colour temperature, emissivity spectral index and column density
for each clump field (the three SDP fields are shown in \cite{Juvela2011}
using {\it Herschel\/} maps). The $N_{\rm{H}_{2}}$ contours correspond to the ticks on the colour bar.
For T and $\beta$, the contours are shown respectively at 2K intervals and for $\beta =1.85$.
}
\label{fig:TB}
\end{figure}

Fig.~\ref{fig:TB} shows the $1^\circ \times 1^\circ$ maps for dust colour
temperature and dust spectral index that were estimated using the
{\it IRAS\/} 100\,$\mu$m data and the three highest frequency HFI channels (see
Section~\ref{sect:T_beta_Nh}) without background removal. 
The three {\it Herschel\/} SDP fields (S8, S9 and S10) are not presented here,
because the analysis of $T$ and $\beta$ was already described in detail 
in \cite{Juvela2011}.

\
Except for the faint (and not so cold) source S2, a clear signature of 
cold dust emission is visible directly on these temperature maps at the
location of the clumps. The colour temperature characterizing the
line-of-sight integrated intensity in the direction of the clump 
is typically about 14\,K, warmer by a few Kelvin than the clump temperature
deduced from its SED after warm background subtraction. 
The clumps are embedded within extended structures in the temperature range
16--18\,K.  For reference, dust in the diffuse medium of the local Galactic
neighbourhood has a temperature of around 17.5\,K \citep{Boulanger1996, Schlegel1998}. 
In several cases (S4, S5, S6 and S7) the cold clumps appear as the coldest spot 
(or one of the coldest spots) along filamentary structures, which are
already colder than the larger scale environment.
The temperature and column density maps are correlated, 
with the coldest structures corresponding, as expected, to the most opaque 
parts of the field at the angular resolution probed by \Planck. The sharp edges 
seen in the intensity maps of the fields S1, S3 and S7 are also clearly
associated with column density sharp transition. This may be the signature of shock 
compression and triggering core formation that should be investigated in further studies. 

We have compared the results on column densities with extinction measurements. 
Extinction is calculated with the NICER method \citep{Lombardi2001}
using stars from the 2MASS catalogue \citep{Skrutskie2006}. 
The $A_V$ maps of the fields are derived assuming an extinction curve that
corresponds to $R_V$=3.1.  Preliminary extinction maps are created using nearby
fields where {\it IRAS\/} 100\,$\mu$m data indicates a low dust column
density.  The $A_V$ maps obtained with this method are shown in the
Appendix (in Fig.~\ref{fig:Av}) at 4.5$\arcmin$ resolution. 
A clear signature is visible, with an increase by about a factor of 2
in $A_V$ toward the clumps.
The $A_V$ values derived (${\sim}\,2$--5 mag) are consistent with the
previous $N_{\rm{H}_2}$ estimates of several times $10^{21}$cm$^{-2}$,
derived from the SEDs.  

In most fields the colour temperature and the spectral index are seen
to be anticorrelated, with high spectral indices being found at the
location of the temperature minima. The parameters $T$ and $\beta$
are known to be connected so that any noise present in the
observations tends to produce a similar anticorrelation. However, the spatial
coherence of the $T$ and $\beta$ maps strongly suggests that the
results of Fig.~\ref{fig:TB} are not caused by statistical noise. 
There is still the possibility of systematic errors, but the
similarity of the results obtained in separate fields makes this
unlikely. A pure calibration error would change the $\beta$ and $T$
values in a systematic way but would not significantly affect the
dispersion of the spectral index values \citep{Juvela2011}. The
line-of-sight temperature variations tend to decrease the observed
spectral indices.
The effect should be stronger toward the cold clump. The fact that $\beta$ 
at those positions is larger than in
the diffuse areas suggests that the emissivity index of the grains
has increased in the cores more than what is visible in the maps.
Therefore, the results support the idea of spatial variation
of grain properties that could result from temperature-dependent  
processes in the dust emission mechanism
\citep[][see\ Paper~I discussion]{Meny2007,Boudet2005, Coupeaud2011a}.

The CO correction to 353\,GHz band fluxes would typically increase the
$\beta$ values by ${\sim}\,10$\% or less, and the effect on the derived
colour temperatures is at most a few 0.1\,K. The main effect is a
change in the general amplitude of the values without major changes in the
morphology of the parameter maps.

\subsection{Analysis using ancillary data}

\subsubsection{{\it Herschel\/} and {\it AKARI\/} maps}
\label{sect:substruct_Herschel}

For all sources we have higher resolution dust continuum data in the
form of {\it Herschel\/} (resolution 37$\arcsec$ or better) and/or
{\it AKARI\/} (58$\arcsec$ or better) maps. These are used: to examine 
internal structure of the sources; to derive physical
parameters of the \Planck\ sources and compare them with the results of
Section~\ref{sect:ResWithPlanck}; and to study the properties of
the structures found at scales below the \Planck\ beam size.

\begin{figure*}
\center
\begin{tabular}{cc}

\includegraphics[width=5.75cm]{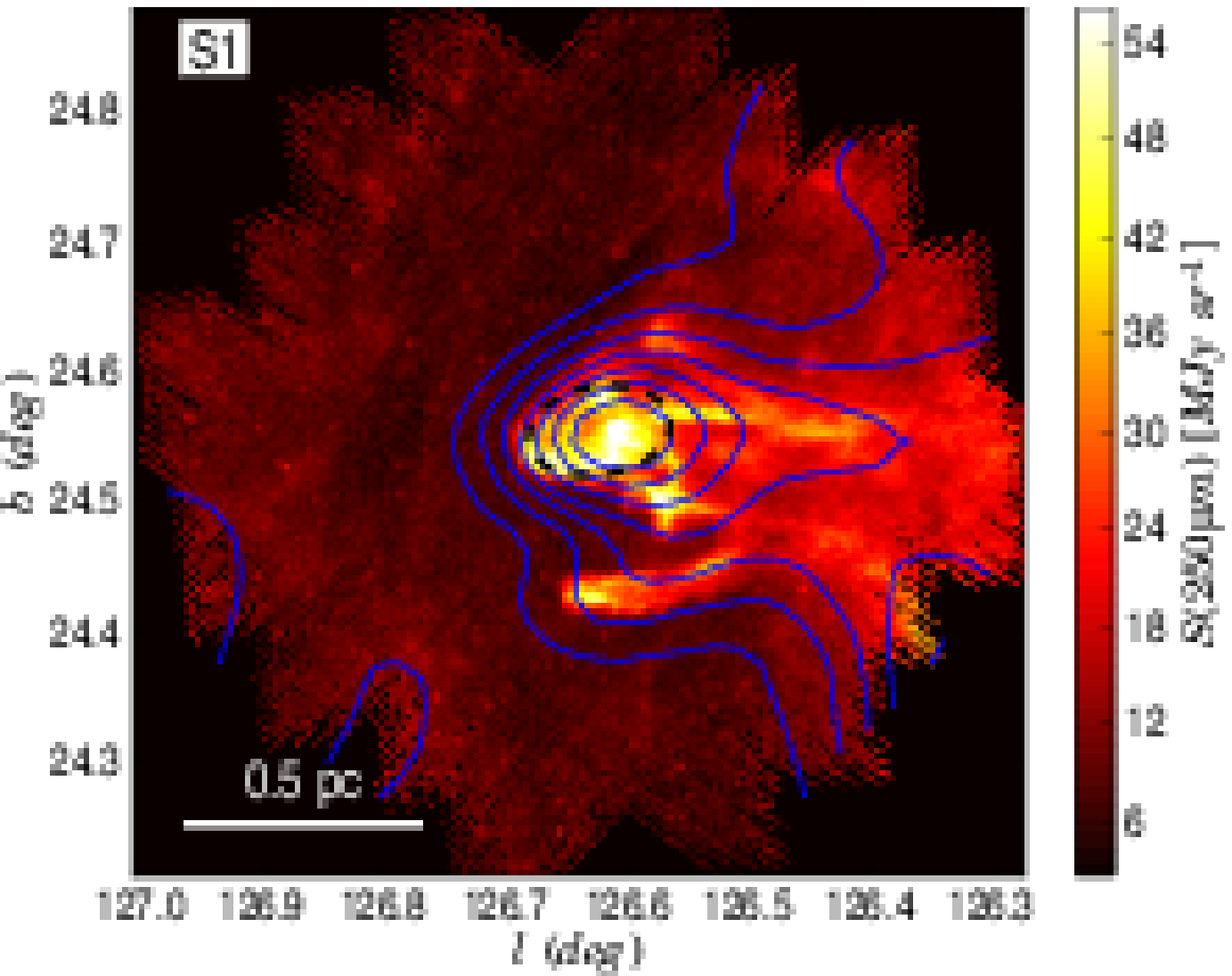}  &
\includegraphics[width=5.75cm]{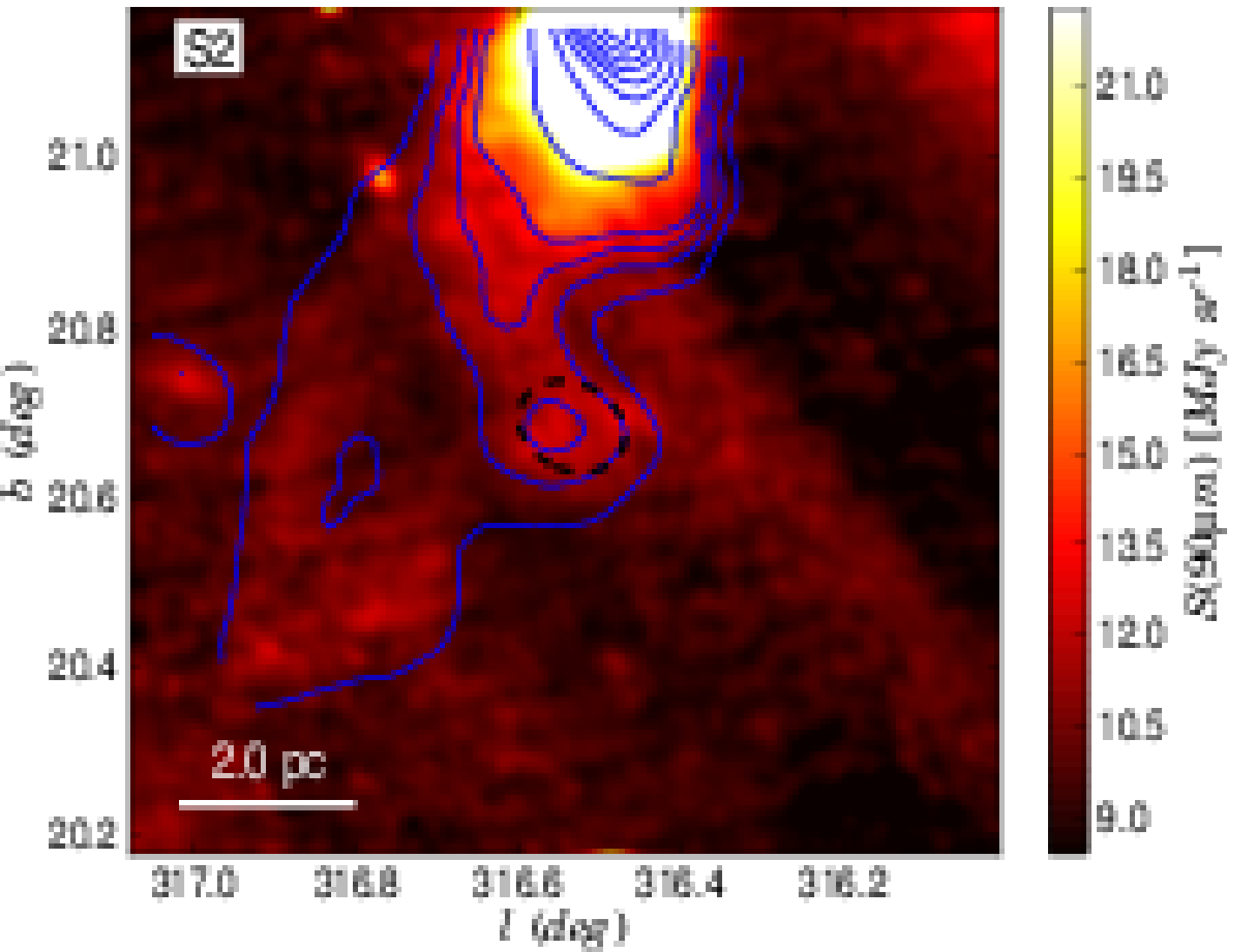} 
     \\
 
\includegraphics[width=5.75cm]{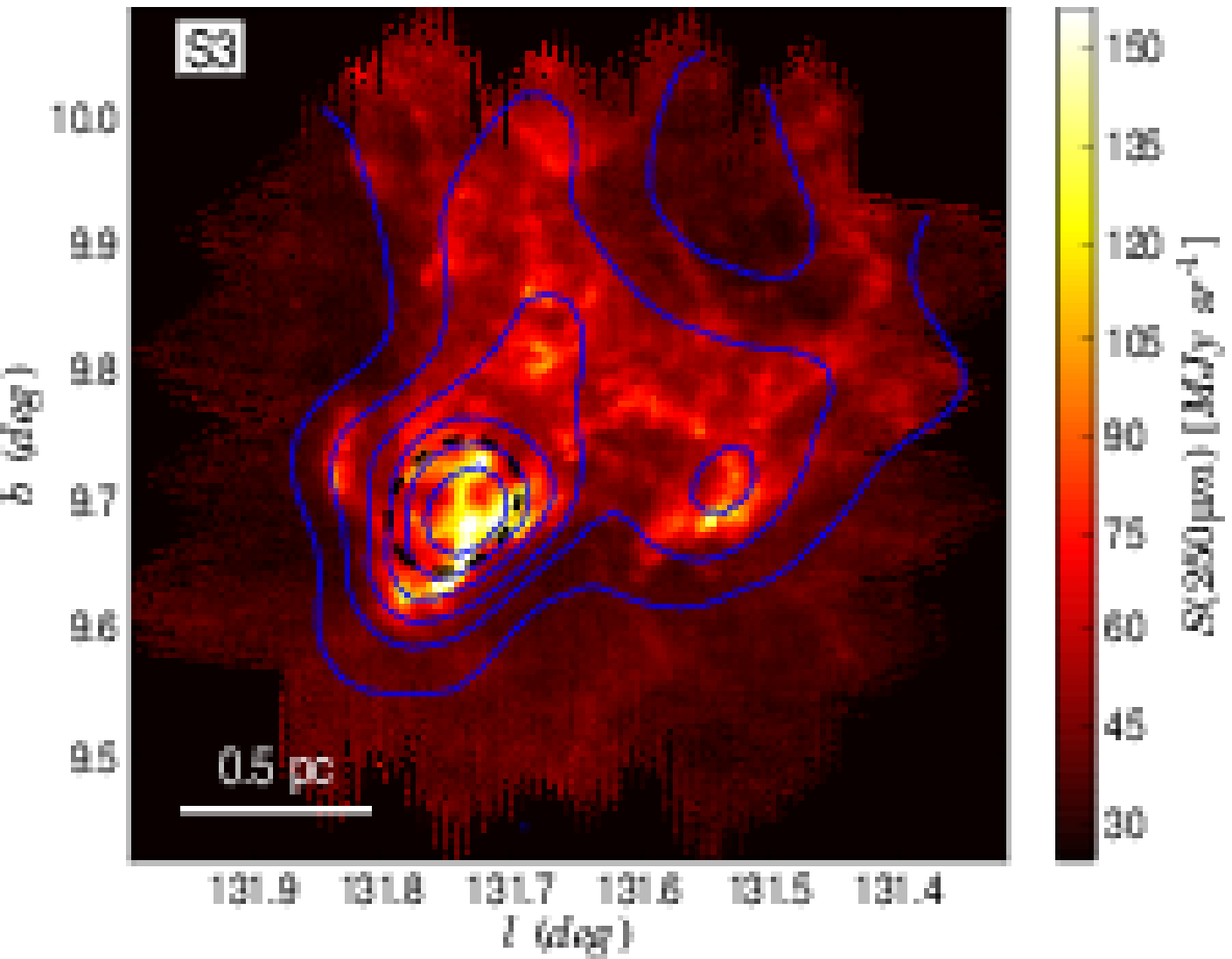}  &
\includegraphics[width=5.75cm]{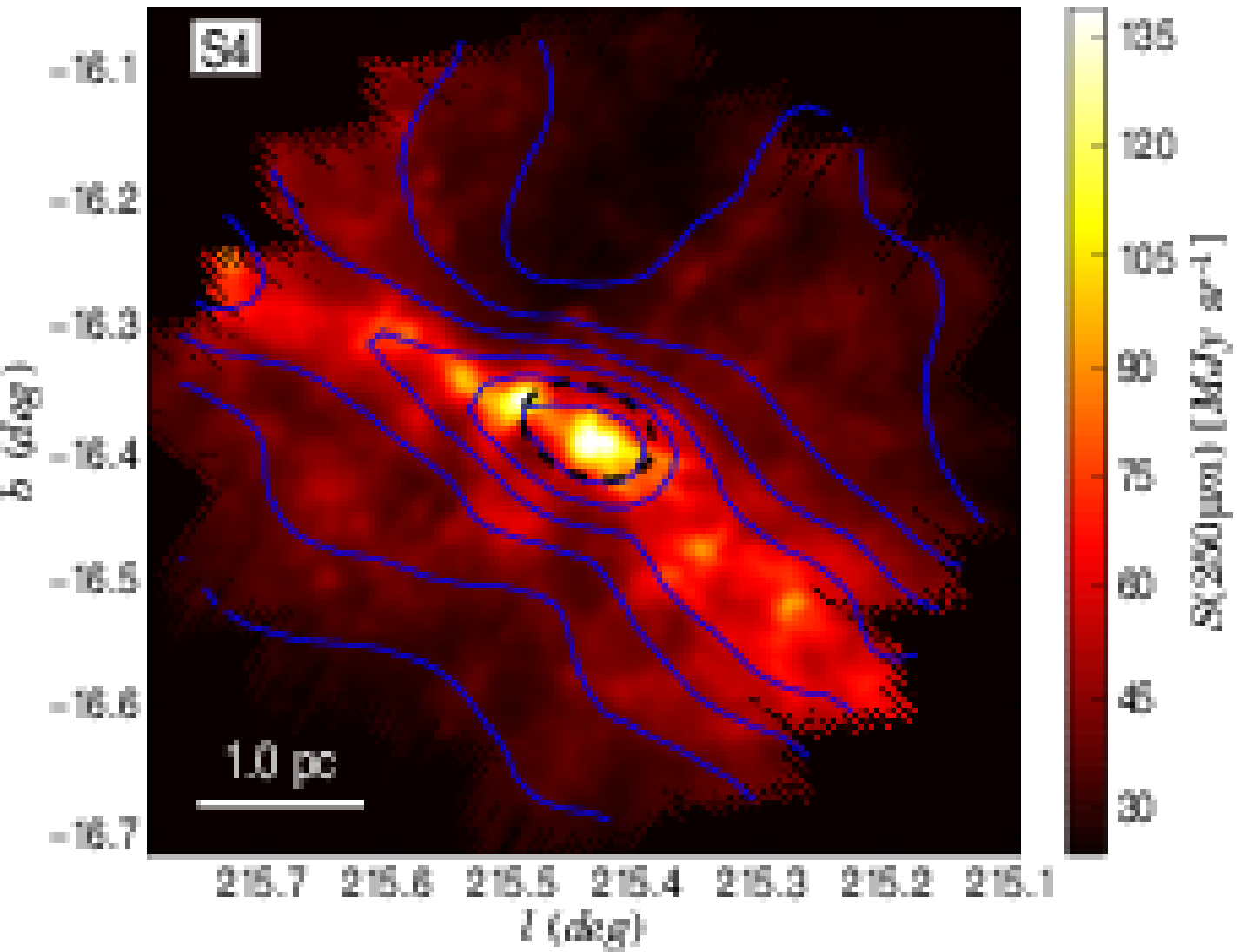} \\
 
\includegraphics[width=5.75cm]{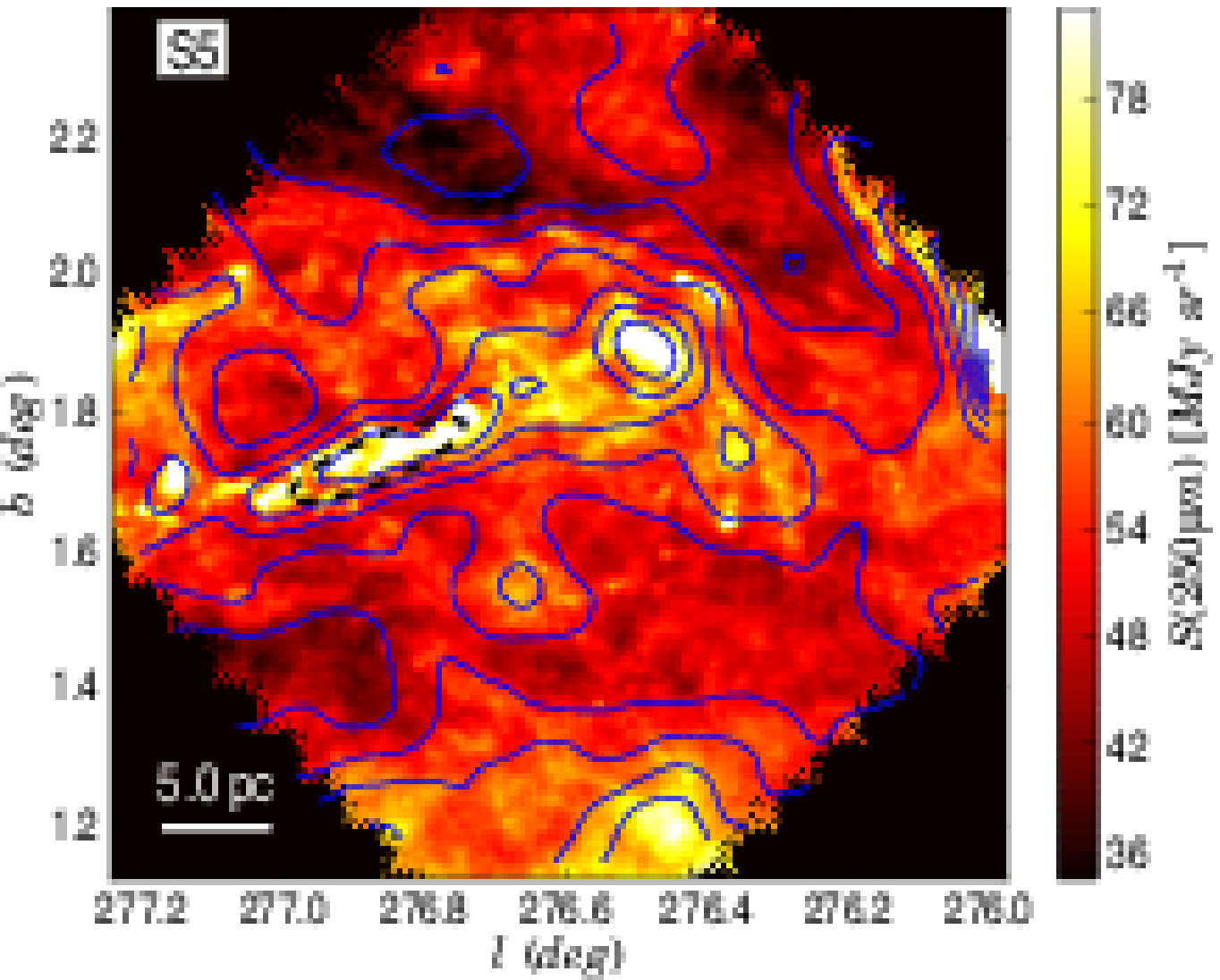}  &
\includegraphics[width=5.75cm]{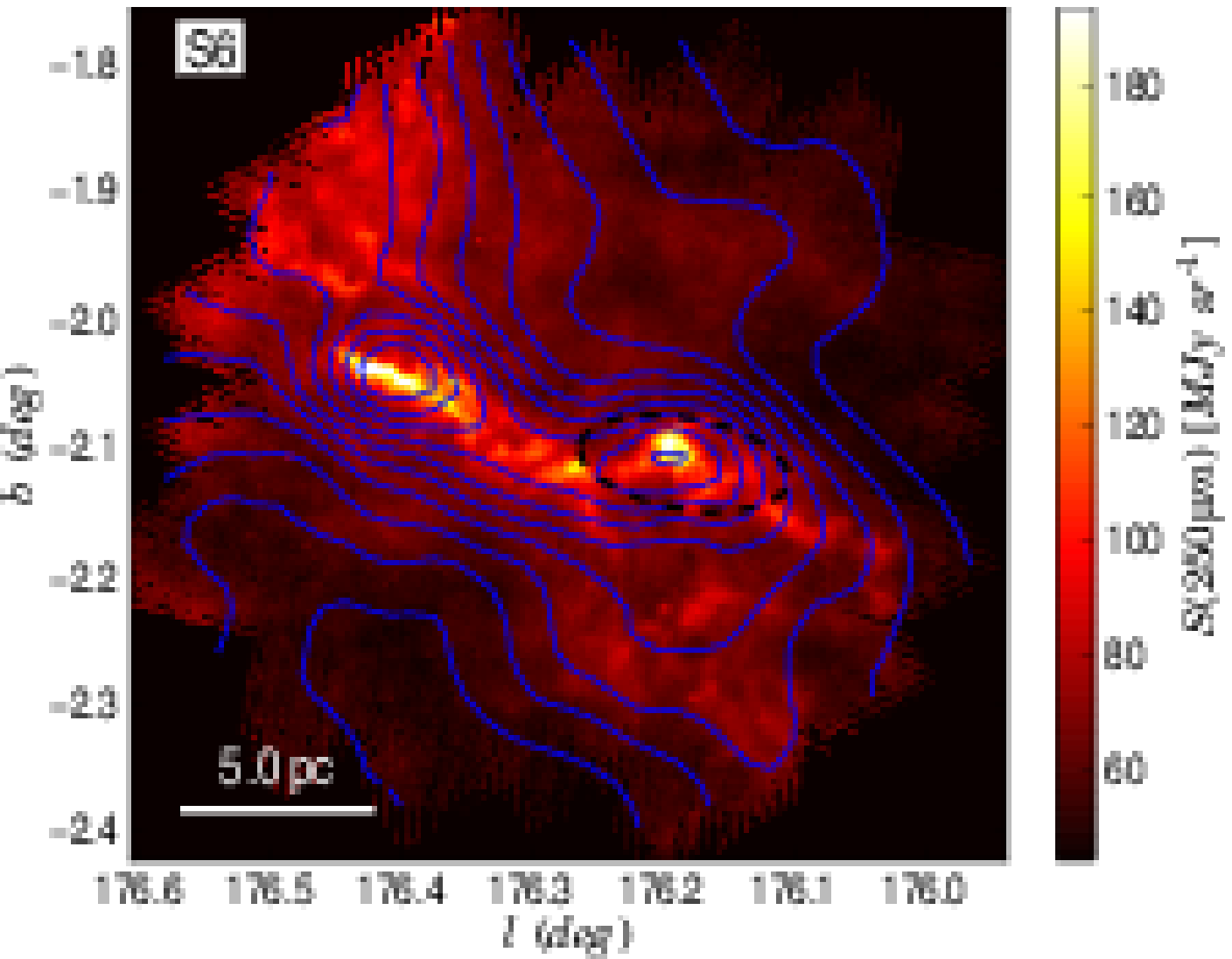} \\

\includegraphics[width=5.75cm]{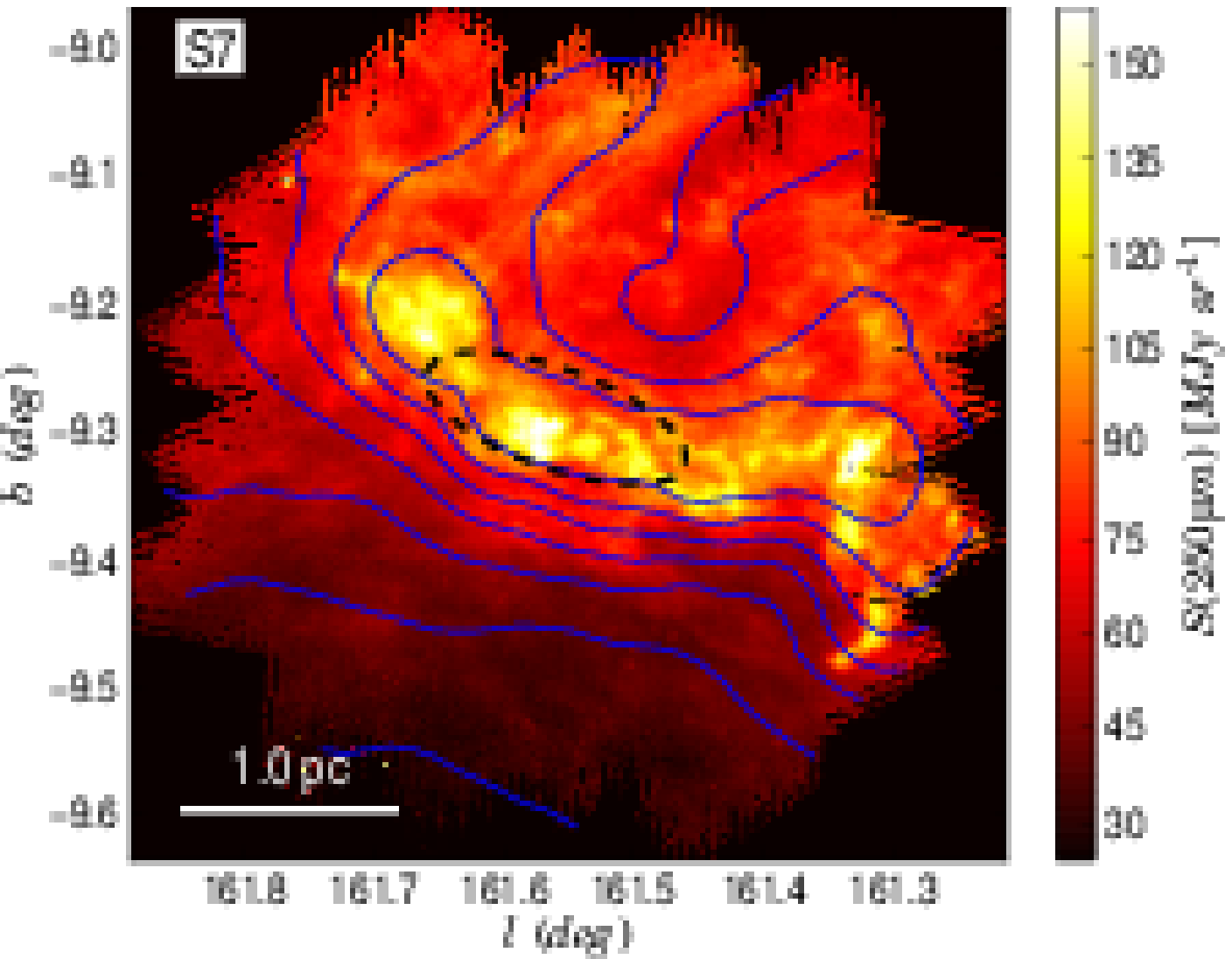}  & 
\includegraphics[width=5.75cm]{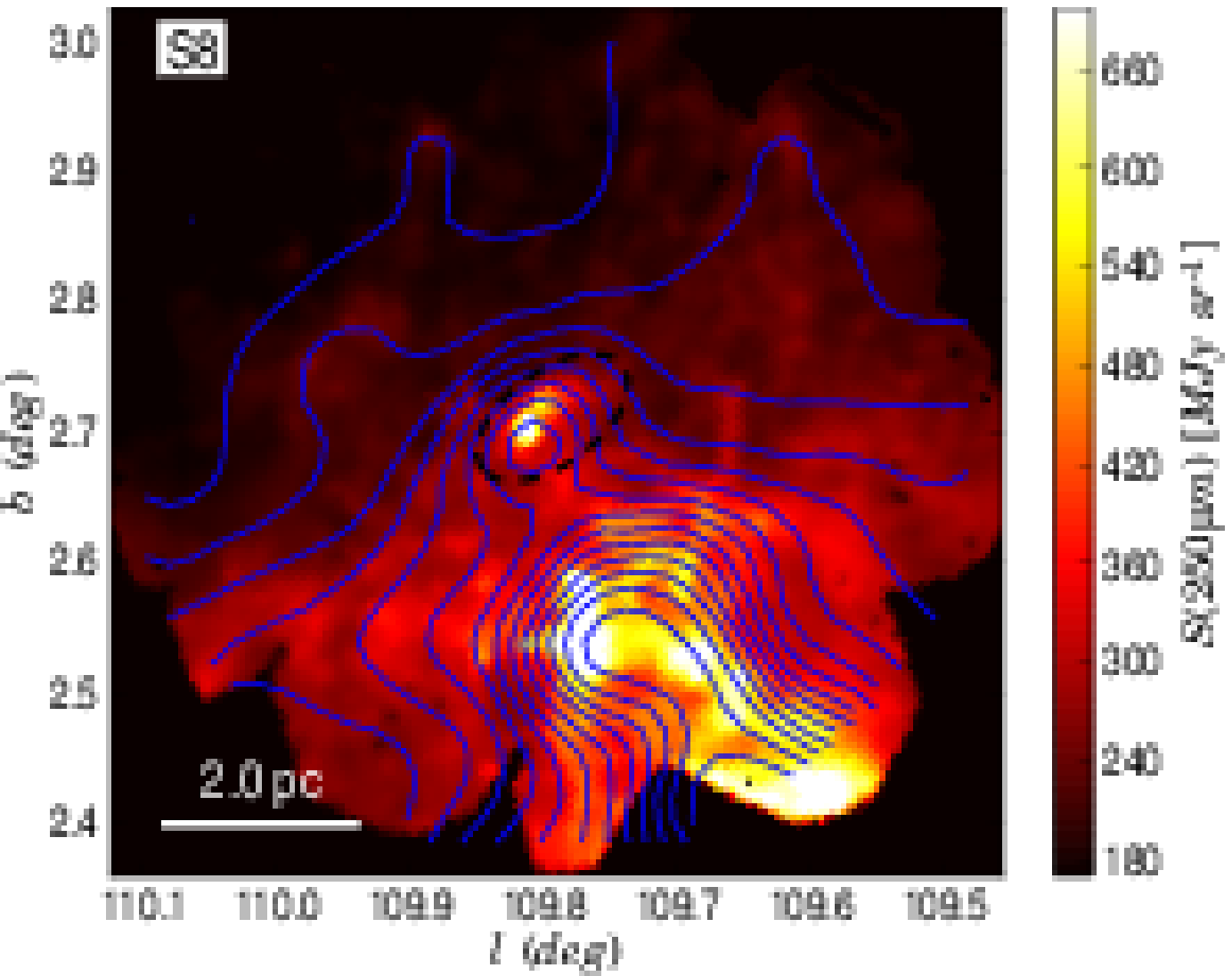}  \\

\includegraphics[width=5.75cm]{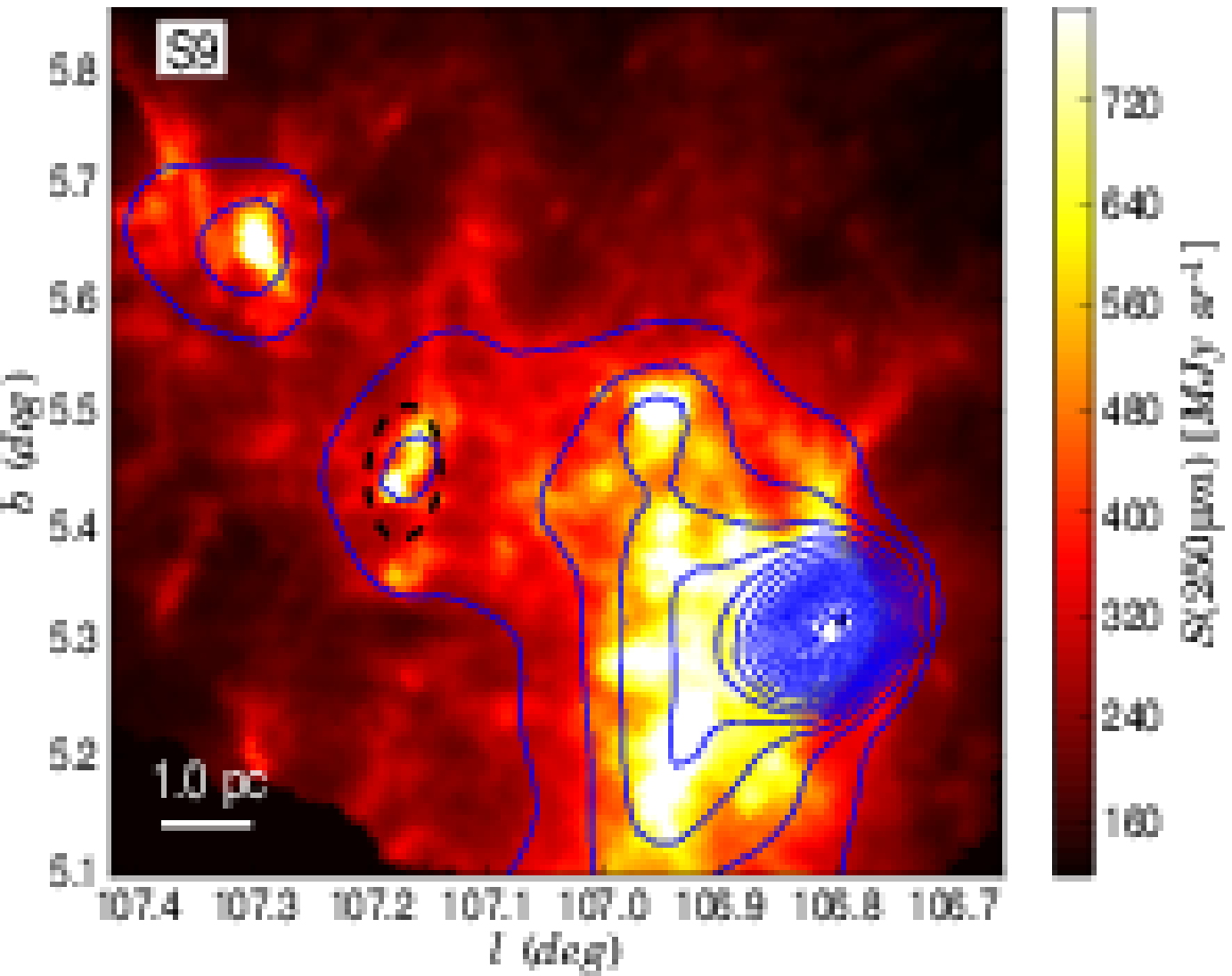}  & 
\includegraphics[width=5.75cm]{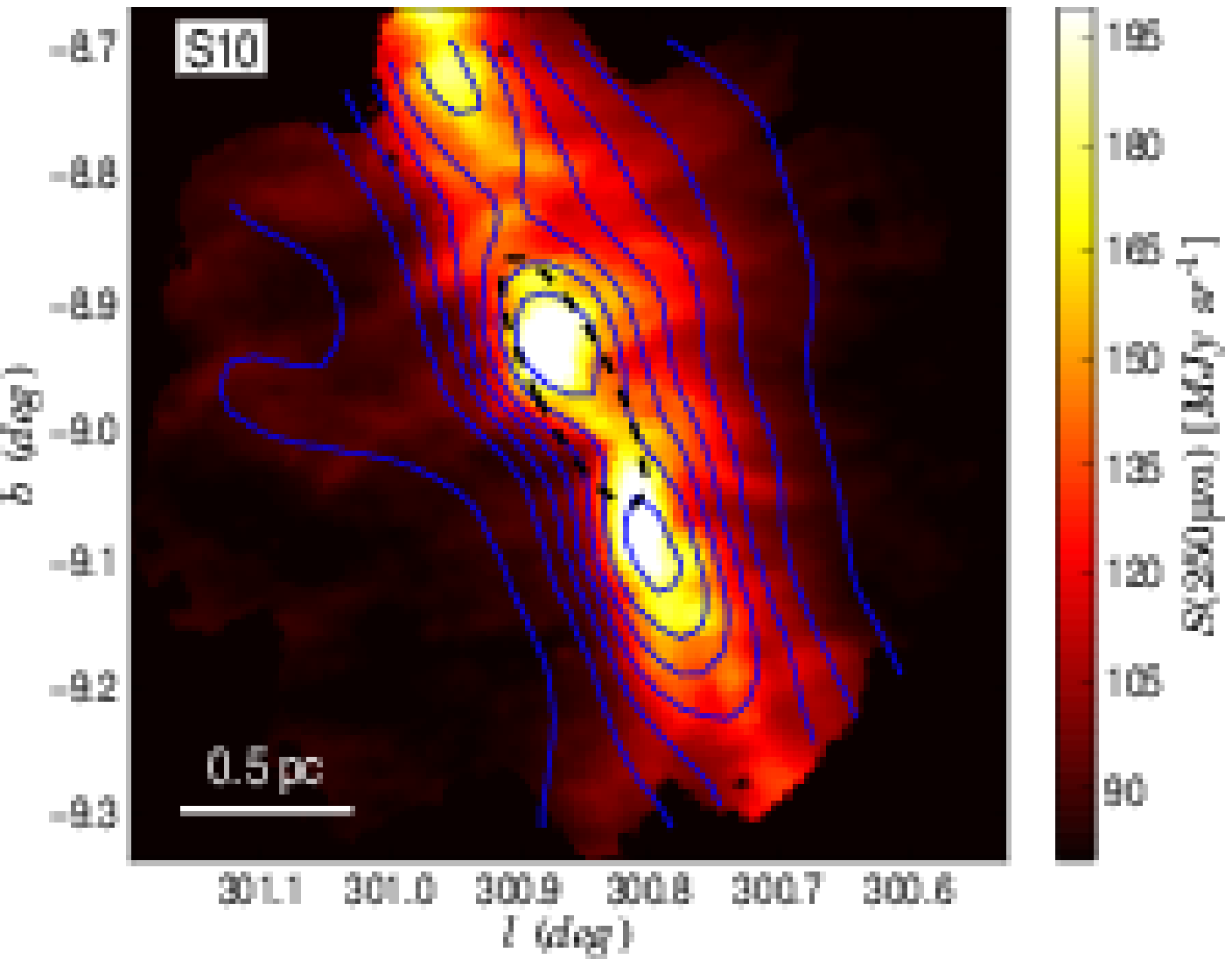}  \\

\end{tabular}
\caption{\Planck\ 857\,GHz surface brightness contours on {\it Herschel\/}
SPIRE maps at 250\,$\mu$m. The source S2 has not been observed with
{\it Herschel\/} and the displayed image corresponds to the {\it AKARI\/}
90\,$\mu$m wide filter. The dashed ellipses correspond to the estimated size
of the \Planck\ cold clump. 
}
\label{fig:Planck-Herschel}
\end{figure*}

\begin{figure*}
\center
\begin{tabular}{cc}

\includegraphics[width=5.75cm]{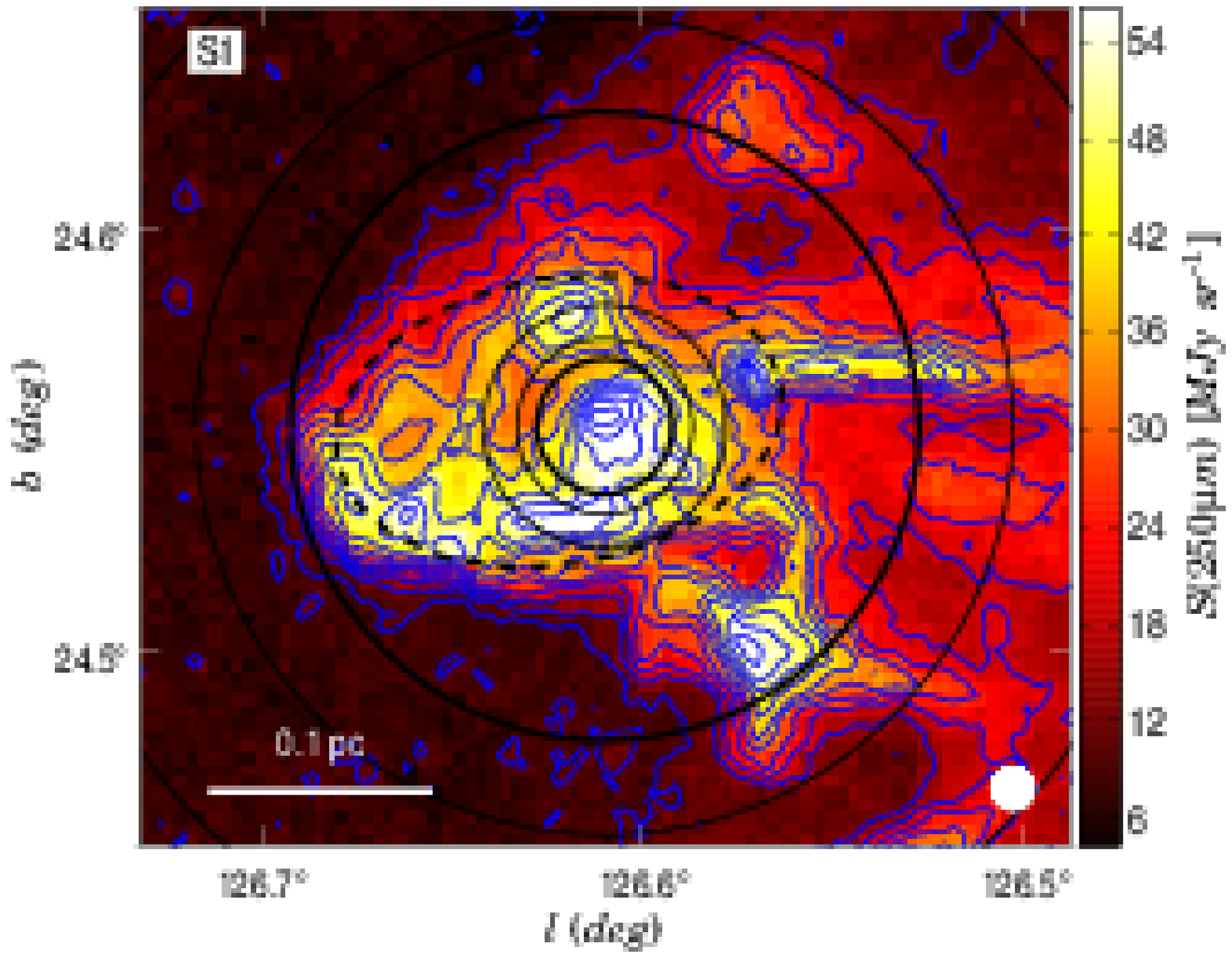}  &
\includegraphics[width=5.75cm]{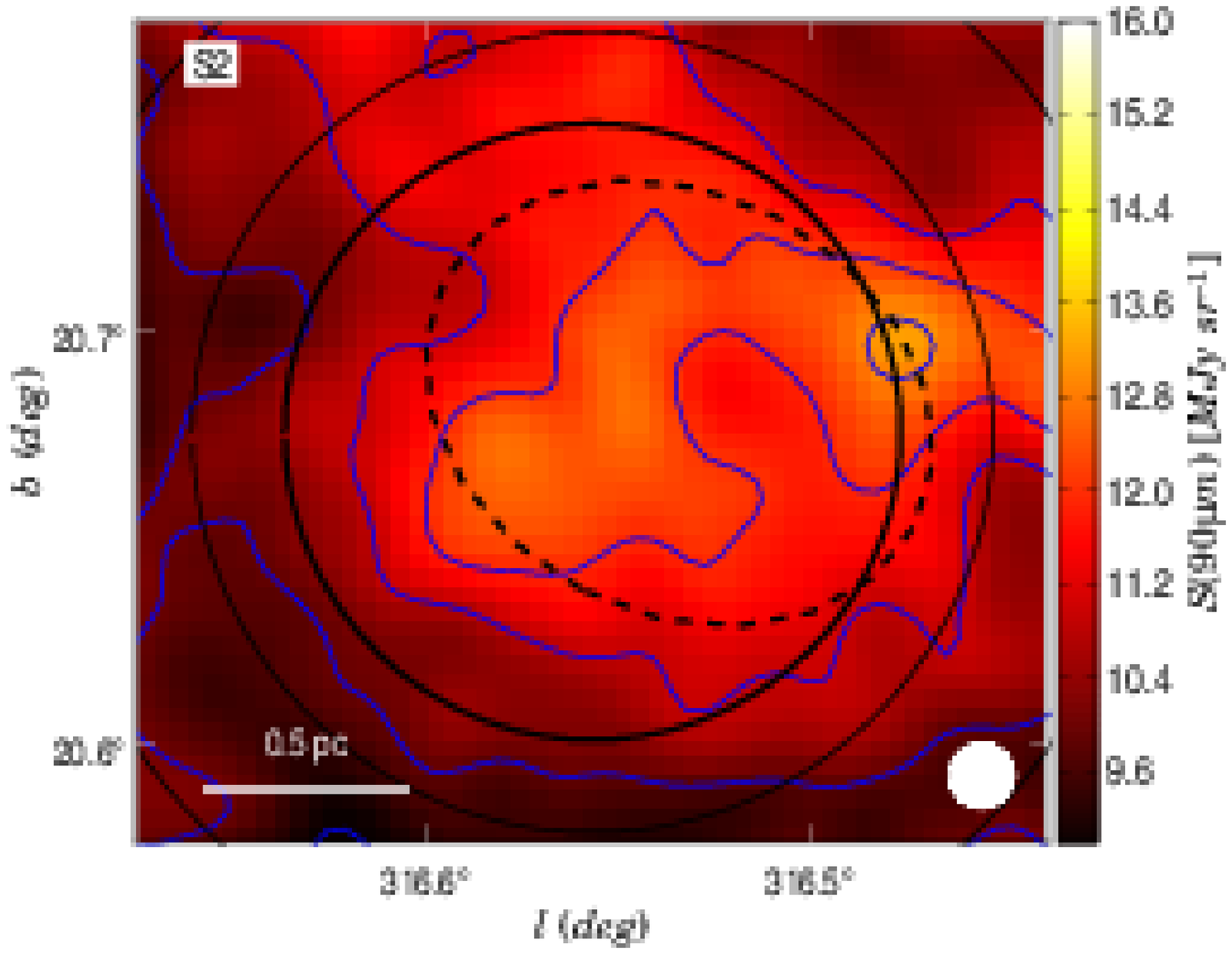} \\
 
\includegraphics[width=5.75cm]{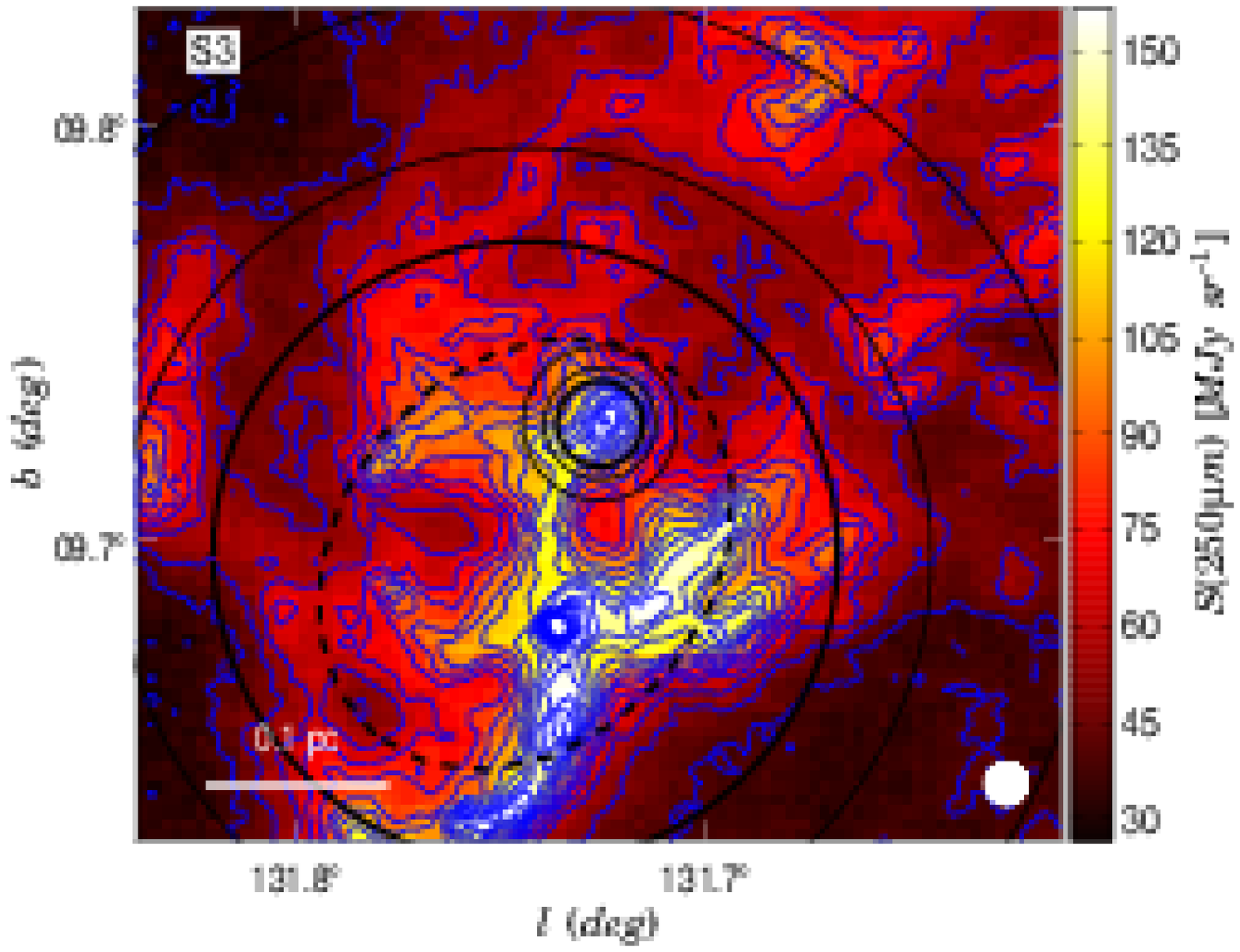}  &
\includegraphics[width=5.75cm]{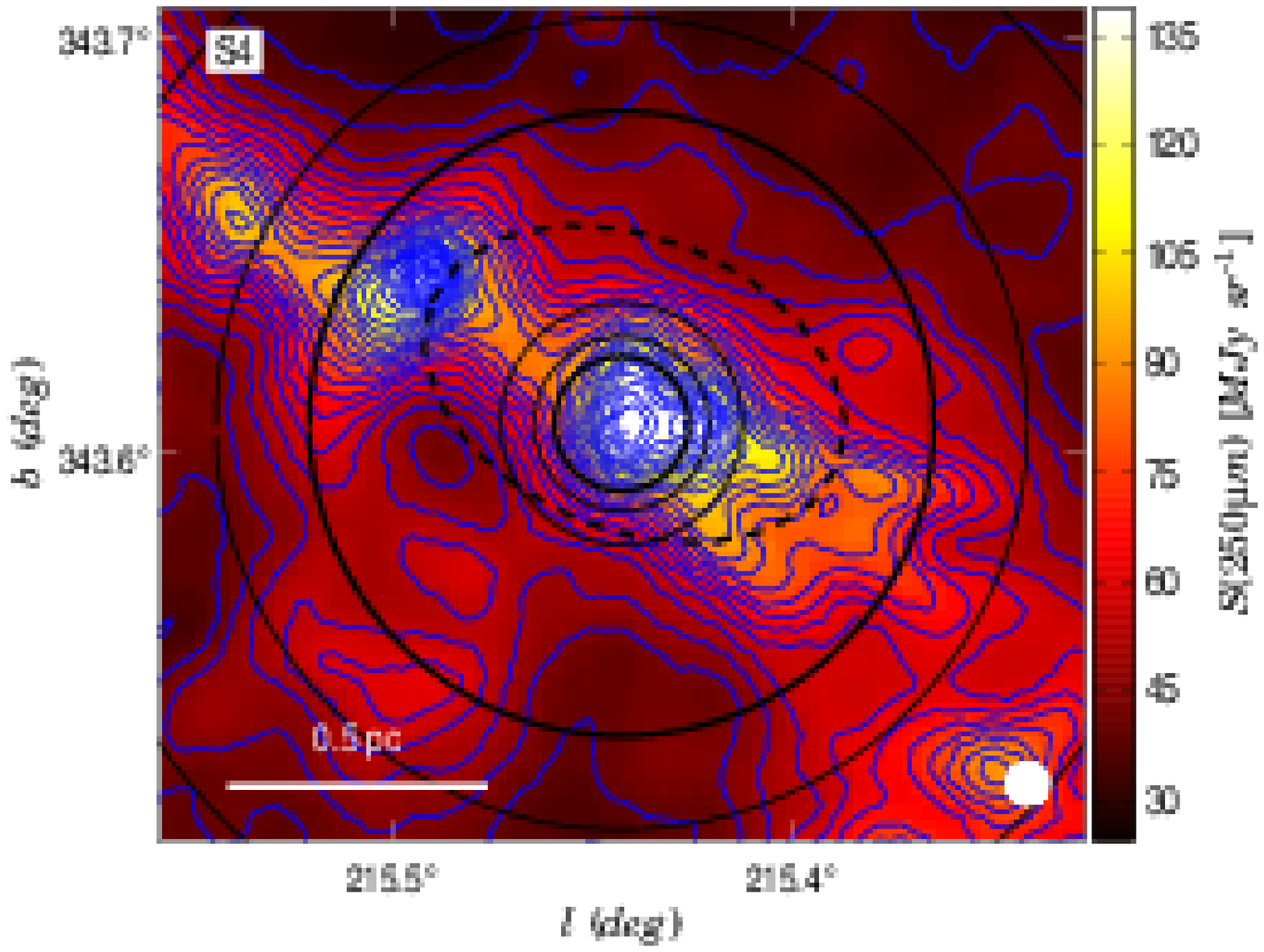} \\
 
\includegraphics[width=5.75cm]{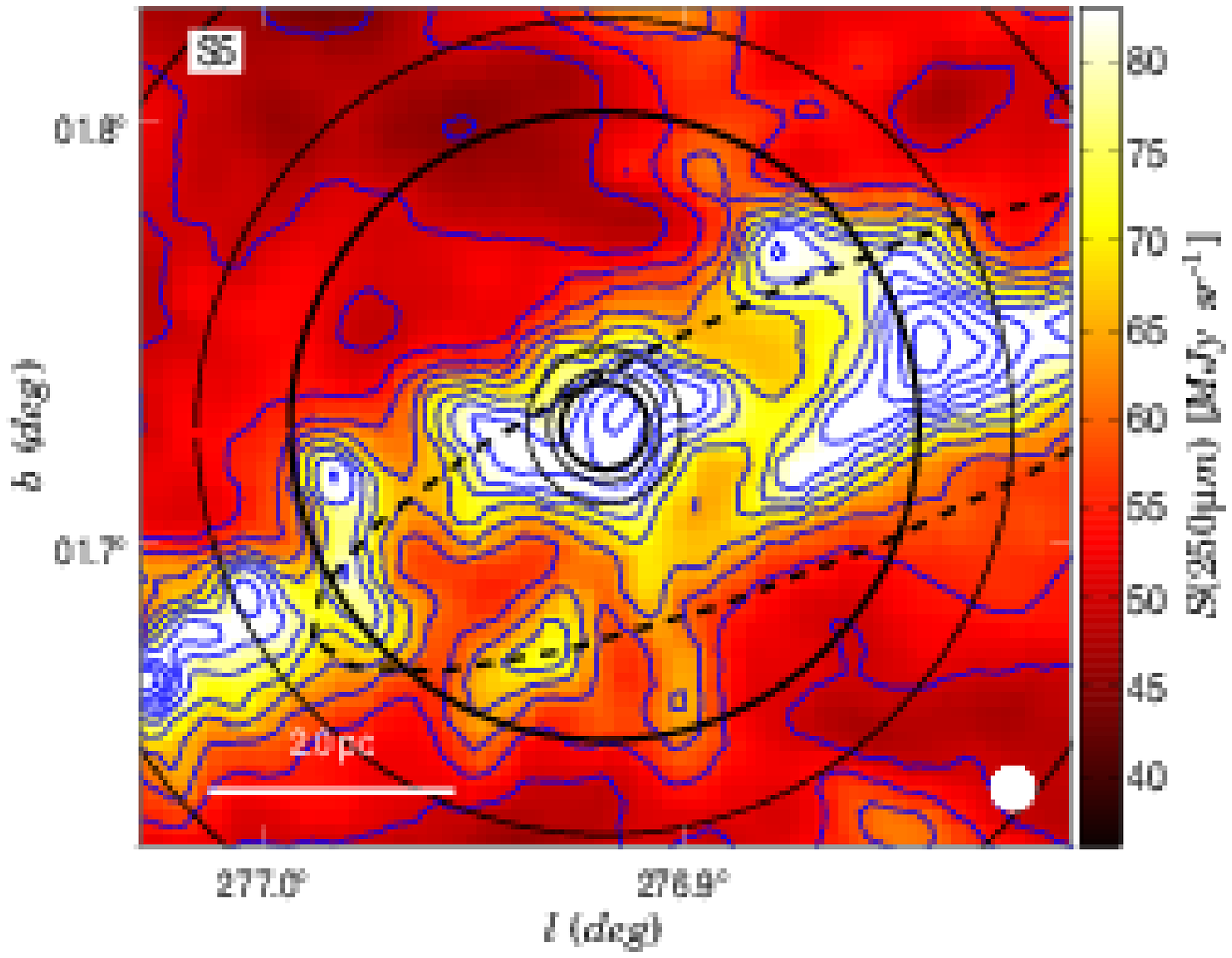}  &
\includegraphics[width=5.75cm]{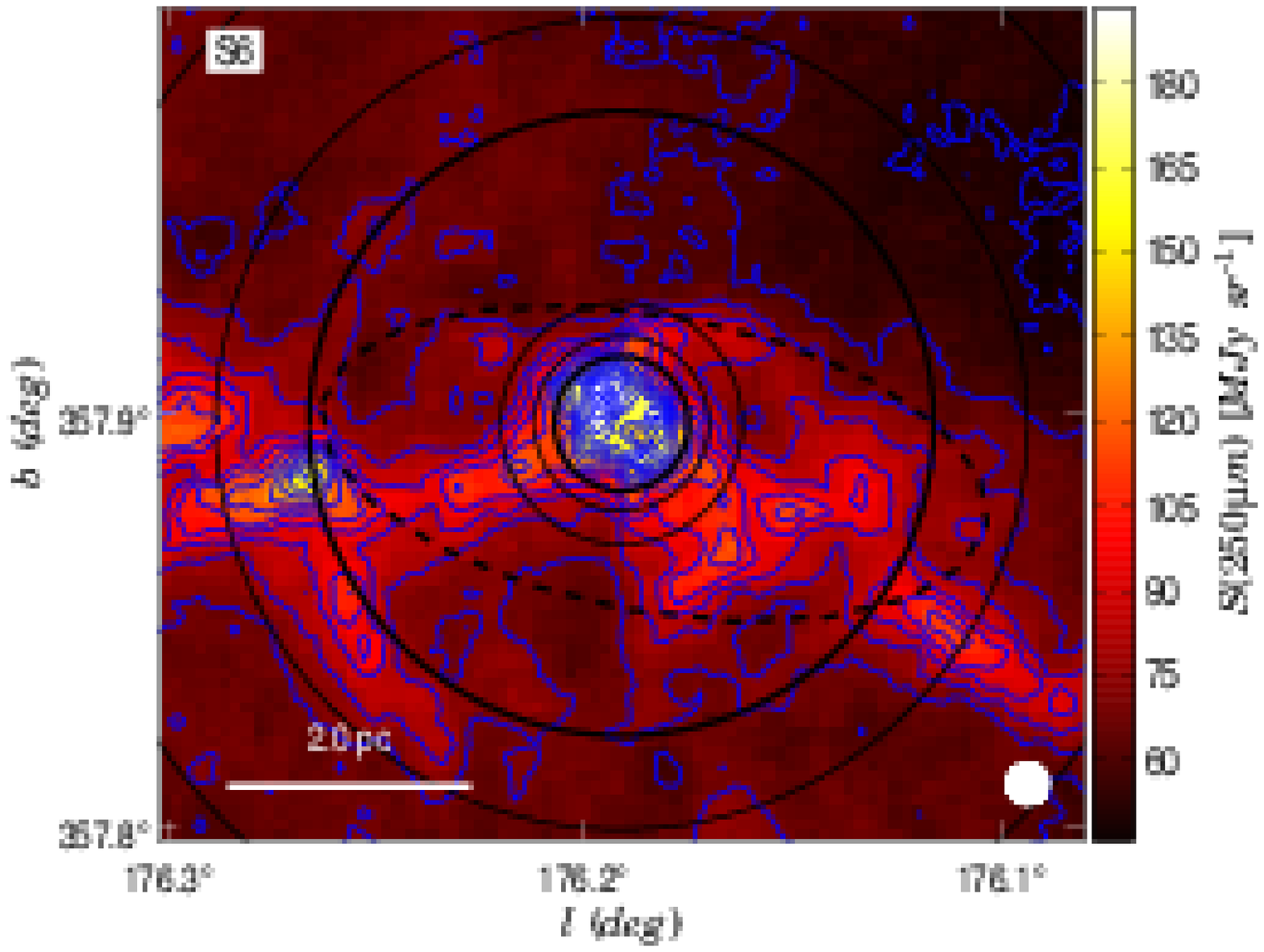} \\

\includegraphics[width=5.75cm]{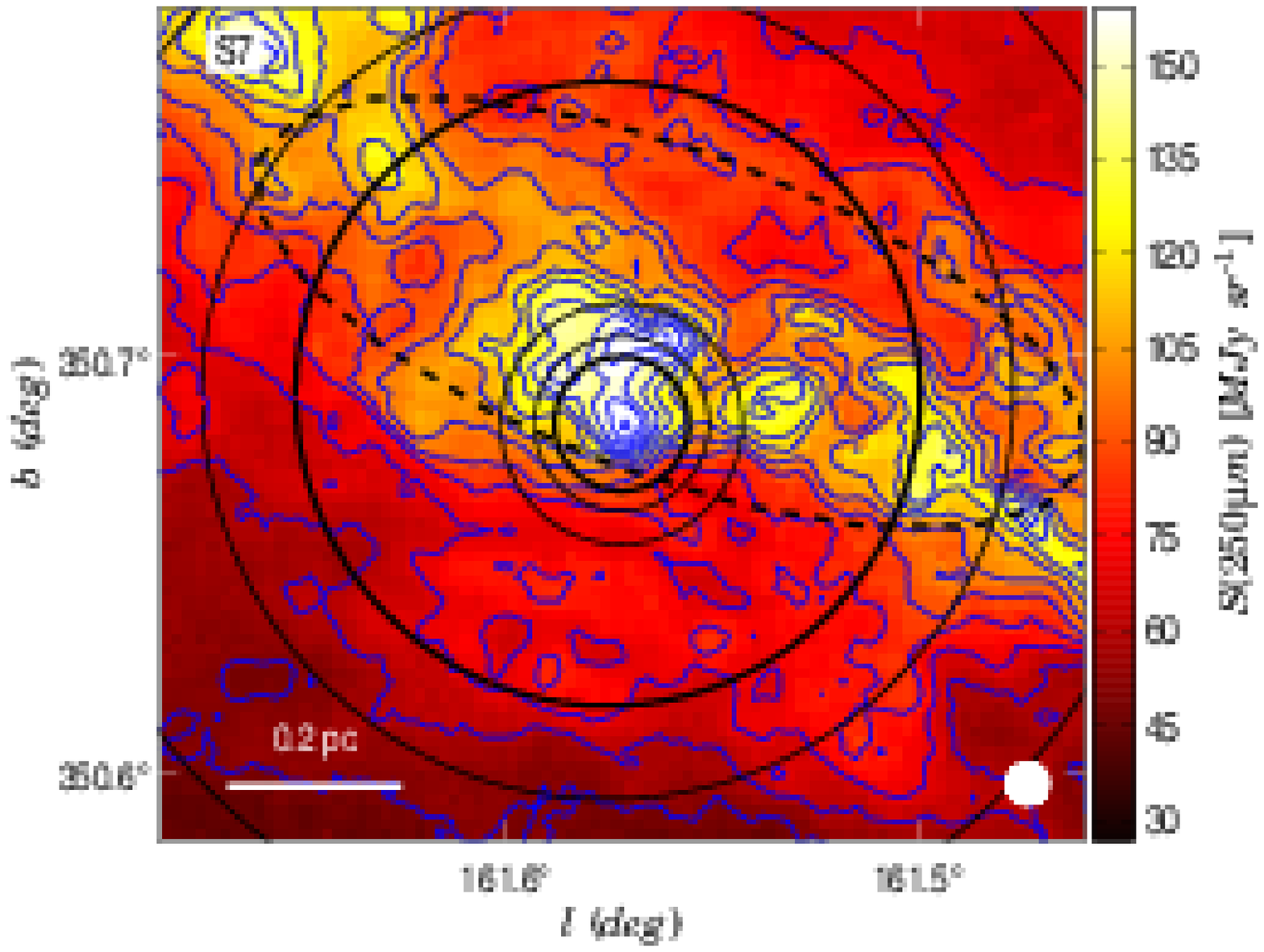}  &
\includegraphics[width=5.75cm]{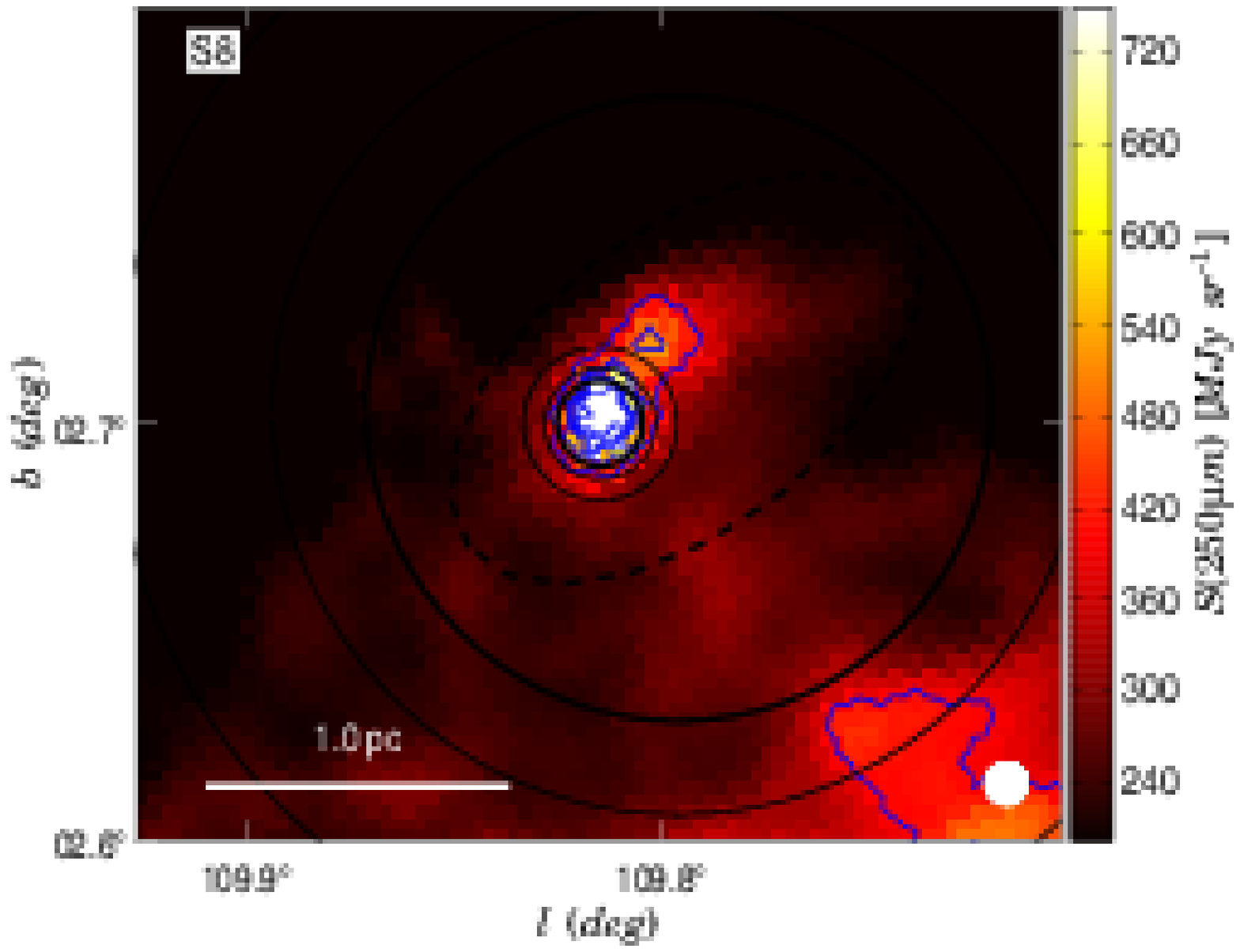} \\
 
\includegraphics[width=5.75cm]{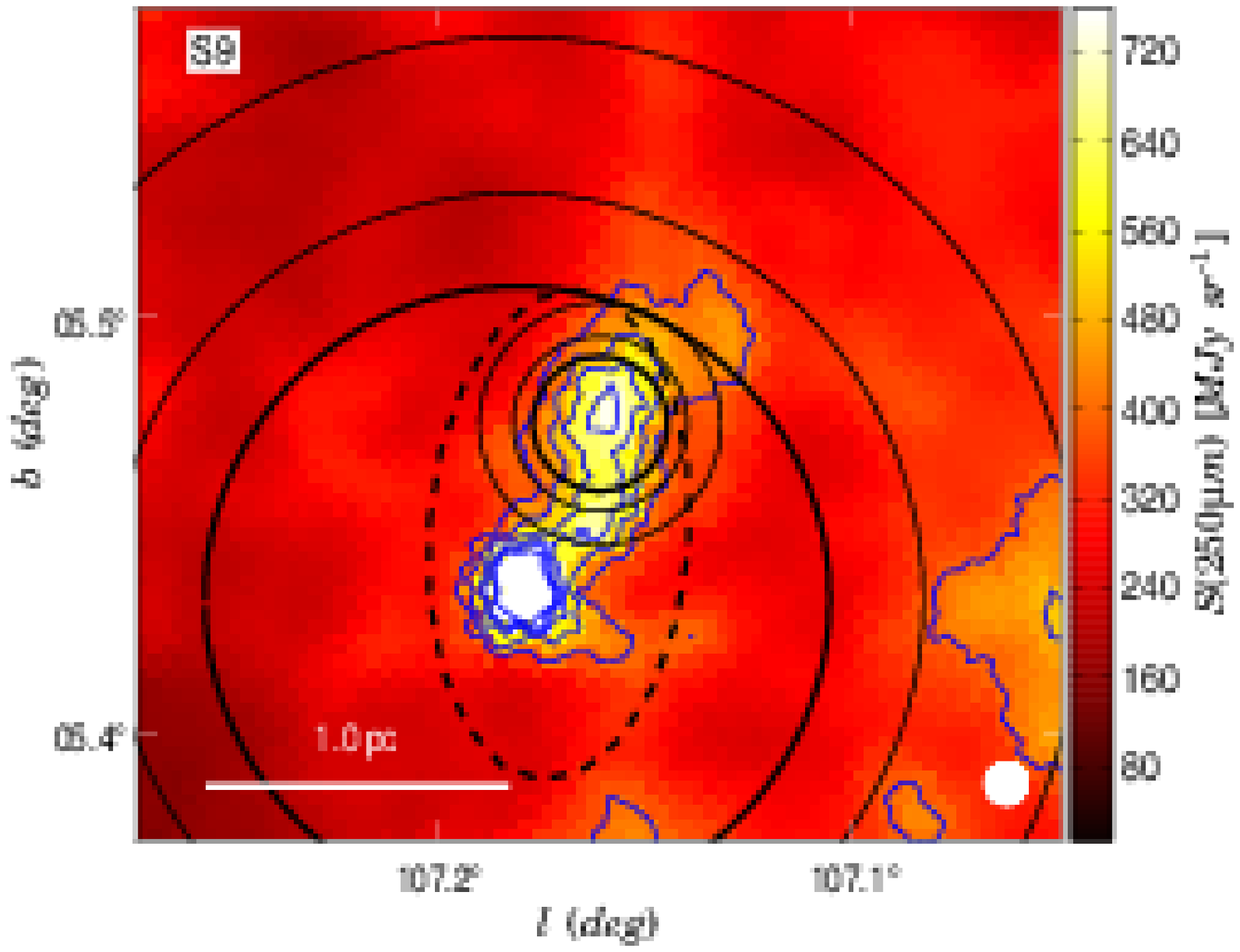}  &
\includegraphics[width=5.75cm]{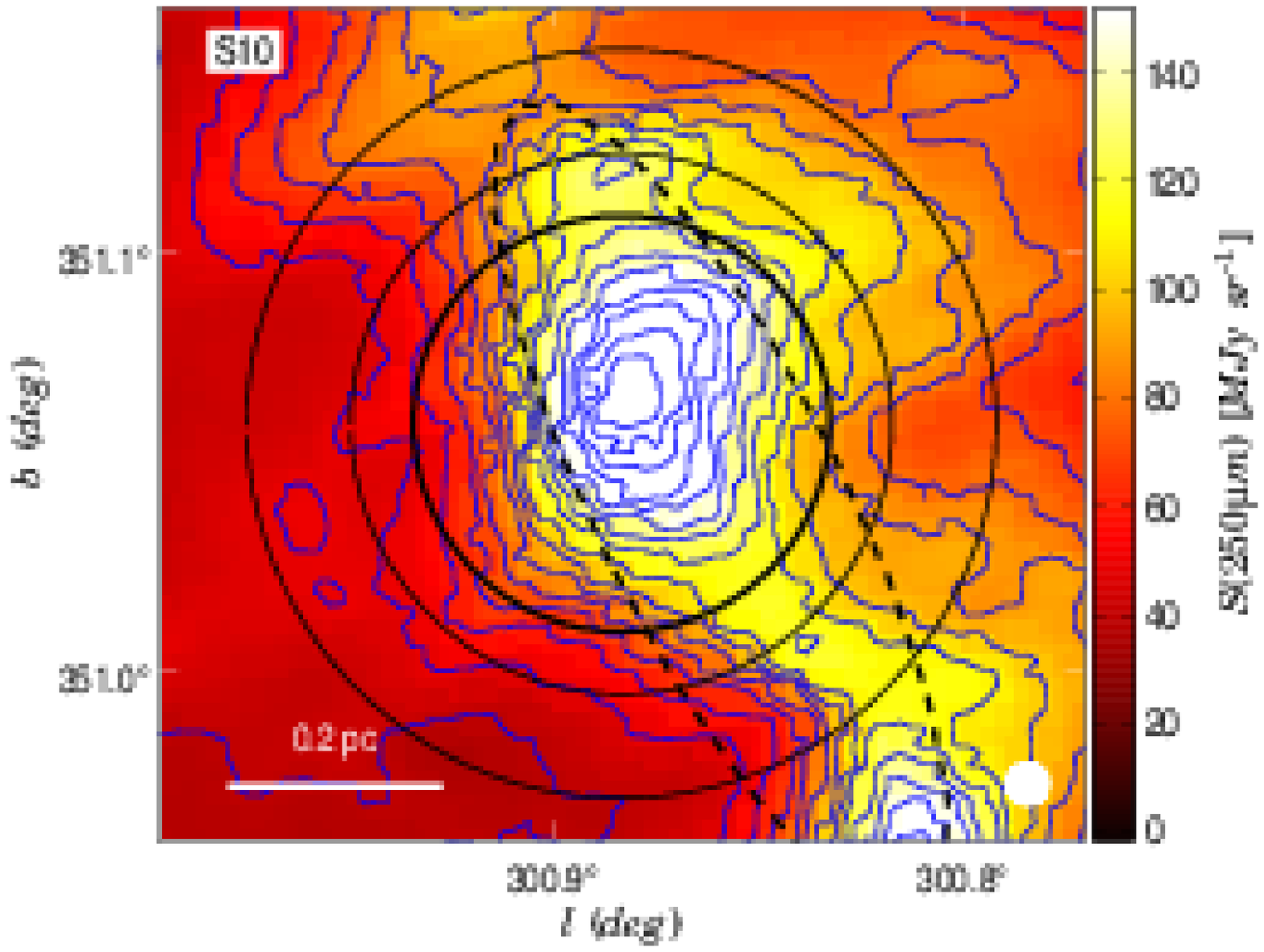} \\

\end{tabular}
\caption{Blow-up of {\it Herschel\/}
SPIRE 250\,$\mu$m maps at the location of the \Planck\ cold clumps 
localised by their elliptic boundary. 
The different circles and the annuli refer to the apertures adopted
for the photometry used in the SEDs.
The large aperture has a $9\arcmin$ diameter, 
and the smaller ones correspond to either 74$\arcsec$ (S3 and S5), 
116$\arcsec$ (S1, S4, S6, S7, S8 and S9), or 360$\arcsec$ (S10).
}
\label{fig:blow-up}
\end{figure*}

Fig.~\ref{fig:Planck-Herschel} shows {\it Herschel\/} 250\,$\mu$m data
(resolution ${\sim}\,18\arcsec$) for all, but one, sources. S2 has
not been observed with {\it Herschel\/} and we show the {\it AKARI\/}
90\,$\mu$m map instead (${\sim}\,39\arcsec$). For comparison, 
the \Planck\ 857 GHz brightness contours are overplotted and ellipses  
show the cold clump sizes as derived 
from elliptical Gaussian fits on the cold
residual maps (cf.\ Paper~I and Section~\ref{sect:Planck_photometry}).
A blow-up of the {\it Herschel\/} 250\,$\mu$m maps 
is shown in Fig.~\ref{fig:blow-up}.

In most fields, a significant amount of structure at scales below the \Planck\
resolution is visible. The Musca field (S10) is an exception. The
{\it Herschel\/} observations there resolve the filament but do not
reveal further structure within the \Planck-detected clump. Interestingly,
this field is among the closest to the Sun.
The other filamentary structures (S4, S5, S6, and S7) break up into
numerous smaller bright knots, the brightest often 
coinciding with the position of the \Planck\ detection.
The two cometary shaped clouds (S1 and S3) similarly harbour a
number of bright smaller knots and narrow filaments that show up 
with a large contrast ($>2$) above
an extended background. 

\subsubsection{Photometry with \Planck, {\it Herschel\/} and {\it AKARI} }

We have performed fits on the SEDs of selected substructures
within the \Planck\ cold clumps to further characterise their inner 
structure. 
The aperture photometry is performed after subtraction of a
background level estimated as
described in Section~\ref{sect:analysis_methods}. The radius of the
largest aperture is 4.5$\arcmin$ (i.e., a diameter of 540$\arcsec$). For
the substructures, we use apertures of either 74$\arcsec$ or
116$\arcsec$ diameter. The locations of these apertures are drawn in
Fig.~\ref{fig:blow-up} and listed in Table~\ref{table:aperture_SED}.
The SED fits involve data at wavelengths longer than 100\,$\mu$m.  This
removes the problem of a possible contribution from transiently heated
grains to the 100\,$\mu$m flux densities, but also reduces the effects of the
line-of-sight temperature variations. 
For the large aperture SEDs, where \Planck\ fluxes are used, 
the 353\,GHz data are corrected for the
CO emission, when available. This correction has 
little effect on the results of the aperture photometry.

\begin{figure}
\center
\begin{tabular}{c}
\includegraphics[width=7cm]{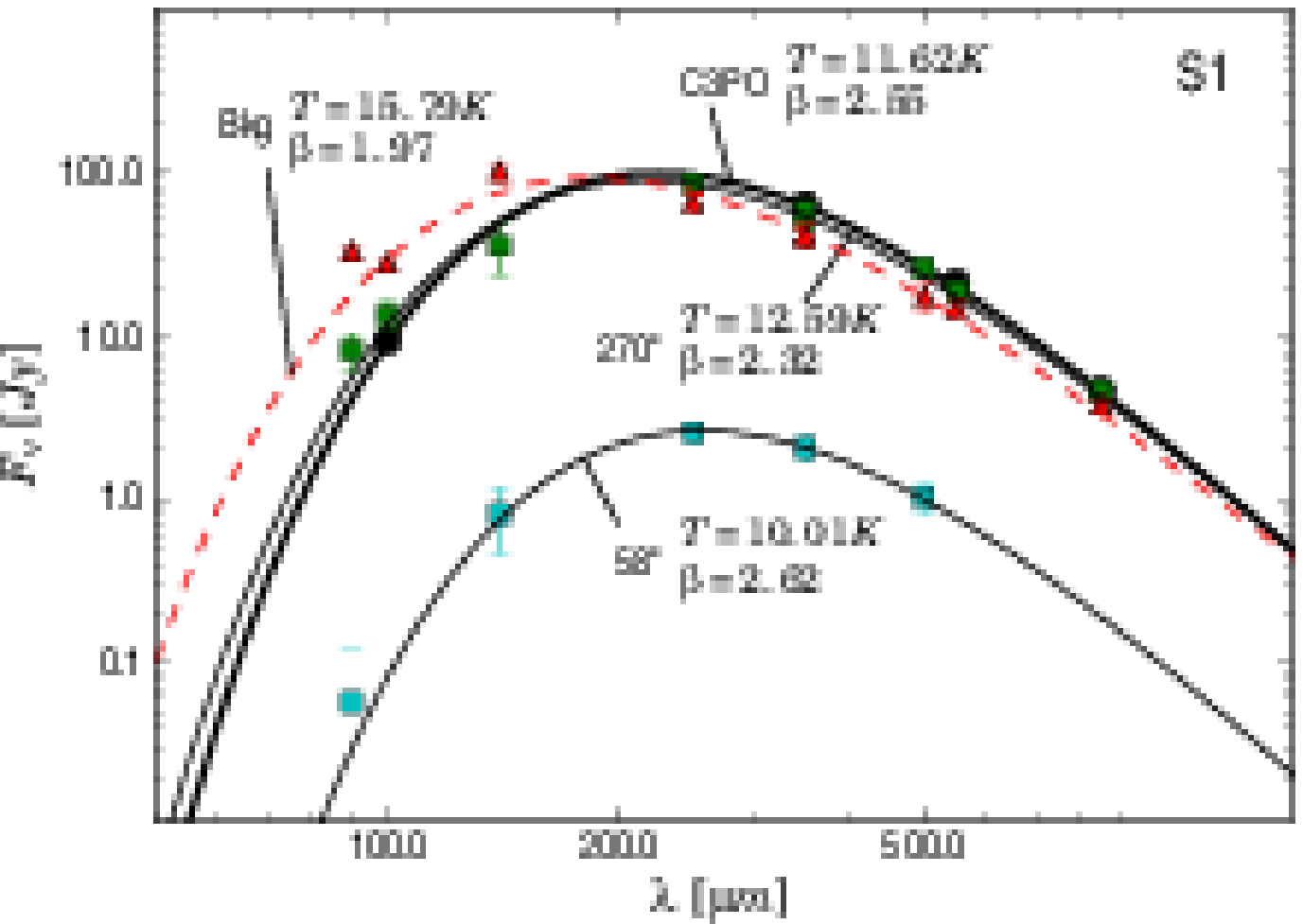} \\

\includegraphics[width=7cm]{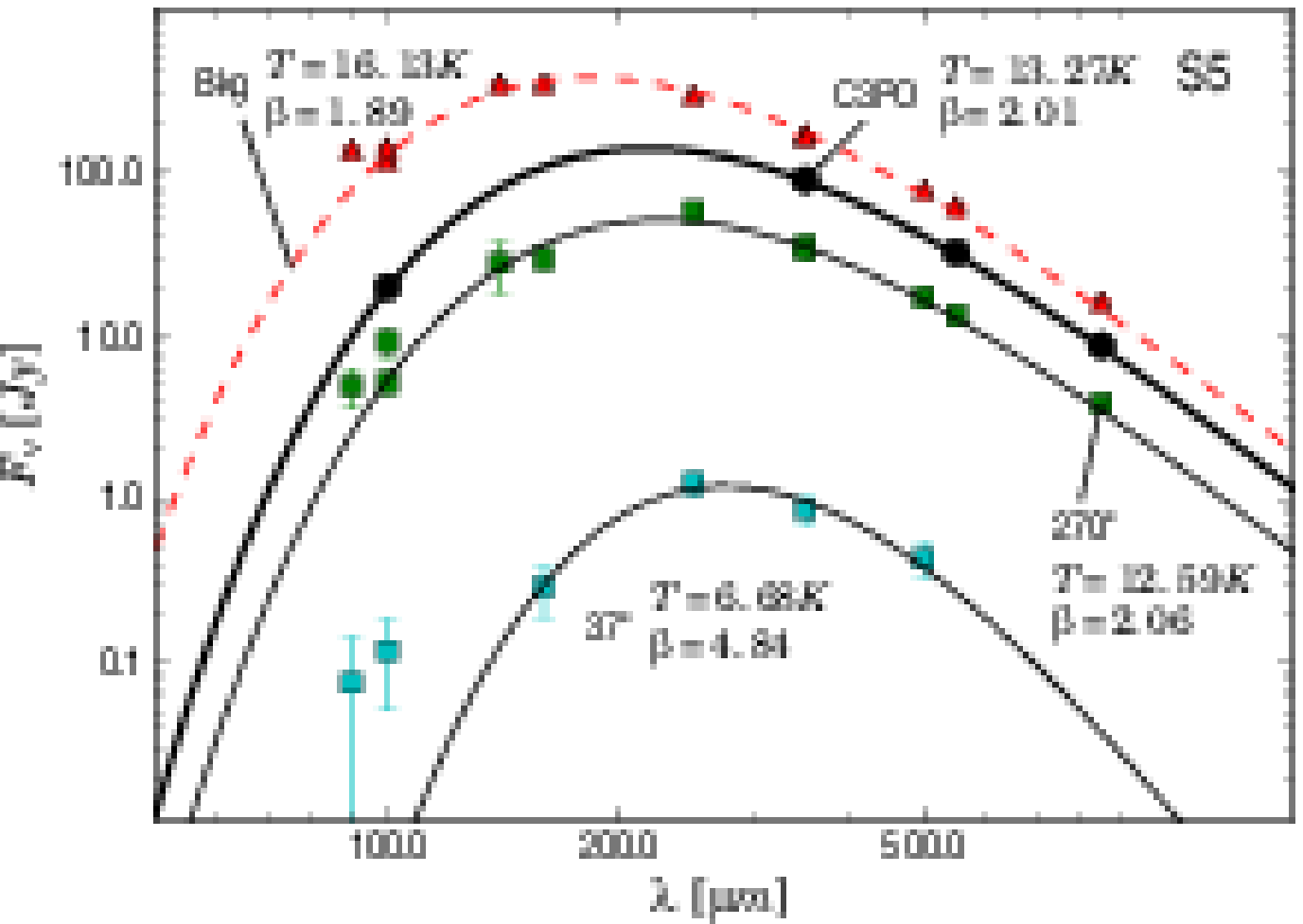} \\

\end{tabular}
\caption[]{
Spectral energy distribution of various components in the S1 (upper panel)
and S5 (lower panel) fields, including: the \Planck\ SED according to the C3PO catalogue (solid circles and thick
solid line), aperture photometry with 270$\arcsec$ and 37$\arcsec$ or 58$\arcsec$ radius aperture sizes (squares and thin
solid lines) and the background SED (triangles and dashed line). The
temperatures and $\beta$ values derived from the modified
blackbody fits are given in the figure (see also Table~\ref{table:aperture_SED}).
The background SED is that of the median surface brightness in the reference
annulus surrounding the larger aperture (see Fig.~\ref{fig:blow-up}). 
}
\label{fig:SED}
\end{figure}

Fig.~\ref{fig:SED} illustrates the various SED estimates for the S1 and S5
fields.  It displays the \Planck\ SED as given in the C3PO catalogue
(therefore corrected for the warm background 
contribution) and the aperture photometry using smaller apertures
on the brightest substructure of the S1 field (Fig.~\ref{fig:blow-up}). 
For the large apertures, the flux densities from \Planck\ and {\it Herschel\/}
are in good agreement with each other; in fact it appears that the
assumed sources of uncertainty may overestimate the errors. 
The temperatures obtained for the fixed apertures
(Table~\ref{table:aperture_SED}) 
are the same within the error bars as those derived from the
\Planck\ and IRIS data only (Table~\ref{table:Phys_param}). 
It is therefore most encouraging that the estimates based {\it only\/}
on \Planck\ and IRIS data do not significantly change when the five
{\it Herschel\/} bands are added that better cover the short wavelength
side of the SED maxima.
The modified blackbody parameters derived from the fits are listed
in Table~\ref{table:aperture_SED}. 

\begin{table*}
\begingroup
\caption{Results from SED fits based on flux densities from aperture
photometry (for $\lambda > $ 100\,$\mu$m data).}
\label{table:aperture_SED}
%\centering
\vspace{-12pt}
\setbox\tablebox=\vbox{
\newdimen\digitwidth
\setbox0=\hbox{\rm 0}
\digitwidth=\wd0
\catcode`*=\active
\def*{\kern\digitwidth}
\newdimen\signwidth
\setbox0=\hbox{+}
\signwidth=\wd0
\catcode`!=\active
\def!{\kern\signwidth}
\halign{#\hfil\tabskip=1.0em&
\hfil#\hfil&
\hfil#\hfil&
\hfil#\hfil&
\hfil#\hfil&
\hfil#\hfil&
\hfil#\hfil&
\hfil#\hfil&
\hfil#\hfil&
\hfil#\hfil\tabskip=0pt\cr
\noalign{\doubleline}
Object &  $l$   &  $b$   & Aperture  & Aperture size & $T_{\rm c}$  &  $\beta$  &
 $N_{\rm{H}_{2}}$  &  $n_{\rm{H}_{2}}$  &  $M$ \cr
&  (deg) &  (deg) & (arcsec)  & (pc)  & (K)  &  
 & (cm$^{-2}$)   &   (cm$^{-3}$)  &  (M$_{\odot}$)  \cr
\noalign{\vskip 4pt\hrule\vskip 3pt}
    S1-1 & 126.61 &  !24.55 & 540 &   0.39 & $12.59\pm0.61$ &  $2.32\pm0.14$ &
 $9.3\times10^{\,20}$ & & \cr
    S1-1 & 126.61 &  !24.55 & 116 &   0.08 & $10.01\pm1.32$ &  $2.62\pm0.54$ &
 $1.9\times10^{\,21}$ & & \cr
    S1-1 & 126.61 &  !24.55 & *74 &   0.05 & $*9.75\pm1.98$ &  $2.67\pm0.92$ &
 $3.0\times10^{\,21}$ &   $2.7\times10^{4}$ &  **0.13 \cr
\noalign{\vskip 4pt\hrule\vskip 3pt}
    S2-1 & 316.56 &  !20.68 & 540 &   1.44 & $15.08\pm0.93$ &  $1.67\pm0.17$ &
 $1.9\times10^{\,20}$ & & \cr
\noalign{\vskip 4pt\hrule\vskip 3pt}
    S3-2 & 131.74 &  !*9.70 & 540 &   0.52 & $11.26\pm0.35$ &  $2.36\pm0.11$ &
 $3.3\times10^{\,21}$ & & \cr
    S3-3 & 131.74 &  !*9.68 & 116 &   0.11 & $*7.09\pm0.74$ &  $3.96\pm0.78$ &
 $5.3\times10^{\,22}$ & & \cr
    S3-4 & 131.73 &  !*9.73 & *74 &   0.07 & $*8.53\pm4.03$ &  $3.10\pm1.03$ &
 $2.0\times10^{\,22}$ &   $1.4\times10^{5}$ &  **1.54 \cr
\noalign{\vskip 4pt\hrule\vskip 3pt}
    S4-1 & 215.44 & $-$16.39 & 540 &  1.18 & $12.23\pm0.49$ &  $2.18\pm0.16$ &
 $1.5\times10^{\,21}$ & & \cr
    S4-1 & 215.44 & $-$16.39 & 116 &  0.25 & $*7.02\pm0.34$ &  $4.40\pm0.42$ &
 $3.9\times10^{\,22}$ & & \cr
    S4-1 & 215.44 & $-$16.39 & *74 &  0.16 & $*7.24\pm0.80$ &  $4.02\pm0.45$ &
 $3.2\times10^{\,22}$ &   $9.8\times10^{4}$ & *12.4* \cr
\noalign{\vskip 4pt\hrule\vskip 3pt}
    S5-4 & 276.92 &   !*1.73 & 540 &  5.24 & $12.59\pm0.28$ &  $2.06\pm0.13$ &
 $6.1\times10^{\,20}$ & & \cr
    S5-4 & 276.92 &   !*1.73 & *74 &  0.72 & $*6.68\pm0.47$ &  $4.84\pm0.11$ &
 $1.7\times10^{\,22}$ &   $1.1\times10^{4}$ & 127.7* \cr
\noalign{\vskip 4pt\hrule\vskip 3pt}
    S6-2 & 176.19 &  *$-$2.10 & 540 & 5.24 & $11.50\pm0.37$ &  $2.31\pm0.11$ &
 $1.6\times10^{\,21}$ & & \cr
    S6-2 & 176.19 &  *$-$2.10 & 116 & 1.12 & $10.78\pm0.77$ &  $1.87\pm0.51$ &
 $7.7\times10^{\,21}$ & & \cr
    S6-2 & 176.19 &  *$-$2.10 & *74 & 0.72 & $10.66\pm0.62$ &  $1.76\pm0.43$ &
 $8.0\times10^{\,21}$ &   $5.4\times10^{3}$ & *60.0* \cr
\noalign{\vskip 4pt\hrule\vskip 3pt}
    S7-2 & 161.57 &  *$-$9.31 & 540 & 0.92 & $13.63\pm0.47$ &  $2.07\pm0.12$ &
 $1.5\times10^{\,21}$ & & \cr
    S7-4 & 161.57 &  *$-$9.32 & 116 & 0.20 & $13.35\pm1.13$ &  $1.48\pm0.41$ &
 $2.4\times10^{\,21}$ & & \cr
    S7-4 & 161.57 &  *$-$9.32 & *74 & 0.13 & $14.21\pm1.12$ &  $1.19\pm0.30$ &
 $1.9\times10^{\,21}$ &   $7.5\times10^{3}$ &  **0.45 \cr
\noalign{\vskip 4pt\hrule\vskip 3pt}
    S8-2 & 109.81 &   !*2.70 &  *74 & 0.29 & $21.02\pm2.81$ &  $1.29\pm0.45$ &
 $5.9\times10^{\,21}$ &   $1.0\times10^{4}$ &  **7.2* \cr
\noalign{\vskip 4pt\hrule\vskip 3pt}
    S9-2 & 107.18 &   !*5.43 &  116 & 0.45 & $19.34\pm2.79$ &  $1.44\pm0.51$ &
 $6.6\times10^{\,21}$ &   $7.1\times10^{3}$ & *19.5* \cr
\noalign{\vskip 4pt\hrule\vskip 3pt}
   S10-1 & 300.88 &  *$-$8.94 & 360 & 0.39 & $11.84\pm0.65$ &  $2.25\pm0.34$ &
 $5.4\times10^{\,21}$ &   $6.7\times10^{3}$ & *12.2* \cr
\noalign{\vskip 4pt\hrule\vskip 3pt}
} }
\endPlancktablewide
\endgroup
\end{table*}

One of the most interesting results is the low  temperatures obtained 
for the SEDs in the 74$\arcsec$ apertures. Except for sources S8 and S9,
they are all significantly smaller than those obtained in the larger
apertures, the lowest values being at about 7\,K.
Therefore, in spite of the large telescope beam size, the sensivity of
\Planck\ is such that the cold clump detection method is actually able
to pick-up and locate (within a few arcmin)
structures of very cold dust that are significantly smaller than the beam.

In sources S8 and S9, the substructures are found to be warmer, with:
$T{\sim}\,19$--21\,K.  As presented in \cite{Juvela2010}, the PACS
$100\,\mu$m maps have revealed the
presence of bright compact sources within the \Planck-detected
clumps.  These objects are likely to be very young stellar objects,
still embedded in a more
extended dense and cold cloud, whose emission is dominating the submillimetre
wavelength range studied with \Planck.  As described in the
Appendix, S8 and S9 are both located close
to active star-forming regions, with S8 having nearby HII regions, and OB
stellar associations.

We note that neither S8 or S9 has any clear counterpart in the $100\,\mu$m
{\it IRAS\/} map (see Fig.~\ref{fig:allnu_sources3}); the emission of the
young stellar objects is not visible, even although this is the same wavelength
as for PACS, probably because the sources are very faint, and diluted with
the other components in the {\it IRAS\/} beam. However, for S8, both the
170\,$\mu$m {\it ISO\/} Serendipity Survey \citep{Stickel2007} and the
{\it AKARI\/} FIS survey at 140\,$\mu$m \citep{Doi2009} showed a faint feature
at this position.  Shorter wavelength data with higher angular resolution
would be interesting to combine for these particular sources.  Among our
sample, the field S8 has been observed by {\it Spitzer}, with the
24$\,\mu$m MIPS data revealing a number of compact sources.
Within the \Planck\
clump, the brightest submillimetre peak (PCC288-A in Juvela et
al. 2010) is seen to contain at least four distinct mid-infrared
sources. These data will be analysed in detail in a future
publication.

We have also estimated the column densities within each aperture, and
the average densities and masses obtained in the smallest aperture.
In most cases, the derived column densities are higher than those
estimated from the \Planck\ clump SEDs.  Moreover, as we found with the
linear mass densities estimated with \Planck\ data, they vary over a factor
of about 30.
These values should probably be considered as lower limits,
because, as shown in
Fig.~\ref{fig:blow-up}, the annulus associated with the small apertures
are close to the source and often include a fraction of signal
associated with the source itself. We have estimated the impact of the
background level on the derived column density by using the
average brightness in the largest annulus instead of the smallest.
The derived column densities are increased slightly, by up to a factor of
2.

\subsubsection{Analysis of molecular line data}\label{sect:gas_properties}

\begin{figure}
%\centering
\includegraphics[width=8cm]{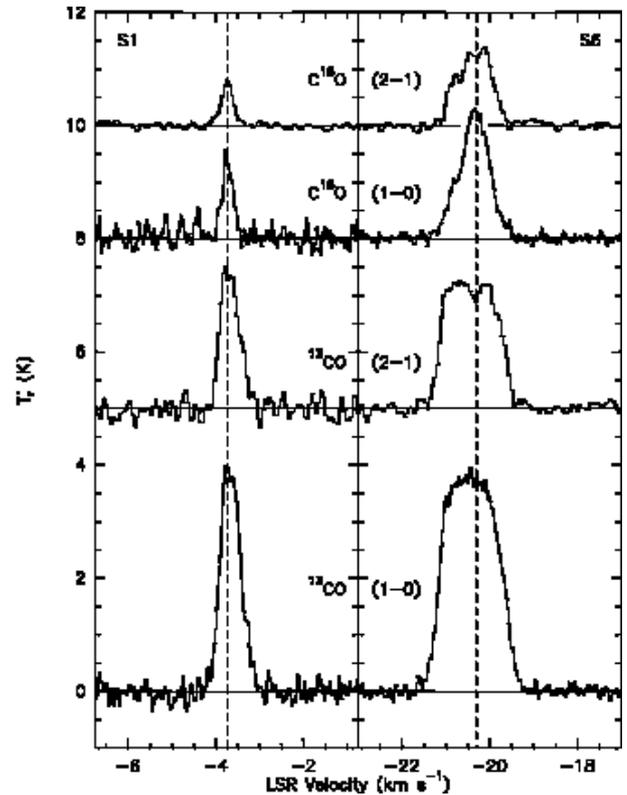}
\caption{
CO spectral lines from the \Planck\ clumps S1 and S6 (taken towards the central positions
given in Table~\ref{table:source_selection}, and with beam apertures in
Table~\ref{30m}. The spectra are in the antenna temperature scale (T$_a^*$).
}
\label{fig:155_CO_spectra}%
\end{figure}

\begin{figure}
%\center
\includegraphics[width=6cm]{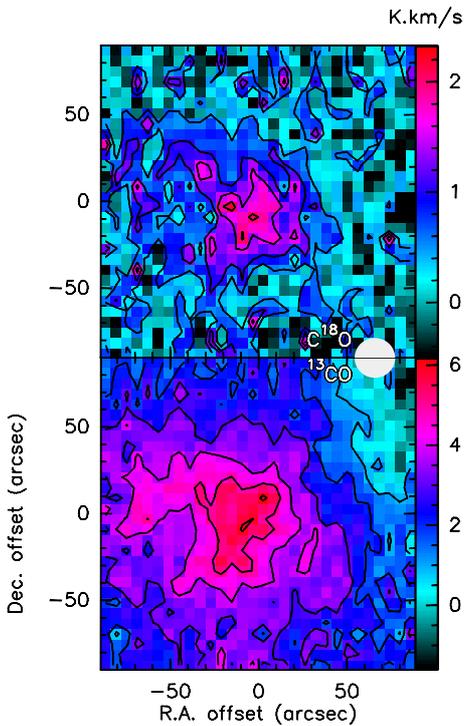}
\caption{
The S6 source spectral line intensity maps for C$^{18}$O $J$=1$\rightarrow$0 (contour steps:0.5\,K\,km\,s$^{-1}$, top image)
and $^{13}$CO $J$=1$\rightarrow$0 (contour steps:1\,K\,km\,s$^{-1}$, bottom image).The white disk indicates
 the 30-m HPBW size at the 2.7 mm wavelength, valid for both maps.
}
\label{fig:6518_spectra}%
\end{figure}

All the \Planck\ clumps observed in $^{13}$CO and C$^{18}$O lines have a clear line detection, 
and the small maps show peaks that coincide with the cold substructures
(Figs.~\ref{fig:155_CO_spectra} and \ref{fig:6518_spectra}). 

The $^{13}$CO linewidths  (Table~\ref{table:line_observations}) 
are typically 1--2\,km\,s$^{-1}$ and those of  the C$^{18}$O lines,
although narrower, range between 0.3 and 1.6\,\kms. 
Several velocity components are present, either seen (in a few \CeiO\ spectra)
as two distinct peaks, or inferred from the non-Gaussian lineshapes, as shown by S6 spectra in Fig.~\ref{fig:155_CO_spectra}.
Assuming that the dust and gas in these cold dense clumps is in thermal
balance, then the gas temperature is close to 10\,K and the sound velocity 
in CO is low,
${\sim}\,0.05$\,\kms. The molecular linewidths are therefore all suprathermal.
It is interesting to compare the level of non-thermal support 
provided by these gas motions to the self-gravity.
This can be done by comparing the gas linear mass density with its
critical (or virial) value, $m_{\rm vir}=2 \sigma^2/G$
which is the largest mass per unit length of 
a self-gravitating cylinder, given the non-thermal support provided by the
internal motions of dispersion $\sigma= \Delta v/ 2.35$ \citep{Fiege2000}.  

The virial linear mass densities inferred from the \CeiO\ 
linewidths of Table~\ref{table:line_observations} range between
10\,\mlin\ for S10, to 
260\,\mlin\ for S8. A crude comparison with the linear mass densities 
estimated at the resolution of the \Planck\ data (Table~2) suggests that
the ratios $m/m_{\rm vir}$ are within a factor of a few of unity.
Therefore, as a first approximation, these cold clumps are
in rough equilibrium between self-gravity and non-thermal support. 
 
The column densities have been estimated from
a single line-of-sight towards each object. When available (in the case of IRAM
observations), we have combined the $J$=1$\rightarrow$0 and 2$\rightarrow$1
transitions to derive the excitation temperature and the density using a
Large Velocity Gradient radiative transfer model \citep{Goldreich1974}. 
In the case of the Onsala and APEX
observations, we only have data for one transition; in
these cases, we assume an excitation temperature of 11\,K
and use the C$^{18}$O line intensity to estimate the column density of
the cores. The molecular hydrogen column densities are calculated assuming a
C$^{18}$O fractional abundance of $[\CeiO]/[\HH]= 10^{-7}$ .

The results are presented in Table~\ref{table:line_observations}. 
The table includes estimates of the
mean density, calculated assuming a core size corresponding to
a diameter of 74$\arcsec$.  The column densities are close  
to the values derived earlier (Table~\ref{table:aperture_SED}) from the dust
continuum data, using the smaller aperture sizes (i.e., 74$\arcsec$ or
116$\arcsec$).  The estimated average densities of
the cores are only of the order of $10^3$\,cm$^{-3}$, which is
lower than the densities found in
nearby dense cores \citep[e.g.,][]{Myers1983}. 
The reason for this is that \CeiO\ is not a good tracer of dense gas
because of the significant degree of CO
depletion onto dust grains. Within a cold core, the depletion could be
almost complete \citep[e.g.,][]{Pineda2010} and so the investigation of this
phenomenon towards the \Planck\ cores is a natural topic for follow-up
studies.

%DS- Give aperture, and any other information which might be helpful for
%interpretting this figure

\begin{table}
\begingroup
\caption{CO line parameters for selected lines-of-sight in seven cold
core fields.}
\label{table:line_observations}  
\vspace{-18pt}
%\centering
\setbox\tablebox=\vbox{
\newdimen\digitwidth
\setbox0=\hbox{\rm 0}
\digitwidth=\wd0
\catcode`*=\active
\def*{\kern\digitwidth}
\newdimen\signwidth
\setbox0=\hbox{+}
\signwidth=\wd0
\catcode`!=\active
\def!{\kern\signwidth}
\halign{#\hfil\tabskip=0.5em&
\hfil#\hfil&
\hfil#\hfil&
\hfil#\hfil&
\hfil#\hfil&
\hfil#\hfil\tabskip=0pt\cr
\noalign{\doubleline}
Target & Telescope & Line &  FWHM & ${n_{\rm H_2}}^{\rm a}$ &
 ${N_{\rm{H}_{2}}}^{\rm b}$\cr
 &           &      &  (km\,s$^{-1}$) & (cm$^{-3}$)  &  (cm$^{-2}$)\cr
\noalign{\vskip 4pt\hrule\vskip 3pt}
S1  & IRAM    & $^{12}$CO $J$=1$\rightarrow$0 & $0.71\pm0.04$ \cr
    & IRAM    & $^{12}$CO $J$=2$\rightarrow$1 & $0.78\pm0.07$ \cr
    & IRAM    & $^{13}$CO $J$=1$\rightarrow$0 & $0.53\pm0.01$ &
 3000*&$3.0\times10^{\,21}$\cr
    & IRAM    & $^{13}$CO $J$=2$\rightarrow$1 & $0.50\pm0.02$ \cr
    & IRAM    & C$^{18}$O $J$=1$\rightarrow$0 & $0.30\pm0.02$ &
 2700*&$4.0\times10^{\,21}$\cr
    & IRAM    & C$^{18}$O $J$=2$\rightarrow$1 & $0.32\pm0.01$ \cr
S3  & Onsala  & $^{12}$CO $J$=1$\rightarrow$0 & $2.64\pm0.03$ \cr
    & Onsala  & $^{13}$CO $J$=1$\rightarrow$0 & $1.71\pm0.04$ \cr
    & Onsala  & C$^{18}$O $J$=1$\rightarrow$0 & $1.22\pm0.08$ &
  $19000$  & $6.7\times10^{\,21}$   \cr
S6  & IRAM    & $^{13}$CO $J$=1$\rightarrow$0 & $1.44\pm0.02$ \cr
    & IRAM    & $^{13}$CO $J$=2$\rightarrow$1 & $1.42\pm0.02$ \cr
    & IRAM    & C$^{18}$O $J$=1$\rightarrow$0 & $0.82\pm0.01$ &
  5000*   &  $1.4\times10^{\,22}$ \cr
    & IRAM    & C$^{18}$O $J$=2$\rightarrow$1 & $0.91\pm0.01$ \cr
%V41_8678
S7$^{\rm c}$ & IRAM & $^{13}$CO $J$=1$\rightarrow$0 & $1.60\pm0.10$ \cr
    & IRAM    & $^{13}$CO $J$=2$\rightarrow$1 & $1.70\pm0.10$ \cr
    & IRAM    & C$^{18}$O $J$=1$\rightarrow$0 & $1.40\pm0.20$ &
  7300*  &  $5.6\times10^{\,21}$  \cr
    & IRAM    & C$^{18}$O $J$=2$\rightarrow$1 & $1.60\pm0.20$ \cr
%% V41-8622 = V31-7535
S8  & Onsala  & $^{13}$CO $J$=1$\rightarrow$0 & $2.06\pm0.07$ \cr
    & Onsala  & C$^{18}$O $J$=1$\rightarrow$0 & $1.82\pm1.07$ &
 $6400*$ & $1.3\times10^{\,22}$ \cr
%V41_6518
S9A & Onsala  & $^{13}$CO $J$=1$\rightarrow$0 & $2.54\pm0.04$ \cr
    & Onsala  & C$^{18}$O $J$=1$\rightarrow$0 & $1.57\pm0.12$ &
  $25000$  &  $1.8\times10^{\,22}$   \cr
S9B & Onsala  & $^{13}$CO $J$=1$\rightarrow$0 & $1.95\pm0.05$ \cr
    & Onsala  & C$^{18}$O $J$=1$\rightarrow$0 & $1.41\pm0.11$ & 
  $23000$ & $1.6\times10^{\,22}$ \cr
%PCC288: map1
S10 & APEX    & $^{12}$CO $J$=2$\rightarrow$1 & $1.39\pm0.02$ \cr
    & APEX    & $^{13}$CO $J$=2$\rightarrow$1 & $0.82\pm0.03$ \cr
    & APEX    & C$^{18}$O $J$=2$\rightarrow$1 & $0.35\pm0.20$ &
  $23000$ & $1.8\times10^{\,22}$  \cr
\noalign{\vskip 4pt\hrule\vskip 3pt}
} }
\endPlancktable
\tablenote a Average density along the line of sight.\par
\tablenote b Gas column density converted to H$_2$ column density 
using the following conversion factors:
$N_{\rm{H}_{2}}$ = 10$^{6} \times N(\thCO)$;
and $N_{\rm{H}_{2}}$= 10$^{7}\times N(\CeiO)$.\par
\tablenote c For this source, the lines are an assembly of three 
components, with strongly asymmetric shapes, which are difficult to analyse
separately.\par
\endgroup
\end{table}

\section{Discussion}\label{sect:discussion}

\subsection{The validity of the cold core detection method}
A perhaps unexpected result is the physical size of the \Planck\ detections.
They may have been anticipated to be compact point sources, the result of 
small cores being diluted within the large beam of
\Planck. However, they are instead found to be significantly
extended and elongated, and embedded in filamentary 
(or cometary) larger-scale structures.  As discussed in previous studies, the
boundary of a core/clump/cloud structure is always difficult to
assess. \cite{Curtis2010} for instance have shown how
much it depends on the method used to identify and extract the core
parameters.

In our case, the detection method used to extract sources from the \Planck\
data is based on the colour signature of the objects, designed to enable us
to detect the cold residuals after the removal of the warmer background.
This results in the discovery of a different, more extended 
cold component with a more complex morphology
than sources found with methods that identify
structures on the basis of surface brightness (e.g., the ``clumpfind''
algorithm of \cite{Williams1994}, or a multi-scale wavelet analysis).
The high sensitivity of \Planck\ helps to better separate the warm
and cold components, particularly in combination with data at somewhat
shorter wavelengths.

The faint, mostly filamentary, emission of cold dust that we detect here
could perhaps be called a ``cold matrix'' linking the
substructures to each other over a broad range of scales.
By combining \Planck\ and {\it Herschel\/} it is possible to probe in detail
the link between 10-pc scale filamentary structures
and sub-parsec scale cold cores. 
Future studies with much larger samples should bring new insights to the
origin of the cores and the mechanisms associated with their formation.

\subsection{The limitations of the SED fitting }

One has to be careful not to over-interpret SED fits based on
a modified blackbody function, with a single
colour temperature and optical depth.  In reality the ISM has a
broad distribution of dust temperatures
and opacities (as well as column densities), within a \Planck\ beam.  
\cite{Shetty2009a} have discussed in detail the biases introduced by
line-of-sight temperature variations and noise in the interpretation  
of such SED fits. Our study suggests that we are not badly contaminated by
these effects for several reasons.  Firstly Planck's
broad spectral coverage, combined with IRIS allows us to
determining both the dust temperature and emissivity spectral index.
And secondly, we stress that
the cold core detection method identifies structures where the submillimetre
SED is dominated by localised dust emission colder than its large scale background.

An indirect indication that the parameters derived from the SEDs fits
are not strongly biased is that 
the physical properties inferred cover the range of values known to 
be those of pre-stellar cores.
Such cores have different properties whether they belong to regions of
low- or high-mass star formation.
Their column densities range between a few times
$10^{21}$\, or $10^{22}$\,\cq\ in nearby star forming regions
 \citep{Motte1998, Kauffmann2008, Enoch2006, Enoch2007,
 Hatchell2005, Curtis2010, Andre2010} to ${\sim}\,10^{23}$\,\cq\ in the cores of 
infrared dark clouds \citep[IRDCs,][]{Simon2006b,Rathborne2010, Peretto2010}.
These IRDCs,
discovered by means of mid-infrared absorption towards the bright
background emission of the Galactic Plane \citep[{\it MSX\/} and ISOGAL
surveys,][]{Egan1998, Perault1996}, are thought to be the sites of formation of
massive stars and star clusters \citep{Rathborne2006}.  
The inferred average densities depend on their size,
which ranges between $<0.1\,$pc for low-mass dense cores of ${\sim}\,1\,$\msol, 
to 0.5\,pc or more for $\sim 10^3$\,\msol\ IRDC cores. 
Our \Planck\ cold clumps sub-sample is not located within the {\it MSX\/} IRDCs spatial distribution, so we 
cannot compare them directly with any IRDC association.
However, the properties we derive fall well within this range. 
It is noteworthy that the \Planck\ cold clumps are elongated structures which
tend to exist within extended filaments
(up to 30\,pc in the present subsample), with thickness up to ${\sim}\,3\,$pc.
The IRDCs identified as extinction peaks in mid-infrared maps have similar
lengthscales and column densities,
but they are usually thinner and therefore have higher densities.    
It is also interesting that the linear mass densities of the filaments in which
the \Planck\ cold clumps are embedded cover the same range as the warm
filamentary structures associated with 
active star-forming regions, i.e., extending up to several 100\, \mlin.
  
The temperatures derived in the sample studied here cover a range (10--14.5\,K)
similar to previous estimates on a number of cold condensations detected using 
multi-wavelength submillimetre observations with the
balloon-borne experiments: PRONAOS, $T\,{\sim}\,12\,$K in
star-forming and cirrus regions \citep{Stepnik2003, Dupac2003,
Bernard1999}; ARCHEOPS, 7--18\,K \citep{Desert2008};
and BLAST, 9--14\,K \citep{Netterfield2009}.
These temperatures are all lower, however, than those measured in the
so-called ``quiescent'' IRDC cores \citep{Rathborne2010} that span
from 17 to 30\,K for a similar range in column density
(0.3--3$\times 10^{22}$\,\cq) and mass (10--$10^3$\,\msol). 
The \Planck\ cold clump population may therefore be representative of
a still earlier stage of evolution of cold dense cores.

\section{Summary and perspectives }\label{sect:conclusion}

We have presented a preliminary analysis of a sample of 10 sources 
from the C3PO catalogue in order to illustrate and better probe 
the nature and properties of the cold objects detected with \Planck.
The sources have been chosen to span a broad range in 
temperature, density, mass and morphology
(inluding filaments and isolated/clustered structures) 
in a variety of environments, from star-forming regions (both remote and
nearby), to high Galactic latitude cirrus clouds. 
The main findings are as follows: 

\begin{enumerate}

\item The sources are significantly larger than the
  \Planck\ beam, with elongated shapes, and appear to
  belong to filamentary structures, with lengths up to 20\,pc;

\item The physical parameters of the sources have been derived from
  SEDs by combining \Planck\ (HFI bands at 857, 545, and 353\,GHz) and IRIS
  data, finding $T\,{\sim}\,10$--15\,K (with a mean value of 12.4\,K),
  $\beta\,{\sim}\,1.8$--2.5 (mean 2.2) and $N_{\rm{H}_{2}}\,{\sim}\,0.8$--$16
  \times10^{21}\,{\rm cm}^{-2}$,
  from which we infer linear mass densities in the range $m=15$--400\,\mlin,
  masses $M\,{\sim}\,3.5$--$1800\,{\rm M}_{\odot}$, bolometric luminosities
  $L\,{\sim}\,1$--300\,\lsol, and $L/M\,{\sim}\,0.1$--0.9\,\lsol/\msol;

\item Except for the faintest source (which lies
  at high Galactic latitude), a clear signature of cold dust emission is
  visible directly in the $1^\circ \times 1^\circ$  maps of dust temperature,
  spectral index and column density, with colour temperatures of typically
  ${\sim}\,14\,$K, surrounded by a warmer extended (often elongated) emission
  at around 16--18\,K;

\item {\it Herschel\/} and {\it AKARI\/} observations at
  higher angular resolution have revealed a rich and
  complex substructure within the \Planck\ clumps, in most cases the
  substructures being colder (down to 7\,K) than the
  \Planck-detected clumps, although in two cases, the substructures are warmer 
  because they harbour compact infrared objects, likely
  protostellar sources at an  early stage;

\item Molecular line observations of 7 of the sources show that 
  all of them are clearly detected
  in $^{13}$CO and C$^{18}$O, $^{13}$CO linewidths
  typically 1--2\,km\,s$^{-1}$, and C$^{18}$O lines always
  narrower (down to ${\sim}\,0.3\,{\rm km}\,{\rm s}^{-1}$) but still clearly
  suprathermal, given the anticipated low temperature of the gas, suggesting
  that the support of \Planck\ cold clumps against self-gravity is dominated
  by non-thermal motions.

\end{enumerate}

Although we have focussed here on a very small sample of 10 clumps, the
results are indicative of what might be expected from the more ambitious
studies which will follow.  We have already shown
that the C3PO list of \Planck\ cold sources contains objects with a wide
variety of physical properties.  These are probably associated with different
evolutionary stages of the star formation process, from quiescent, cold and
starless clumps, through prestellar stages to very young
protostellar objects still embedded in their cold surrounding cloud.
Forthcoming papers will present a more detailed analysis of the
sources using {\it Herschel\/} data combined with \Planck\ for larger and more
stattistically robust samples.  More detailed physical modelling will allow
for characterisation of
clustering and fragmentation within the \Planck\ clumps, addressing the
question of their evolutionary stage, along with the study of stability of
the starless substructures.

The first unbiased all-sky catalogue of cold objects provided by
\Planck\ offers the opportunity to investigate the properties of
the population of Galactic cold objects over the entire sky.  The full
sample will include objects at the very early stages of evolution,
in a variety of large-scale environments, and in particular, outside the
well-known molecular complexes, at high latitude or at large distances
within the Galactic Plane.  For this purpose, dedicated follow-up observations
are needed in both higher resolution continuum mode and spectroscopy,
which is the objective of the ''Galactic
Cold Cores'' key programme, planning a follow-up with PACS and SPIRE of
about 150 \Planck\ cold clumps. In parallel, similar and/or complementary
follow-up studies will be possible on the basis of the Early Cold Core
catalogue which has been delivered to the astronomical community,
providing a robust sub-sample of
the C3PO catalogue, with more than 900 \Planck\ cold clumps
distributed over the whole sky.

\bibliographystyle{aa}
%\bibliography{biblio_v2.1,planck_bib} 
\bibliography{biblio_v4}

\appendix

\section{Selected sources, illustration of detection method, source
elongation and observational parameters}

%\section{The selected sources} \label{sect:selected_sources}

We list below the main features of the selected fields as known based on
previous studies.

{\bf S1} lies in a tenuous high-latitude cloud, MCLD 126.5+24.5,
located at the border of the Polaris Flare, a large molecular cirrus cloud 
in the direction of the north celestial pole \citep{Heithausen1990},
at an estimated distance of 150\,pc \citep{Bensch2003}. The cloud 
was first noted from the POSS prints by \cite{Lynds1965} as the small
reflection nebula LBN 628. 
Its cometary globule shape appears similar to what is usually found in active
star formation regions, although this nebula is far from any such region.
\cite{Boden1993} have studied the gas properties of the densest part of the
cloud, observing  
transition lines of the dense tracors $^{12}$CO, $^{13}$CO, H$_{2}$CO, and 
NH$_{3}$. The ammonia core (${\sim}\,0.14\,{\rm pc} \times 0.07\,$pc)
corresponds to $0.2{\rm M}_\odot$ and 
a density of $4000$cm$^{-3}$. Although this density is high compared to the
cirrus-like environment, \cite{Boden1993} find that both the cloud and the
core are probably gravitationally unbound.  
The line ratios found are consistent with shocked gas being compressed by the
North Celestial Pole \ion{H}{i} loop \citep{Kramer1998}. 

{\bf S2} is embedded in the tail (${\sim}\,30^\prime$ south) of the
well-studied cometary globule CG12. 
This isolated globule is located at high Galactic latitude
($l=316.5^\circ, b=21.2^\circ$), with a $1^\circ$ long nebular tail
nearly perpendicular to the Galactic Plane. 
Its distance of $550\,$pc has been determined using a photometry-extinction
method by \cite{Maheswar2004}. Molecular studies using
$^{12}$CO  \citep{vanTill1975, White1993, Yonekura1999},
C$^{18}$O \citep{Haikala2006}, H$_{2}$CO \citep{Goss1980}
and NH$_{3}$\citep{Bourke1995} reveal star formation activity, with the 
presence of a highly collimated bipolar outflow in the head of the cloud.
CG12 seems to be a rare example of triggered star formation at relatively
large Galactic height \citep{Maheswar2004}.

{\bf S3} is located in the constellation of Cassiopeia and has no
counterpart in the SIMBAD database.
On the sky, the closest Lynds catalogue sources are
LDN~1358 and LDN~1355, which are, respectively, at distances of 114 and
$118\arcmin$. These Lynds sources are associated with the Cepheus
flare at a distance of 200$\pm$50\,pc \citep{Obayashi1998, Kauffmann2008}. The
\Planck\ detection is associated with an {\it AKARI\/} FIS bright source
catalogue source that has a cirrus type spectrum \citep{Yamamura2008}.

{\bf S4} is in the Orion region, within a dark cloud that was
mapped by \cite{Dobashi2005} in extinction,
derived from Digitized Sky Survey images. The source is located six
degrees east of the M42 nebula, in the Dobashi et al.\ cloud 1490,
close to its clump P43. It has not been the subject of any dedicated studies,
so far. In higher resolution extinction images derived
from the colour excess of 2MASS stars, cloud 9711 is seen to
reside within a narrow filament. In the \Planck\ data, the filament is
clearly visible, but remains unresolved. The source itself is a
compact, slightly elongated clump that stands out just as well in the
individual \Planck\ frequency maps as in the cold residual map. The
Nanten CO data show a peak at the same position, with a line intensity
close to 20\,K\,km\,s$^{-1}$. 

{\bf S5} is a filamentary structure located near the Vela Molecular
Ridge (hereafter VMR), a giant molecular cloud complex in the outer
Galaxy. \cite{Murphy1991} mapped the VMR in CO, but the
mapping did not cover S5. However, \cite{Otrupcek2000} carried out
CO observations of the dark cloud DCld 276.9+01.7, with angular extent of
$16\arcmin$ by $3\arcmin$, and this is interposed on the filament.
The detected CO line is rather strong ($T_{\rm A}^{*}=6.1\,$K), with a
velocity (relative to the local standard of rest) of
${\sim}\,2\,{\rm km}\,{\rm s}^{-1}$. 
We derive a kinematic distance for the cloud of ${\sim}\,2\,$kpc.   
This is consistent with the distance estimate of the VMR: \cite{Liseau1992}
give a photometric distance of ($0.7\pm 0.2$)\,kpc for parts A, C and D of the
VMR and 2\,kpc for part B, using the nomenclature established by 
\cite{Murphy1991} and agreeing with their results. 
Thus we consider it is likely that \Planck-4110 is part of the VMR.
Several studies have been published 
on the VMR young star-forming region, for example, 
\cite{Olmi2009} found 141 BLAST cores (starless and proto-stellar) in
the Vela-D region.

{\bf S6} has been detected in the anti-centre direction ($l$, $b$=176.18,
$b=-2.11$), at a distance estimated to 2\,kpc using the
extinction signature method by \cite{Marshall2006}. 
The filament-shaped core seen with \Planck\ does not coincide with any known
object, but is likely associated with the Perseus arm. 

{\bf S7} has been observed by \cite{Ungerechts1987} as part of their large
CO survey of Perseus, Taurus and Auriga. It is included in their 12th area
(see their Table~1), covering $41.8\,{\rm deg}^2$, with a 
distance of 350\,pc. For this large region, they estimate a virial mass that
is four times higher than the CO mass, showing that this region is out of
equilibrium.  The \Planck\ source is not associated with any known objects
in the SIMBAD database..

{\bf S8, S9} and {\bf S10} correspond with
{\it Herschel\/} SDP sources PCC288, PCC550 and PCC249, respectively.
These fields were selected in September 2009 from the cold core detections in
the First Light Survey of \Planck, in order to perform the first follow-up
observations of our {\it Herschel\/} key-program during the 
{\it Herschel\/} SDP.  The fields were mapped with both PACS and SPIRE
(with map sizes from $18^\prime$ to $50^\prime$), and 
the observations have been described in detail in dedicated papers
\citep{Juvela2010,Juvela2011}.

Field S8 is located in Cepheus, at the interface between the Cepheus F
molecular cloud \citep{Sargent1977} and the young stellar group Cep OB3b
\citep{Jordi1996}, which is one of the youngest nearby 
stellar groups. It has been suggestsed that this region could show
direct evidence for star formation triggered by the OB association. 
S9 is an active star formation region close the S140 \ion{H}{ii}
region. \Planck\ has provided two detections, PCC288-P1 and PCC288-P2
\citep[see][]{Juvela2010}, with the southern source PCC288-P2
being the colder one. The two cores have also been identified in
CS $J$=1$\rightarrow$0 mapping of the area by \cite{Tafalla1993}. 
The field S10 is part of the Musca cloud filament, where \Planck\ shows
two secure detections of cold cores ($T\,{\sim}\,11\,$K).
The cores have been studied previously in $^{13}$CO 
and C$^{18}$O by \cite{VilasBoas1994}.

%\section{HFI maps of the sources} \label{sect:HFImaps}

\begin{figure*}
\center
\begin{tabular}{ccccc}

Band & {\tiny S1\qquad\quad} & {\tiny S2\qquad\quad} & {\tiny S3\qquad\quad} &
 {\tiny S4\qquad\quad}  \\
 $100\,\mu$m &
 \includegraphics[width=4cm]{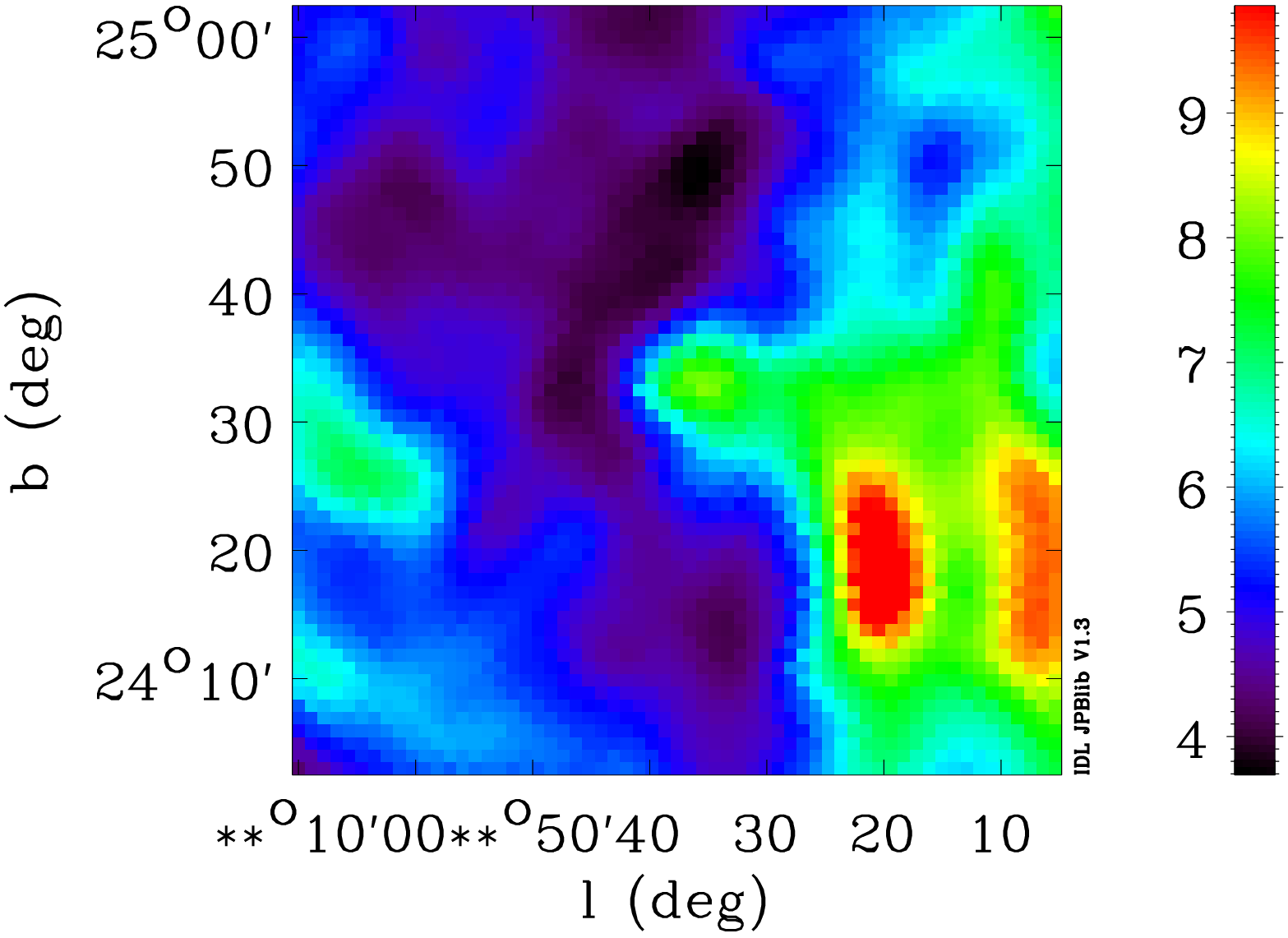} &
 \includegraphics[width=4cm]{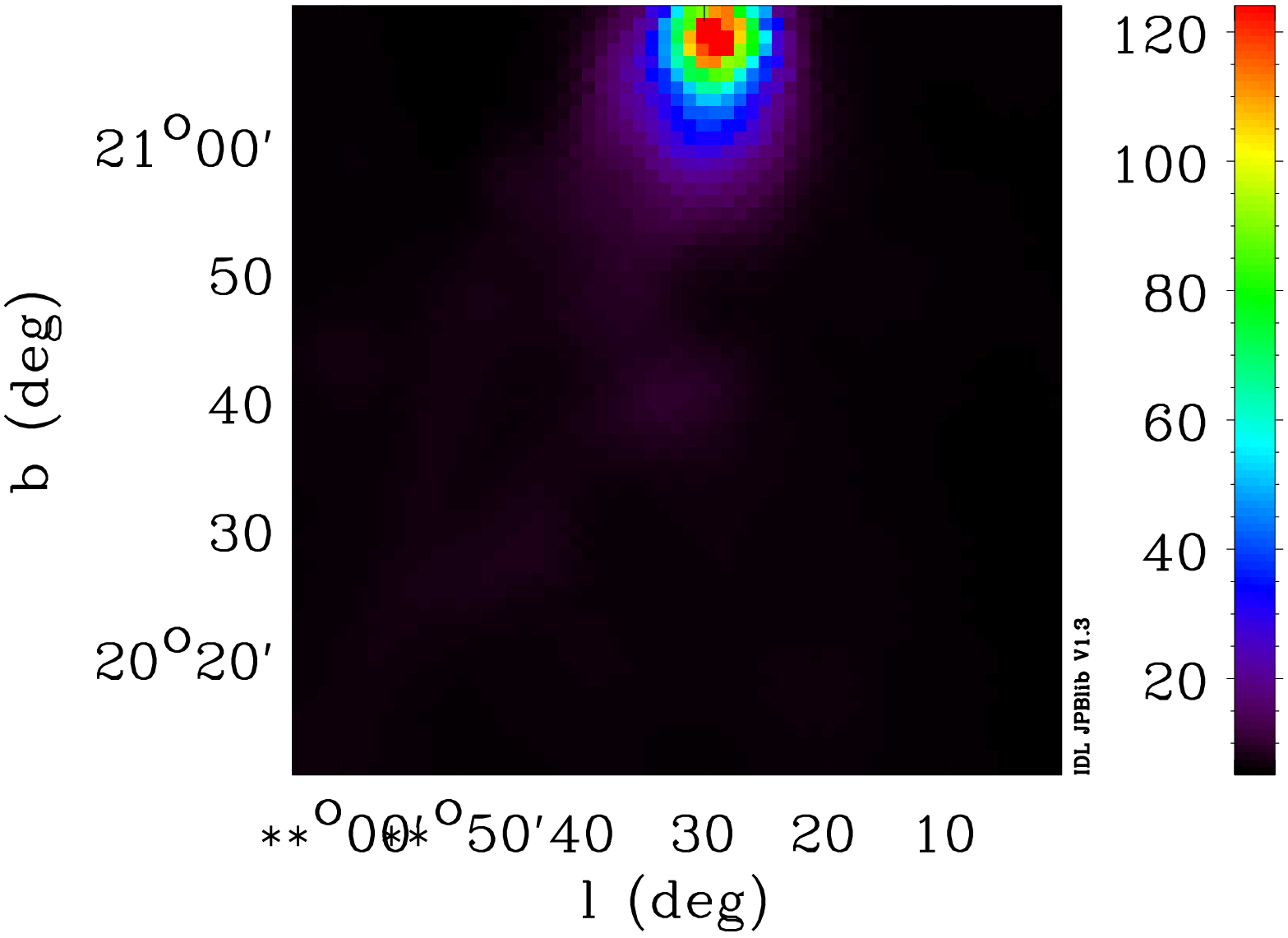} &
 \includegraphics[width=4cm]{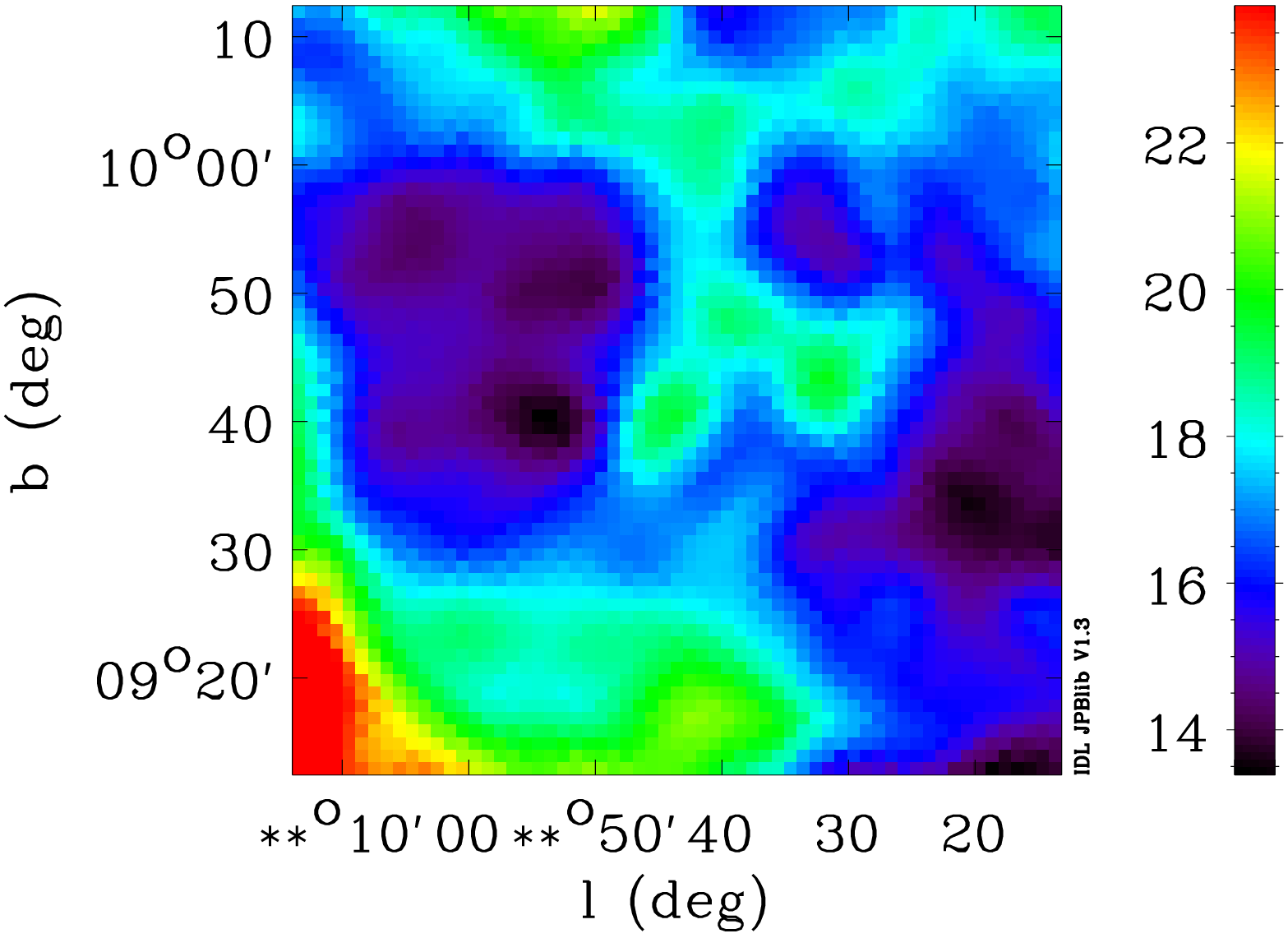} &
 \includegraphics[width=4cm]{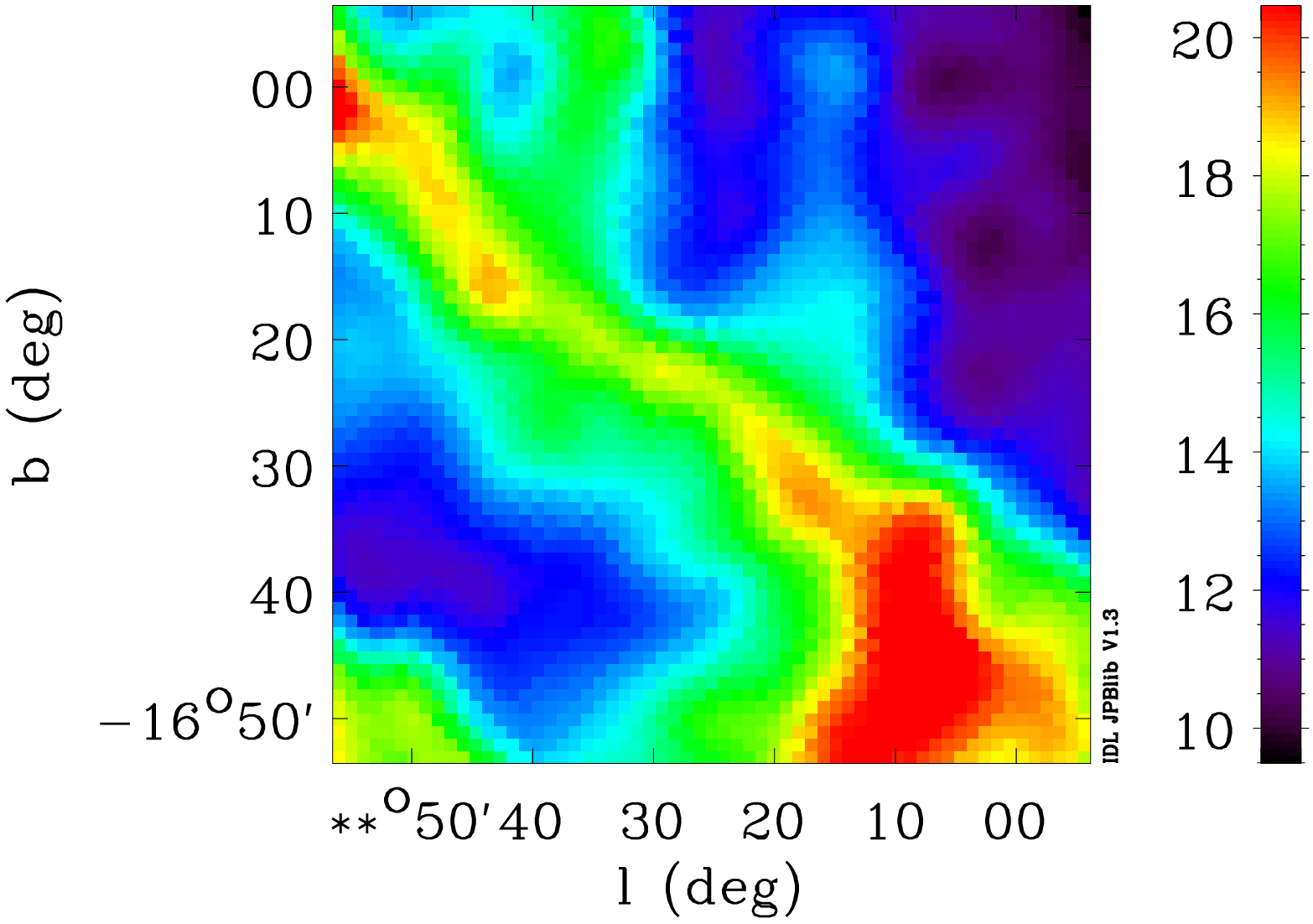} \\

 857\,GHz &
 \includegraphics[width=4cm]{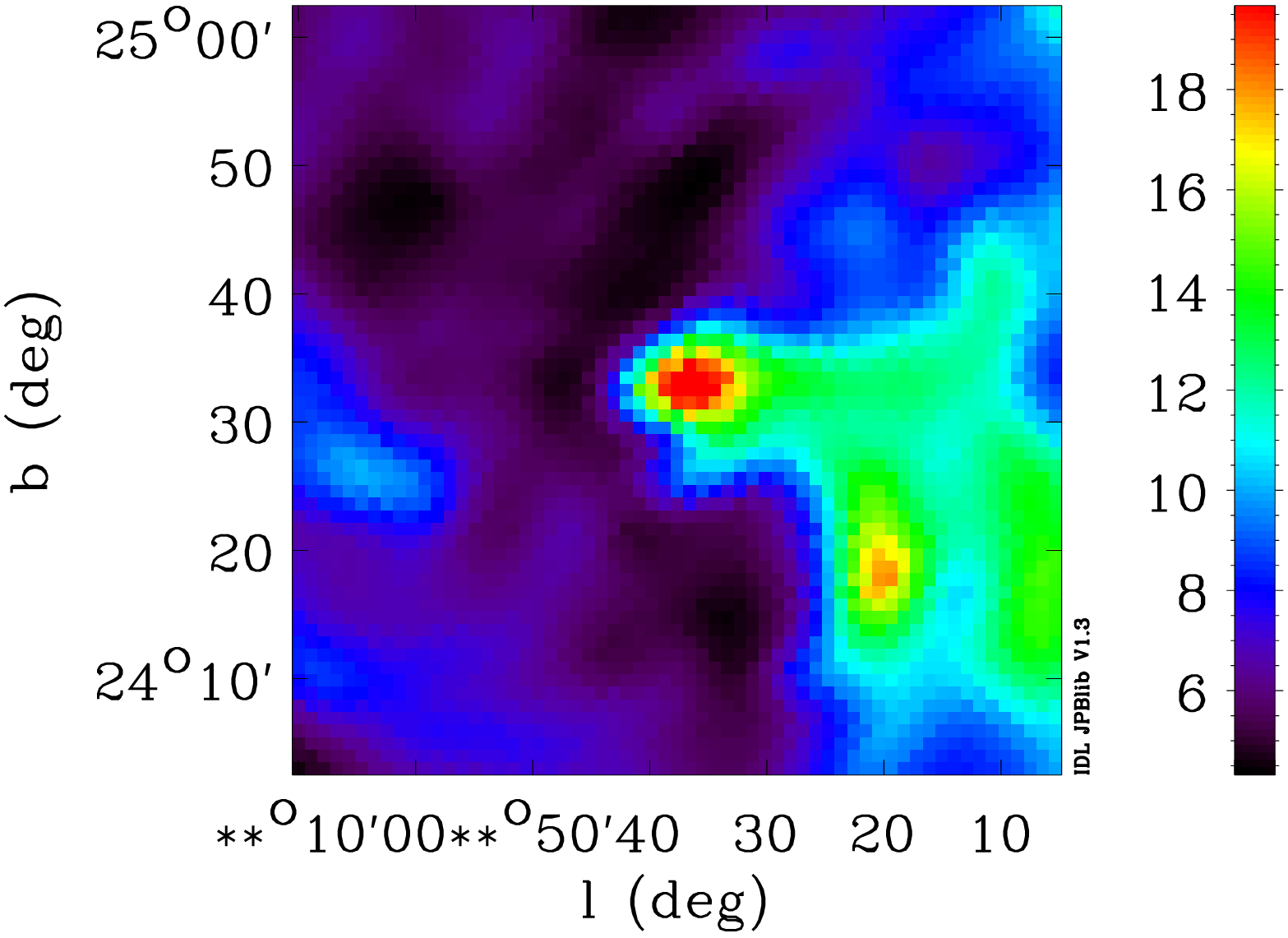} &
 \includegraphics[width=4cm]{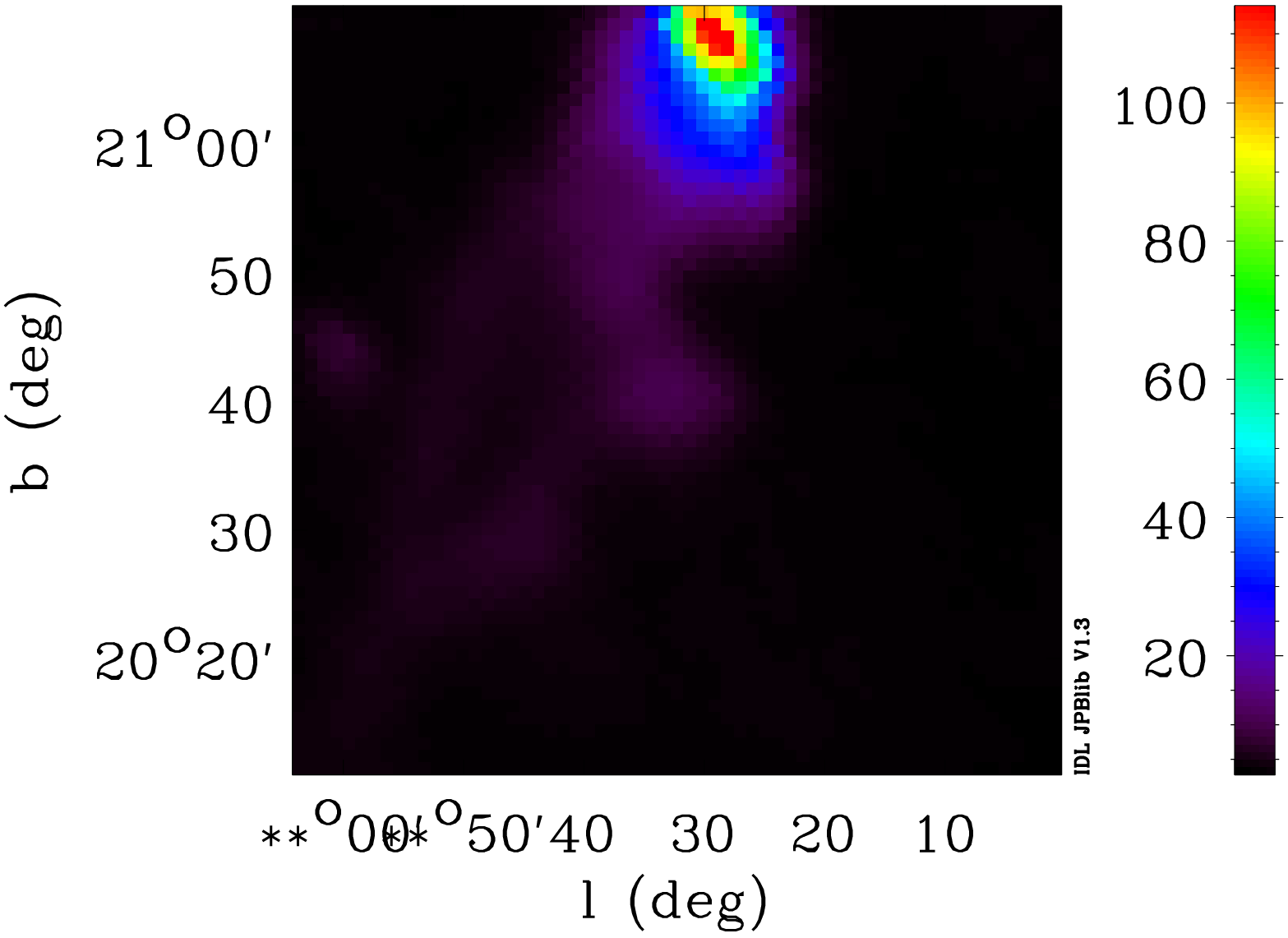} &
 \includegraphics[width=4cm]{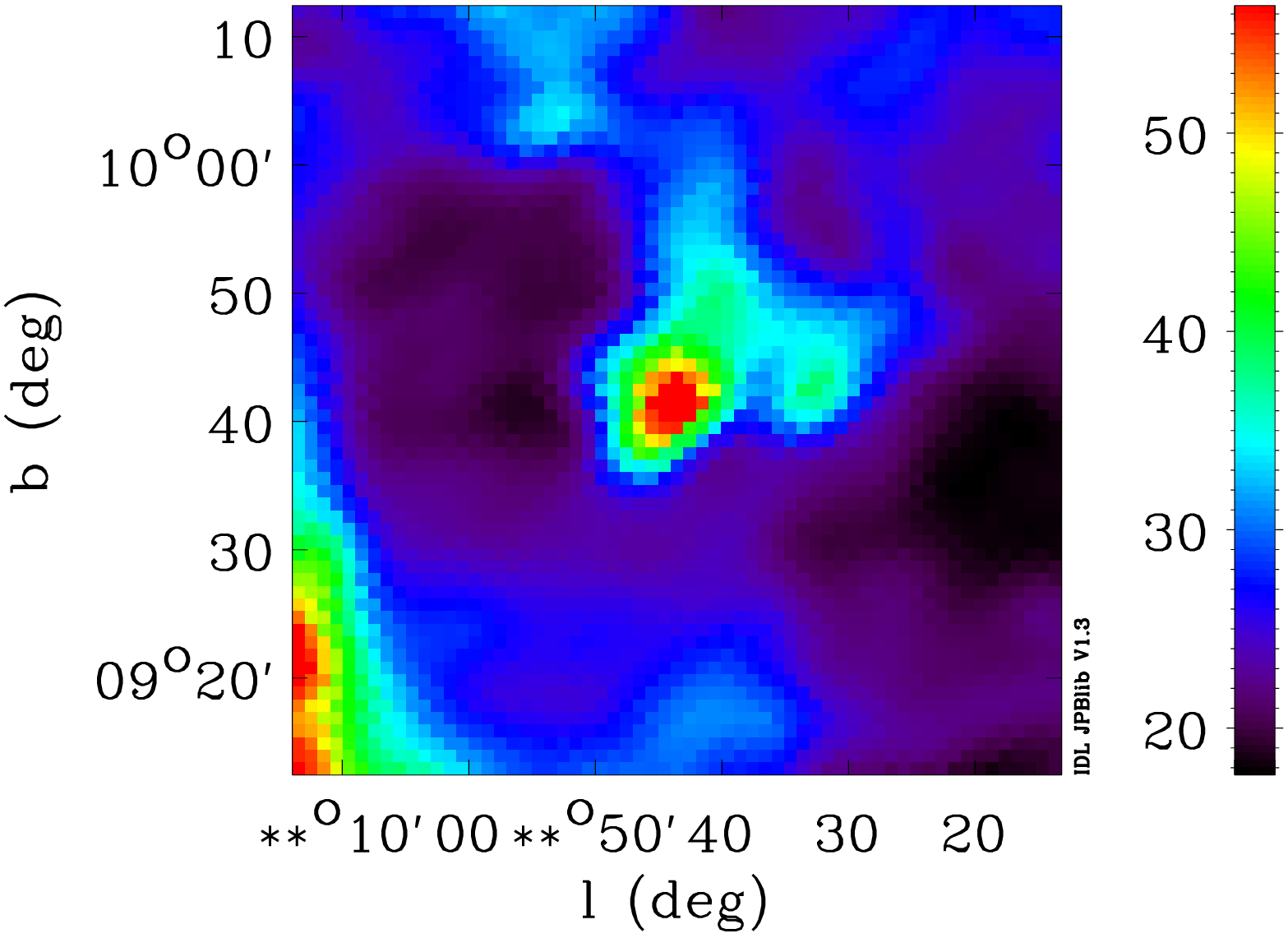} &
 \includegraphics[width=4cm]{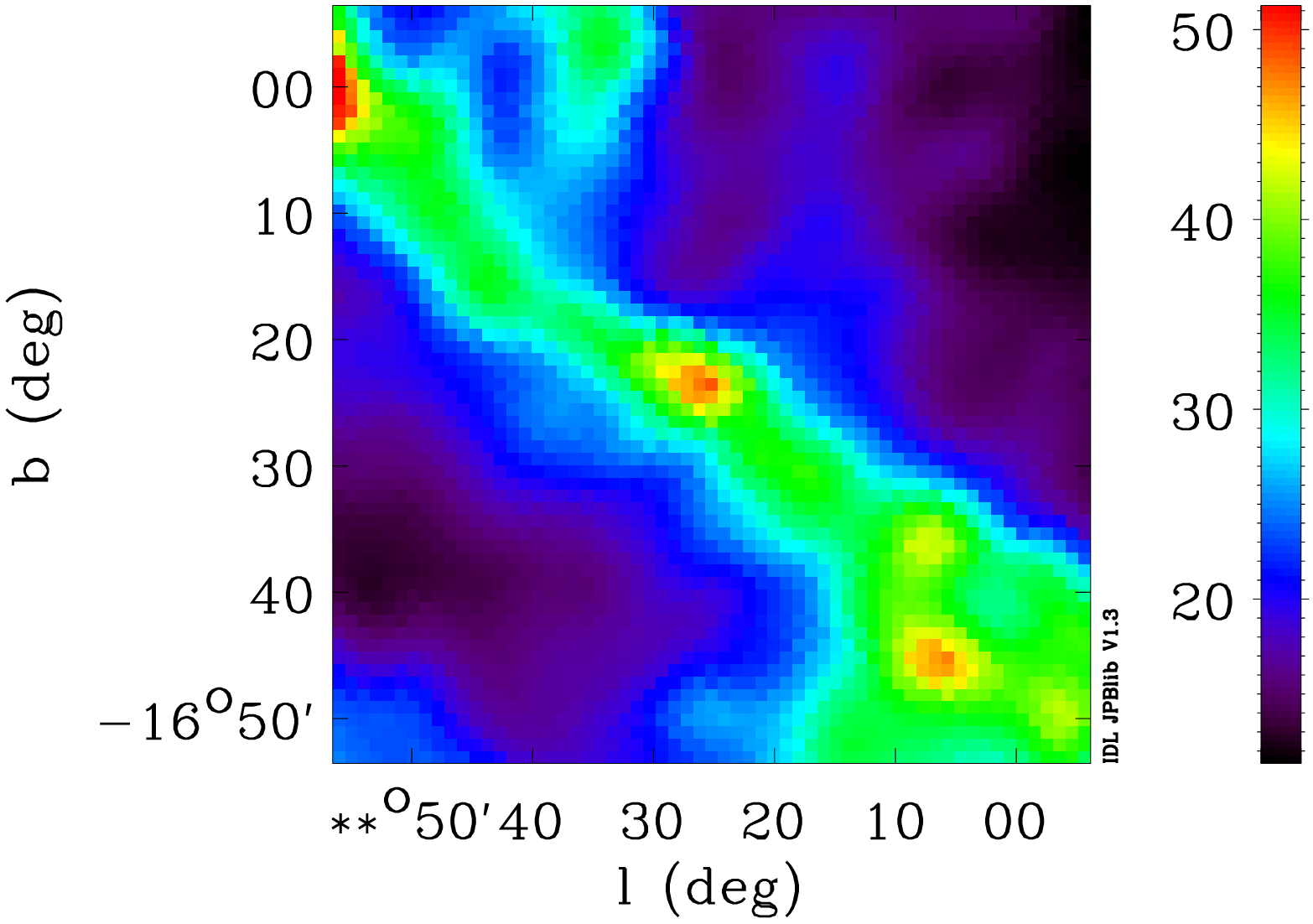} \\

\tiny{Cold residual} &
 \includegraphics[width=4cm]{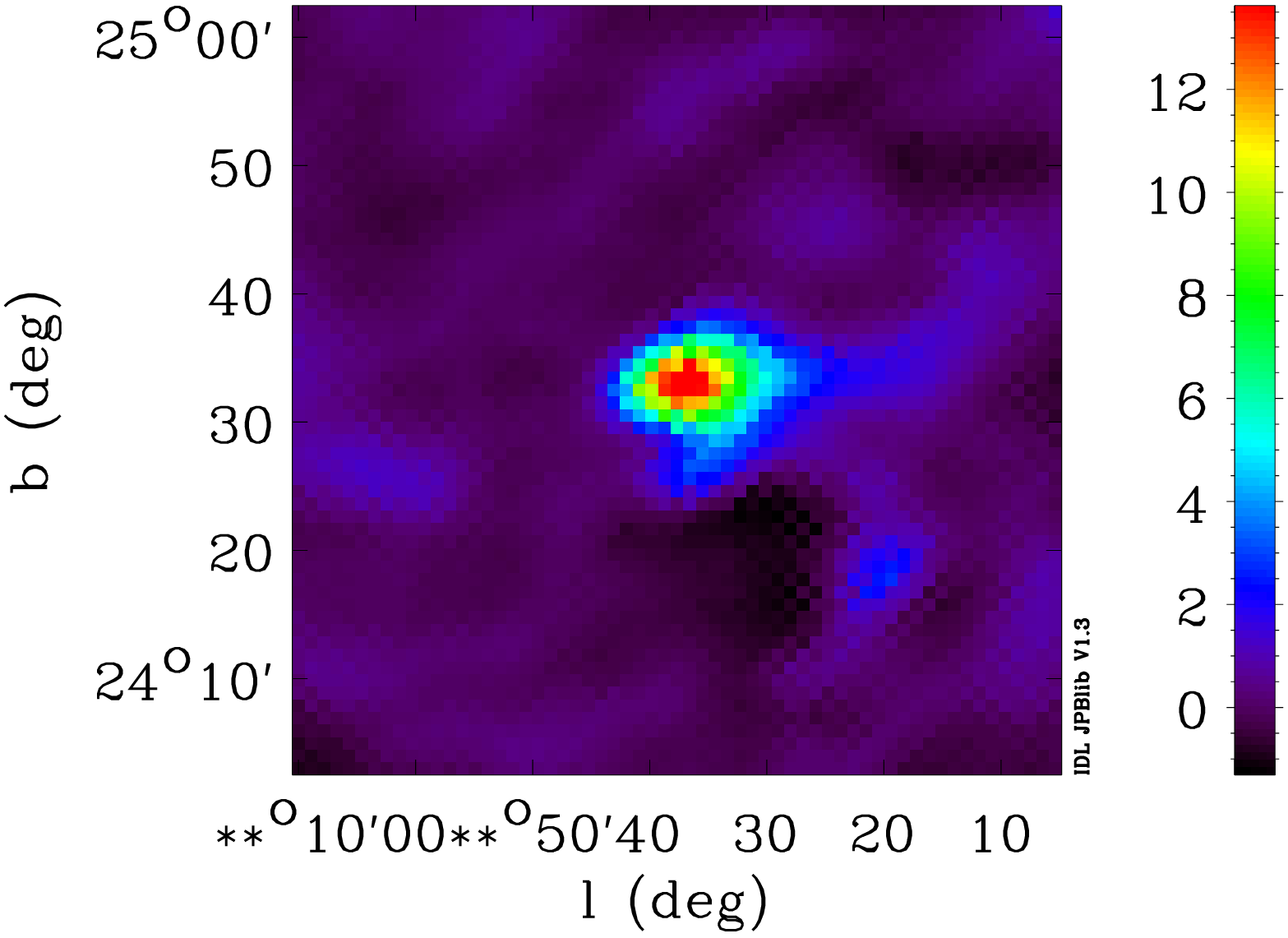} &
 \includegraphics[width=4cm]{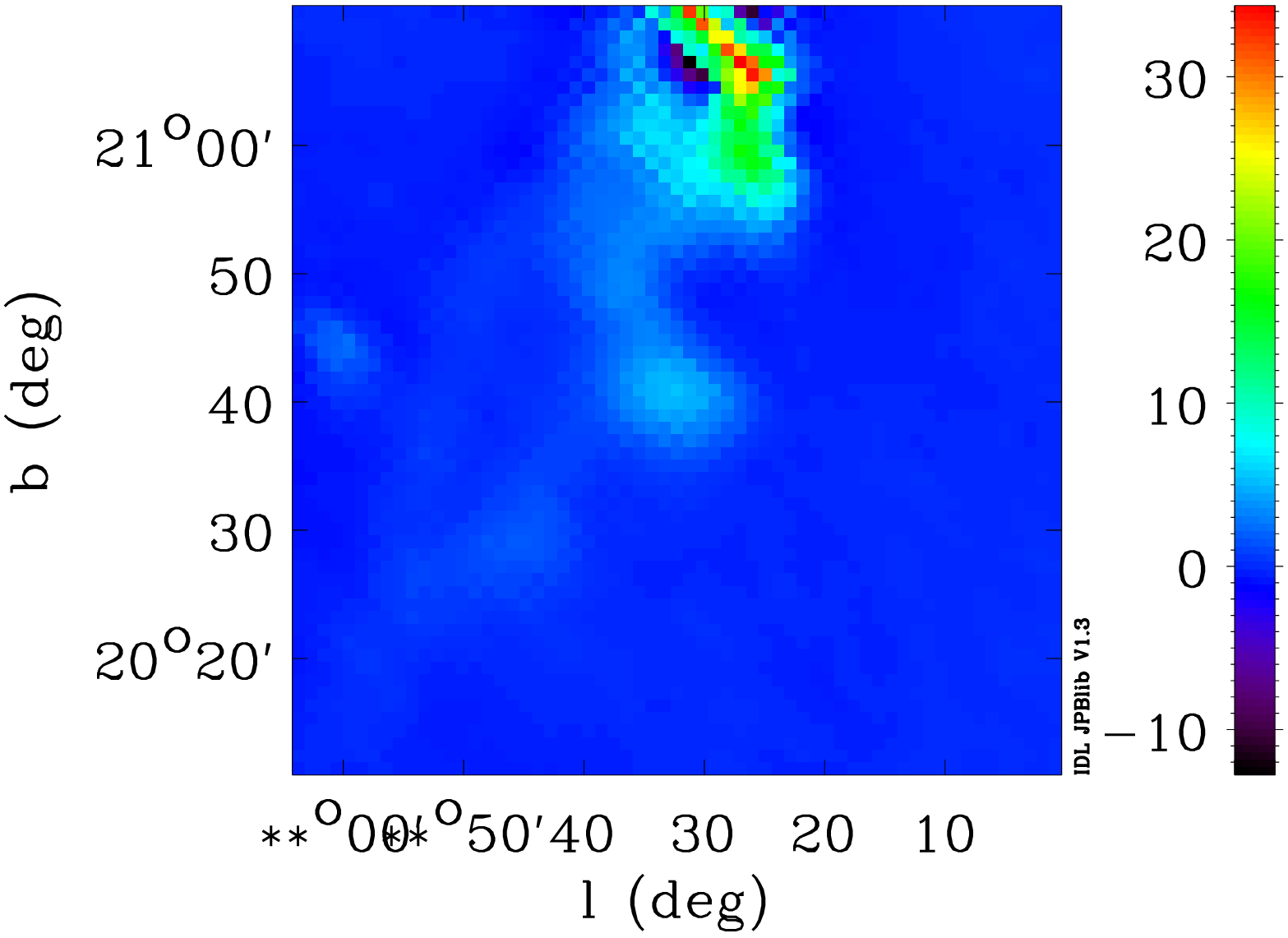} &
 \includegraphics[width=4cm]{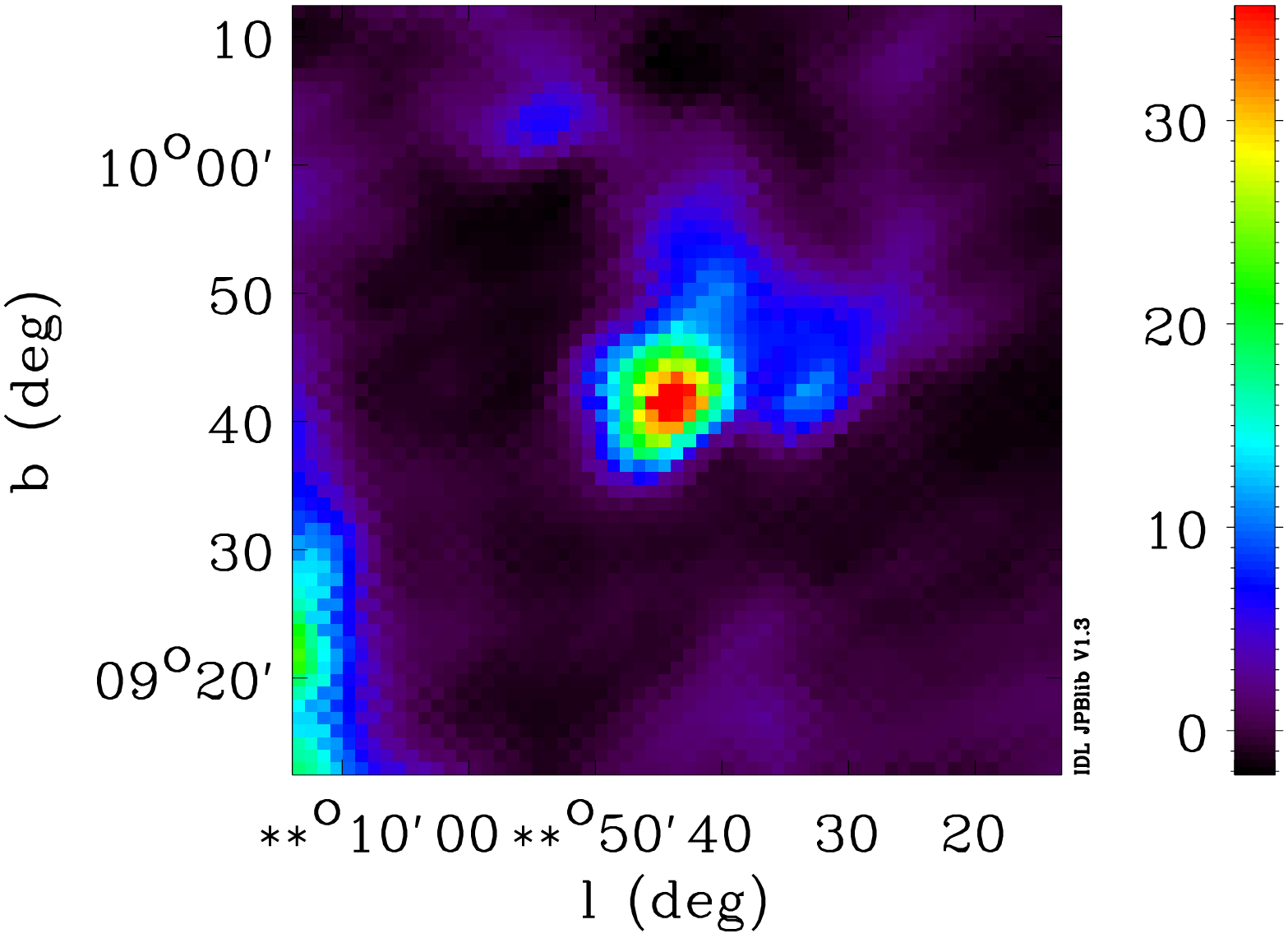} &
 \includegraphics[width=4cm]{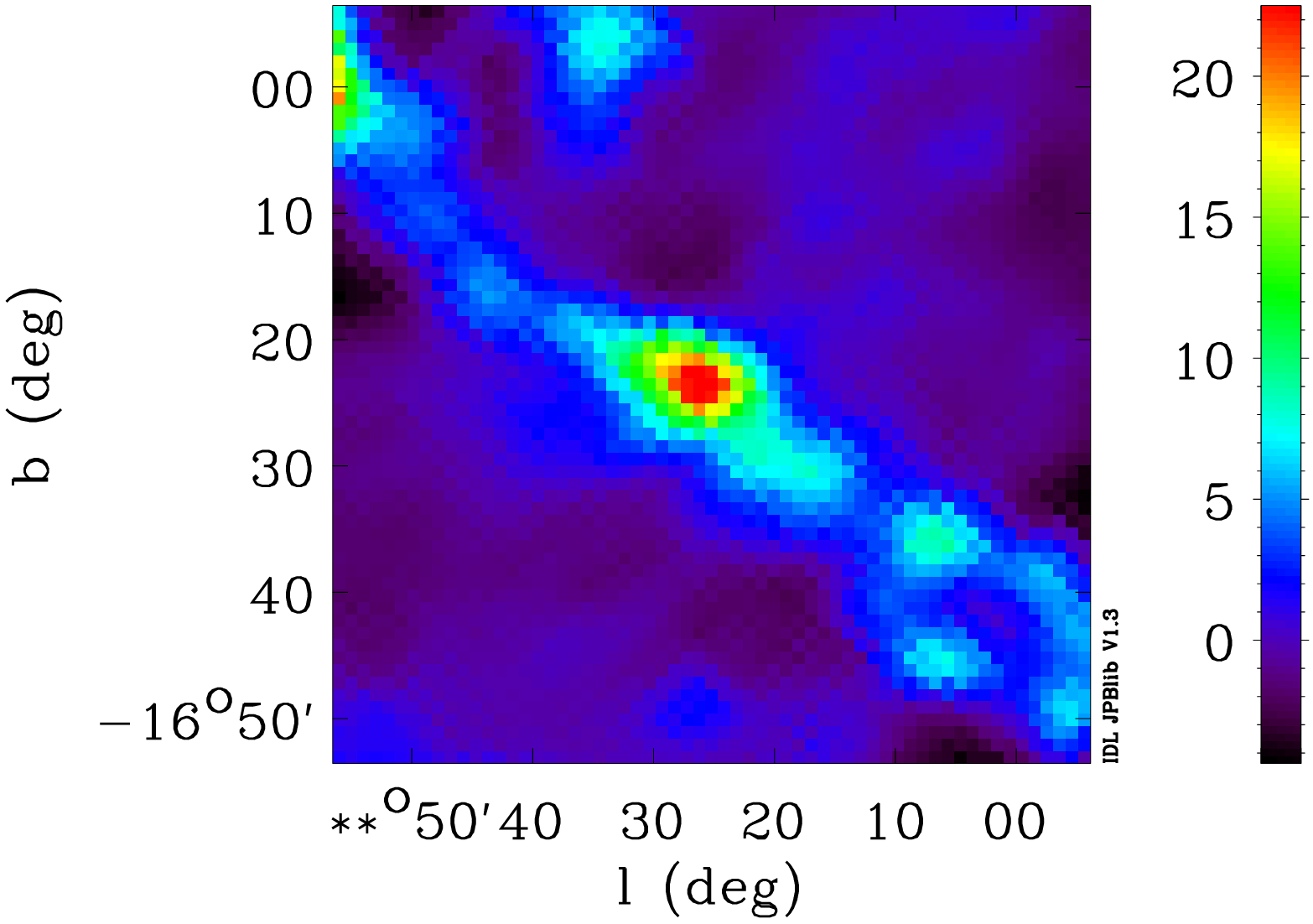} \\

 545\,GHz &
 \includegraphics[width=4cm]{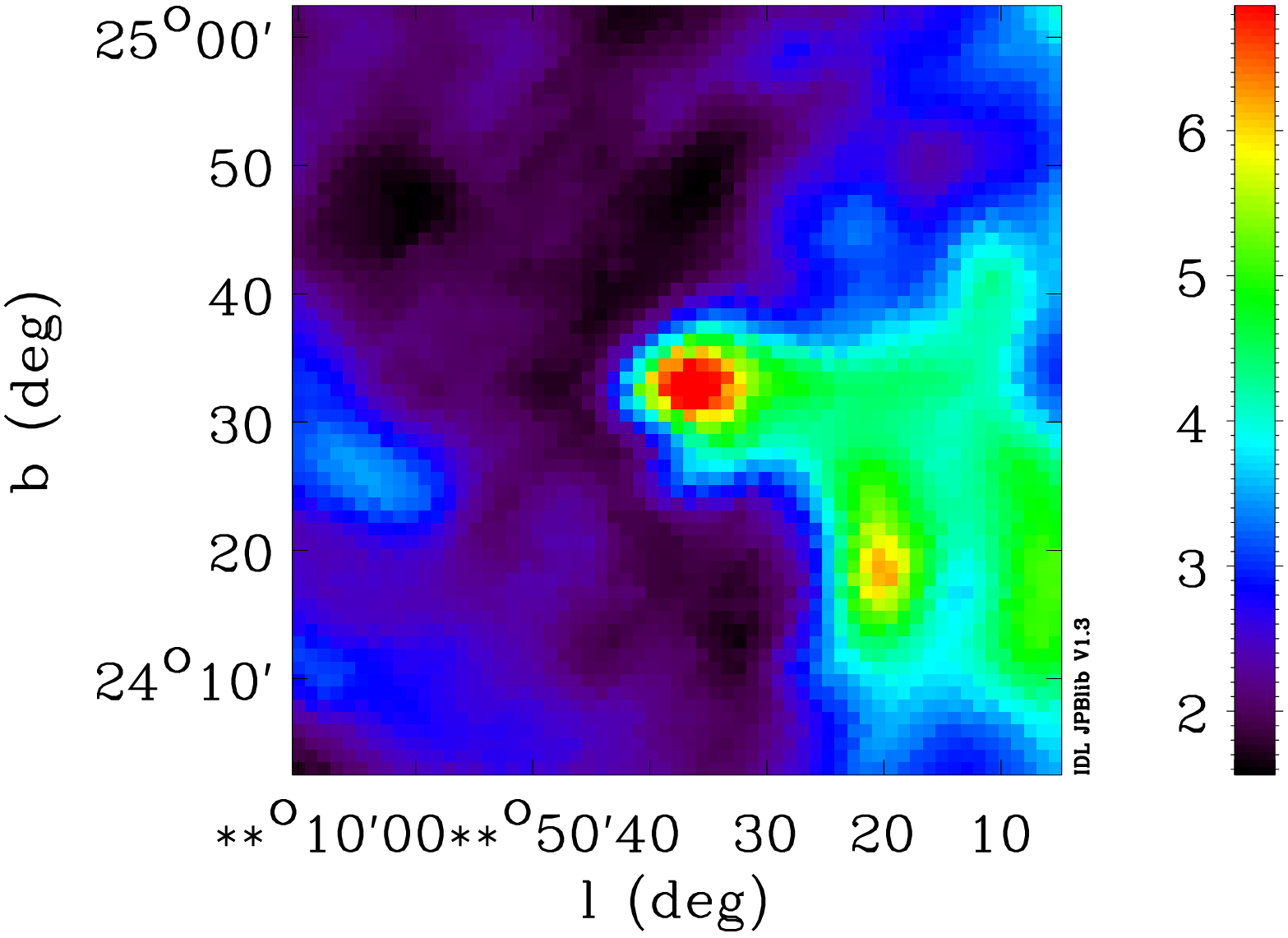} &
 \includegraphics[width=4cm]{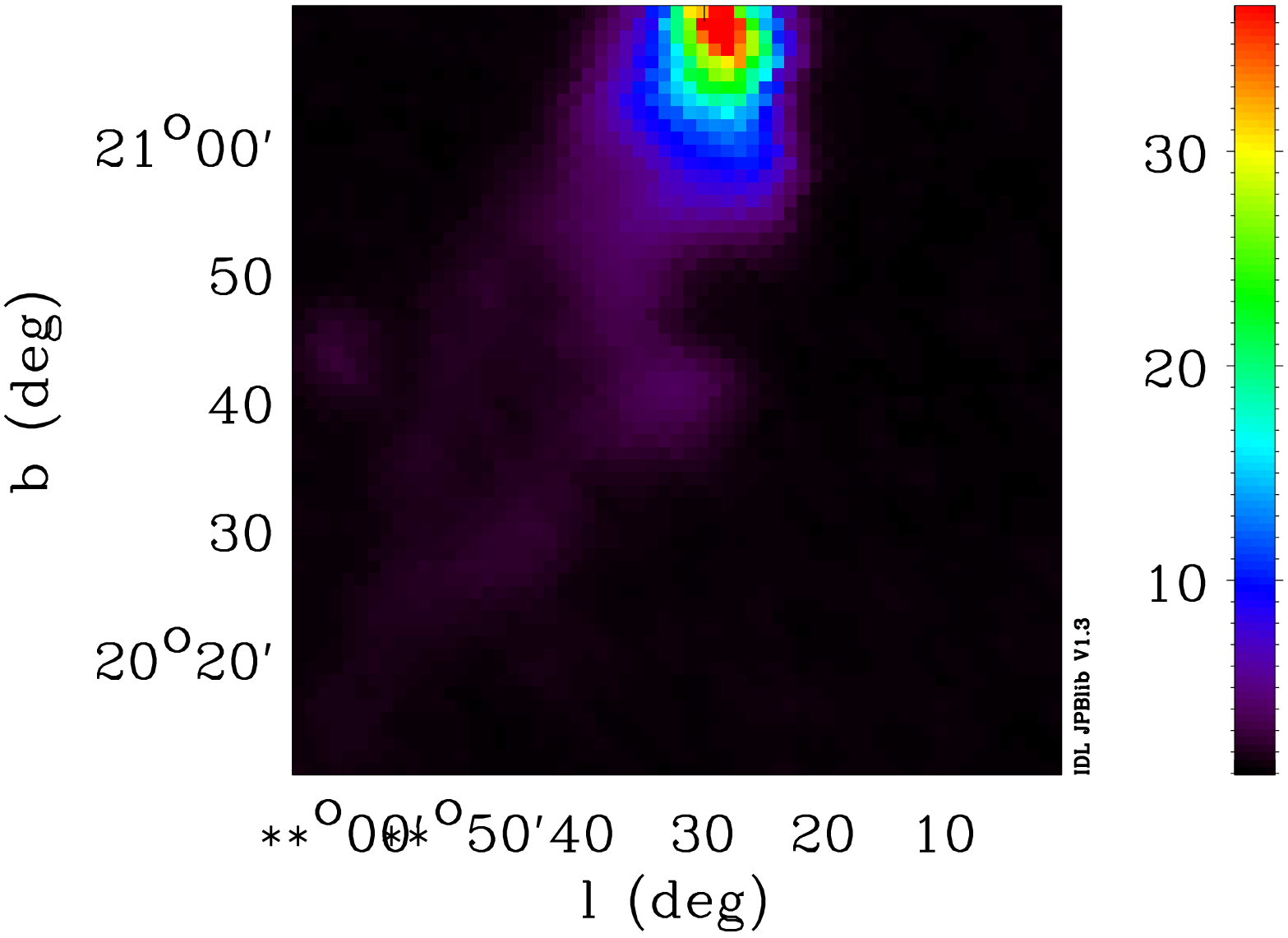} &
 \includegraphics[width=4cm]{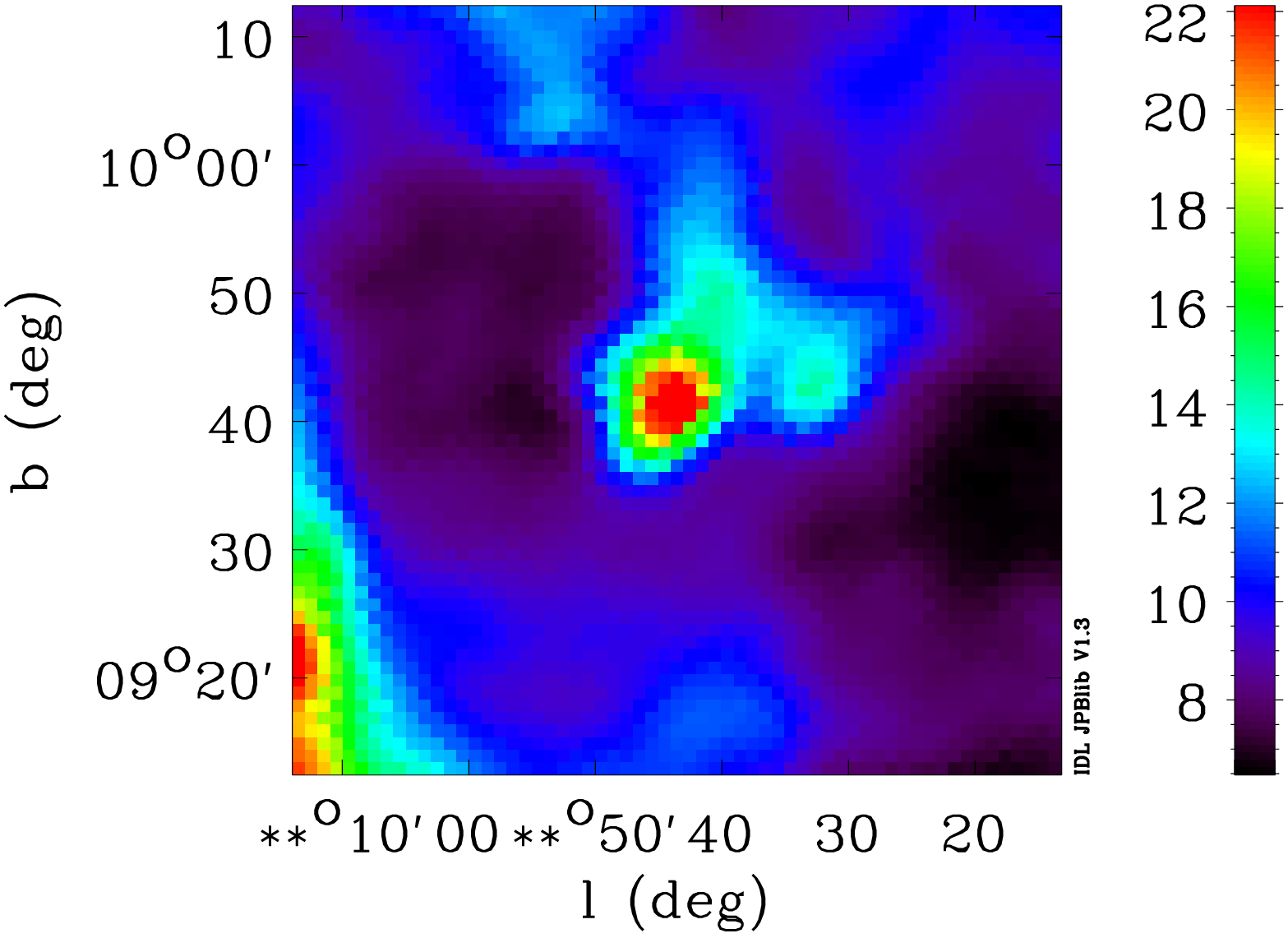} &
 \includegraphics[width=4cm]{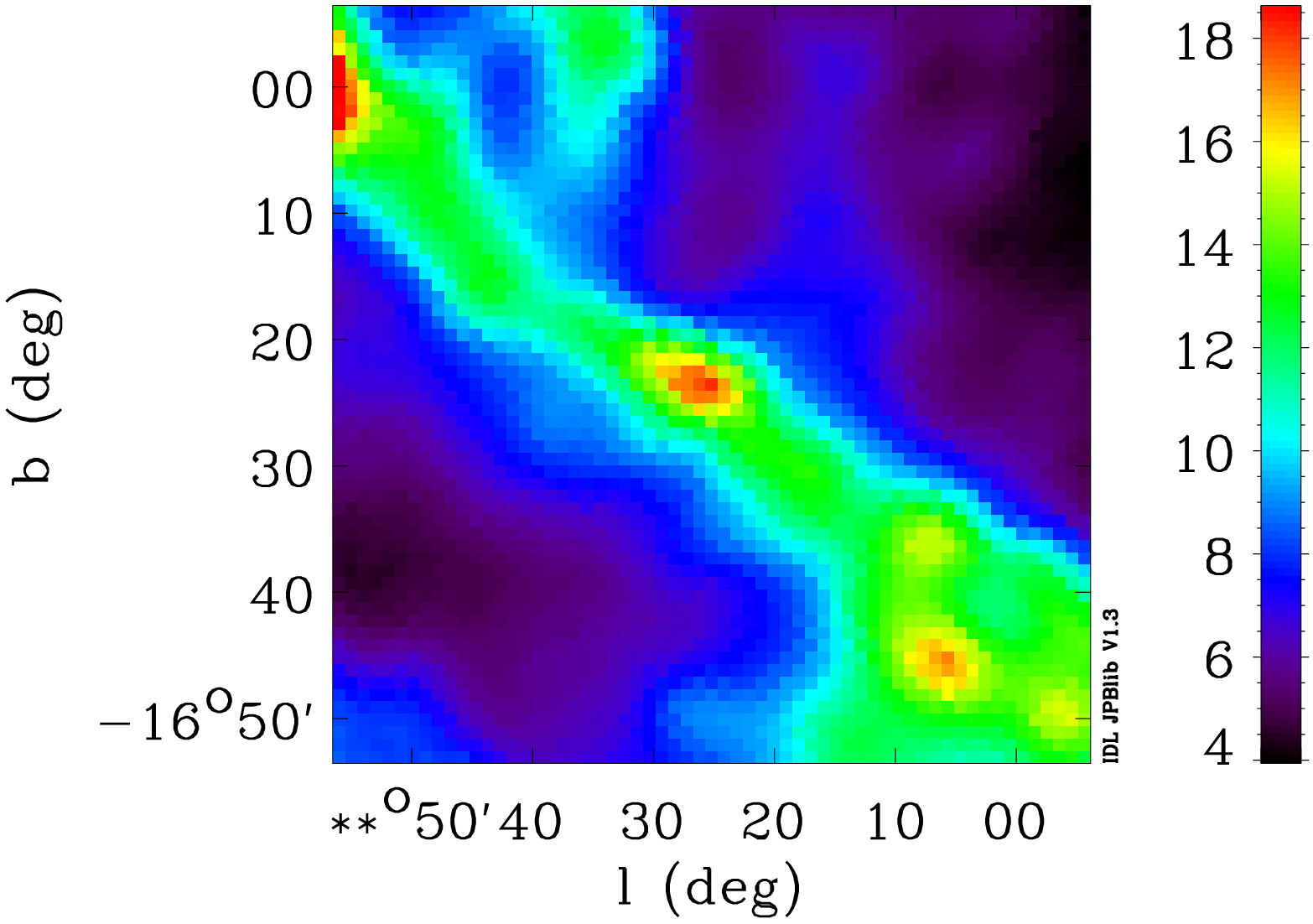} \\

 353\,GHz &
 \includegraphics[width=4cm]{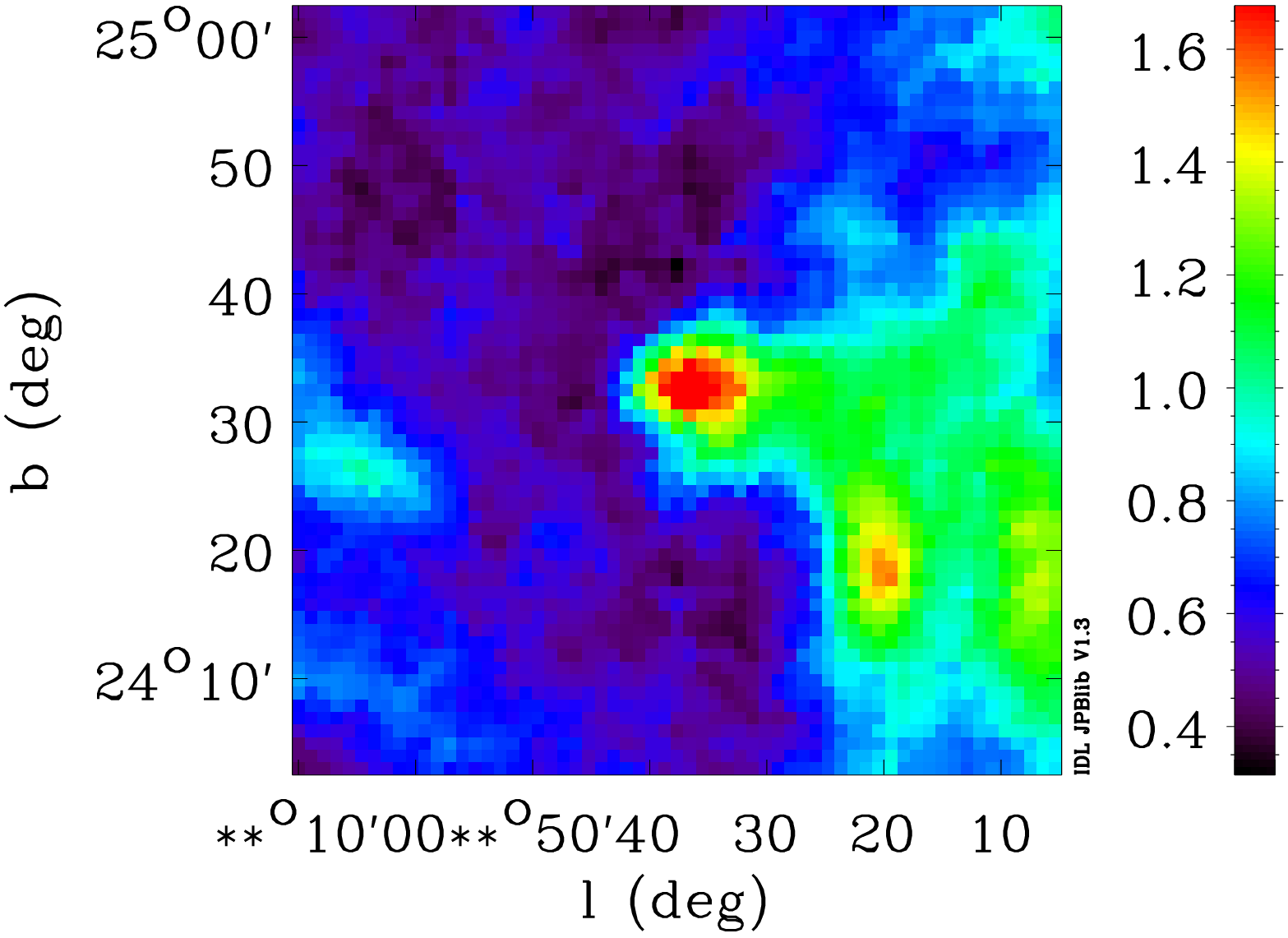} &
 \includegraphics[width=4cm]{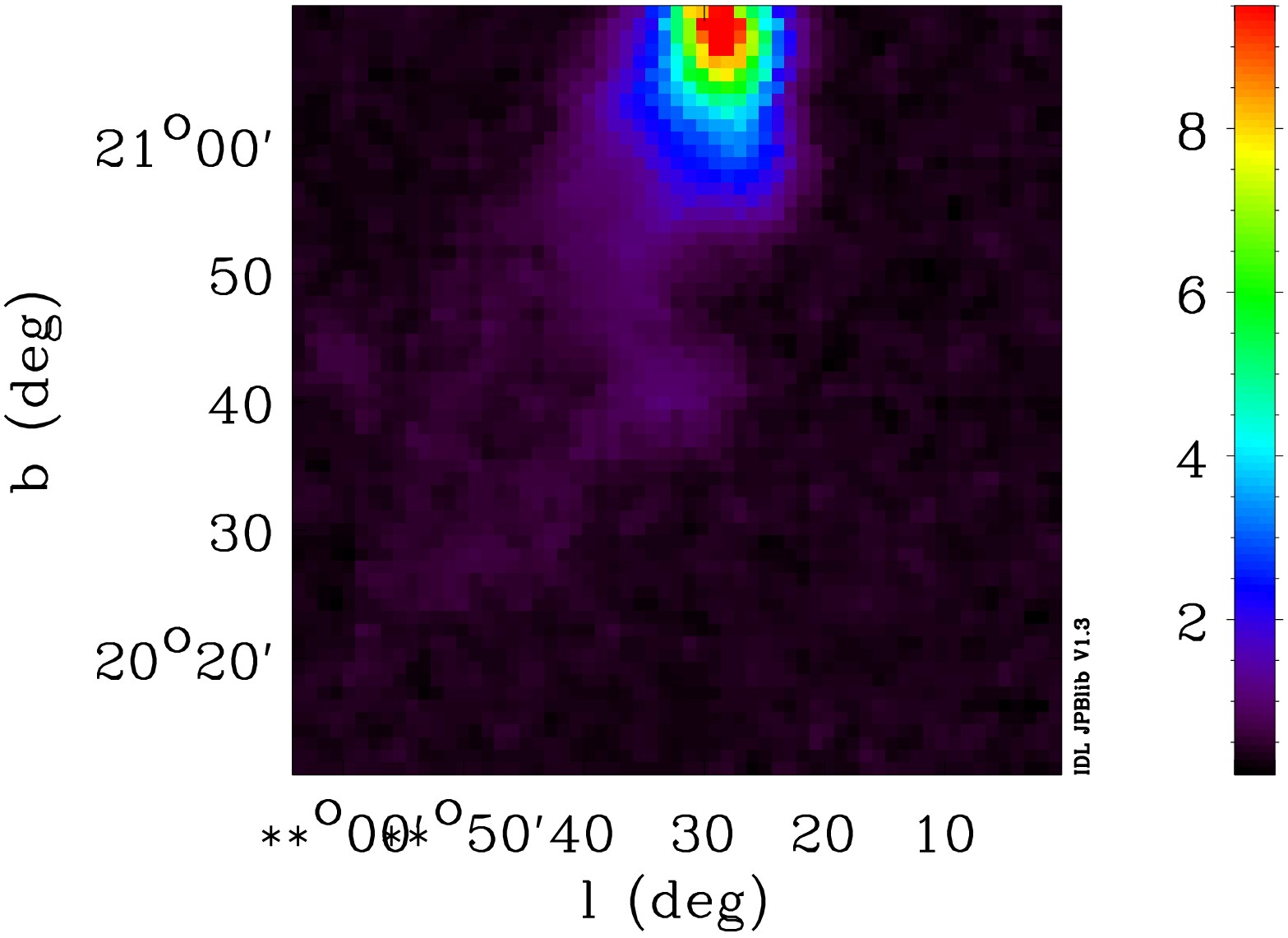} &
 \includegraphics[width=4cm]{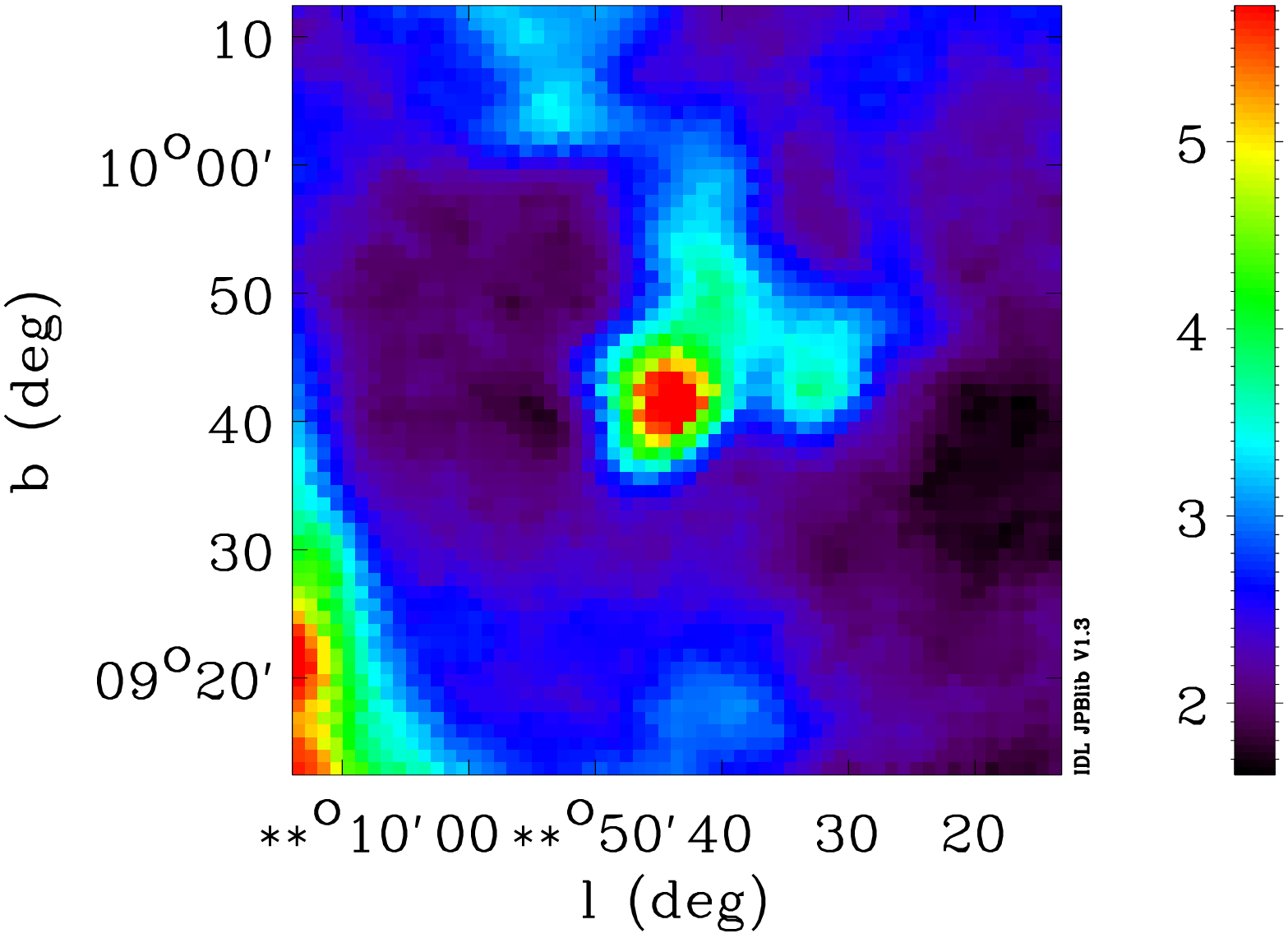} &
 \includegraphics[width=4cm]{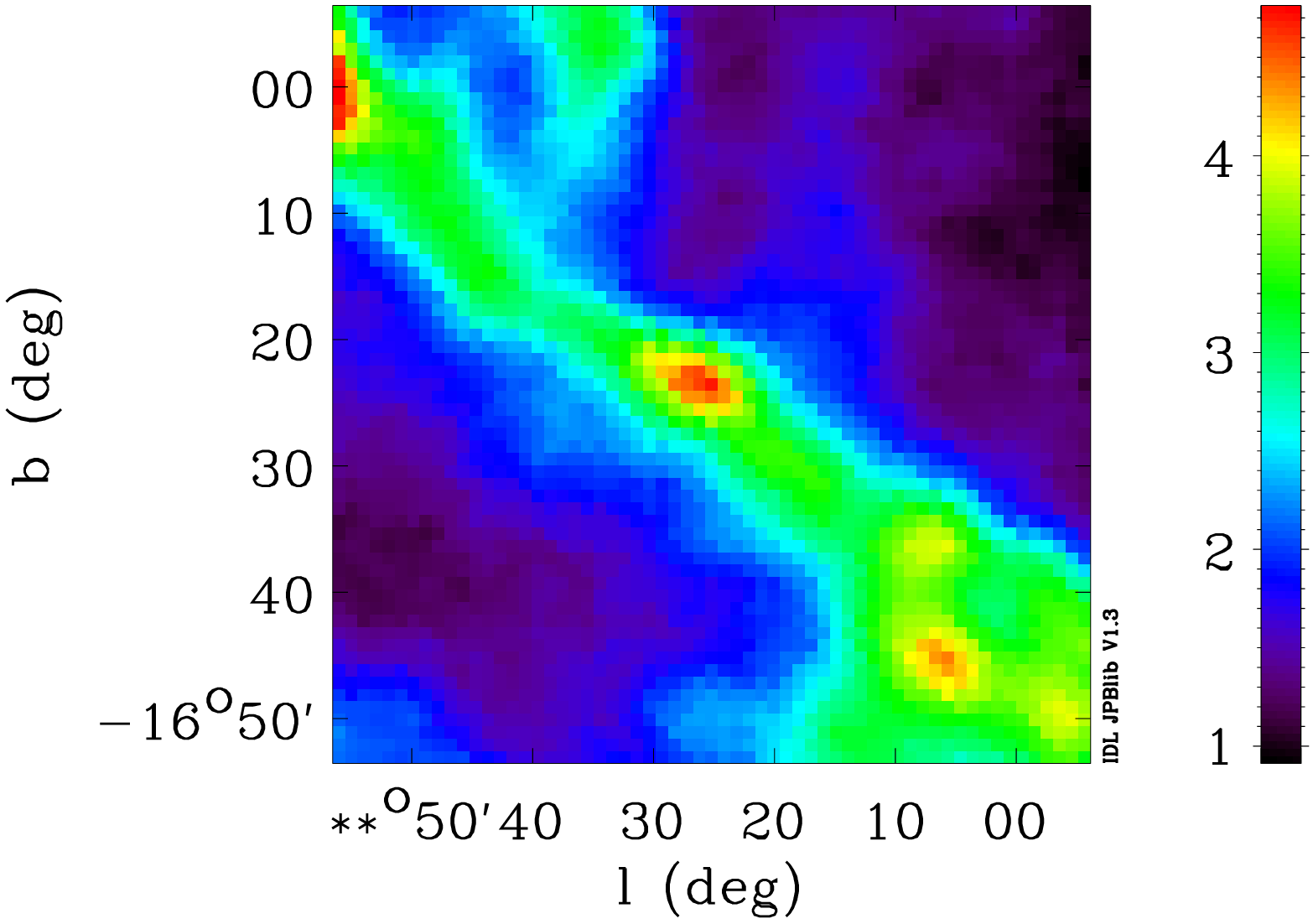} \\

 217\,GHz &
 \includegraphics[width=4cm]{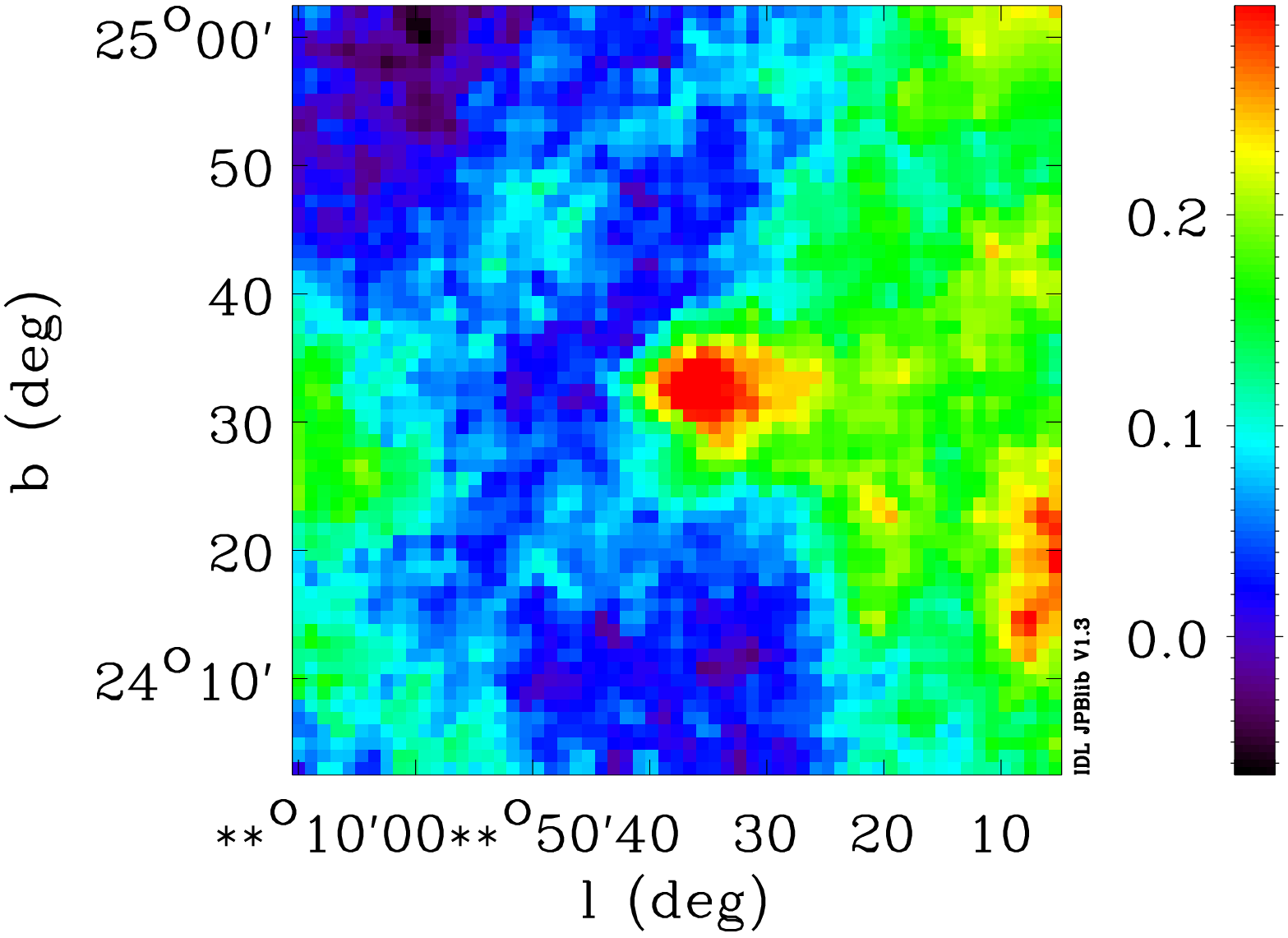} &
 \includegraphics[width=4cm]{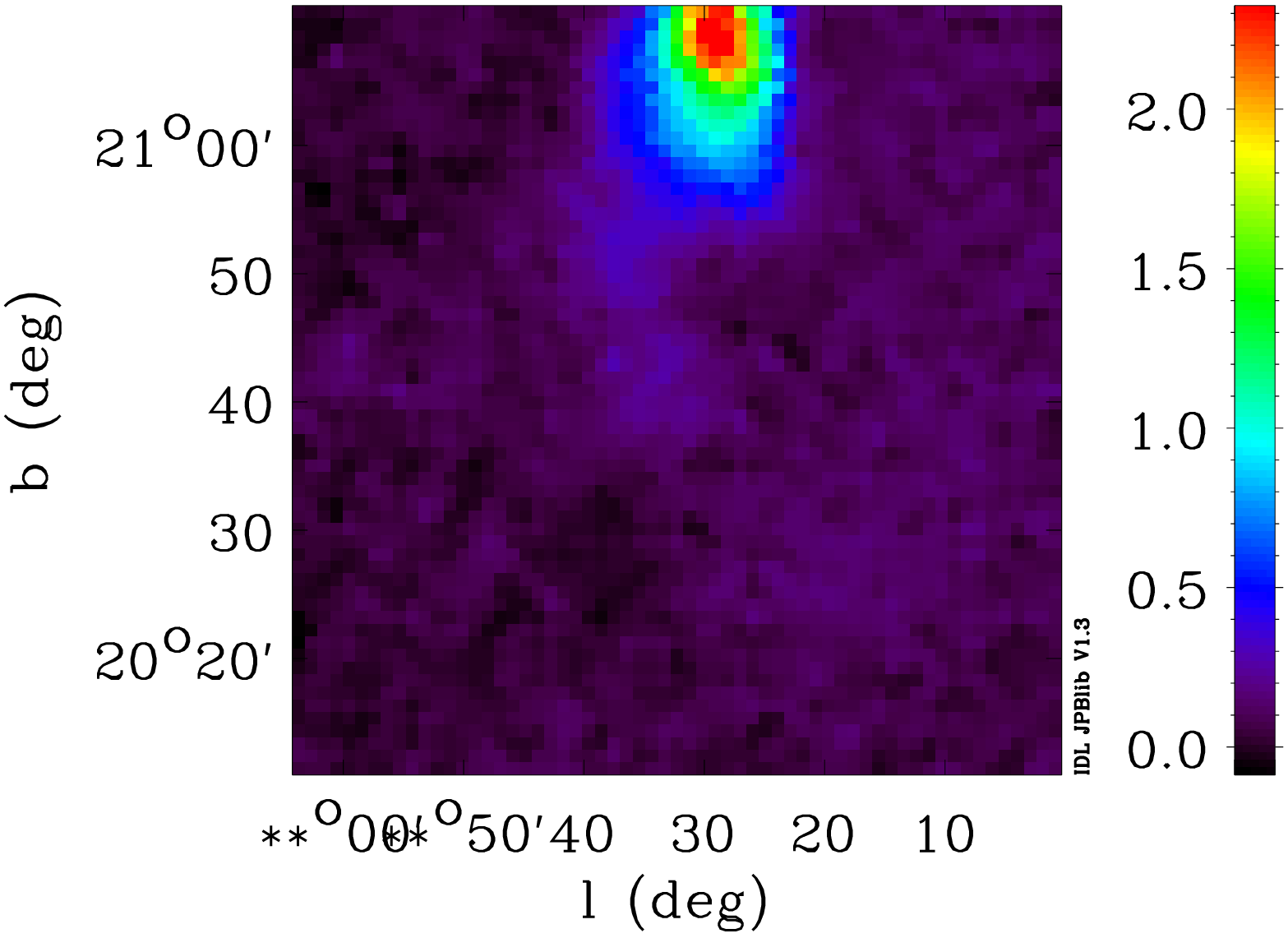} &
 \includegraphics[width=4cm]{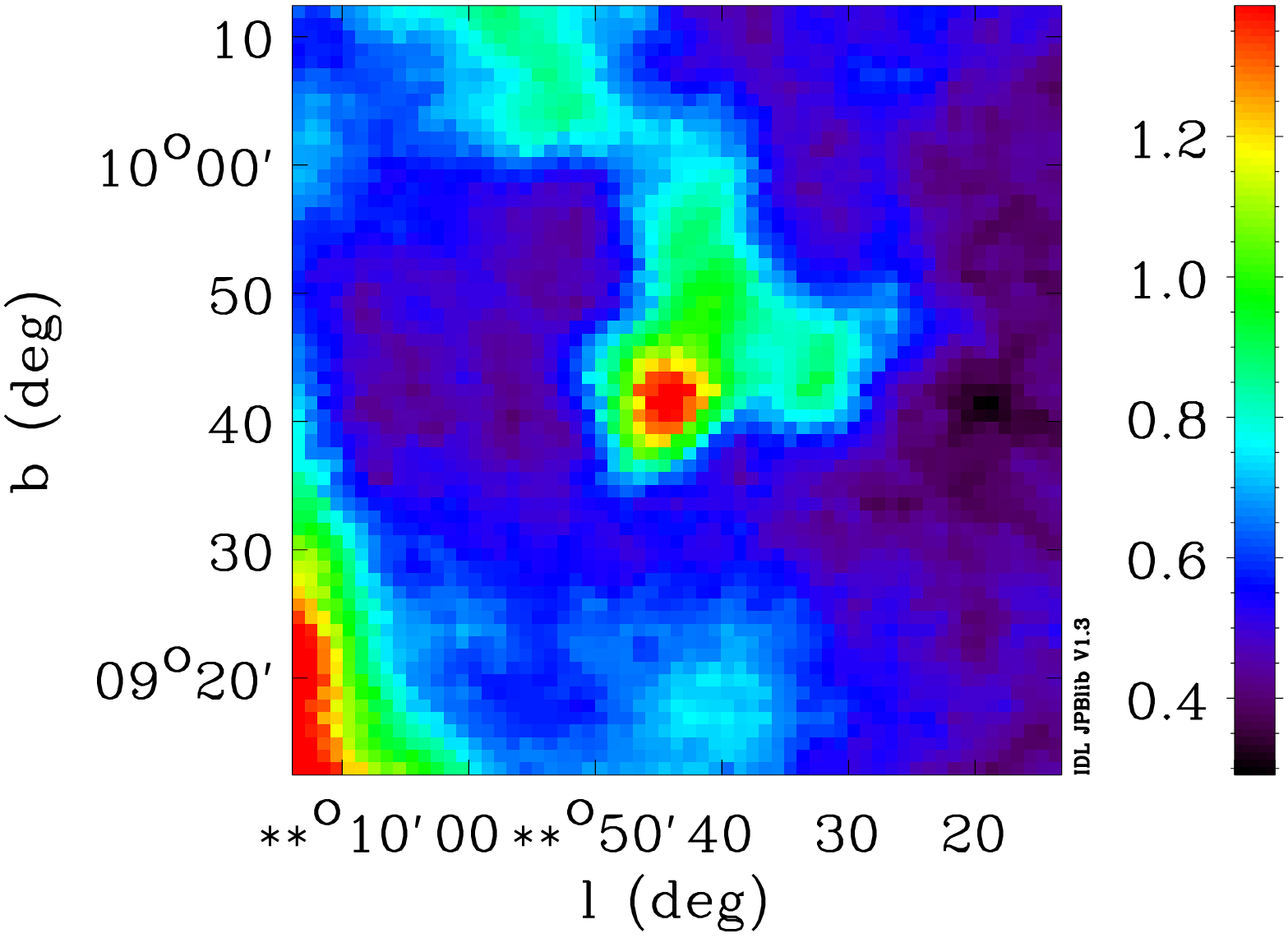} &
 \includegraphics[width=4cm]{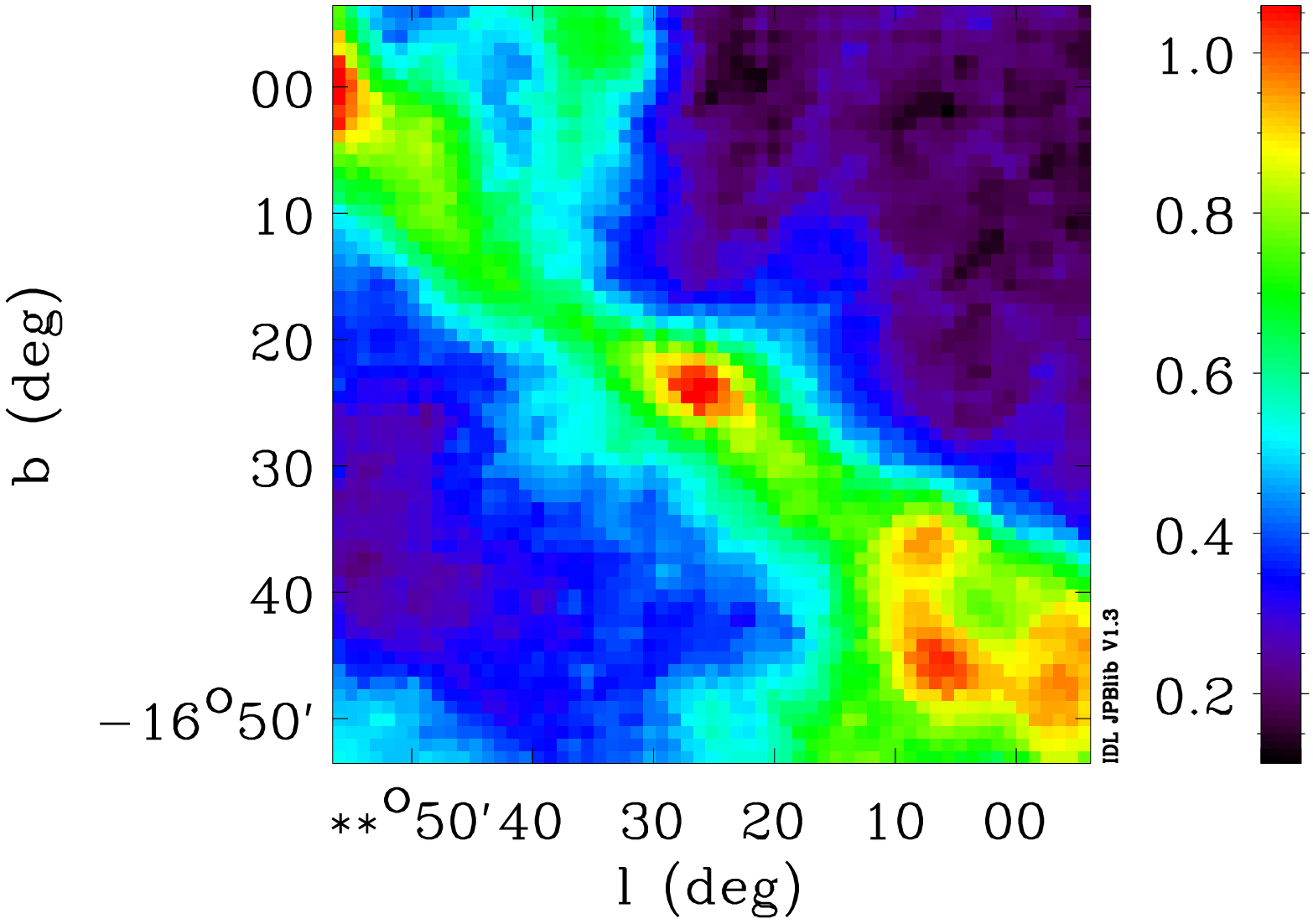} \\

 143\,GHz &
 \includegraphics[width=4cm]{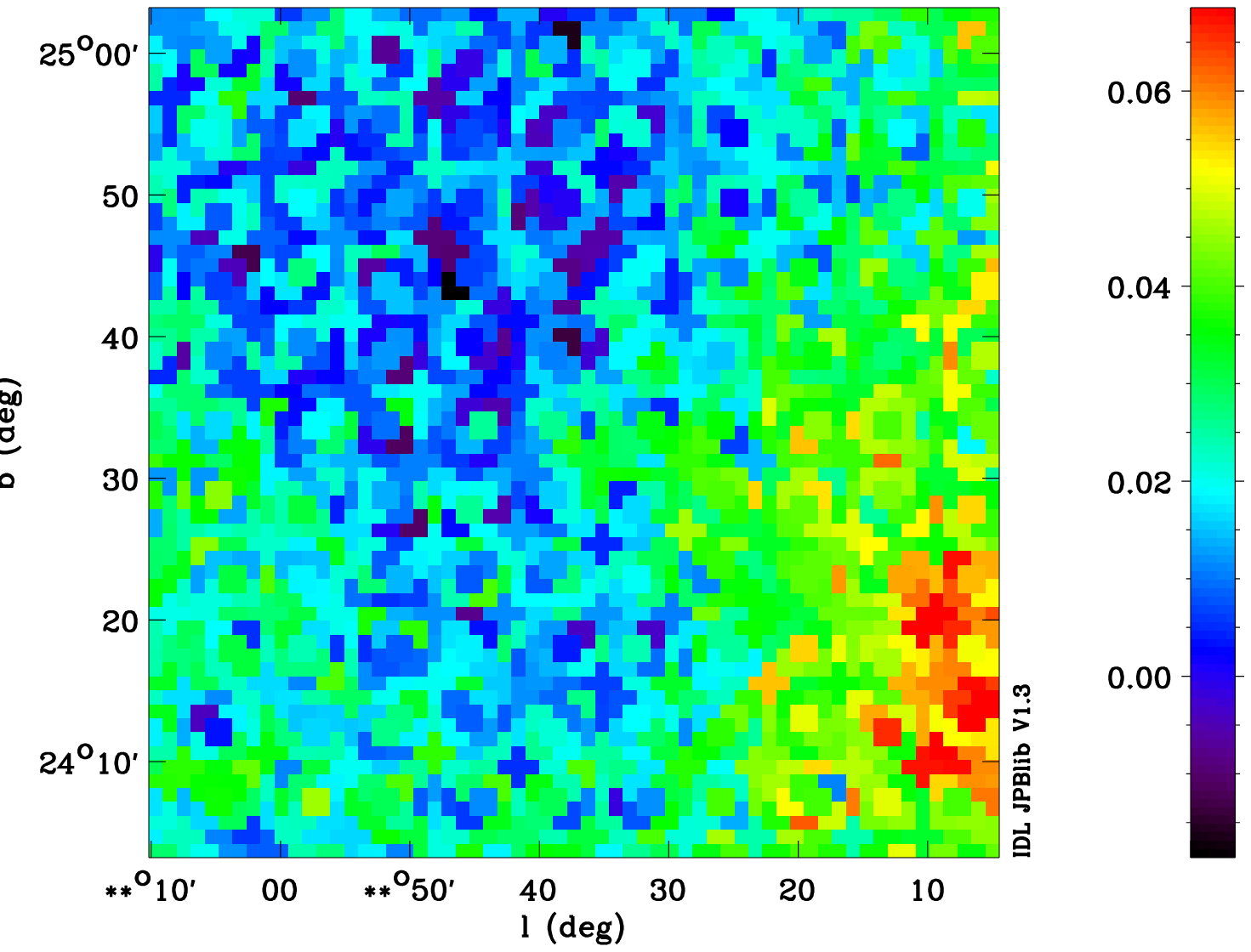} &
 \includegraphics[width=4cm]{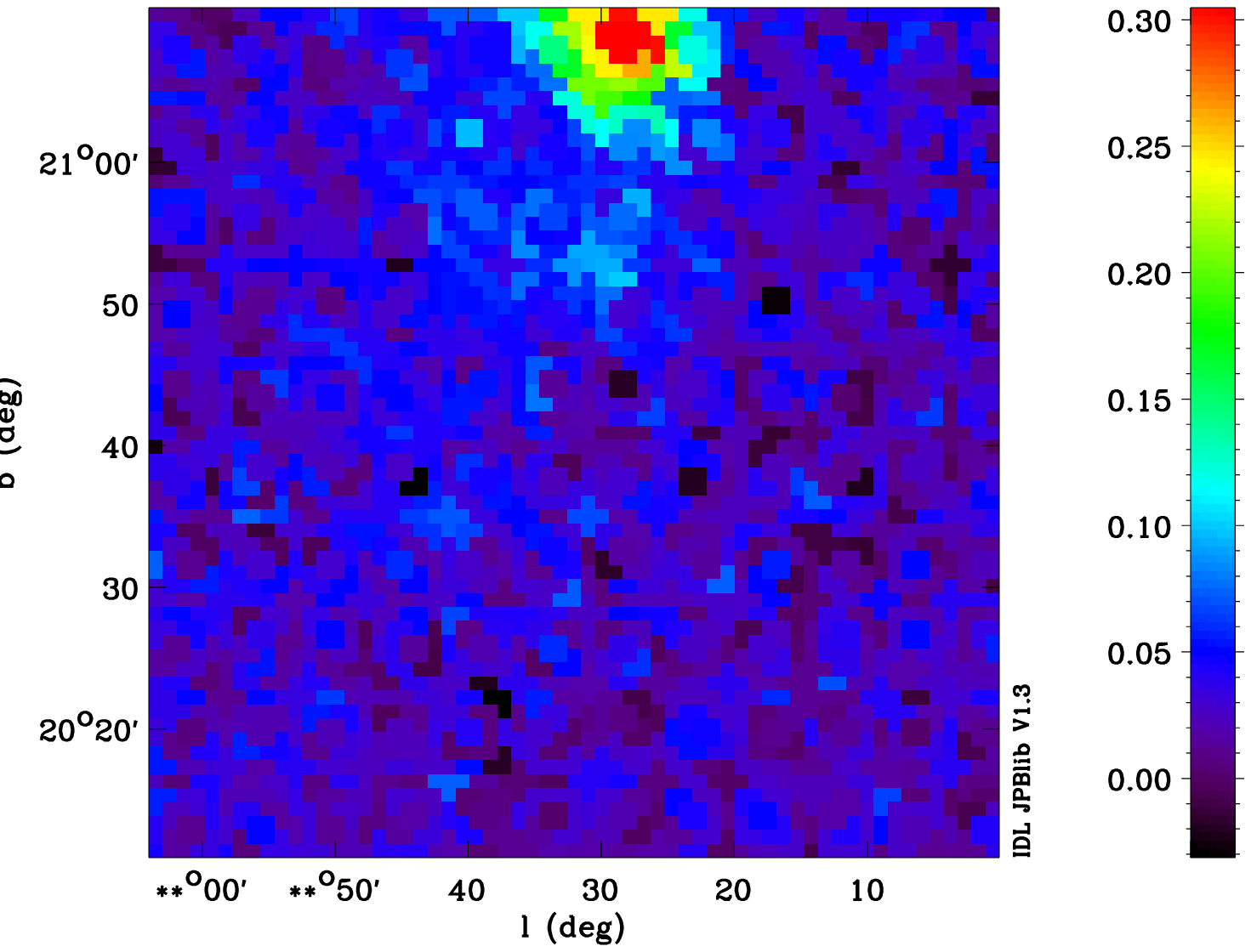} &
 \includegraphics[width=4cm]{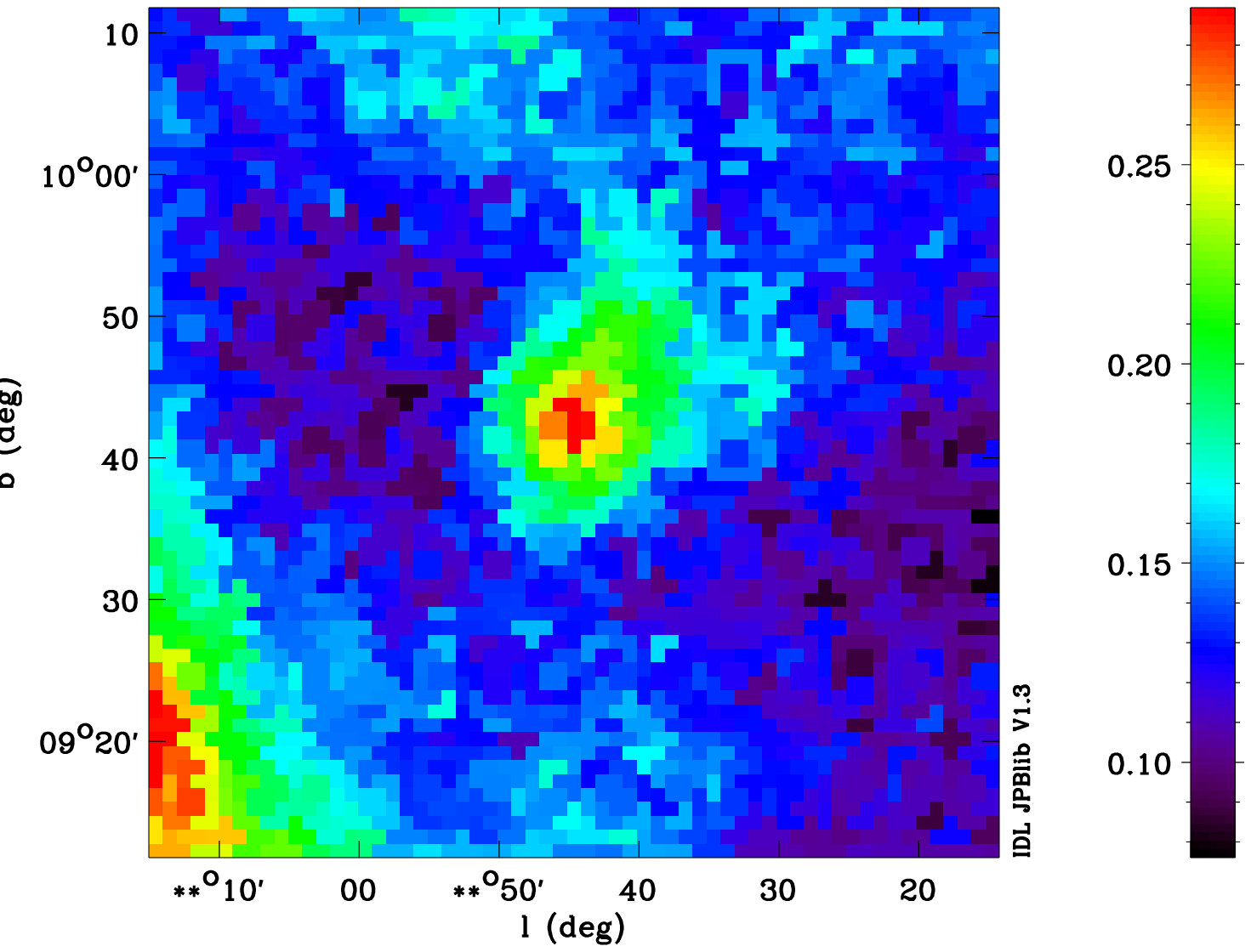} &
 \includegraphics[width=4cm]{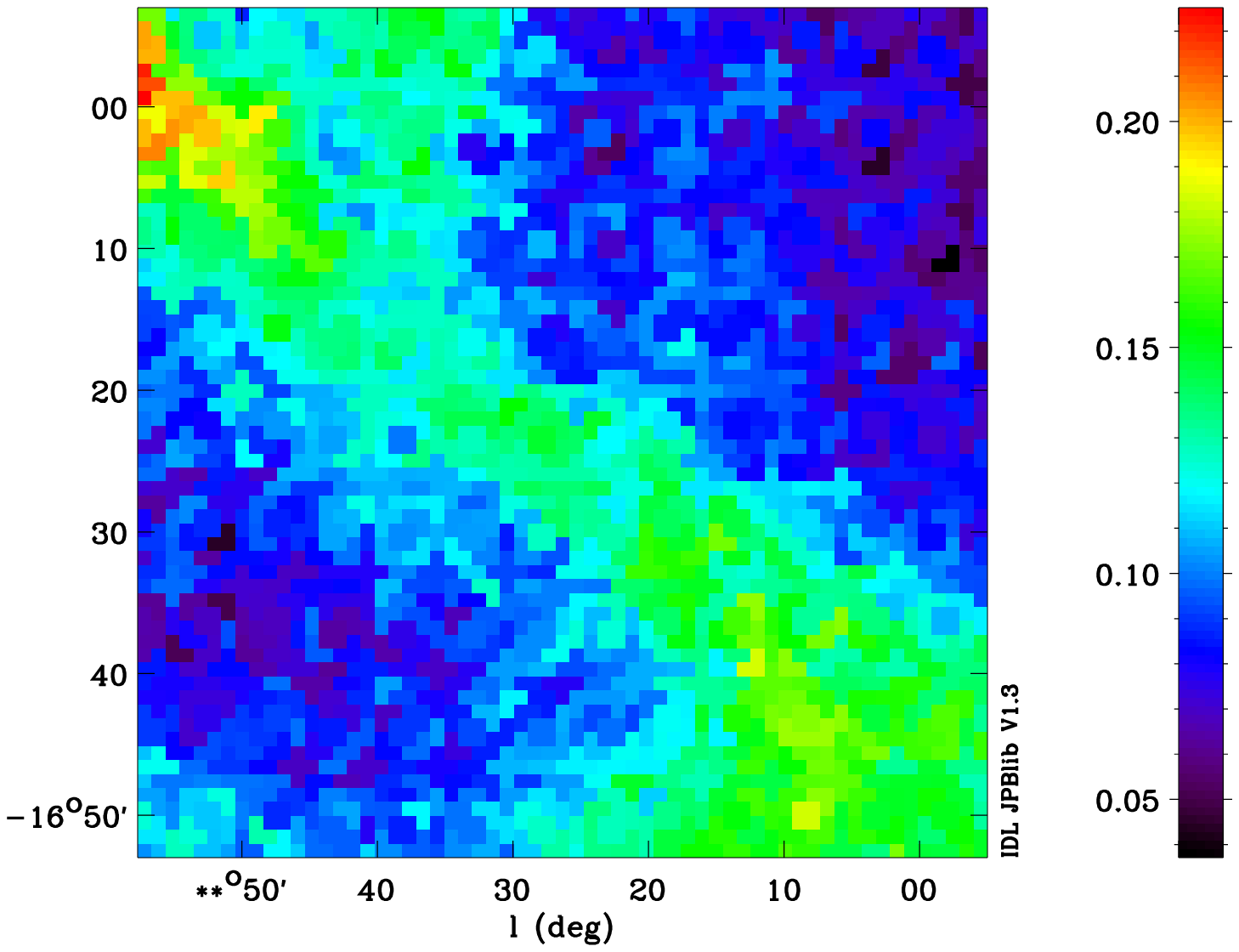} \\

 100\,GHz &
 \includegraphics[width=4cm]{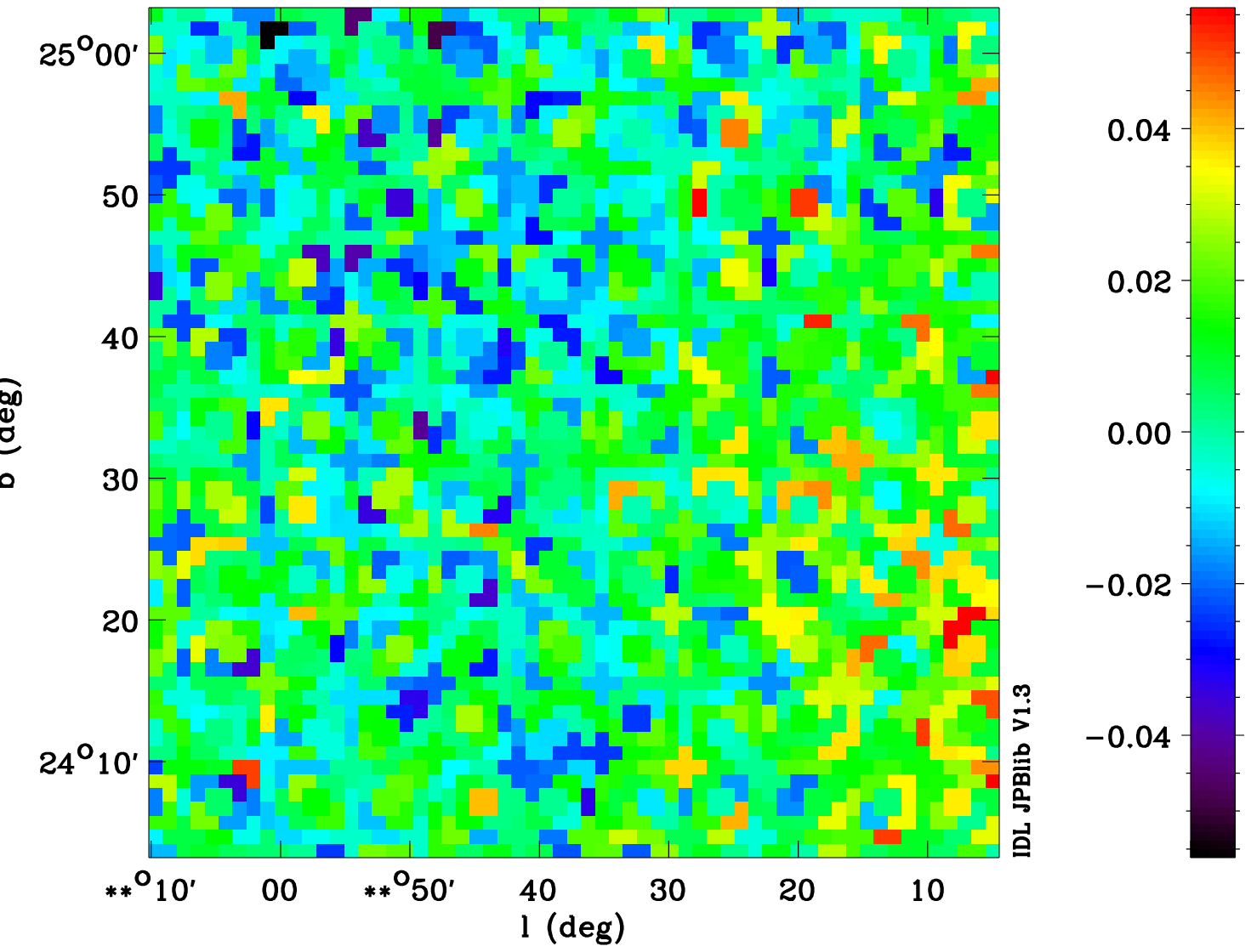} &
 \includegraphics[width=4cm]{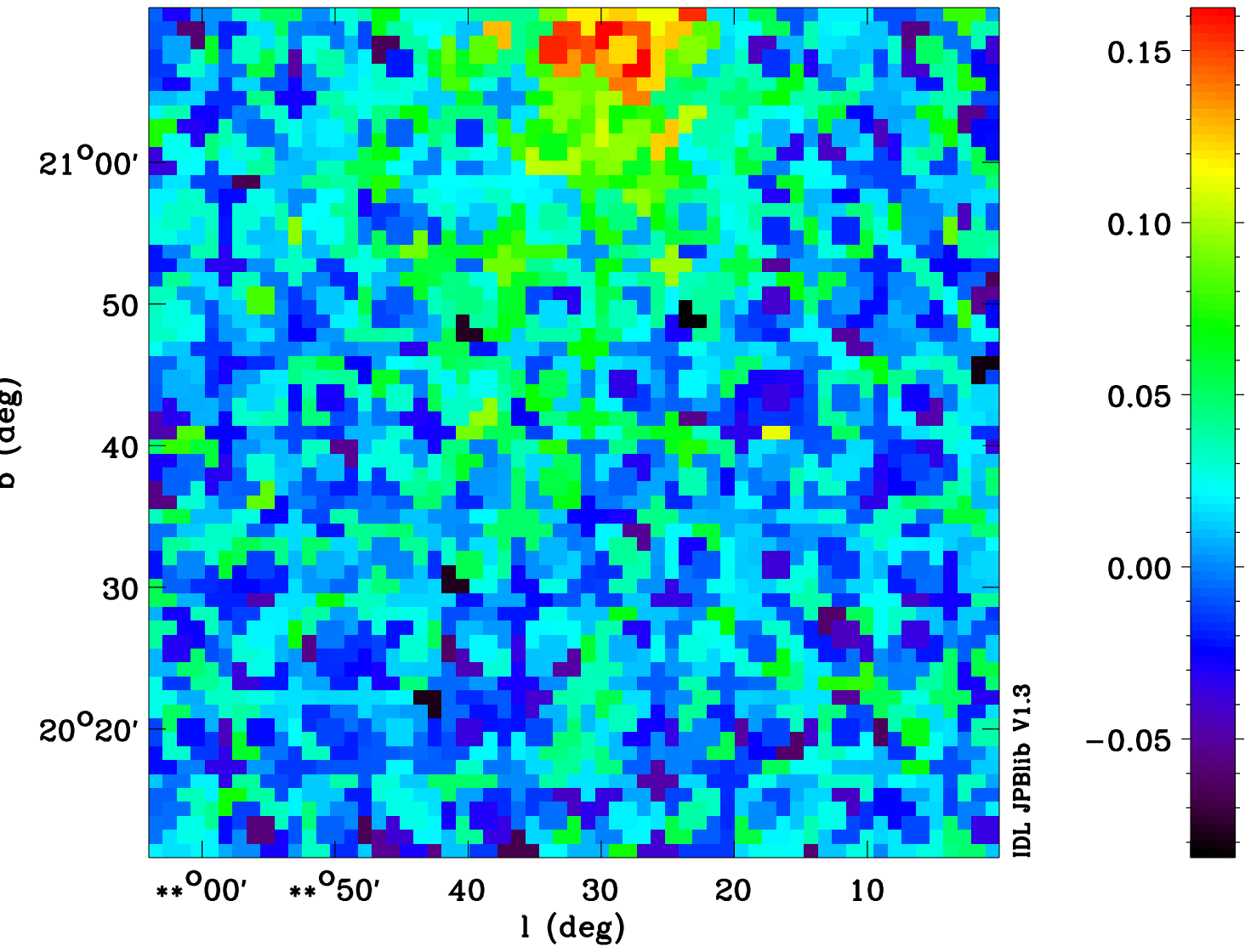} &
 \includegraphics[width=4cm]{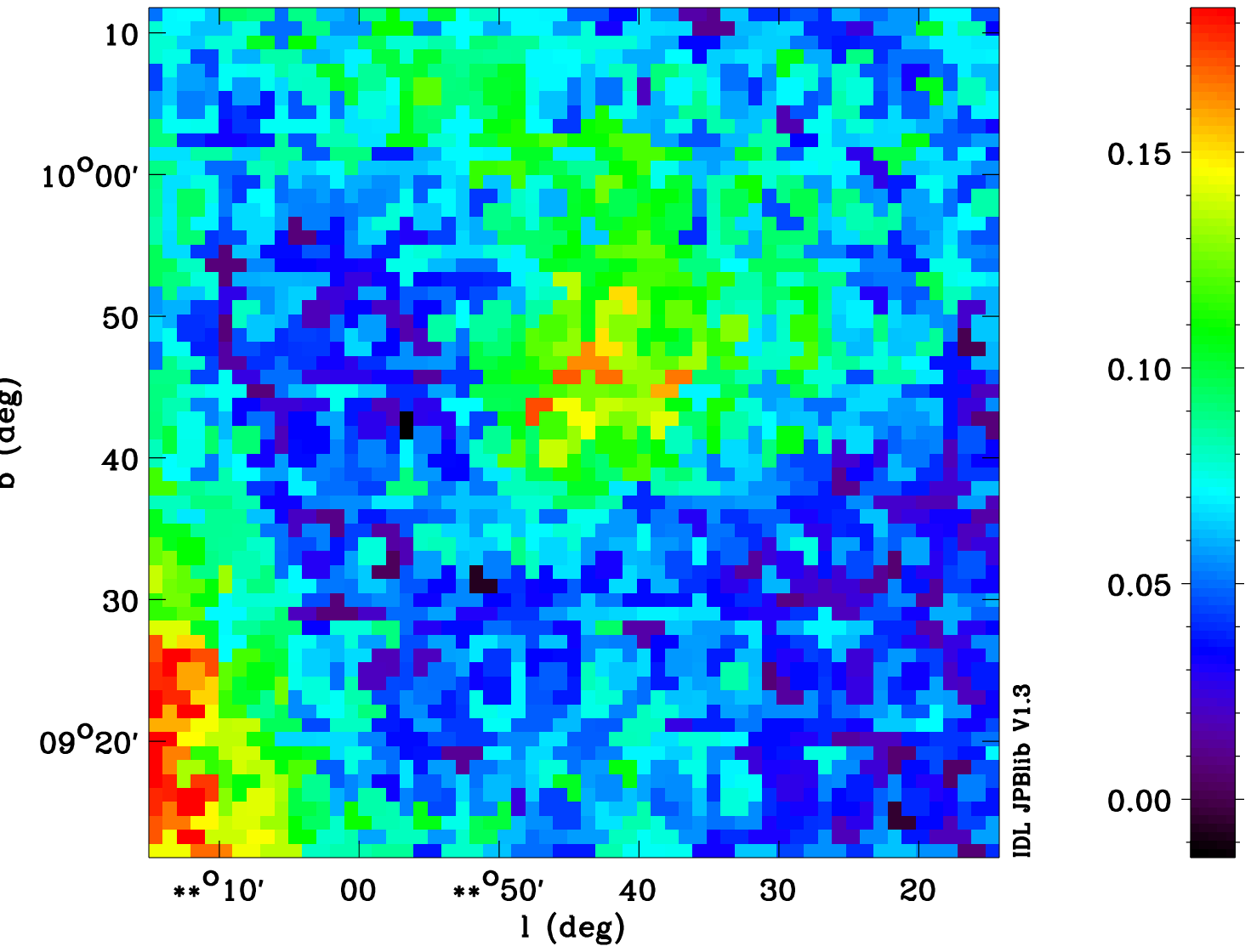} &
 \includegraphics[width=4cm]{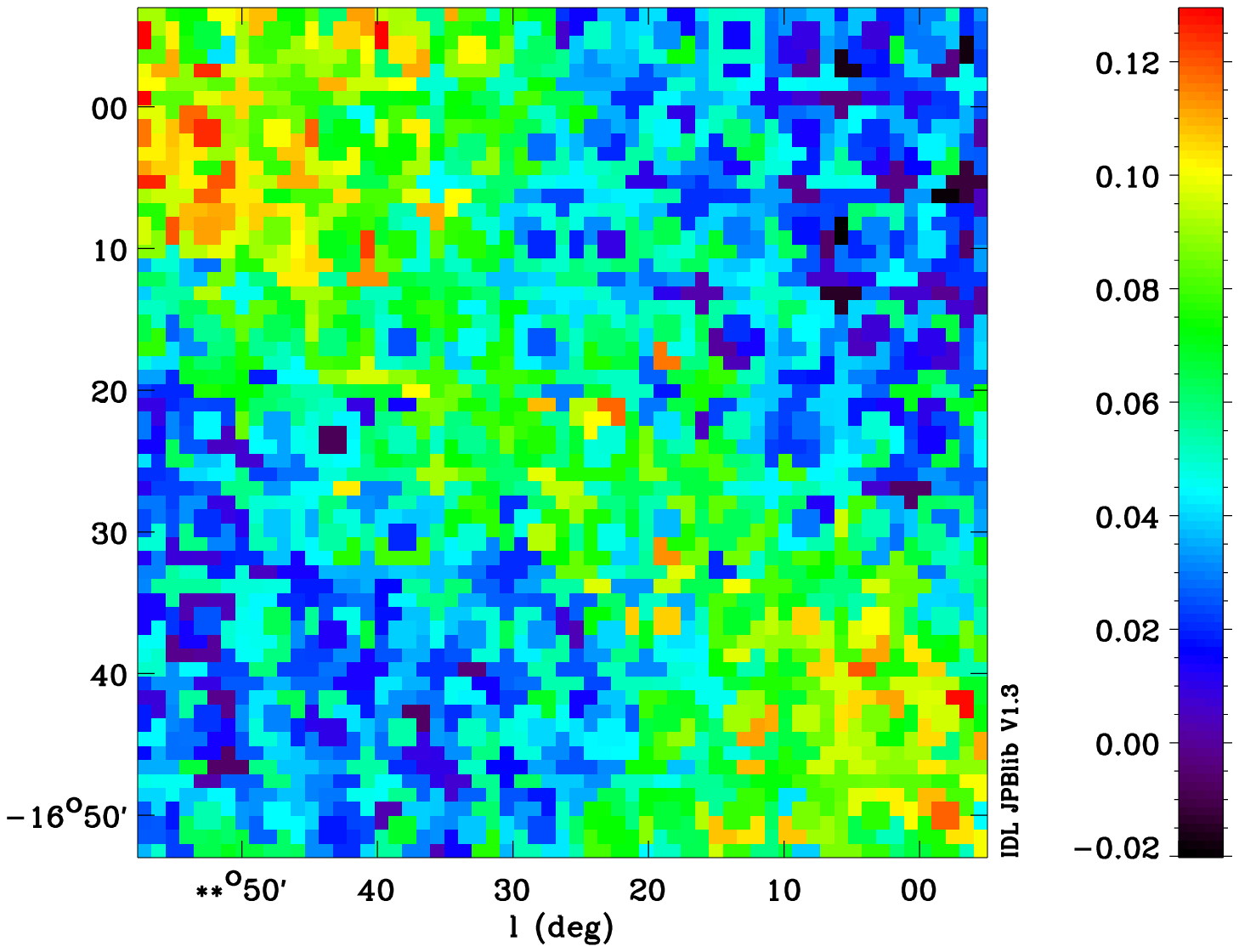} \\

\end{tabular}
\caption{Multi-band emission maps of the sources S1, S2, S3, and S4.}
\label{fig:allnu_sources1}
\end{figure*}

\begin{figure*}
\center
\begin{tabular}{ccccc}

Band & {\tiny S5\qquad\quad} & {\tiny S6\qquad\quad} & {\tiny S7\qquad\quad} &
 {\tiny S8\qquad\quad}  \\
 $100\,\mu$m &
 \includegraphics[width=4cm]{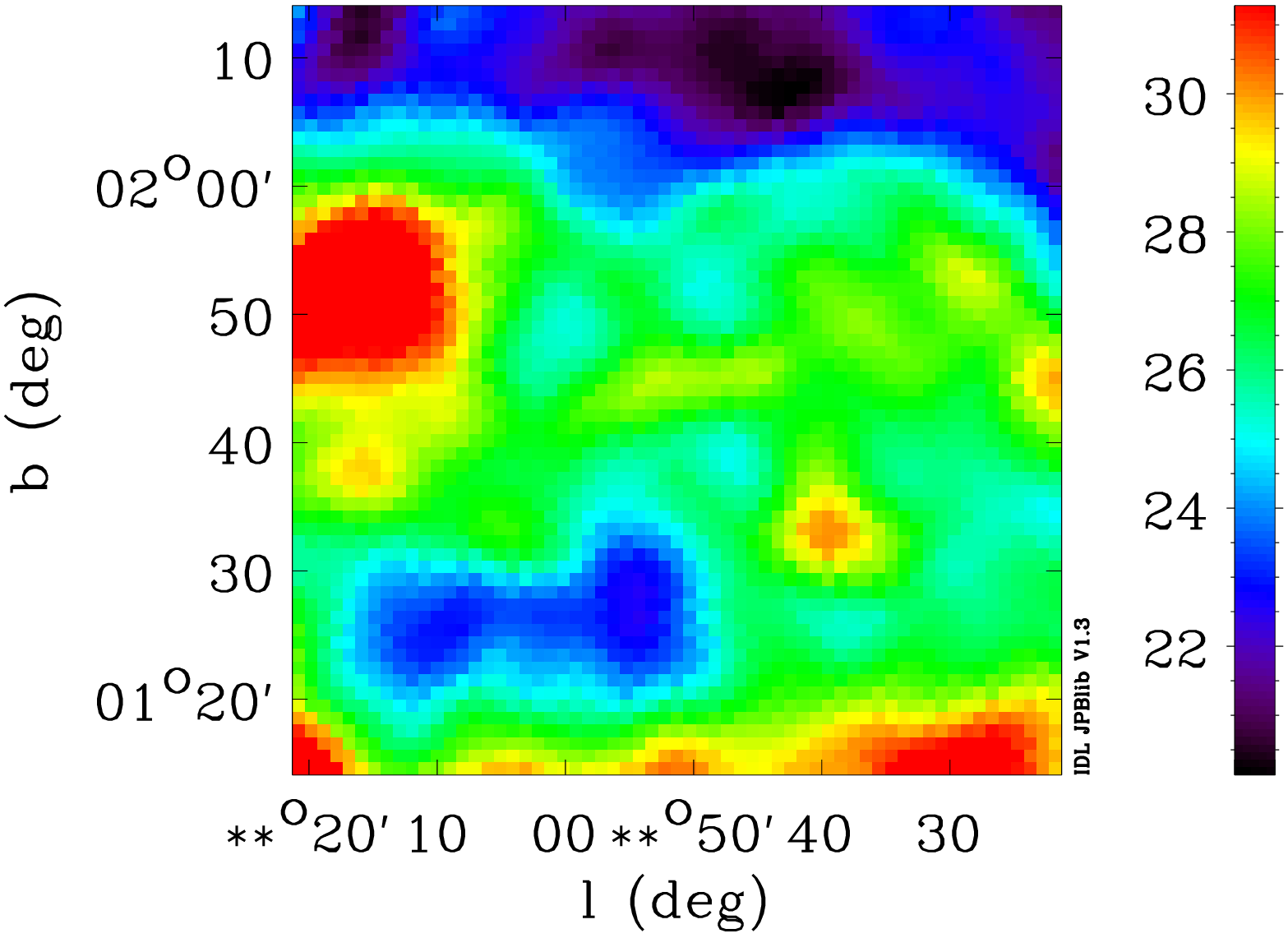} &
 \includegraphics[width=4cm]{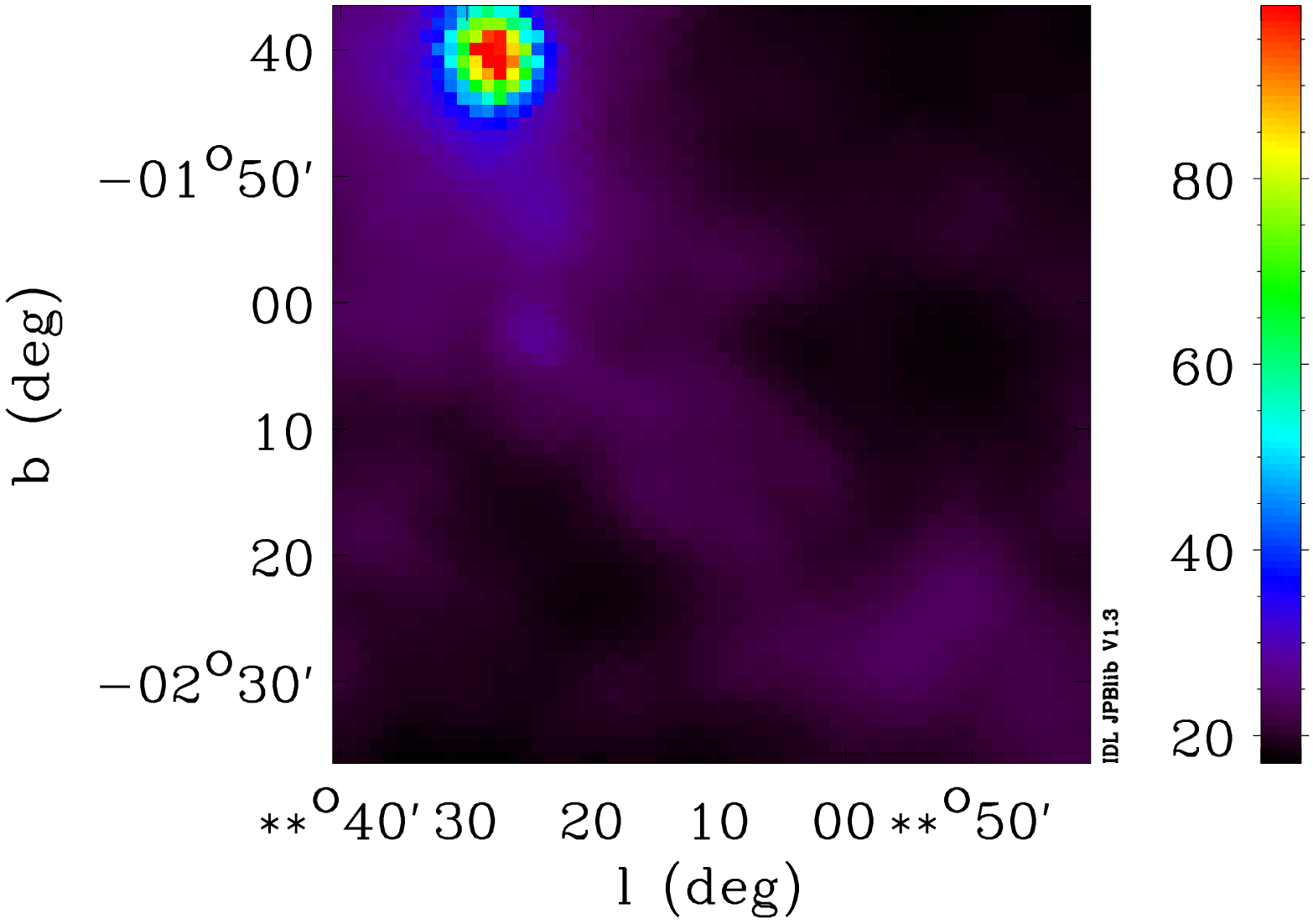} &
 \includegraphics[width=4cm]{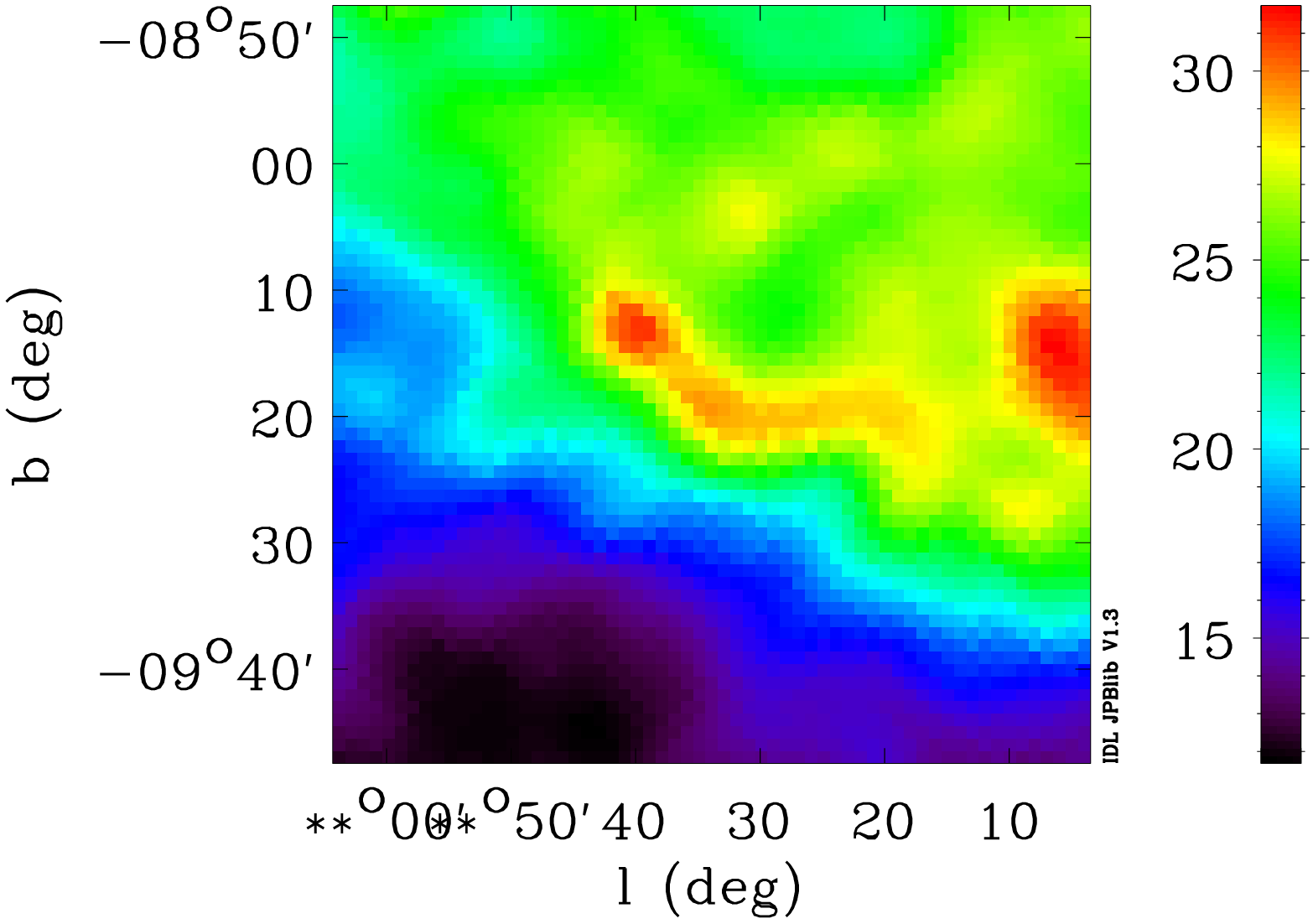} &
 \includegraphics[width=4cm]{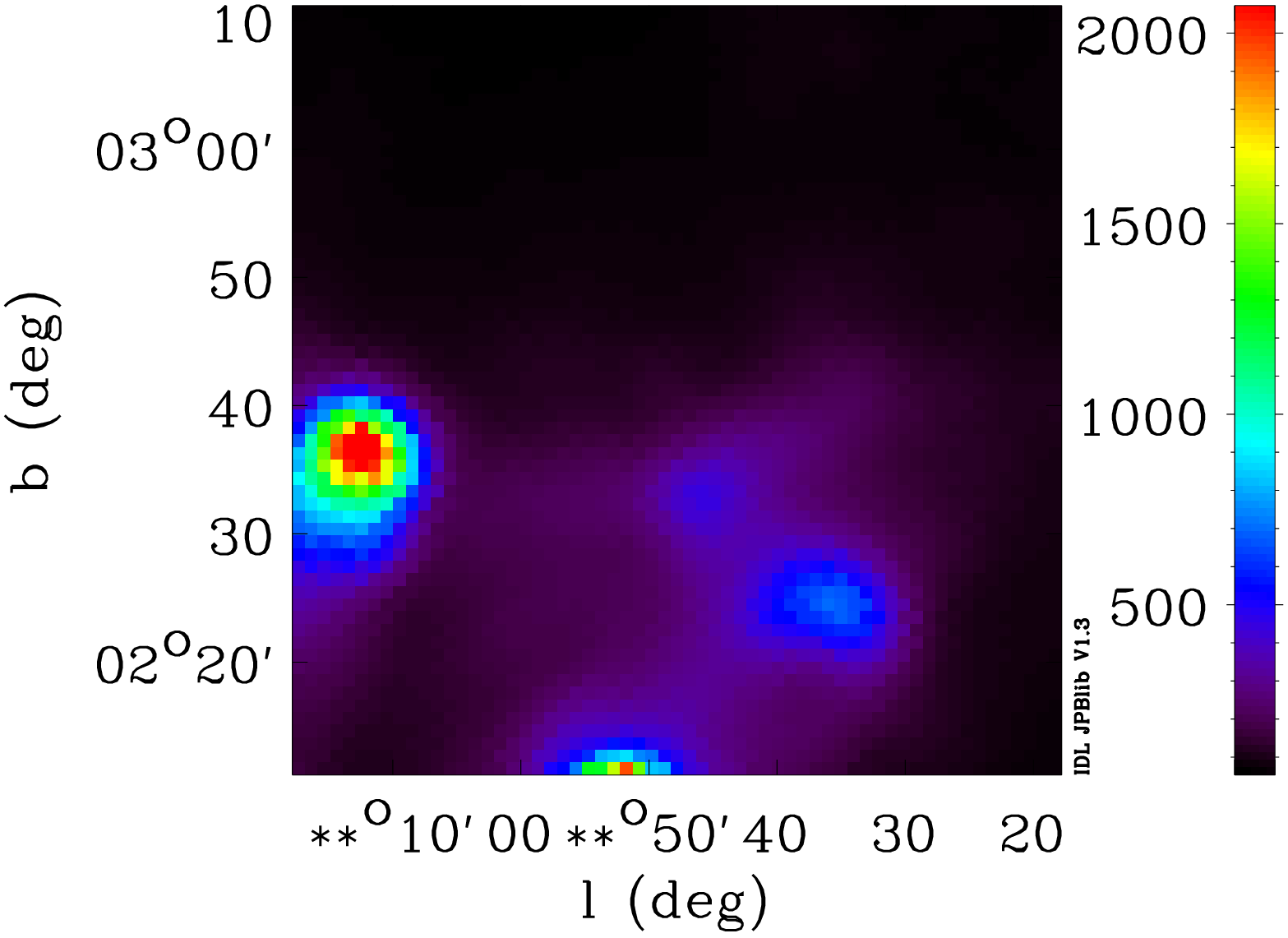} \\

 857\,GHz &
 \includegraphics[width=4cm]{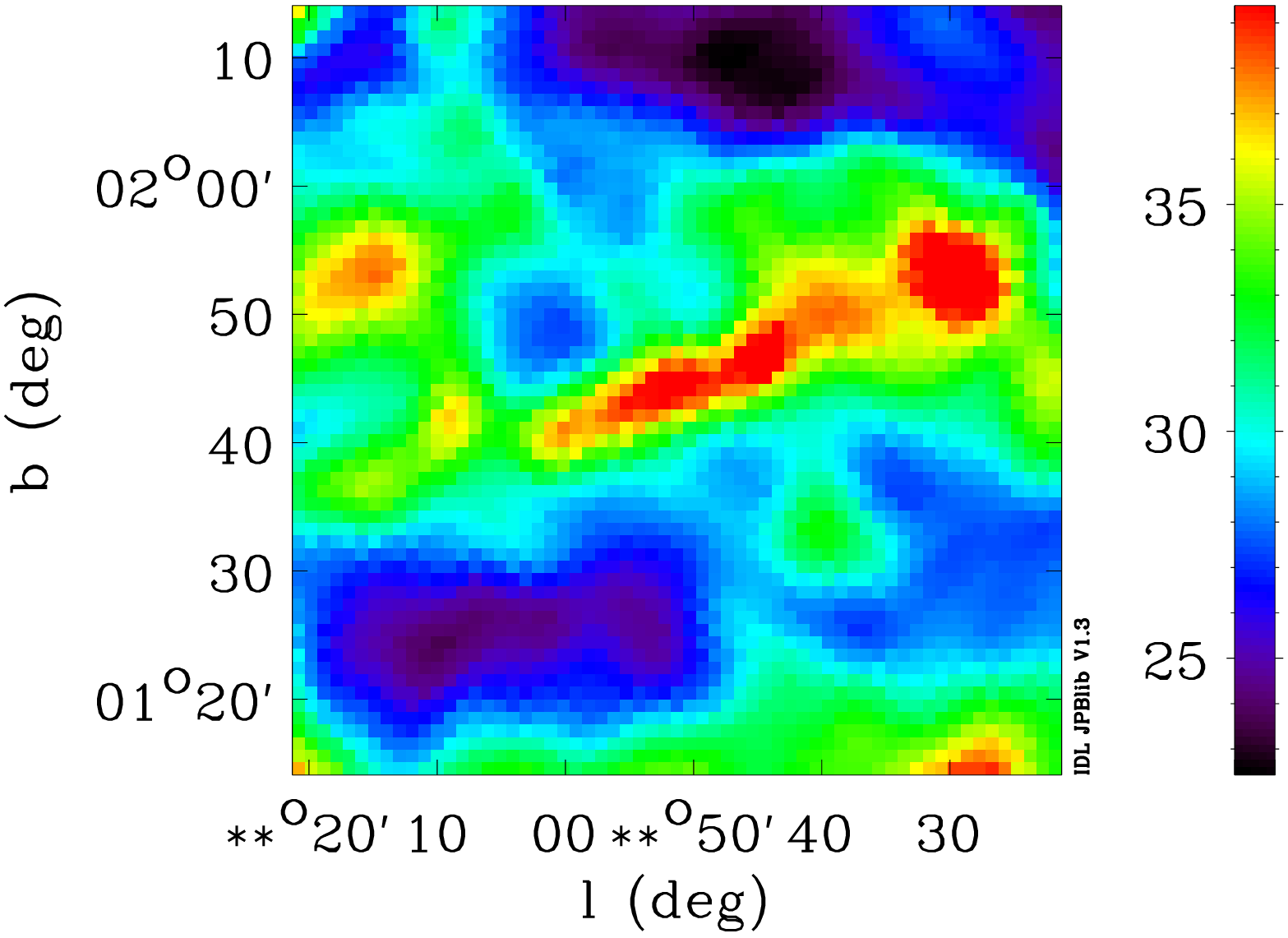} &
 \includegraphics[width=4cm]{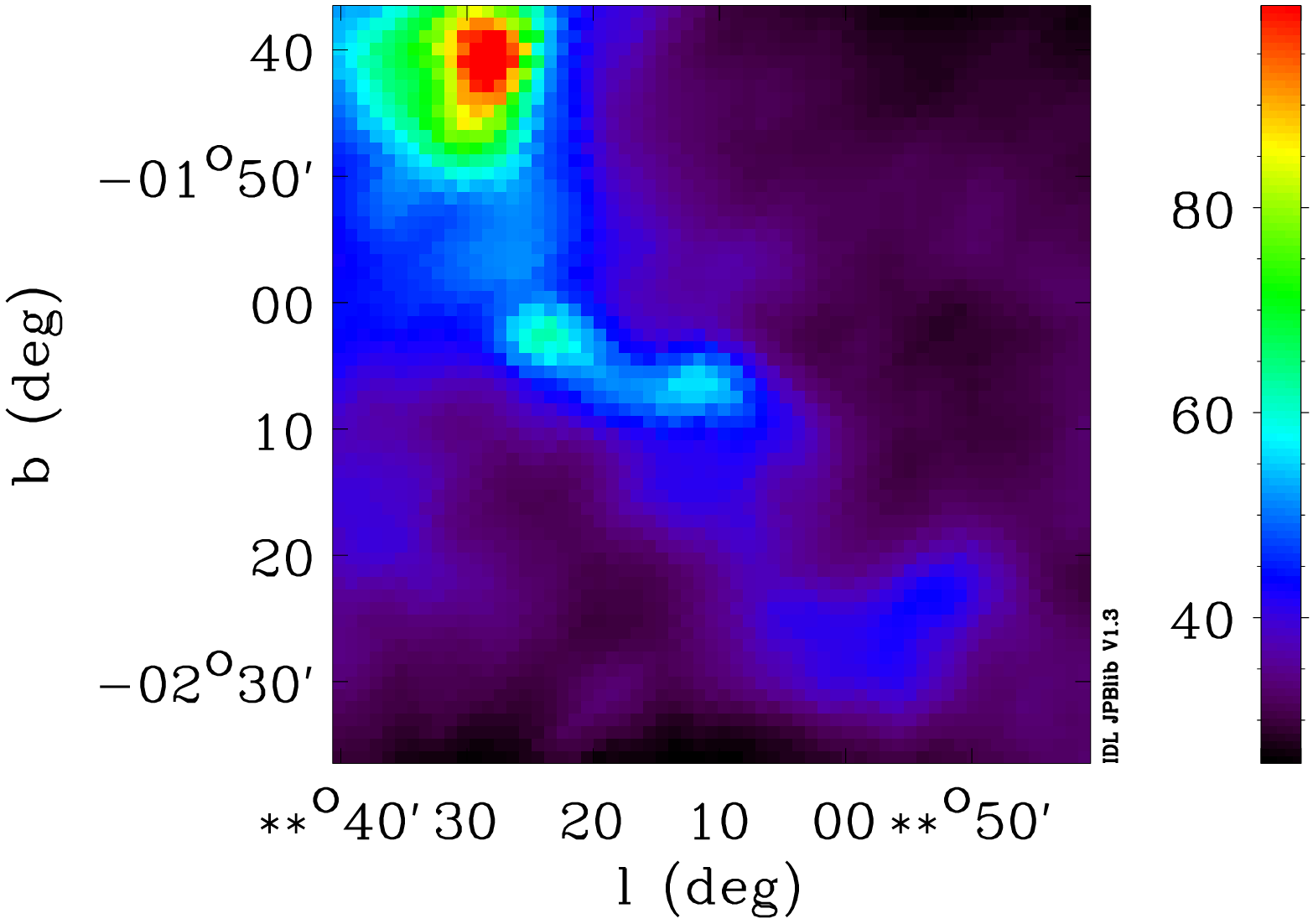} &
 \includegraphics[width=4cm]{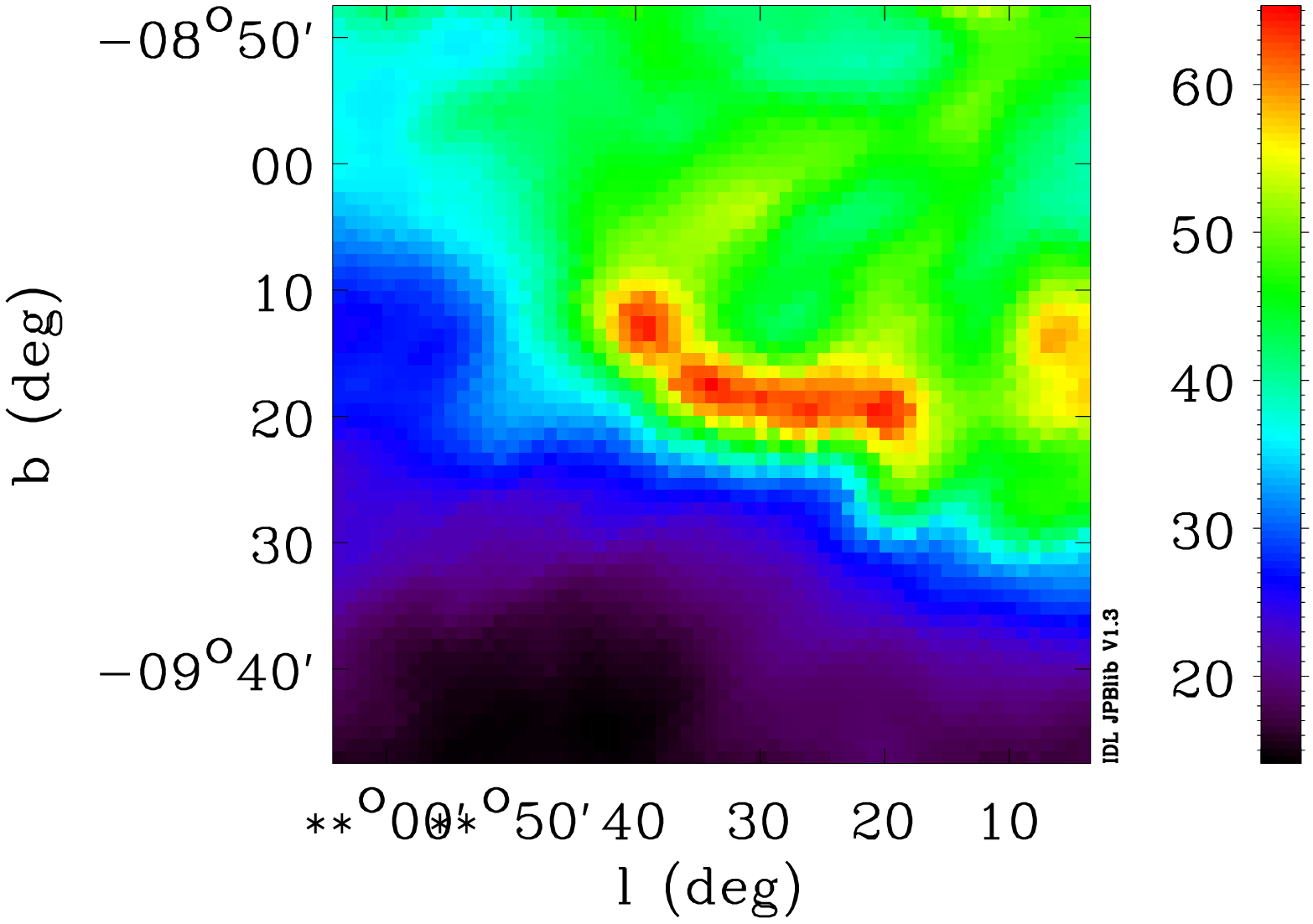} &
 \includegraphics[width=4cm]{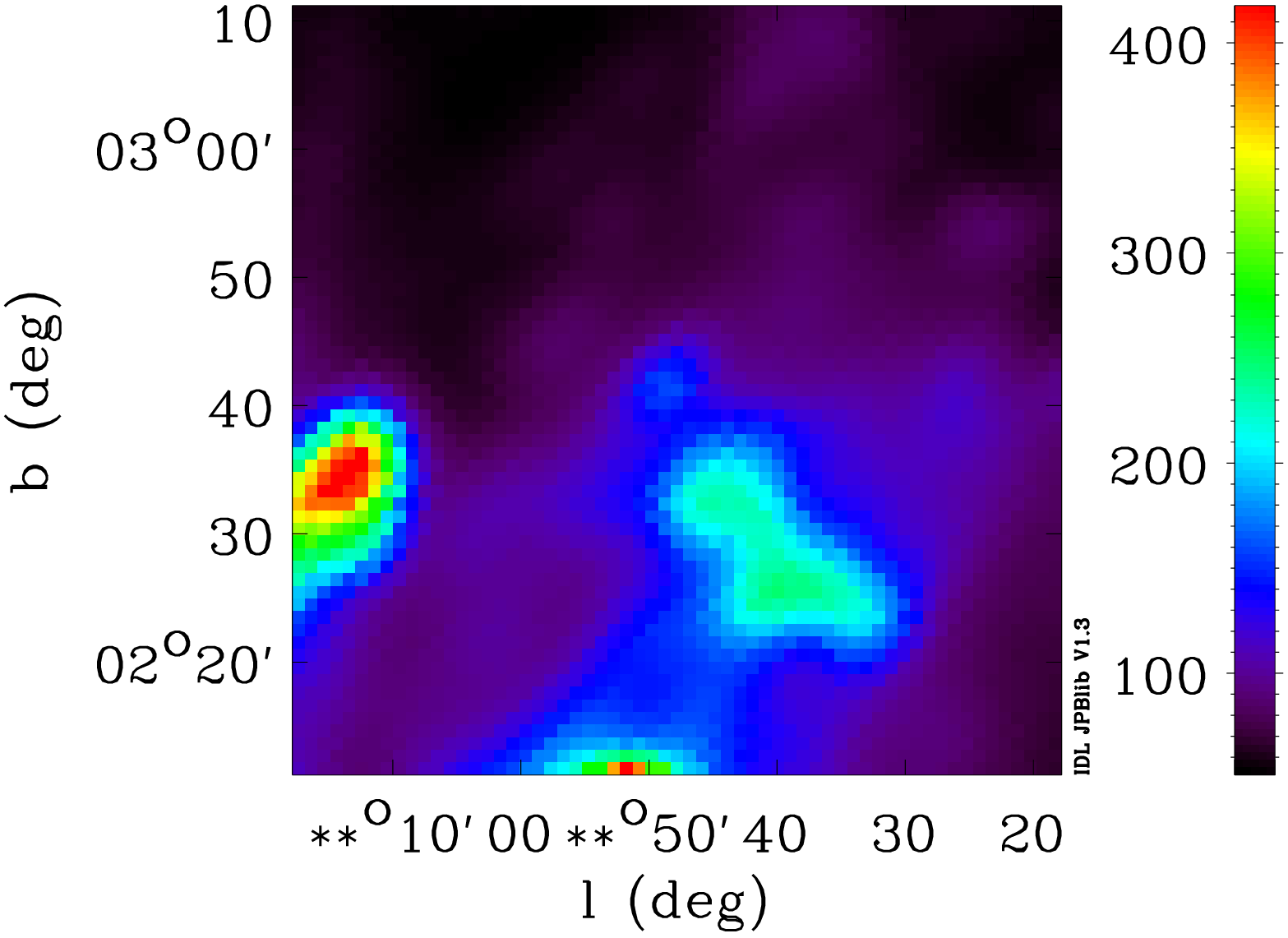} \\

\tiny{Cold Residual} &
 \includegraphics[width=4cm]{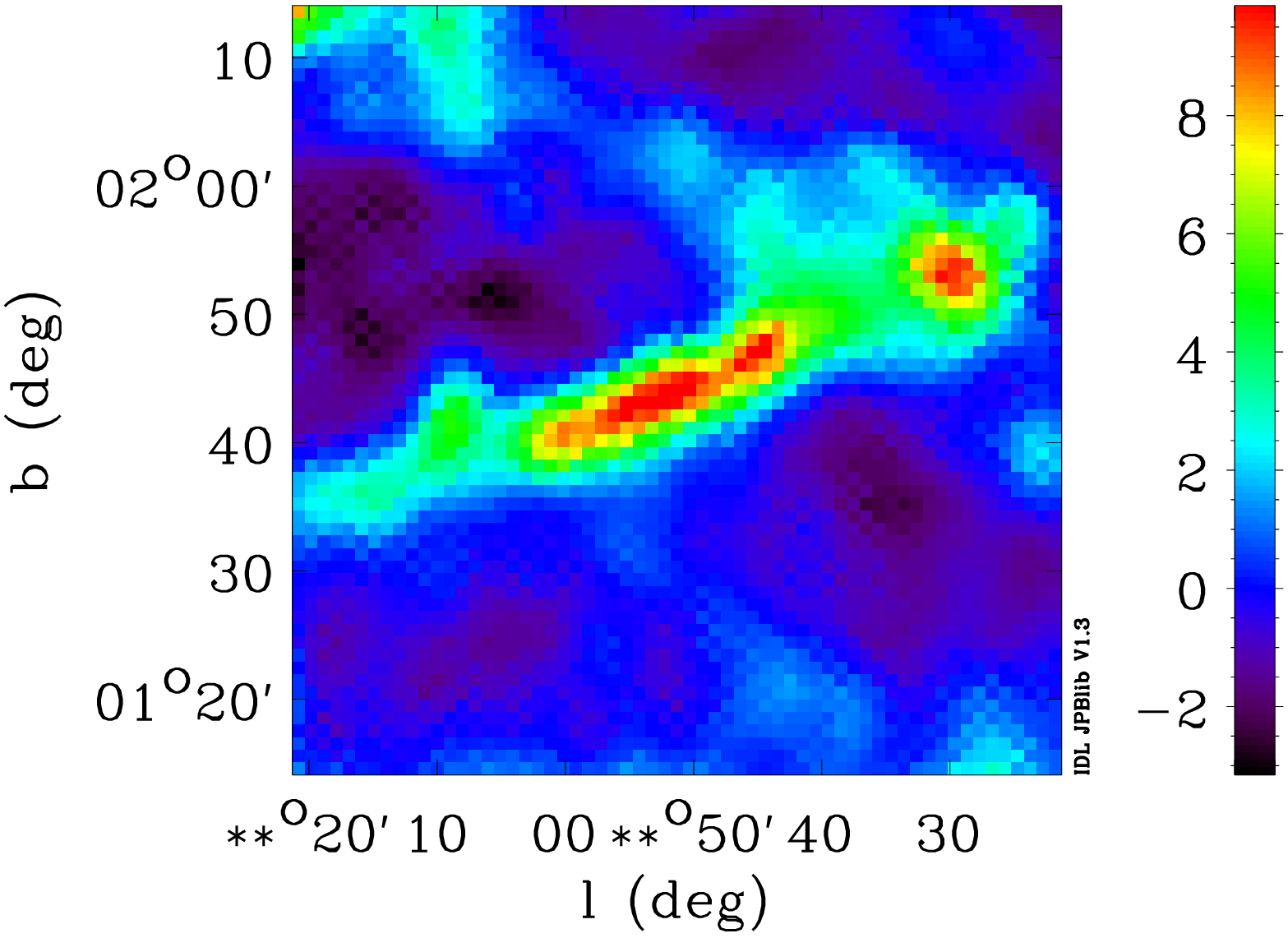} &
 \includegraphics[width=4cm]{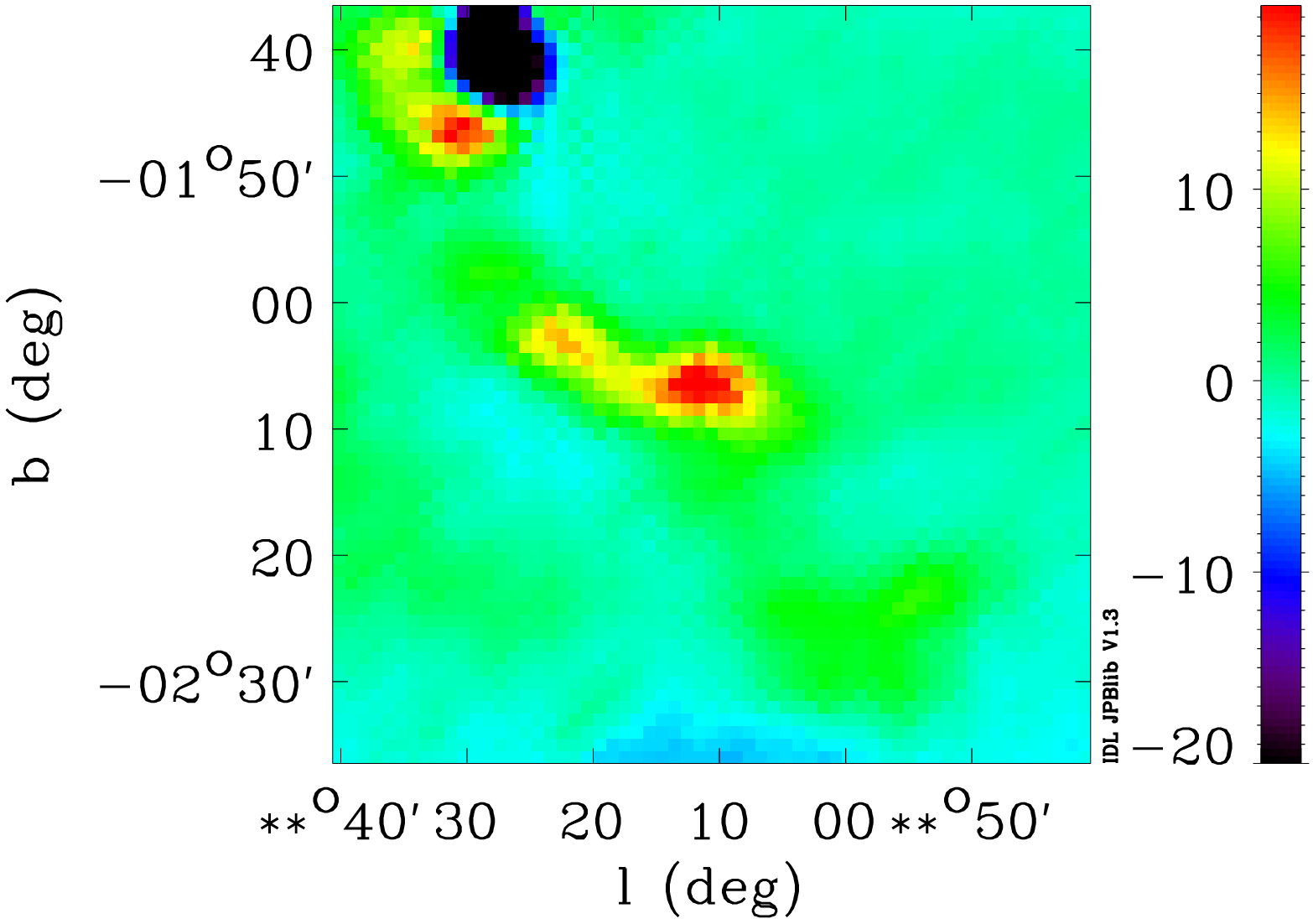} &
 \includegraphics[width=4cm]{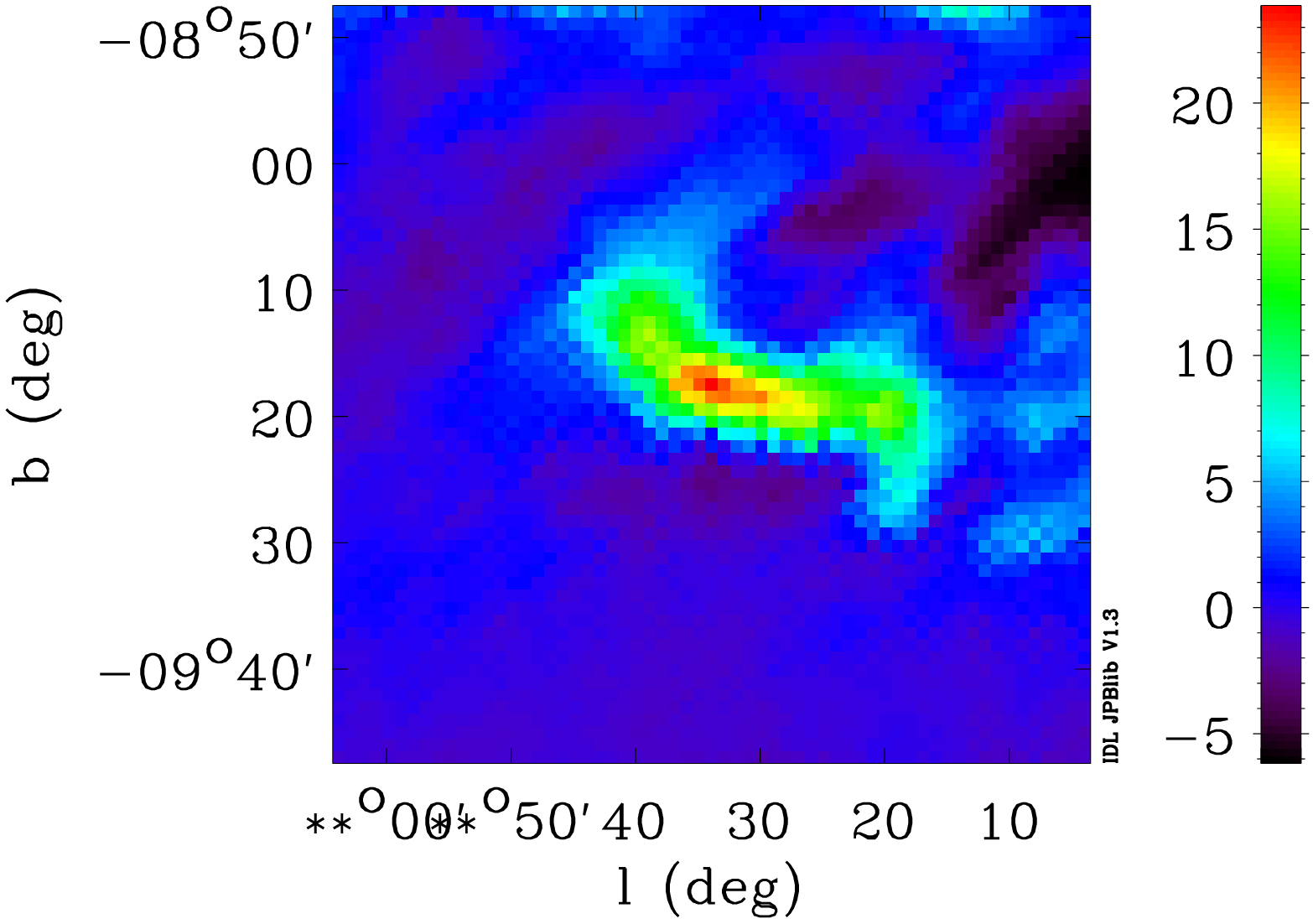} &
 \includegraphics[width=4cm]{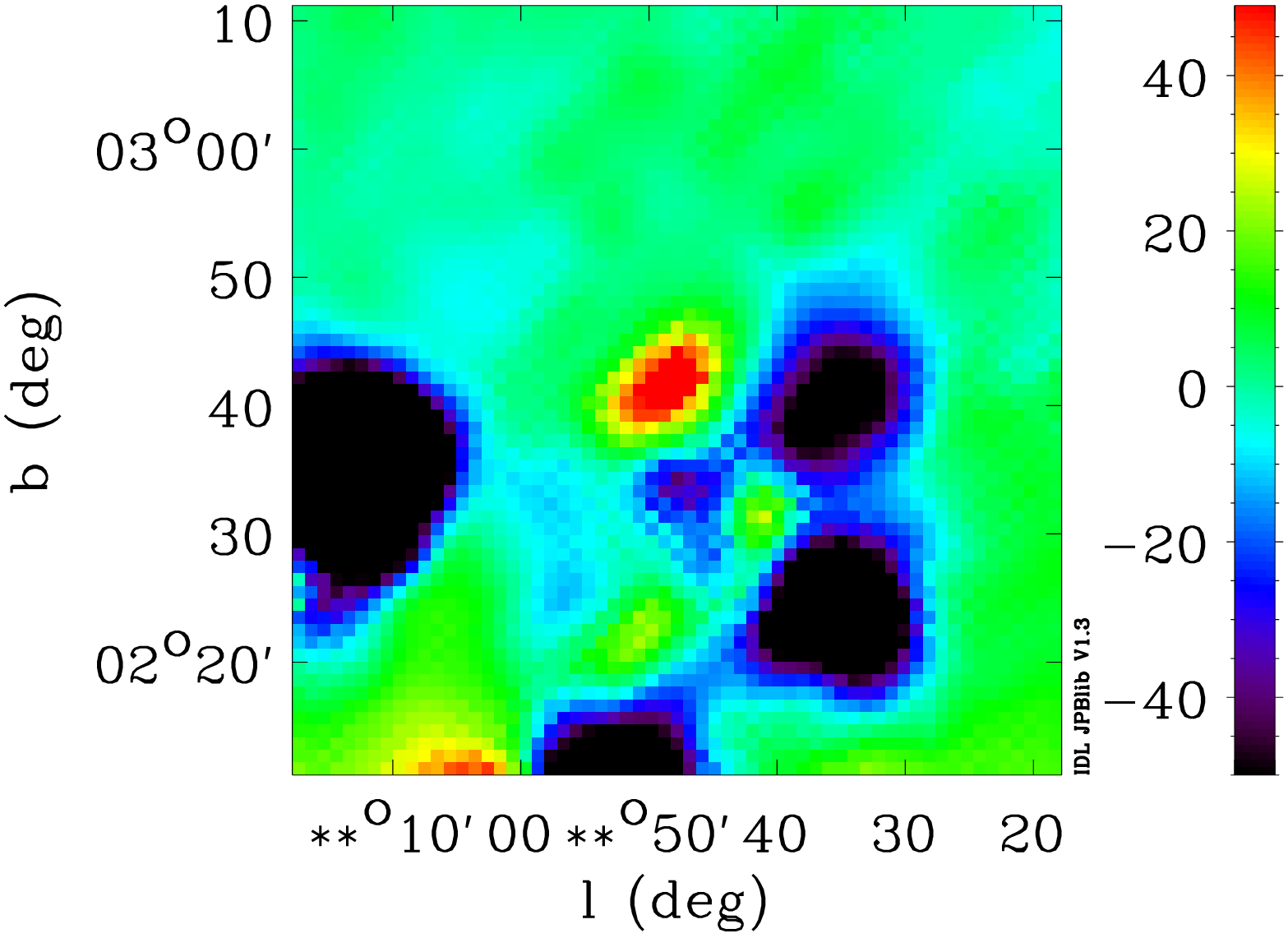} \\

 545\,GHz & \includegraphics[width=4cm]{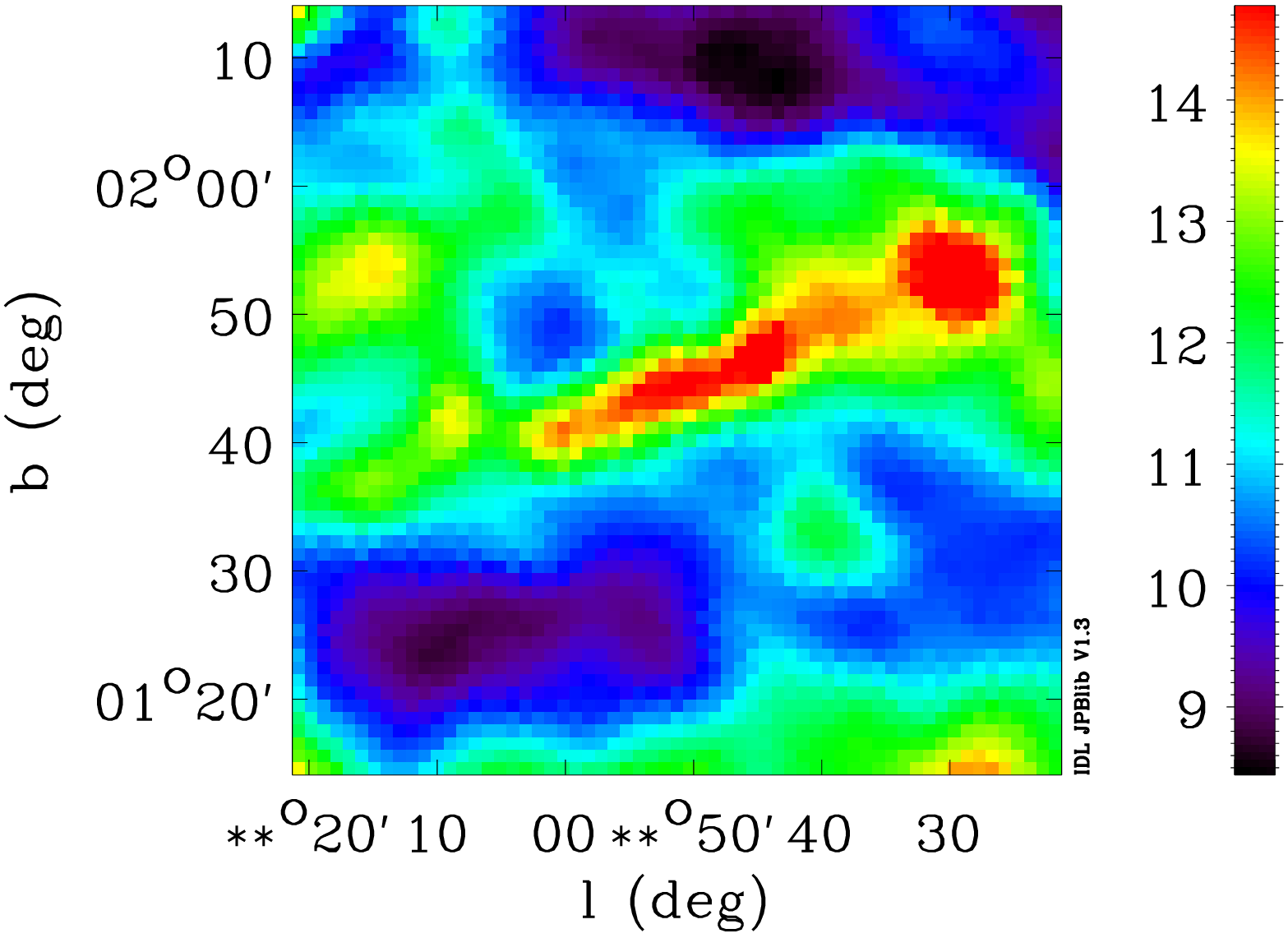} &
 \includegraphics[width=4cm]{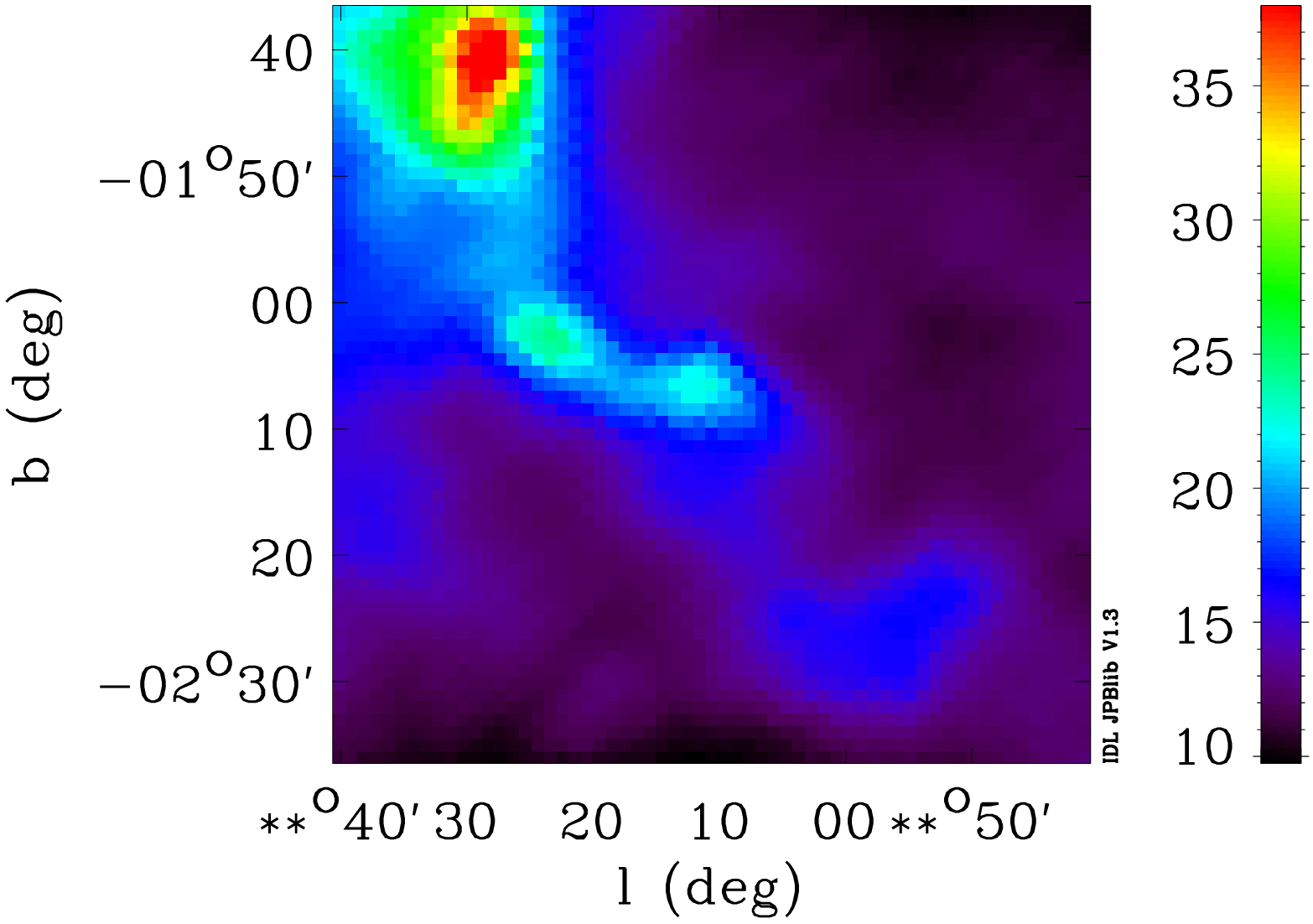} &
 \includegraphics[width=4cm]{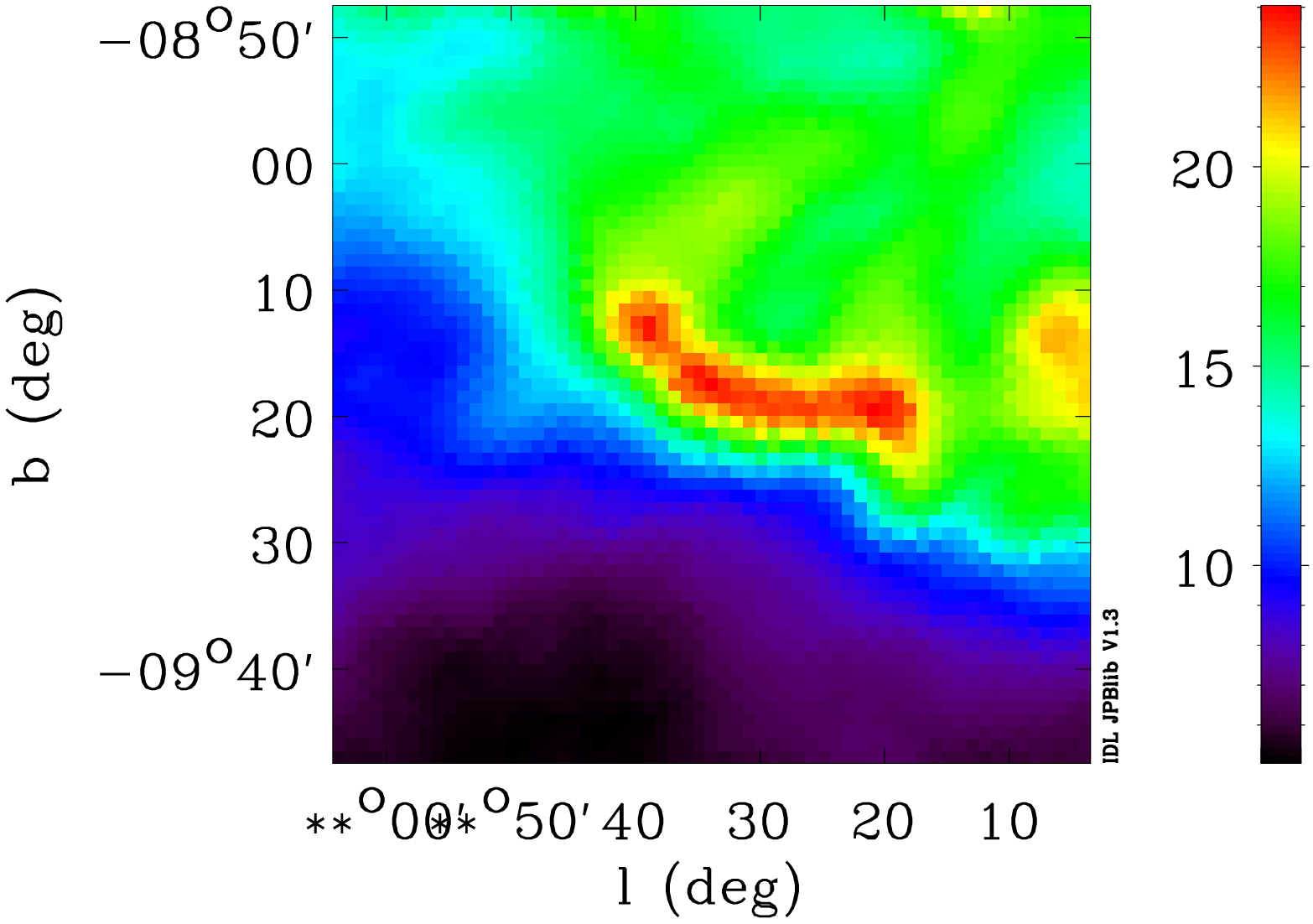} &
 \includegraphics[width=4cm]{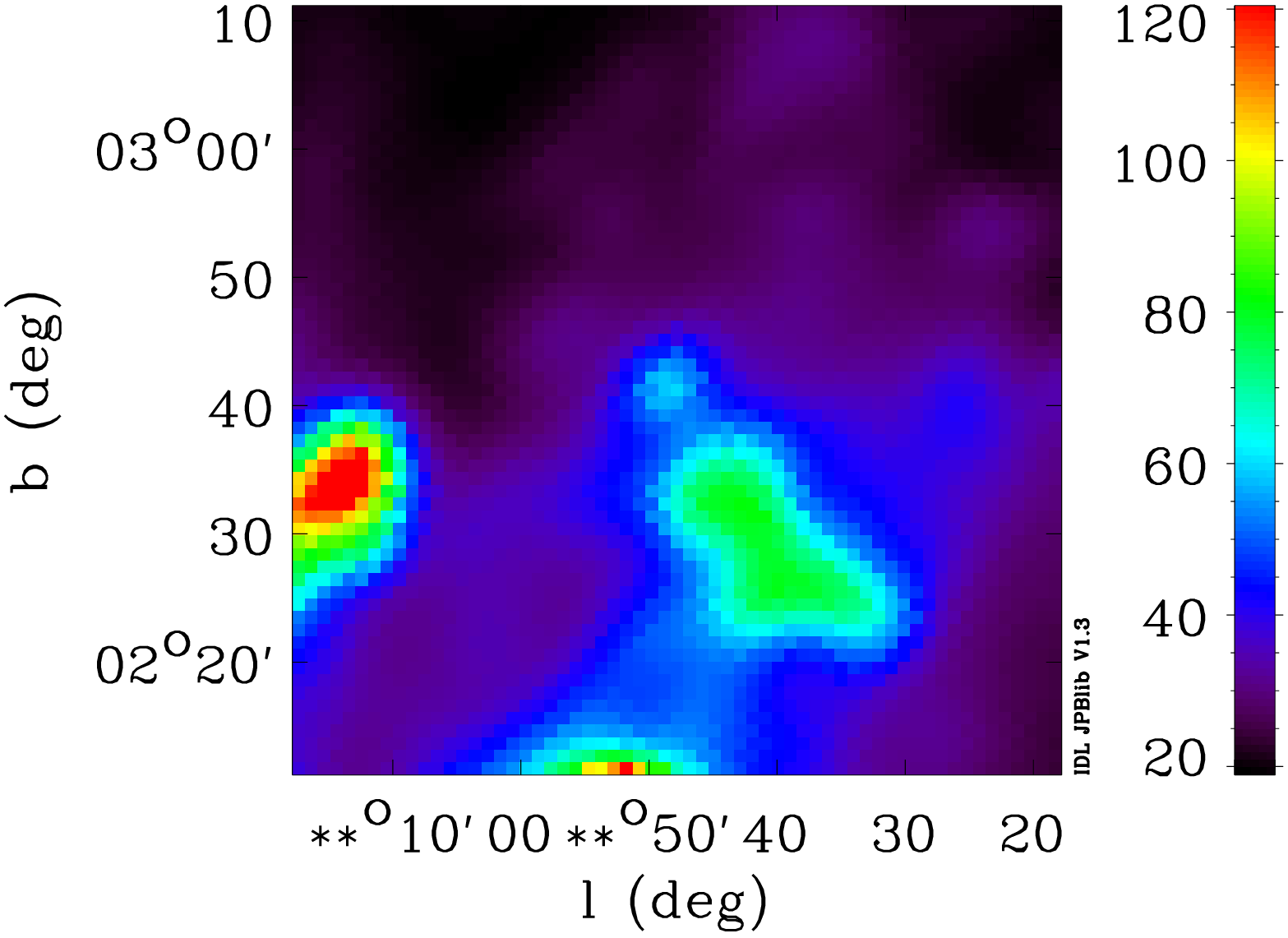} \\

 353\,GHz &
 \includegraphics[width=4cm]{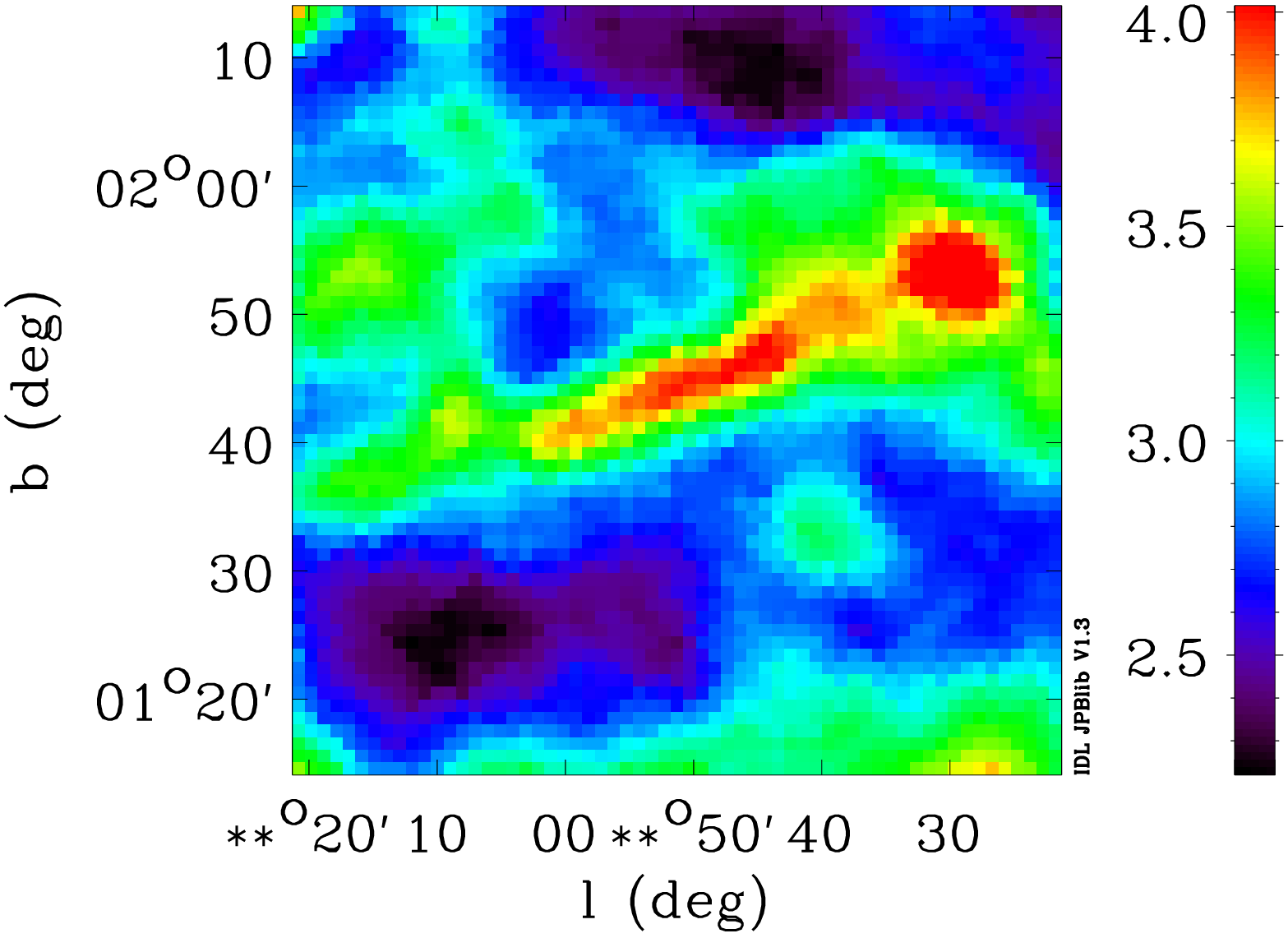} &
 \includegraphics[width=4cm]{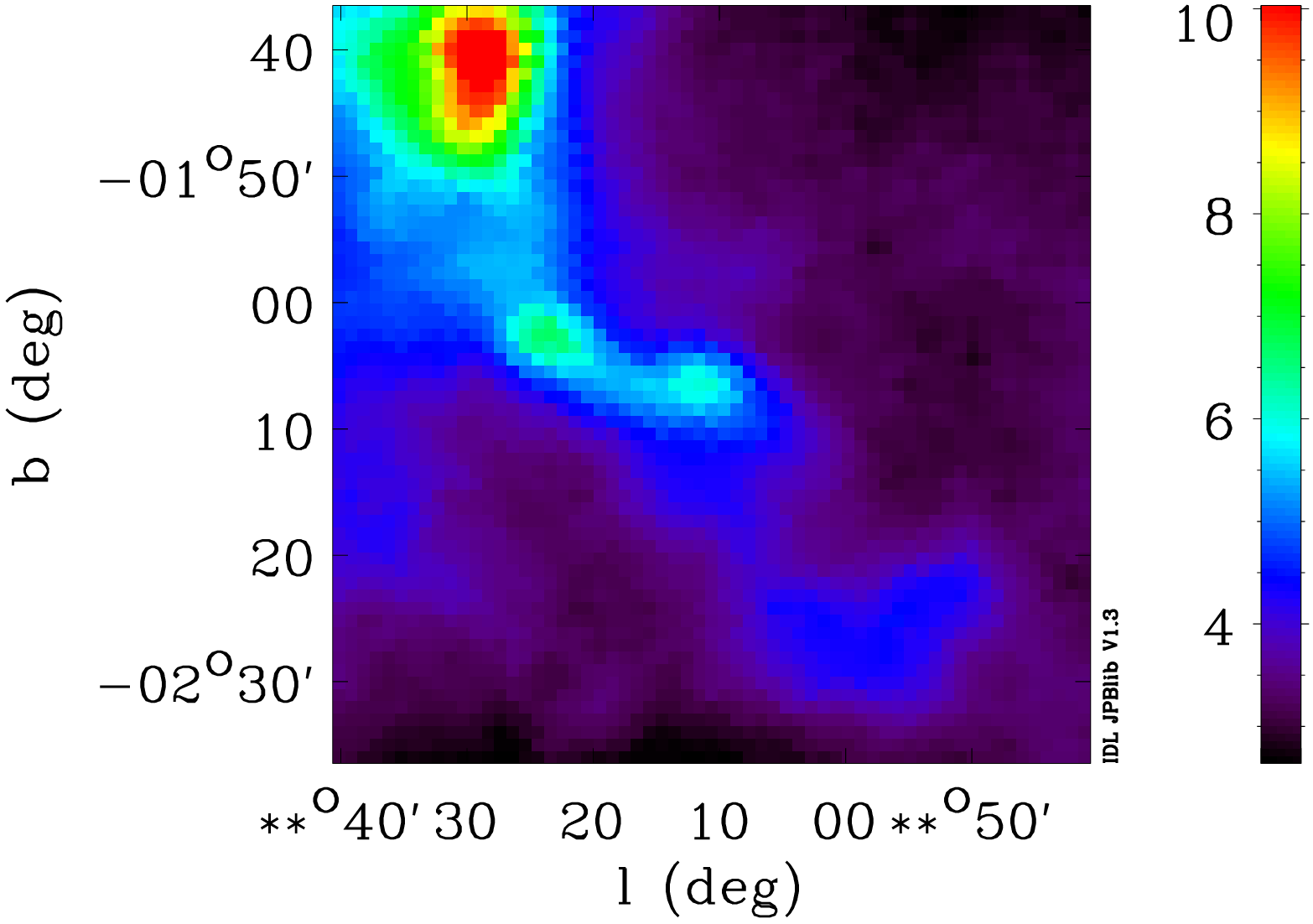} &
 \includegraphics[width=4cm]{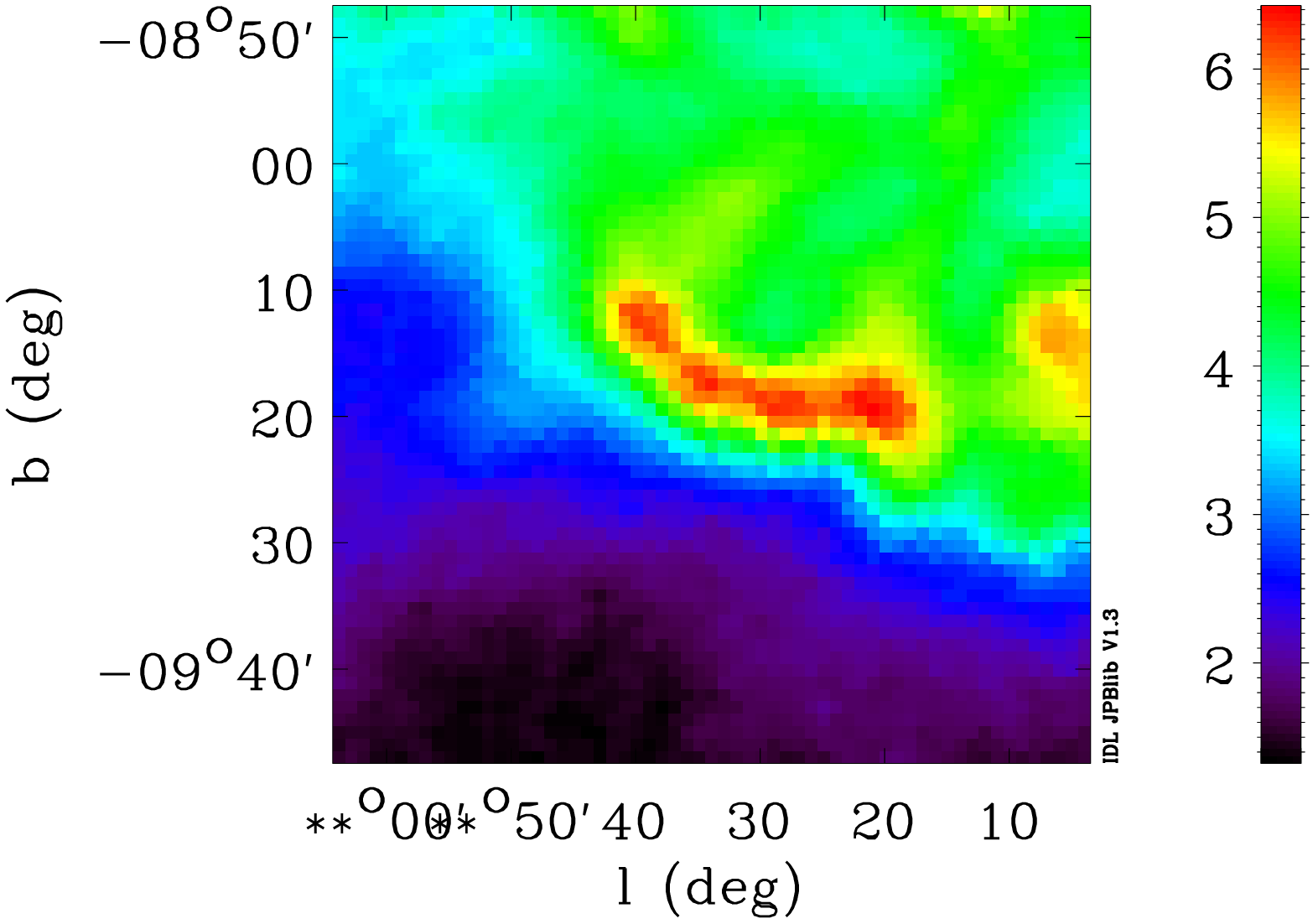} &
 \includegraphics[width=4cm]{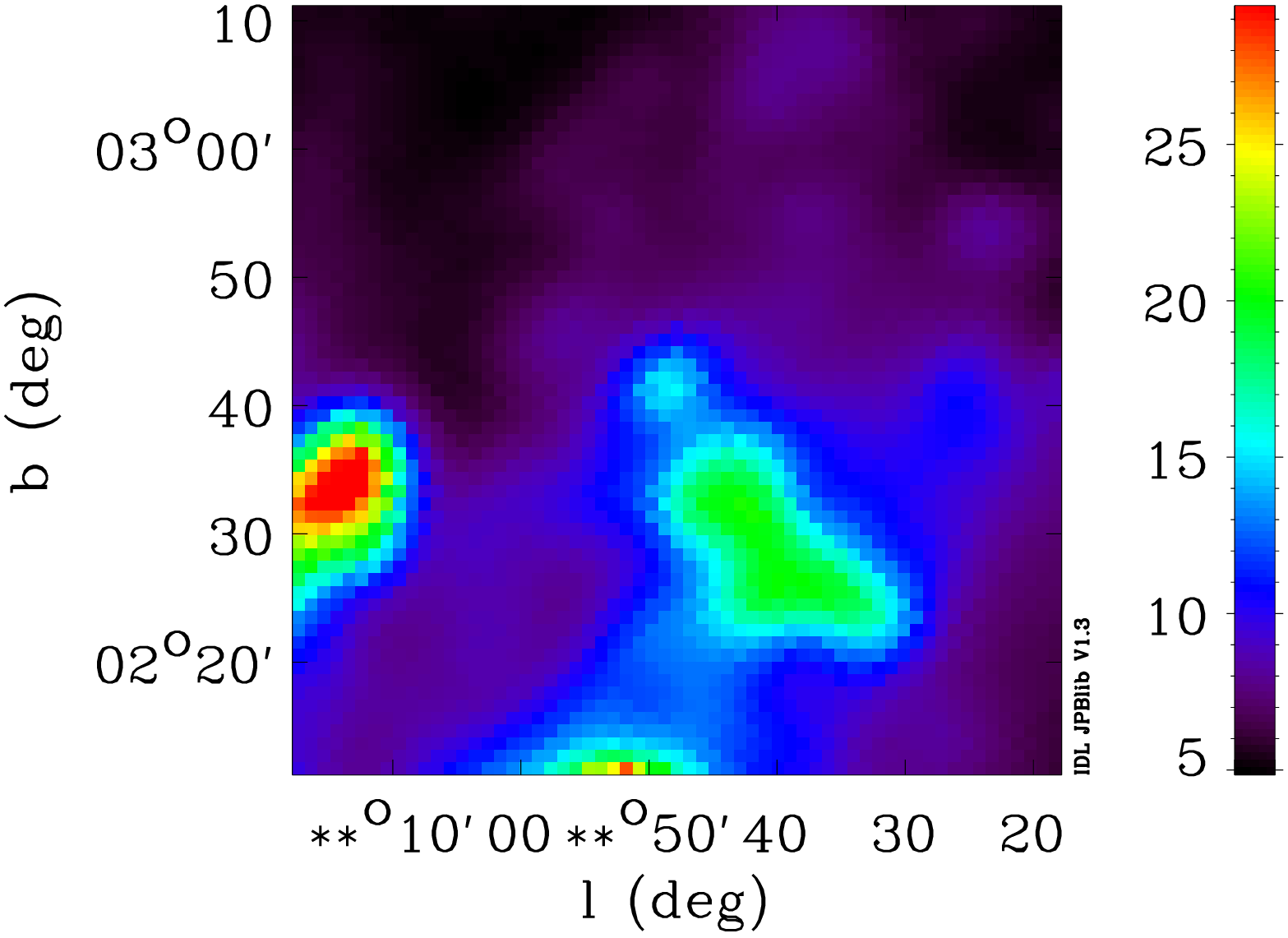} \\

 217\,GHz &
 \includegraphics[width=4cm]{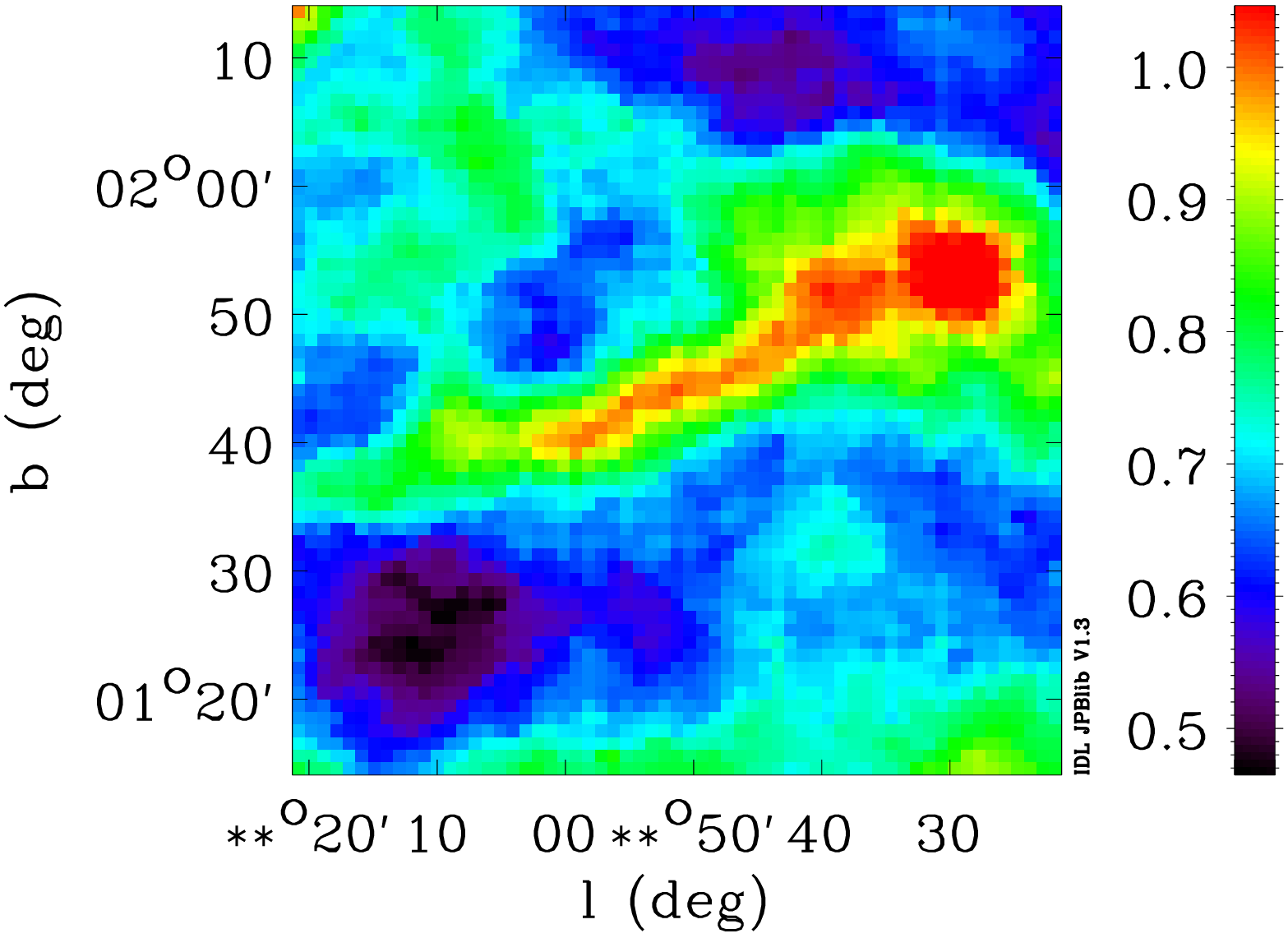} &
 \includegraphics[width=4cm]{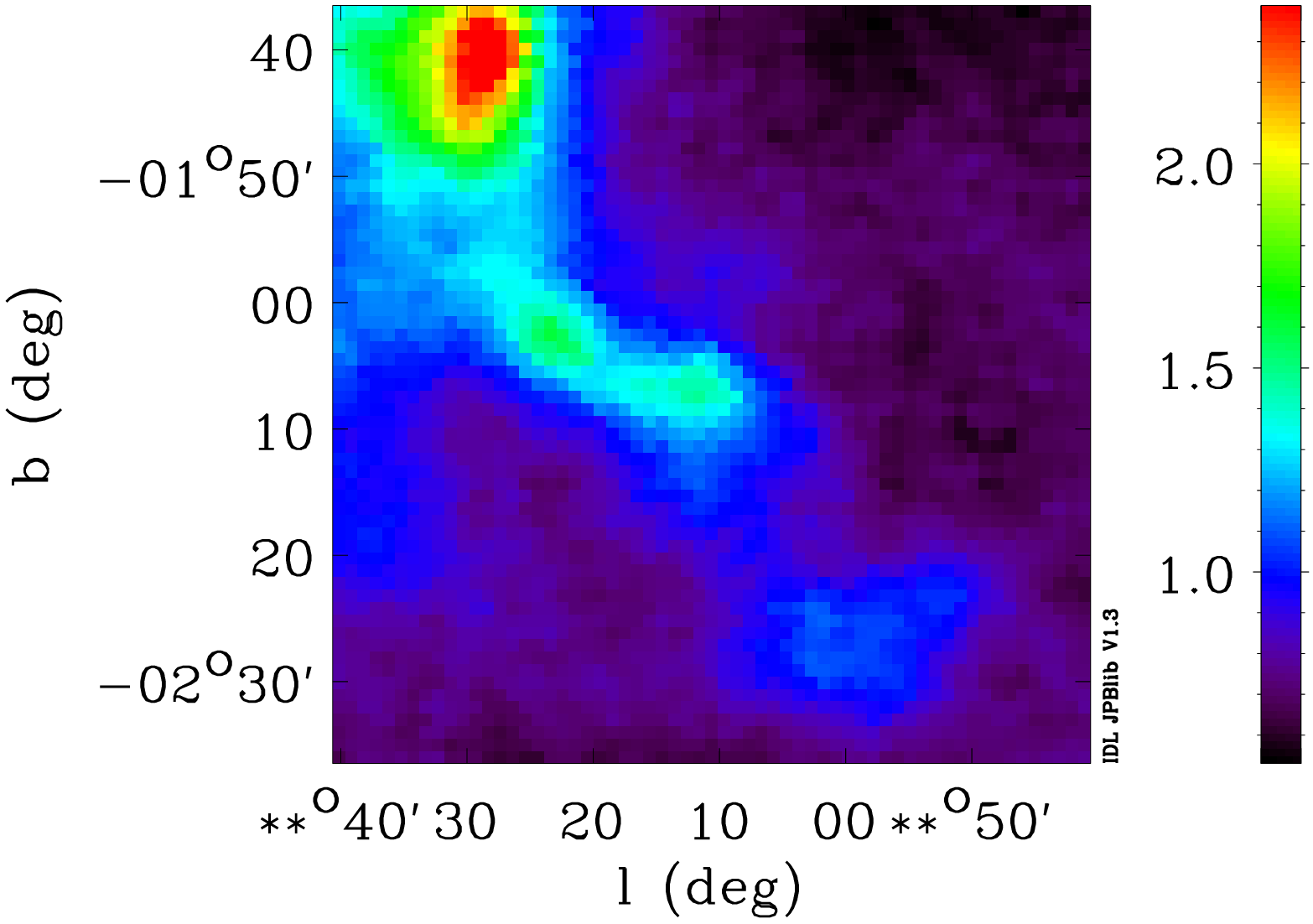} &
 \includegraphics[width=4cm]{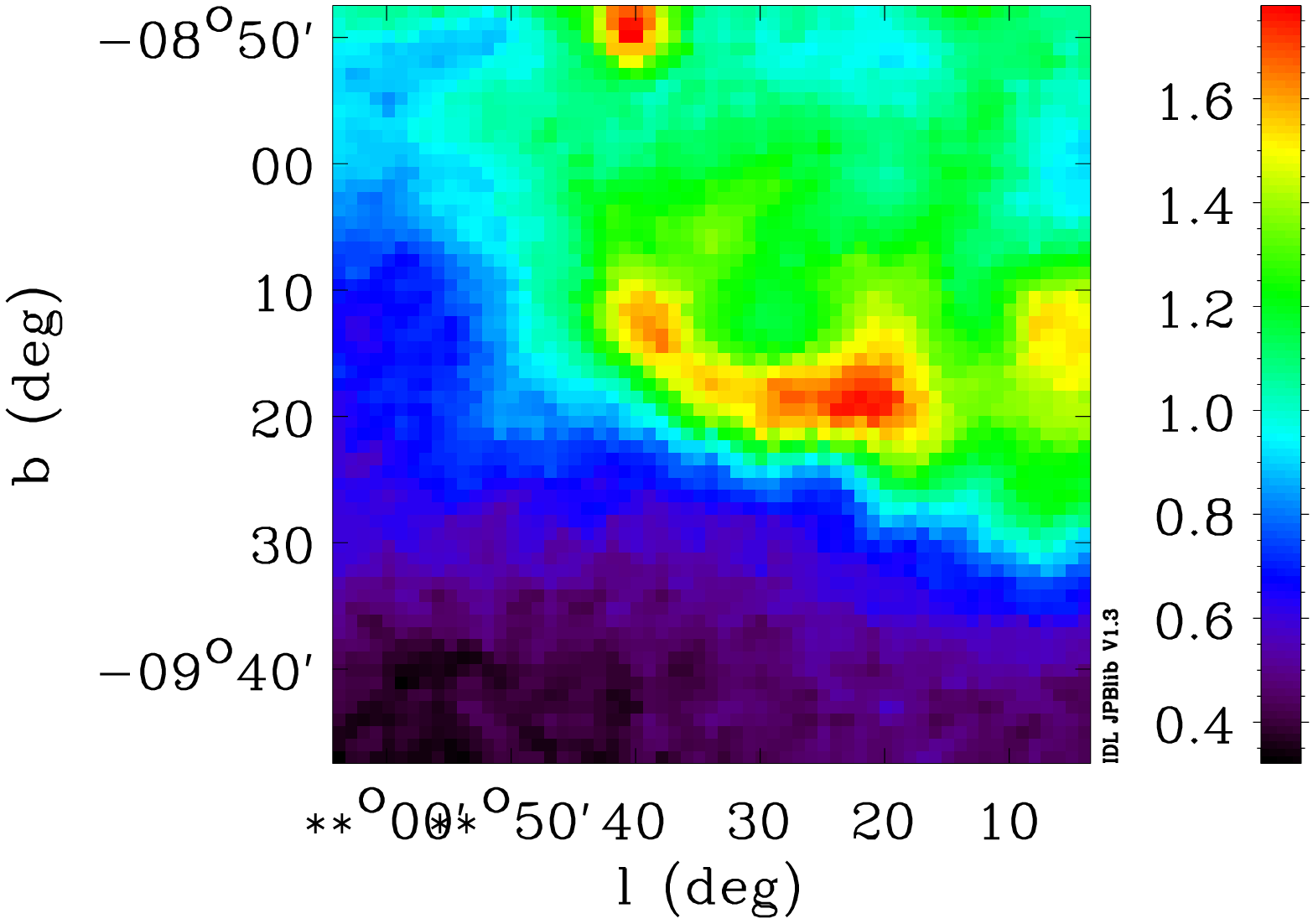} &
 \includegraphics[width=4cm]{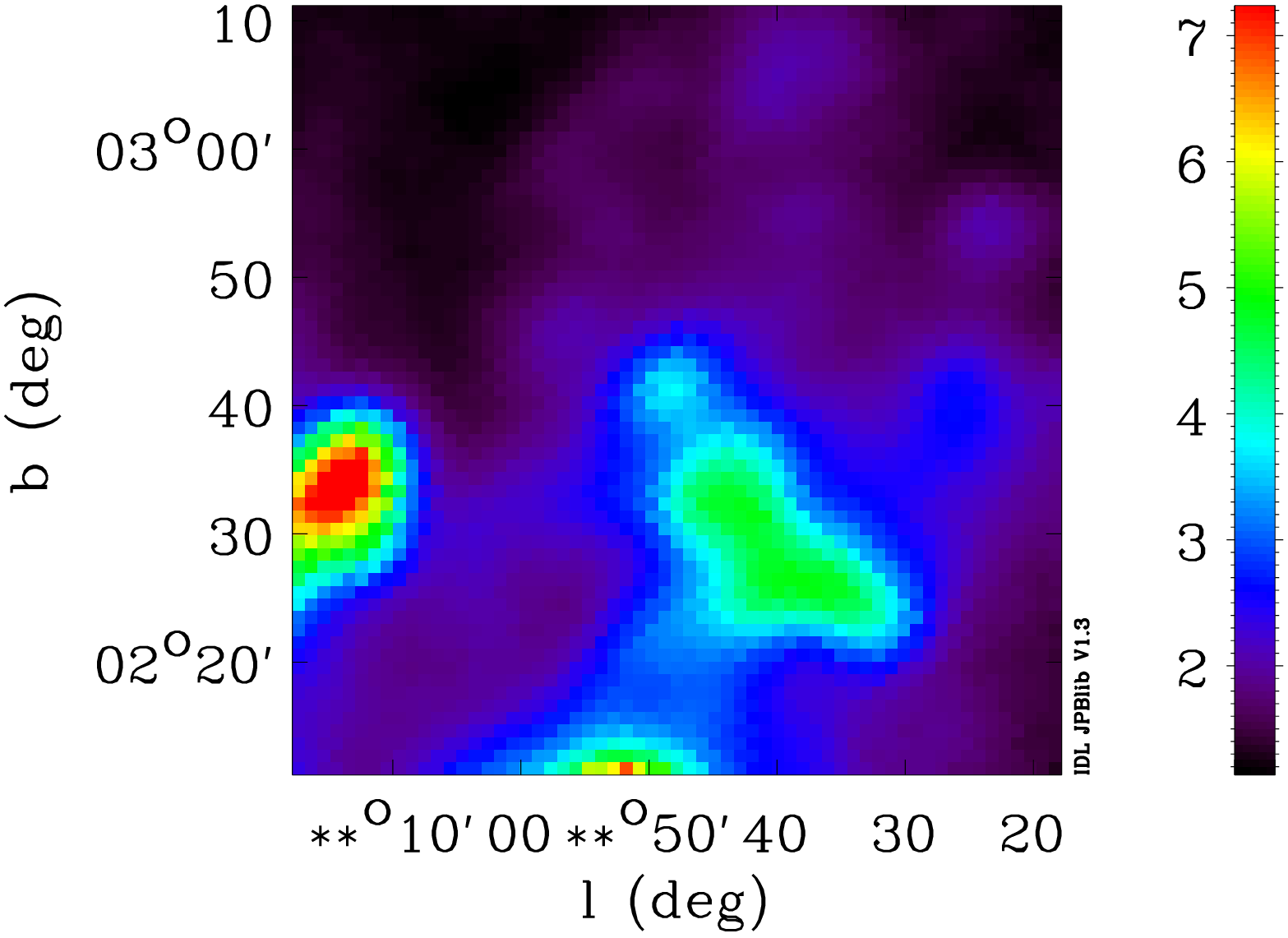} \\

 143\,GHz &
 \includegraphics[width=4cm]{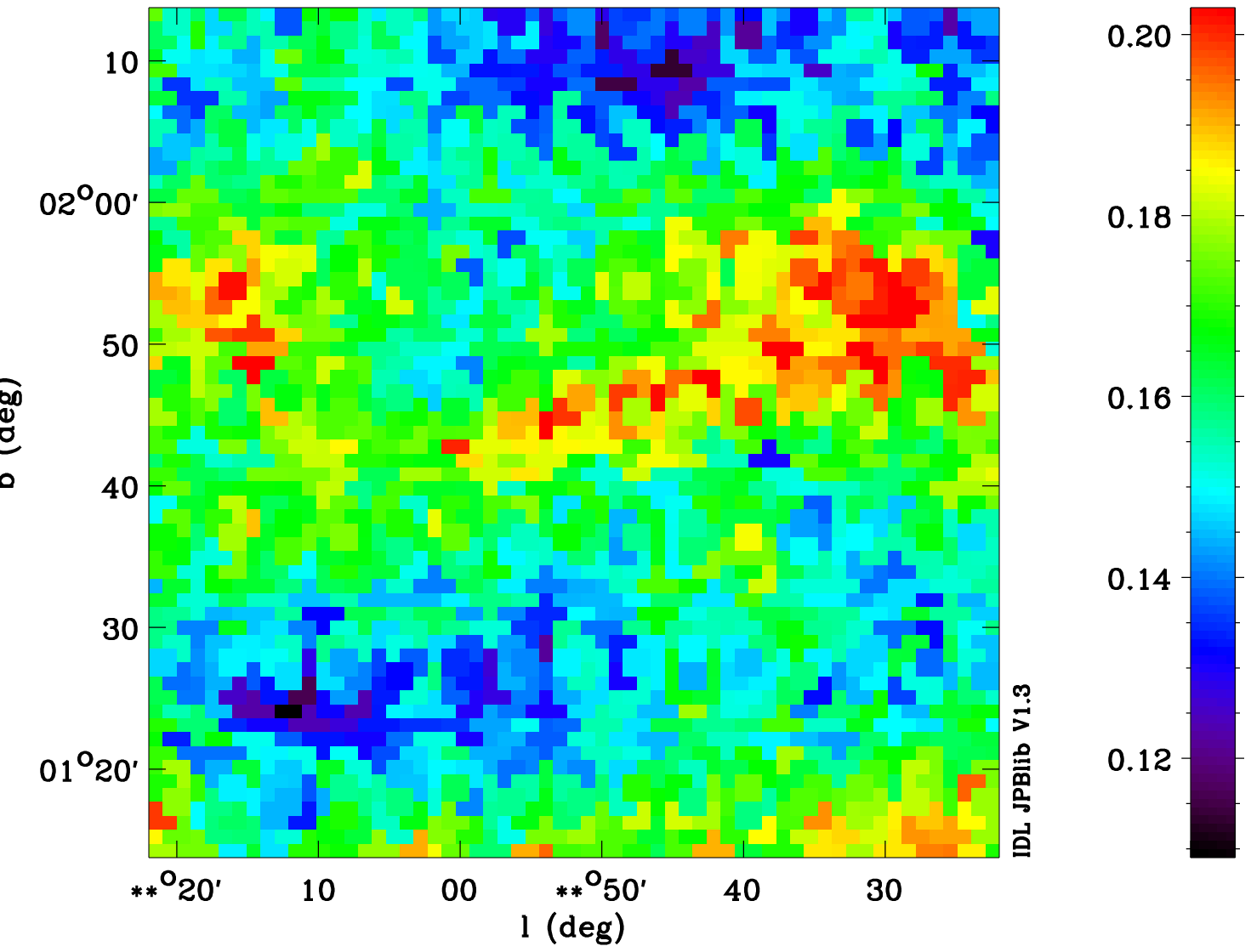} &
 \includegraphics[width=4cm]{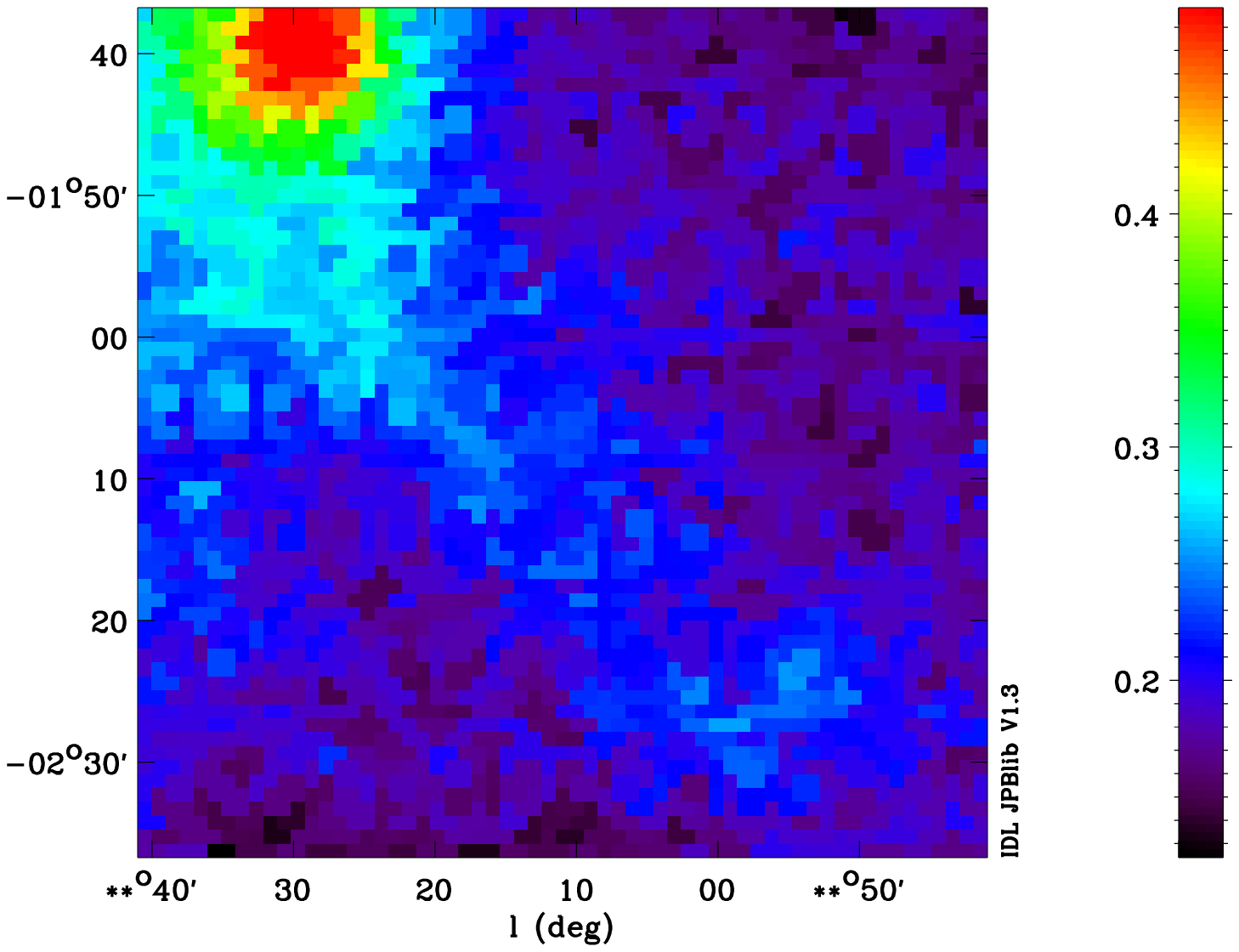} &
 \includegraphics[width=4cm]{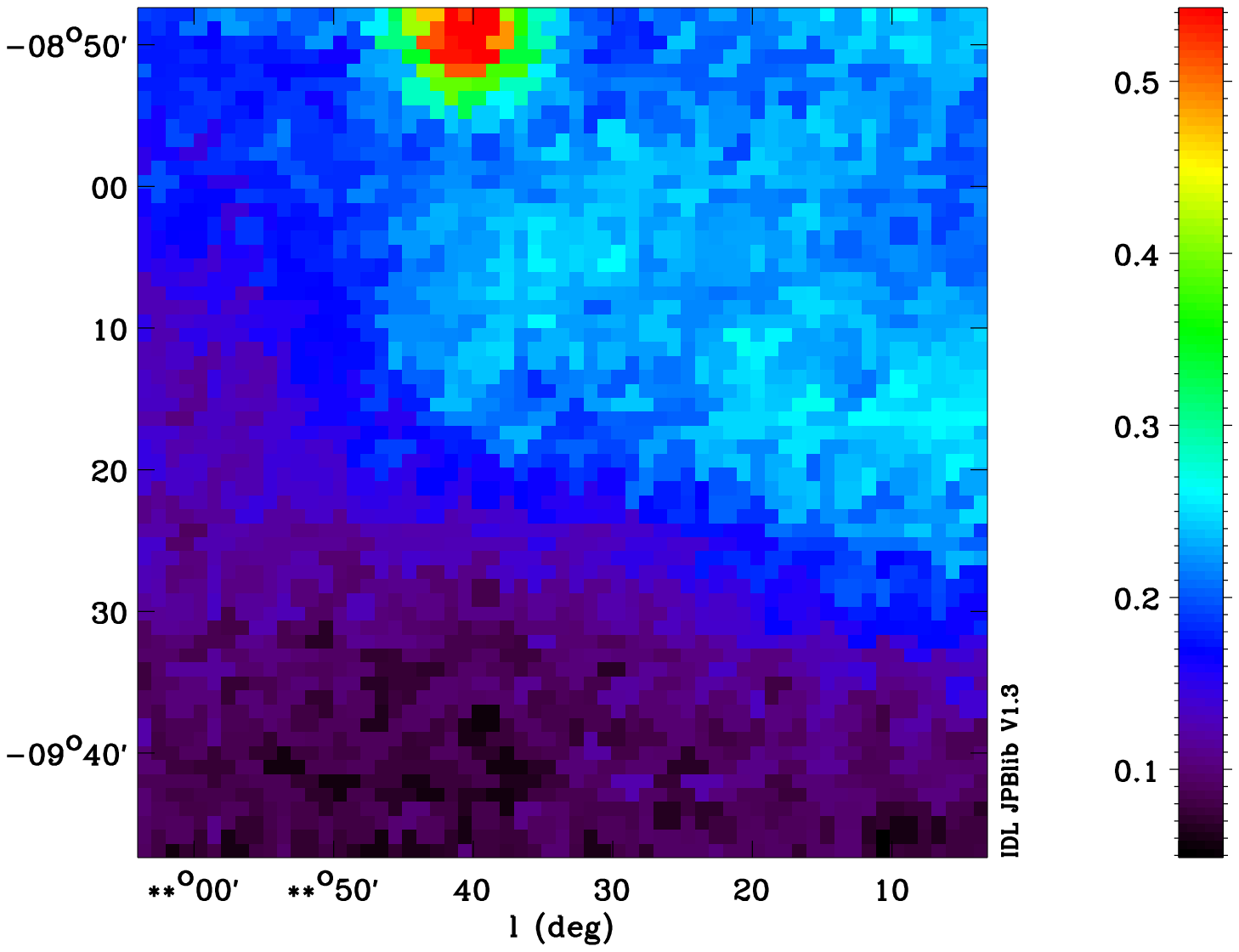} &
 \includegraphics[width=4cm]{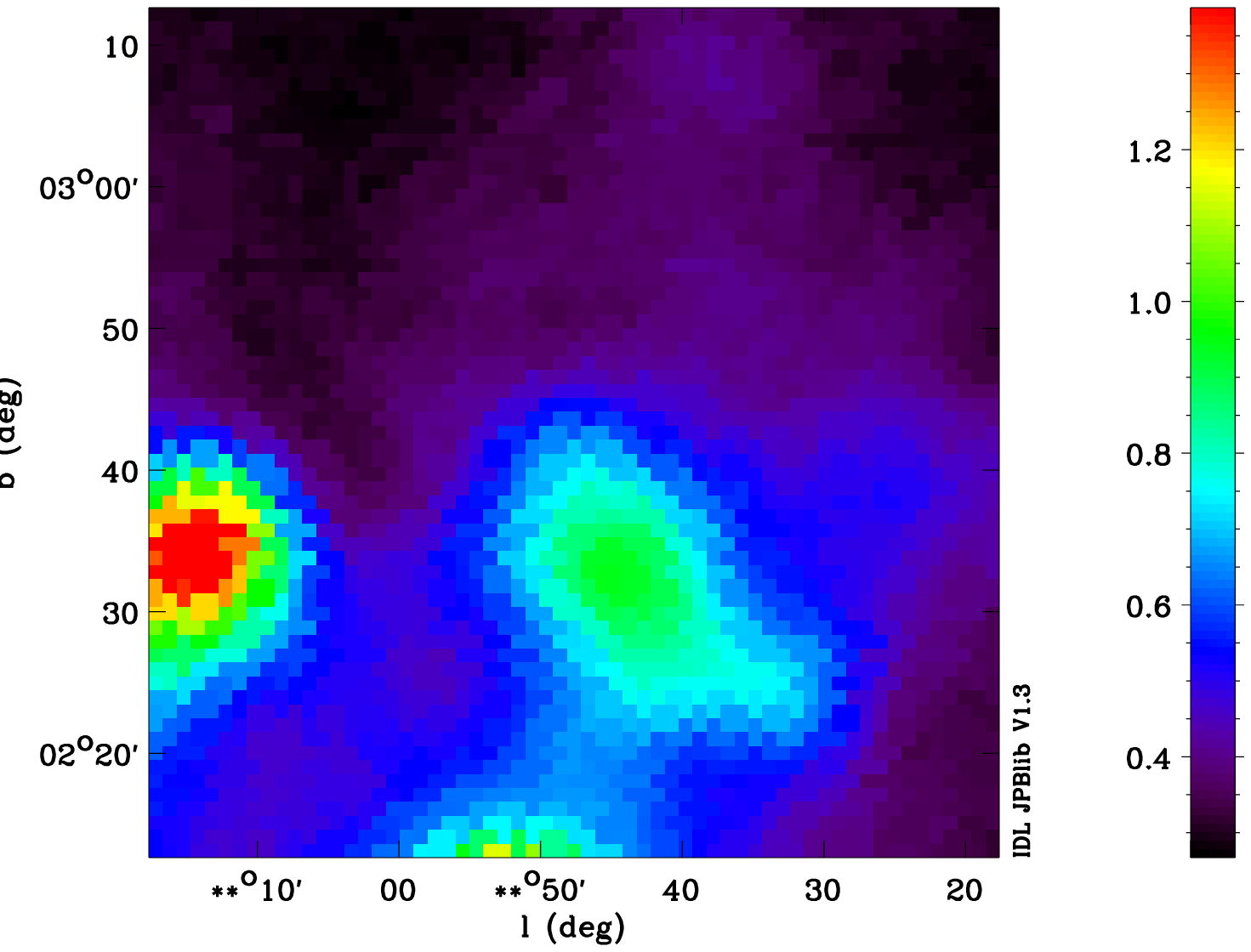} \\

 100\,GHz  &
 \includegraphics[width=4cm]{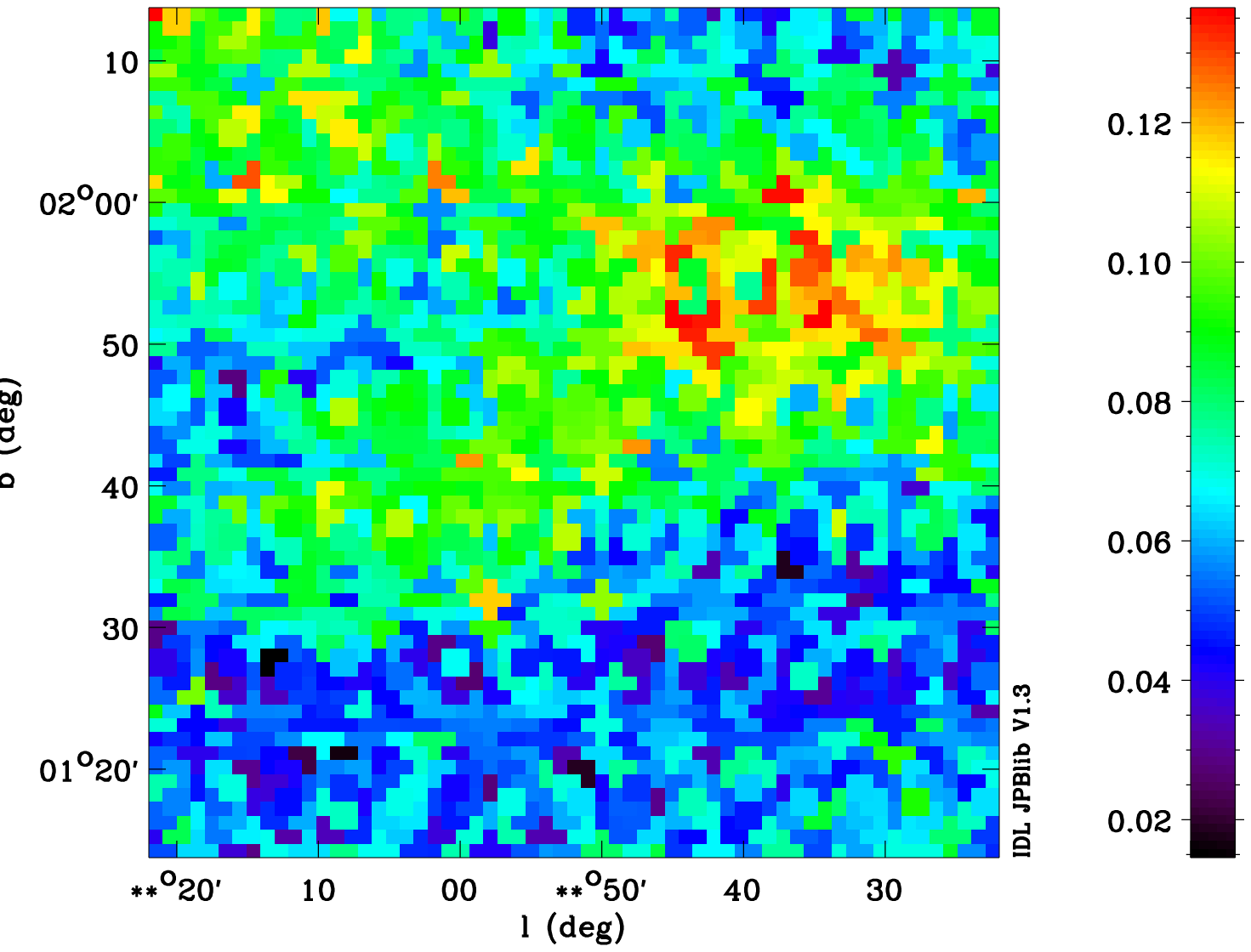} &
 \includegraphics[width=4cm]{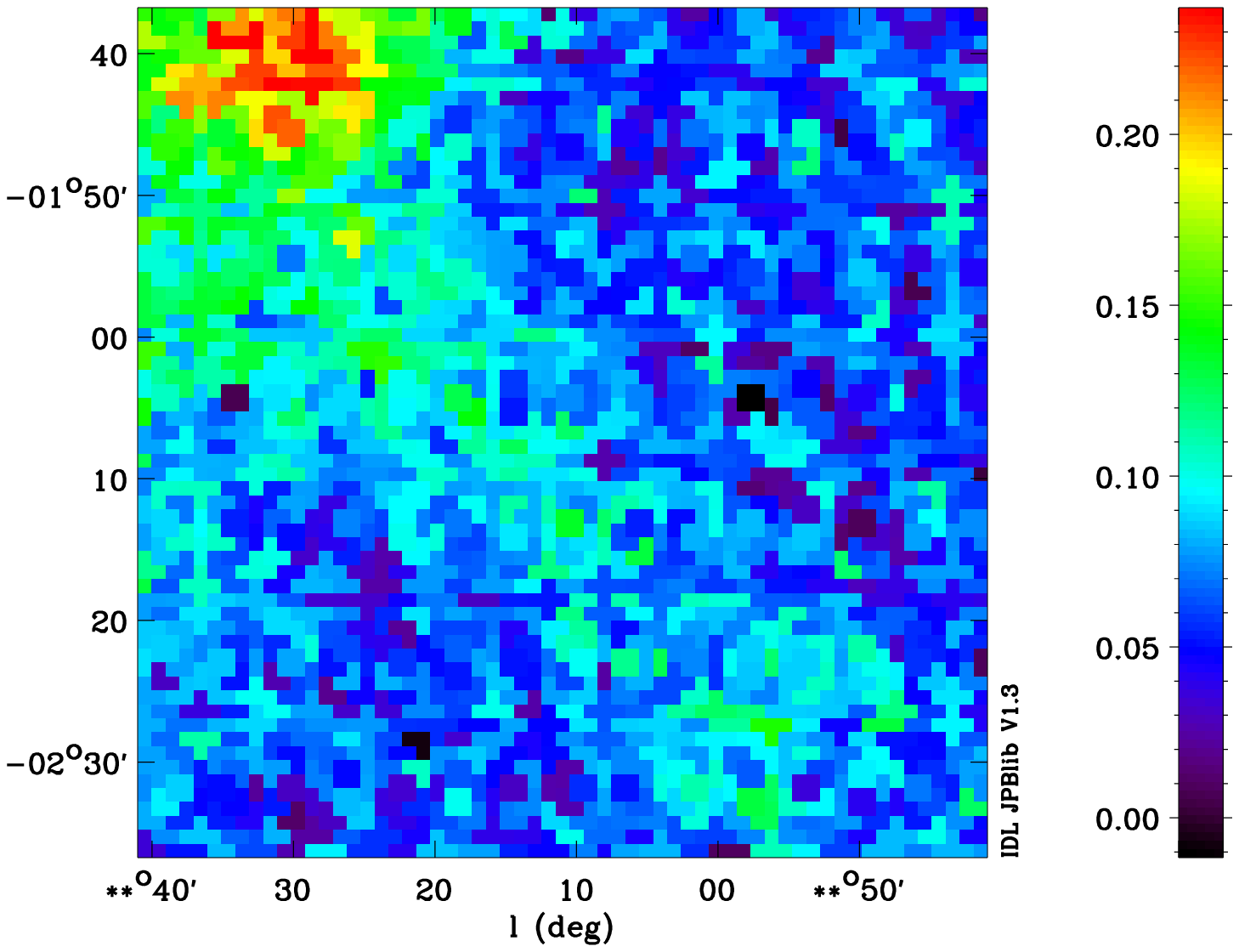} &
\includegraphics[width=4cm]{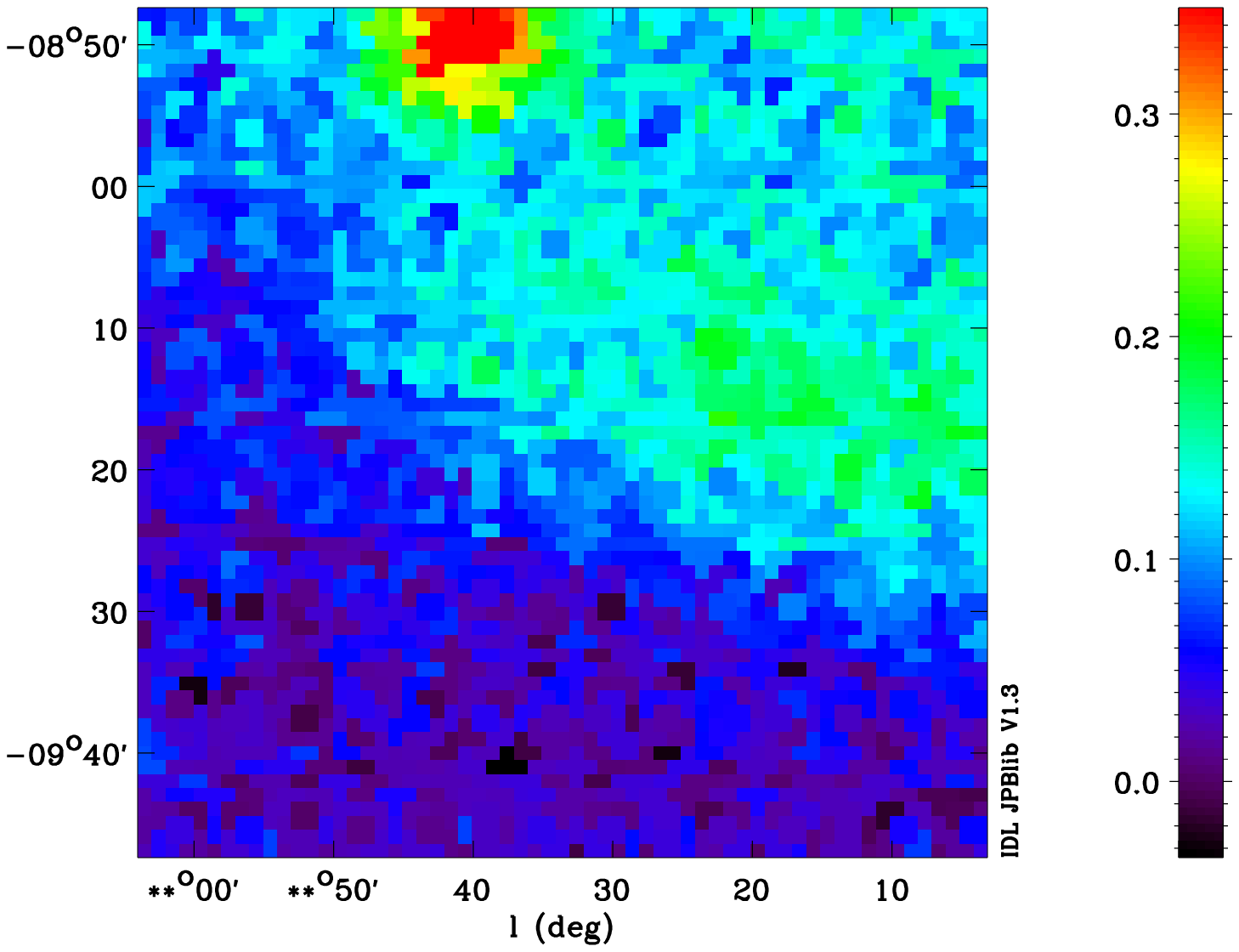} &
 \includegraphics[width=4cm]{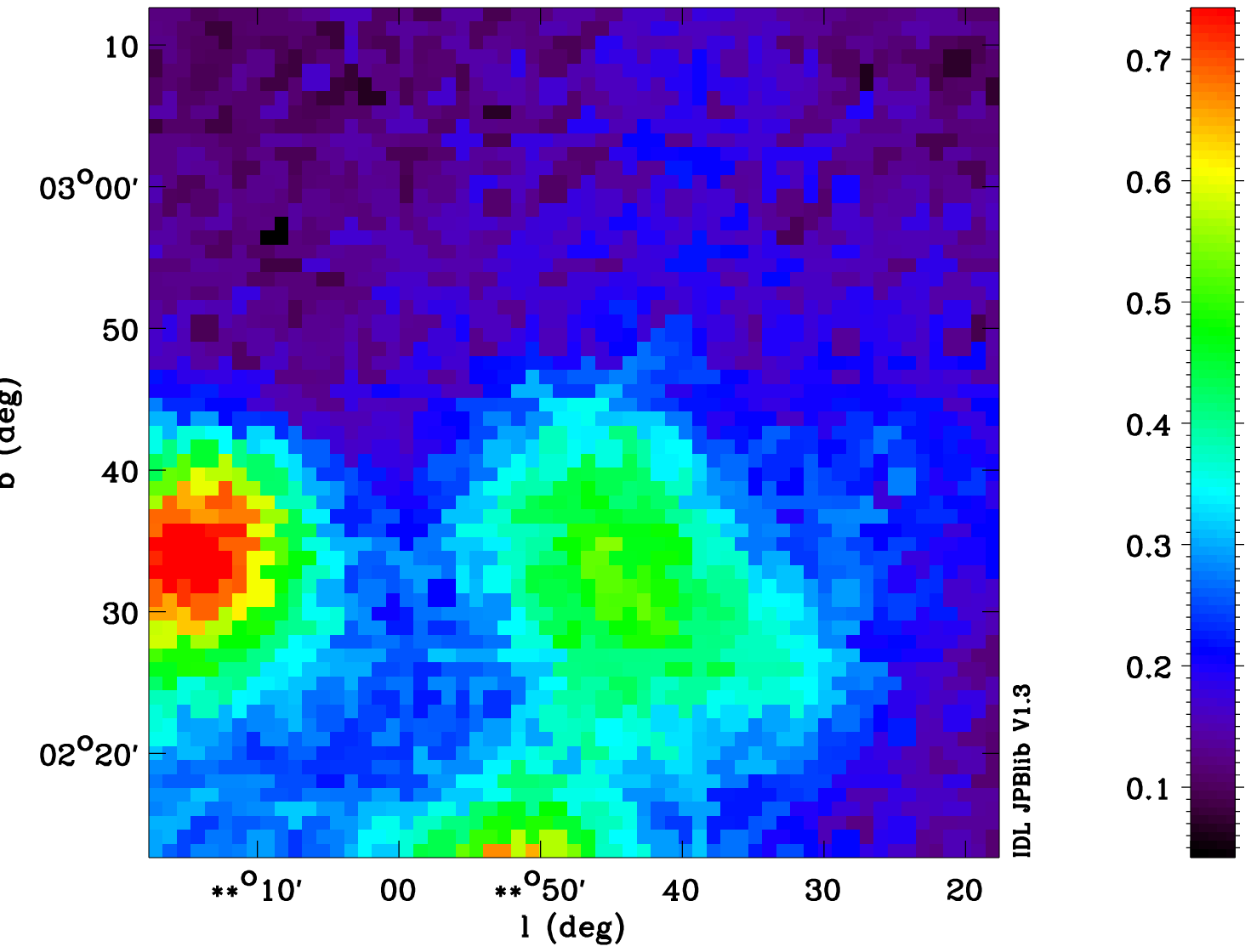} \\

\end{tabular}
\caption{Multi-band emission maps of the sources S5, S6, S7, and S8.}
\label{fig:allnu_sources2}
\end{figure*}

\begin{figure*}
\center
\begin{tabular}{ccc}

Band & {\tiny S9\quad\quad} & {\tiny S10\qquad\quad} \\
 $100\,\mu$m &
 \includegraphics[width=4cm]{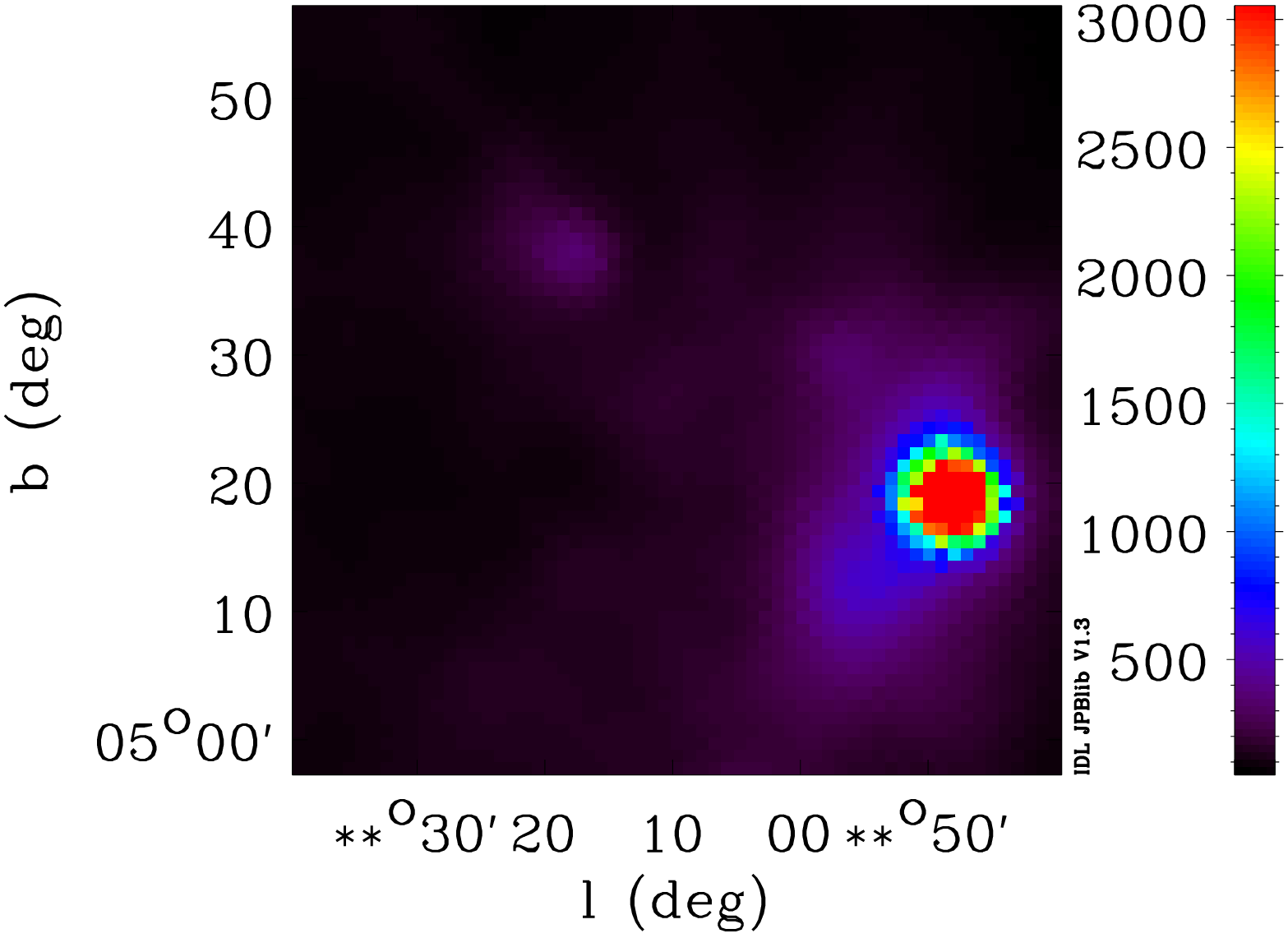} &
 \includegraphics[width=4cm]{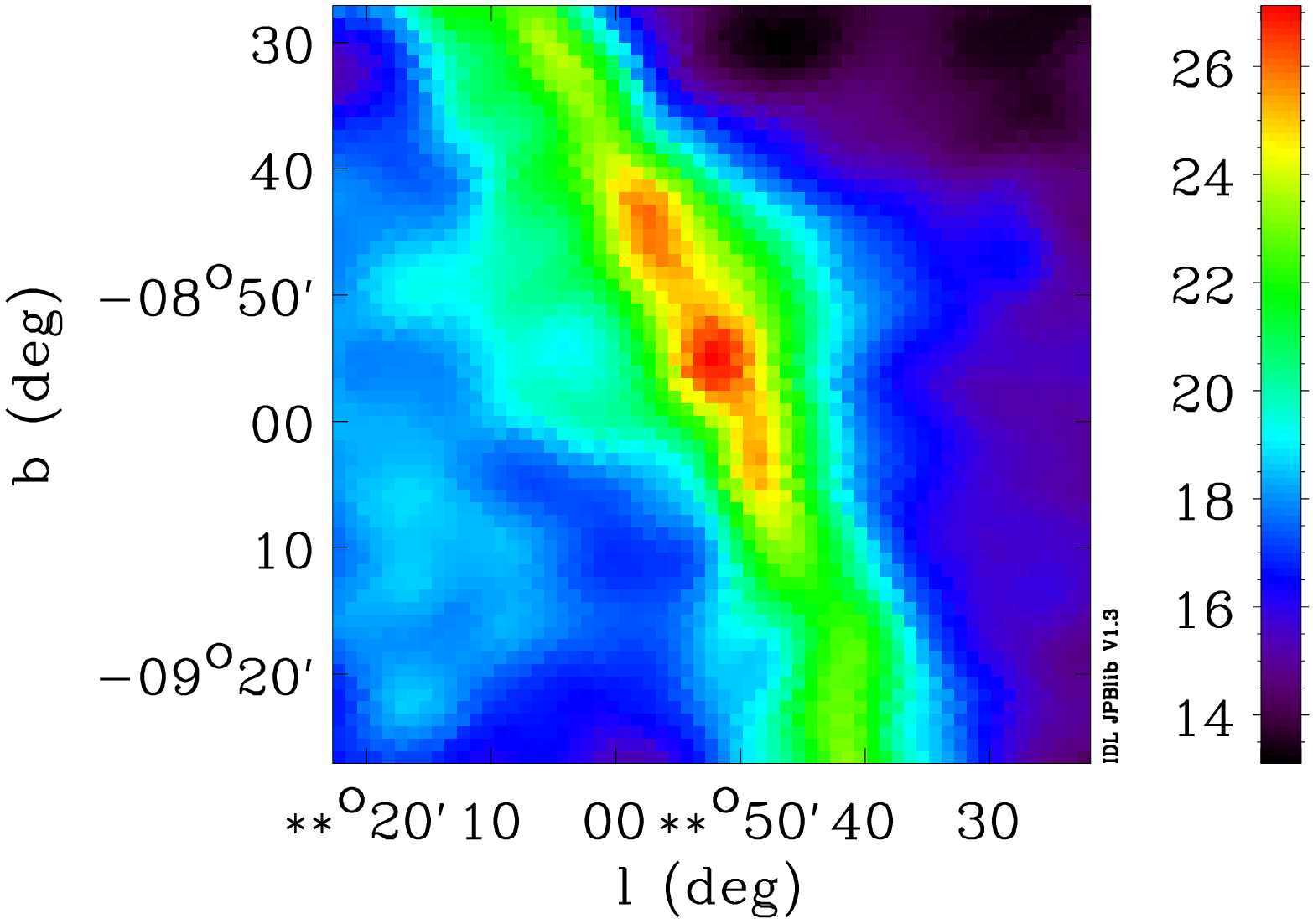} \\

 857\,GHz &
 \includegraphics[width=4cm]{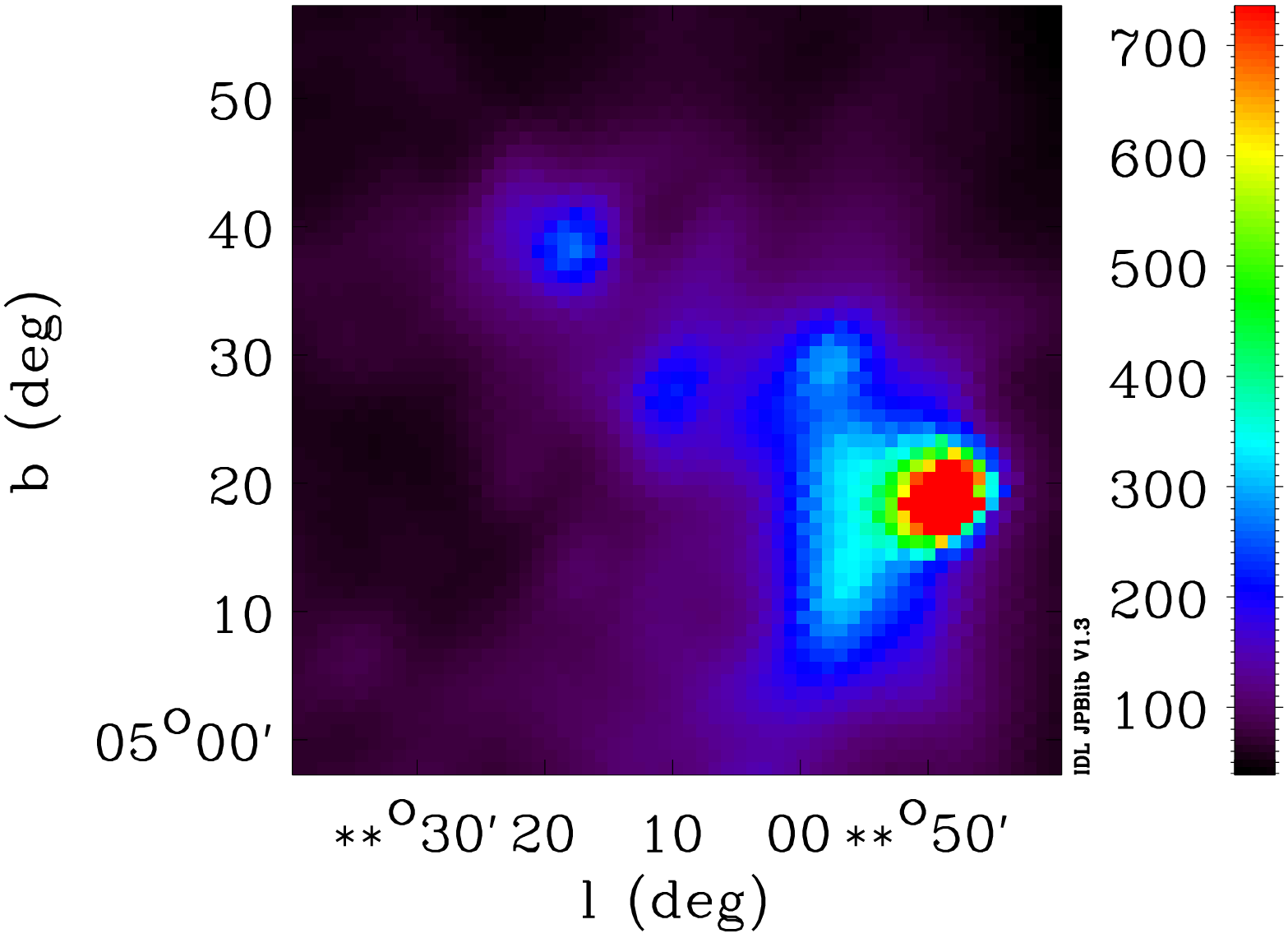} &
 \includegraphics[width=4cm]{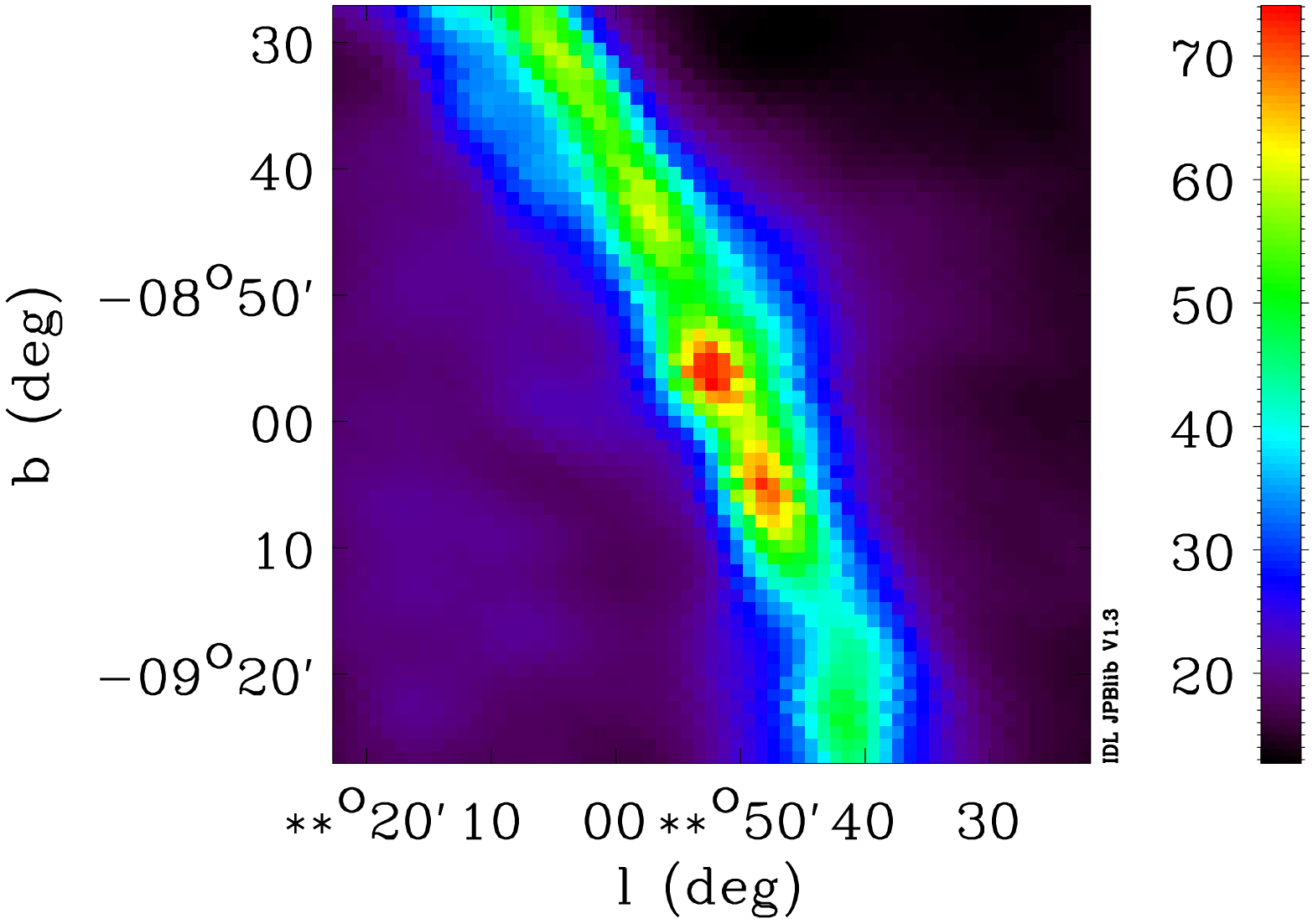} \\

\tiny{Cold Residual} &
 \includegraphics[width=4cm]{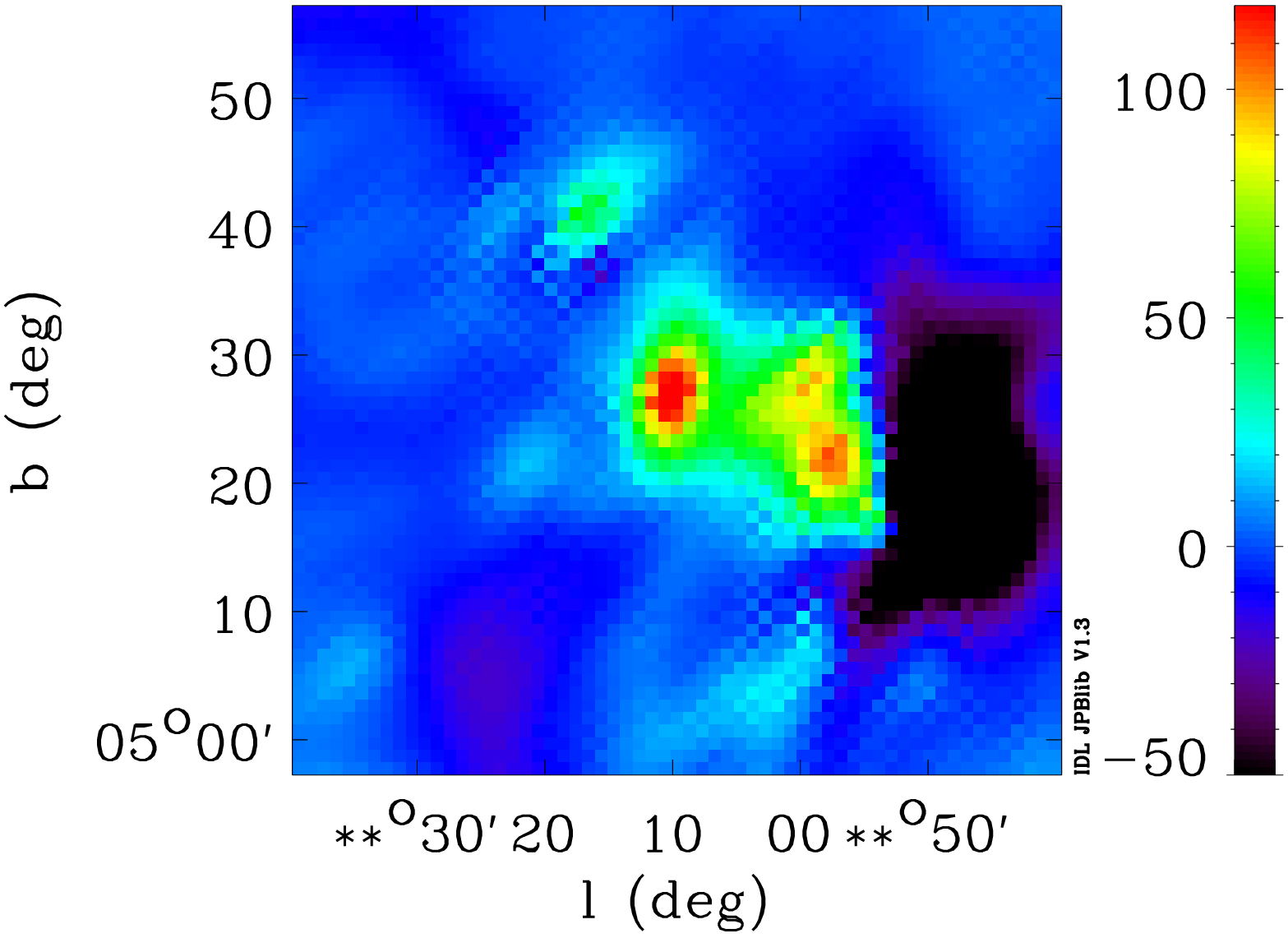} &
 \includegraphics[width=4cm]{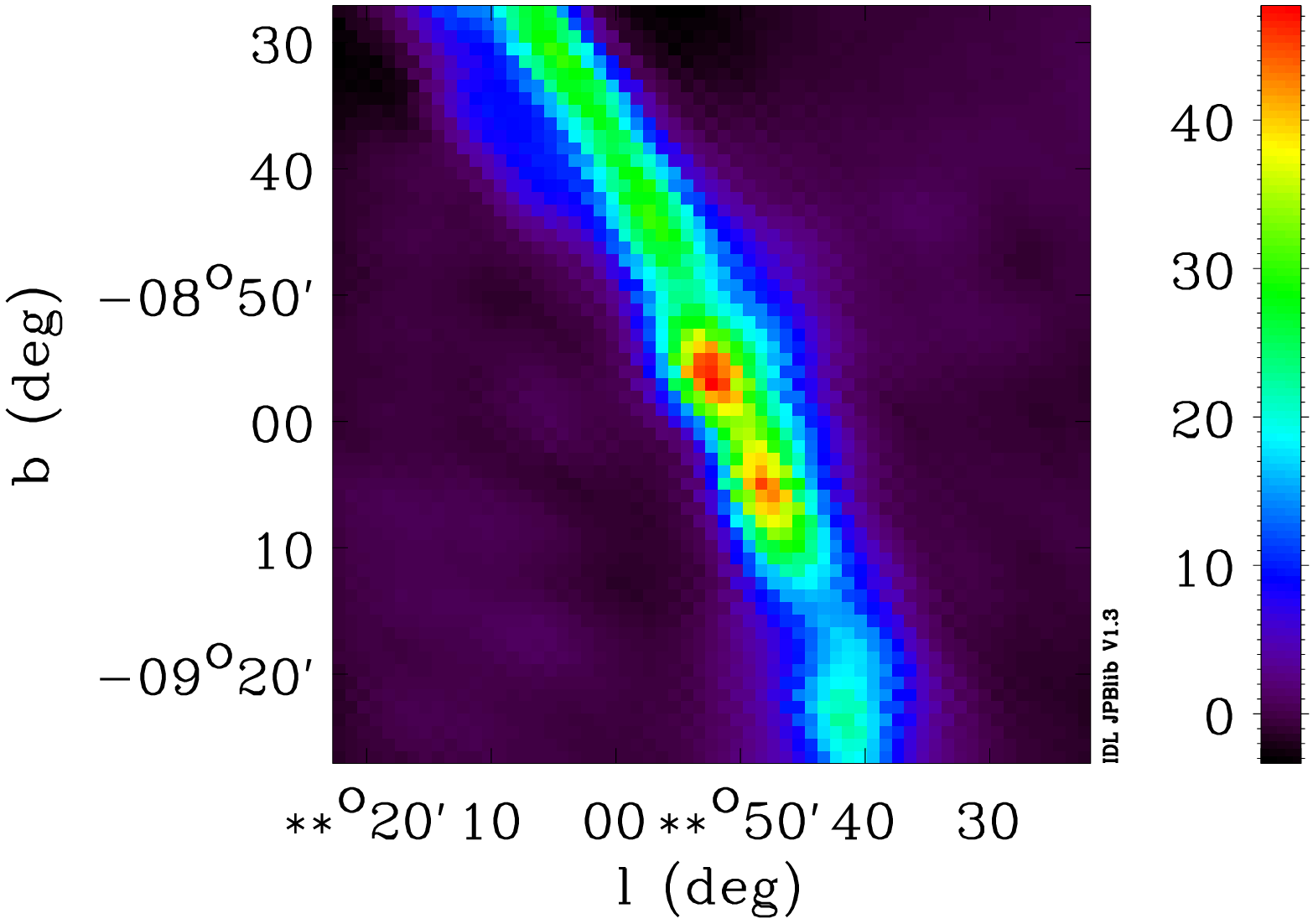} \\

 545\,GHz &
 \includegraphics[width=4cm]{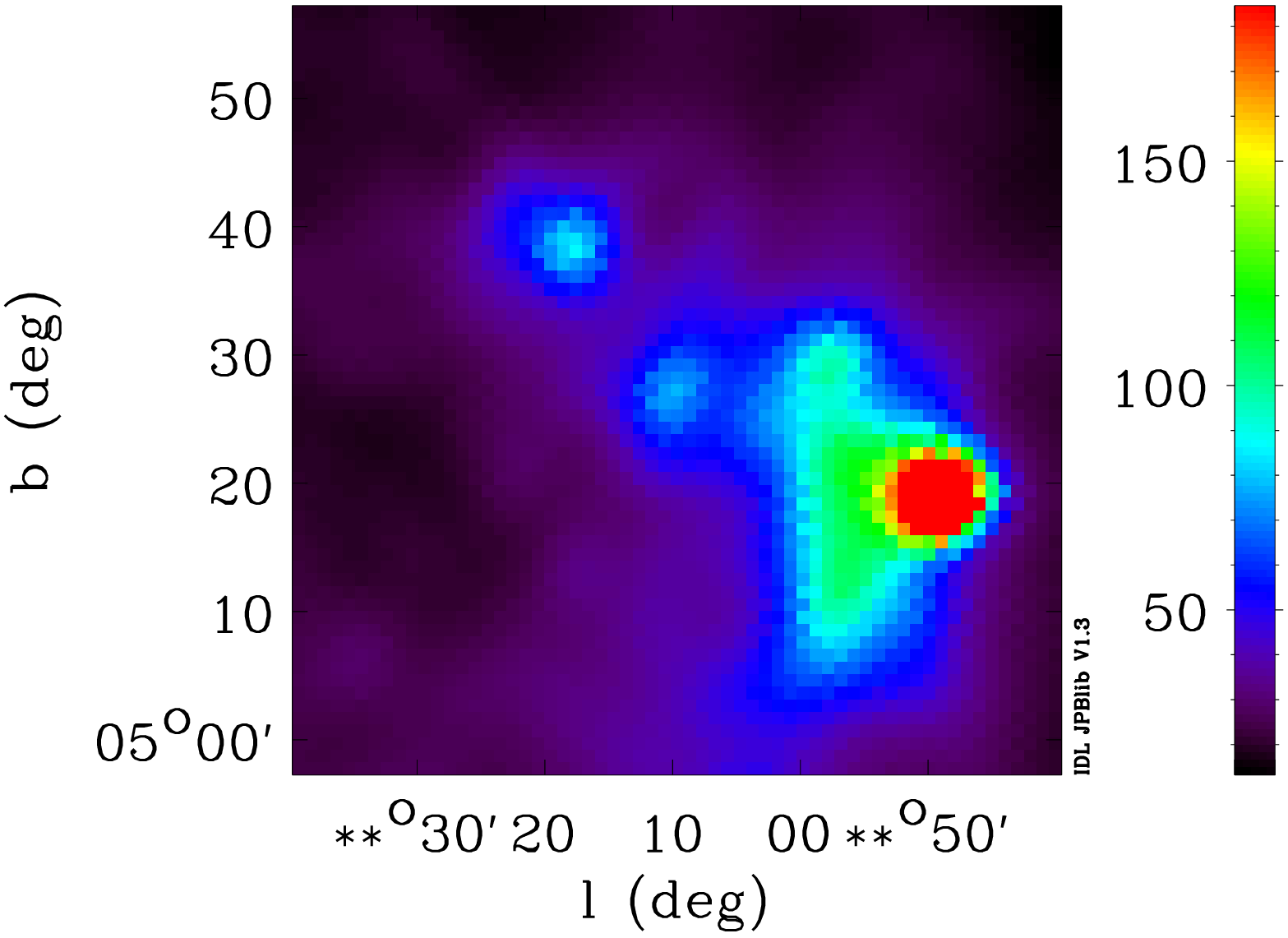} &
 \includegraphics[width=4cm]{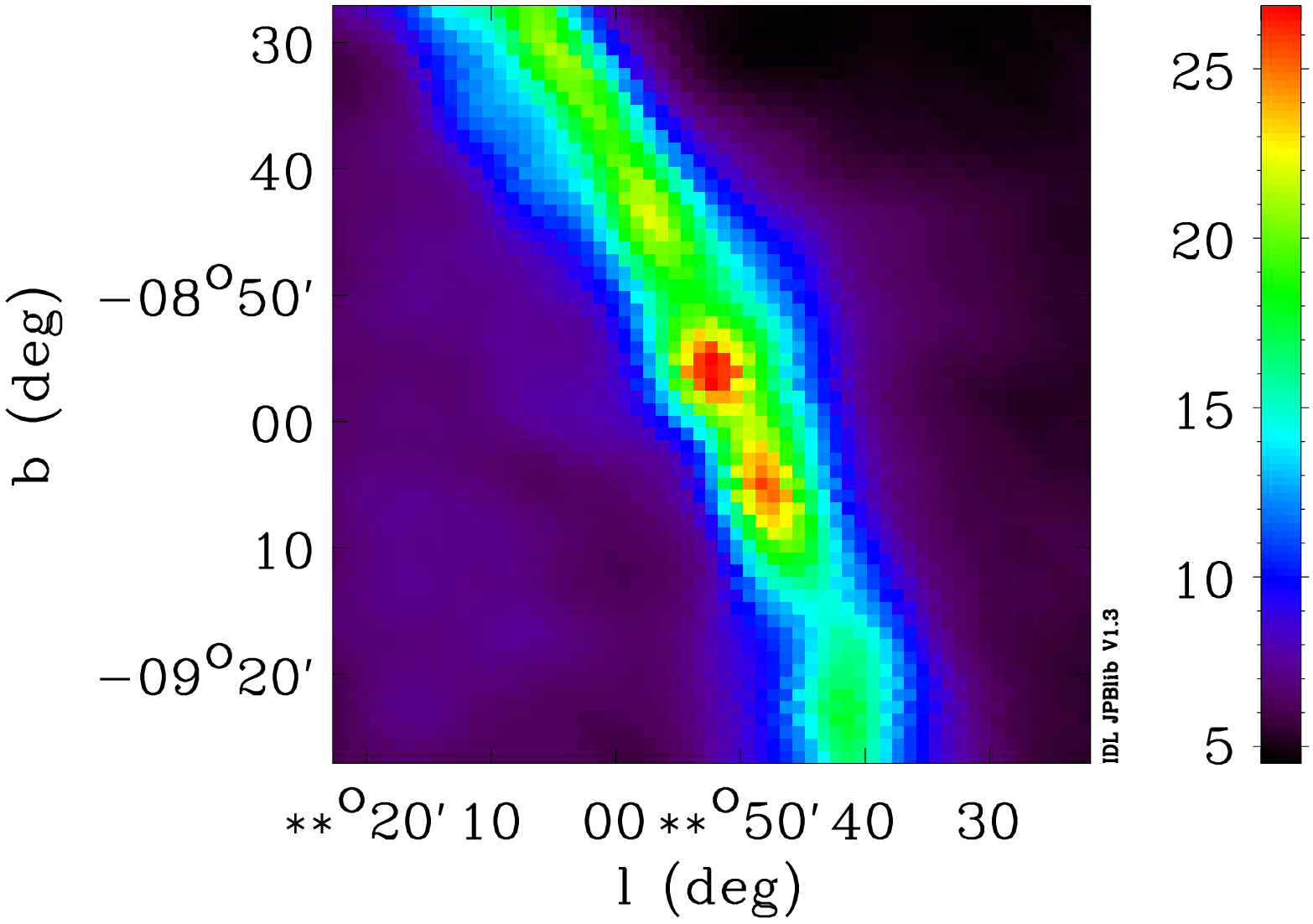} \\

 353\,GHz &
 \includegraphics[width=4cm]{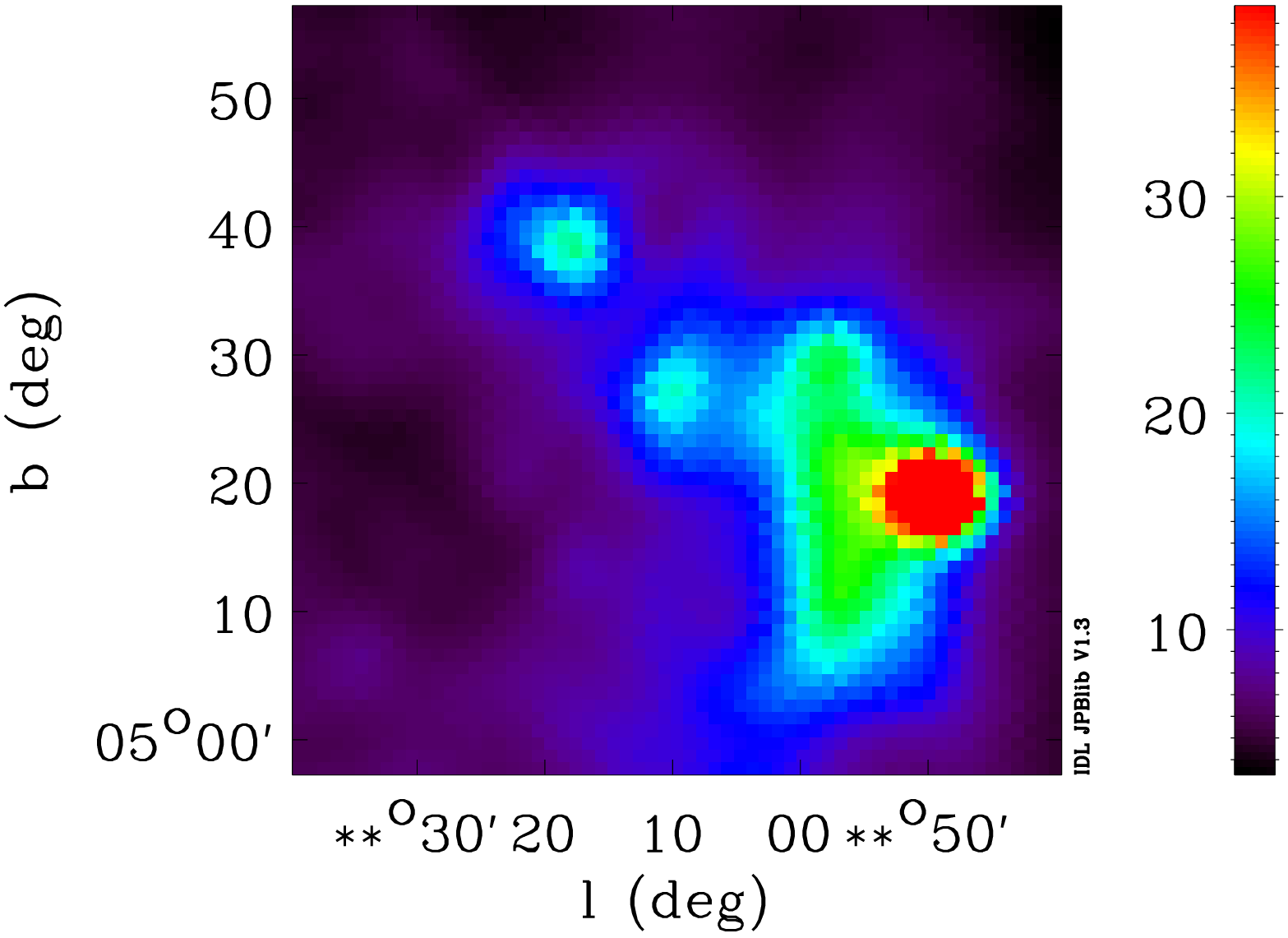} &
 \includegraphics[width=4cm]{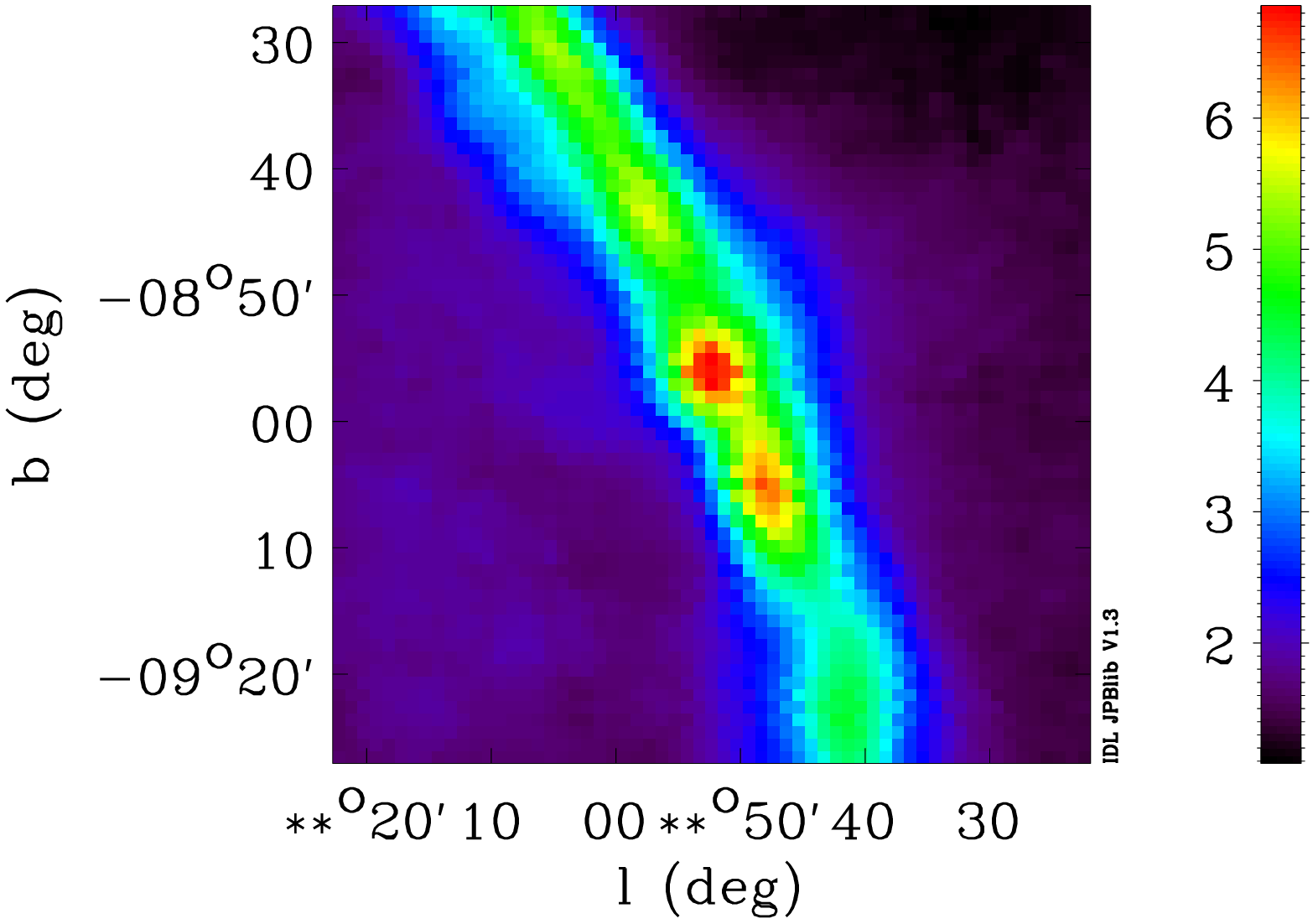} \\

 217\,GHz &
 \includegraphics[width=4cm]{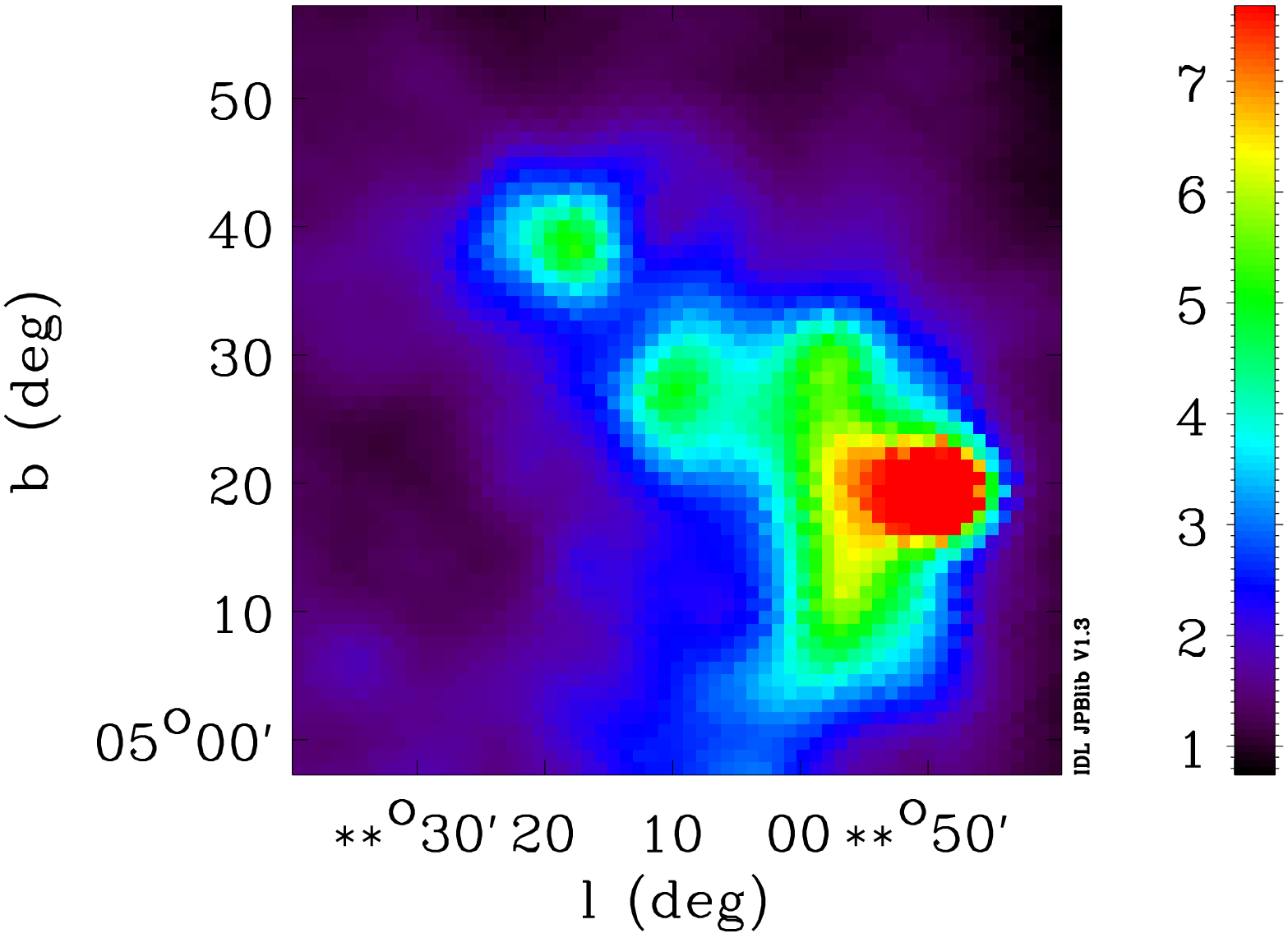} &
 \includegraphics[width=4cm]{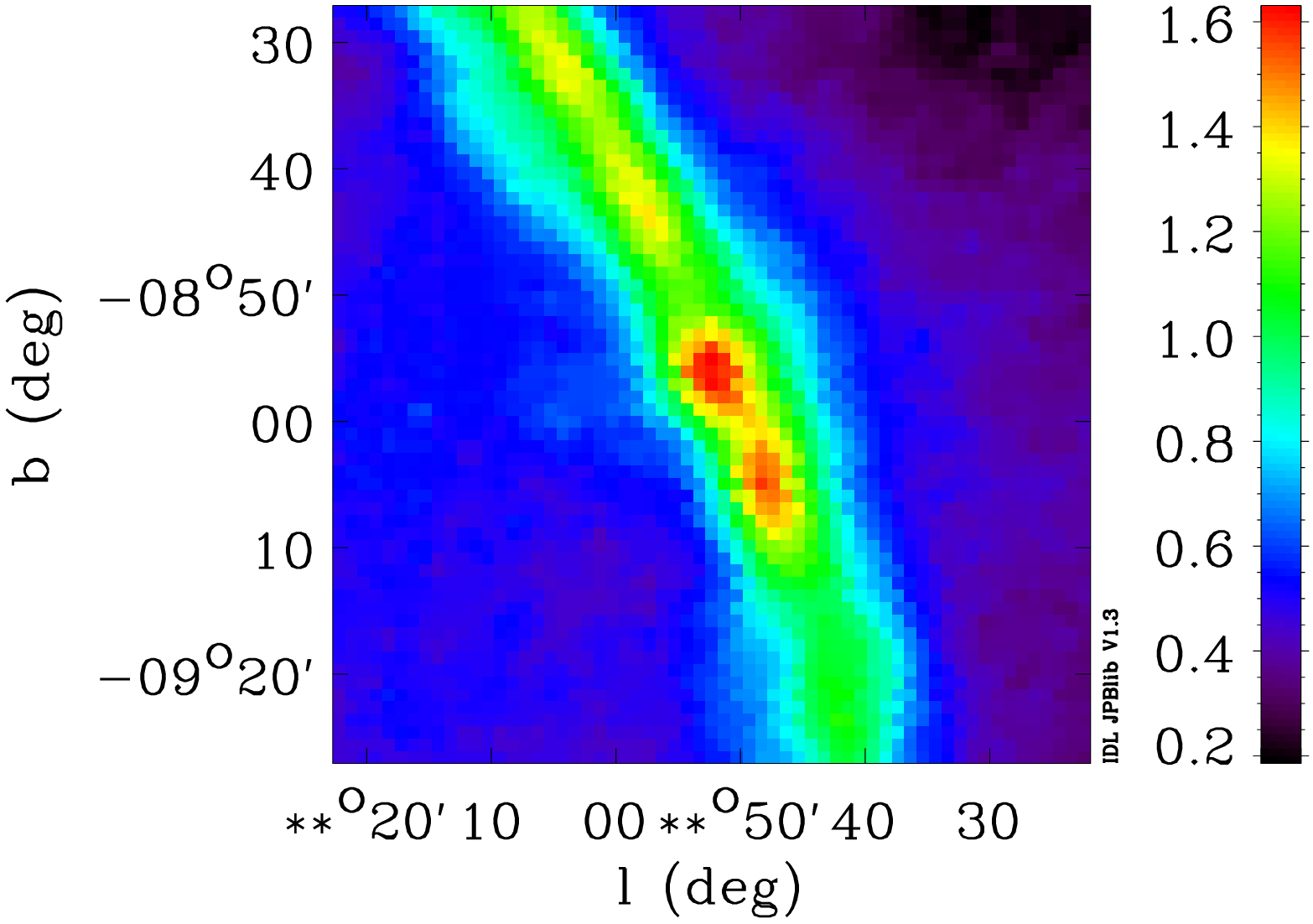} \\

 143\,GHz &
 \includegraphics[width=4cm]{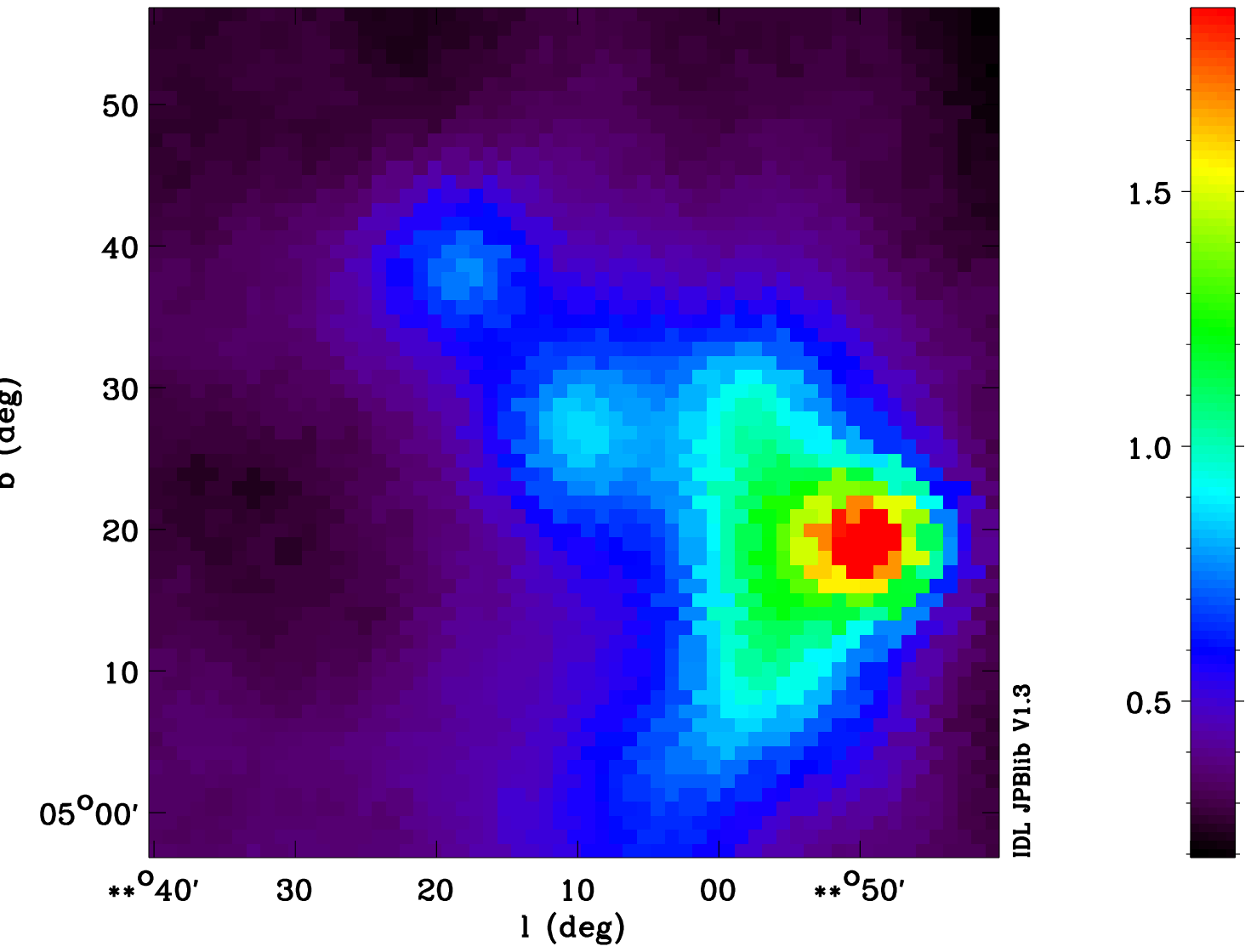} &
 \includegraphics[width=4cm]{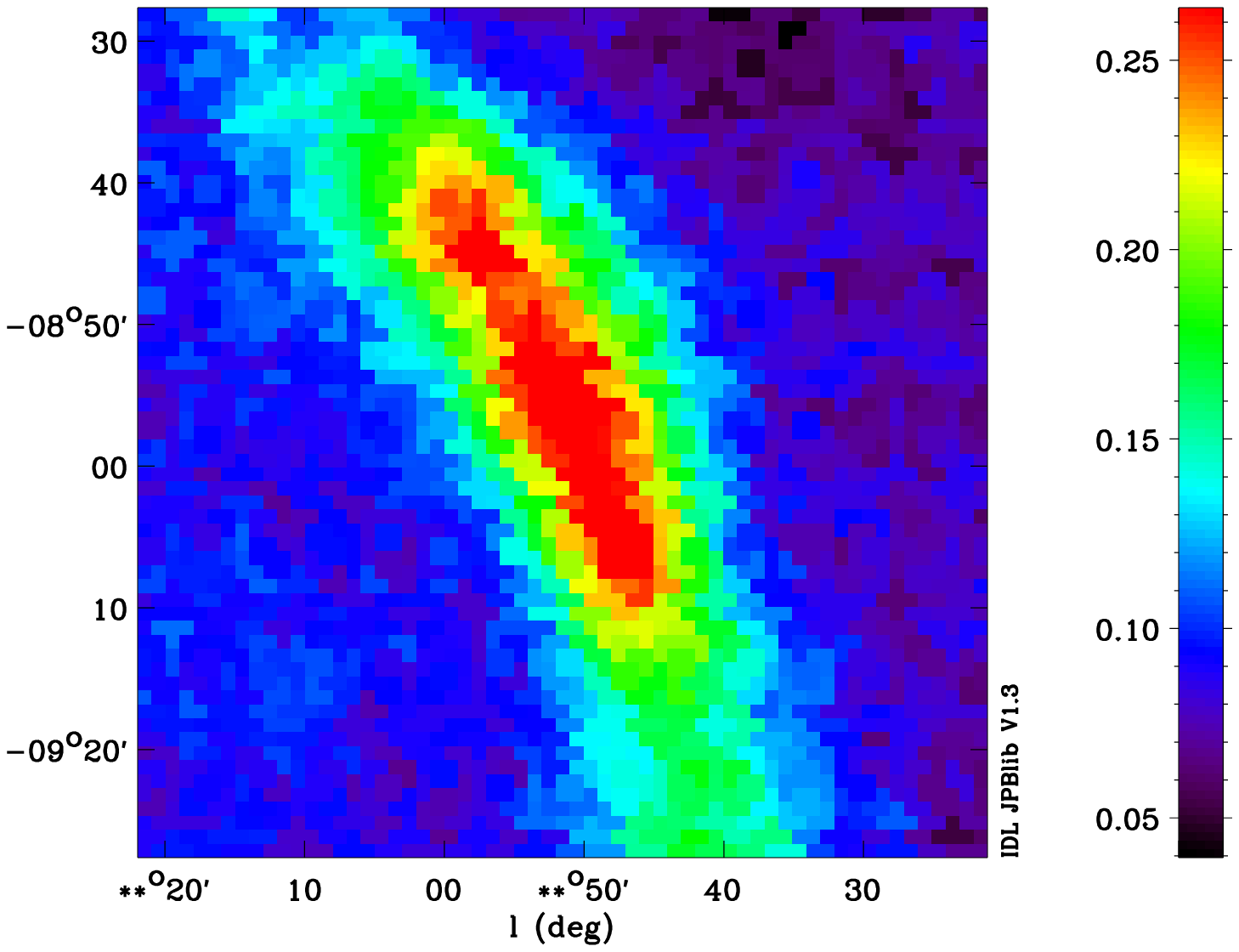} \\

 100\,GHz &
 \includegraphics[width=4cm]{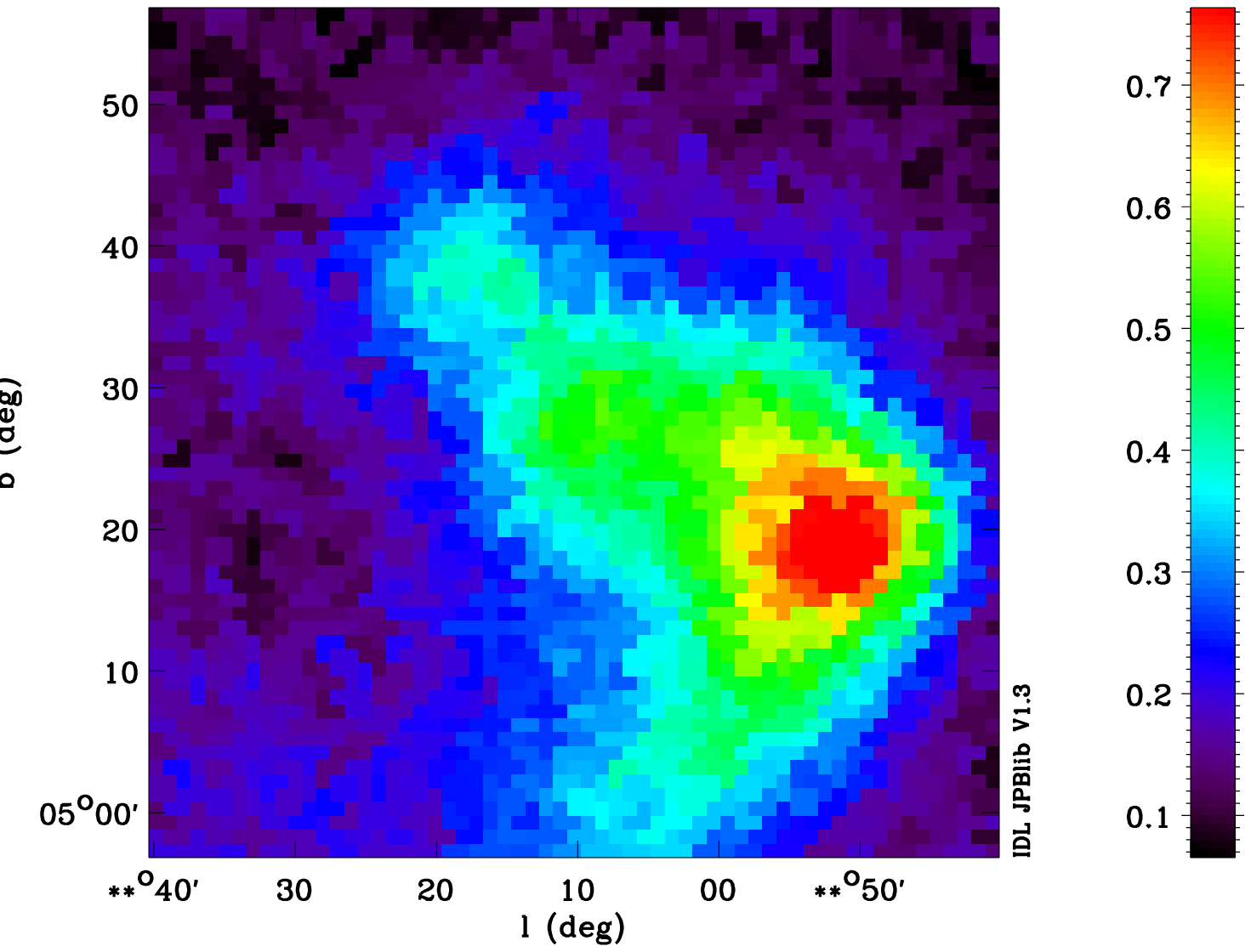} &
 \includegraphics[width=4cm]{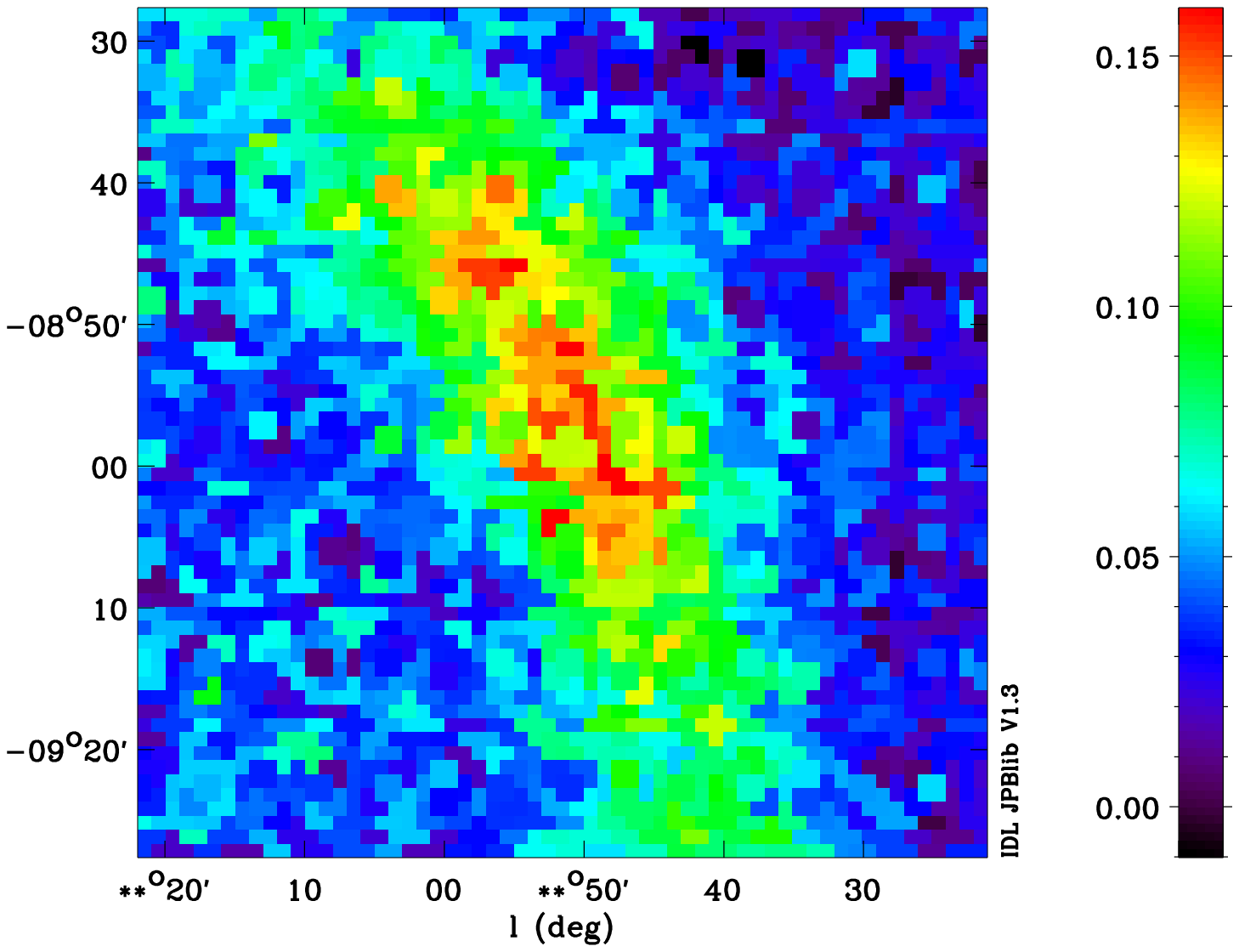} \\

\end{tabular}
\caption{Multi-band emission maps of the sources S9 and S10.}
\label{fig:allnu_sources3}
\end{figure*}

%\section{Illustration of the detection method} \label{sect:detection_method}

\begin{figure*}
\center
\includegraphics[width=18cm]{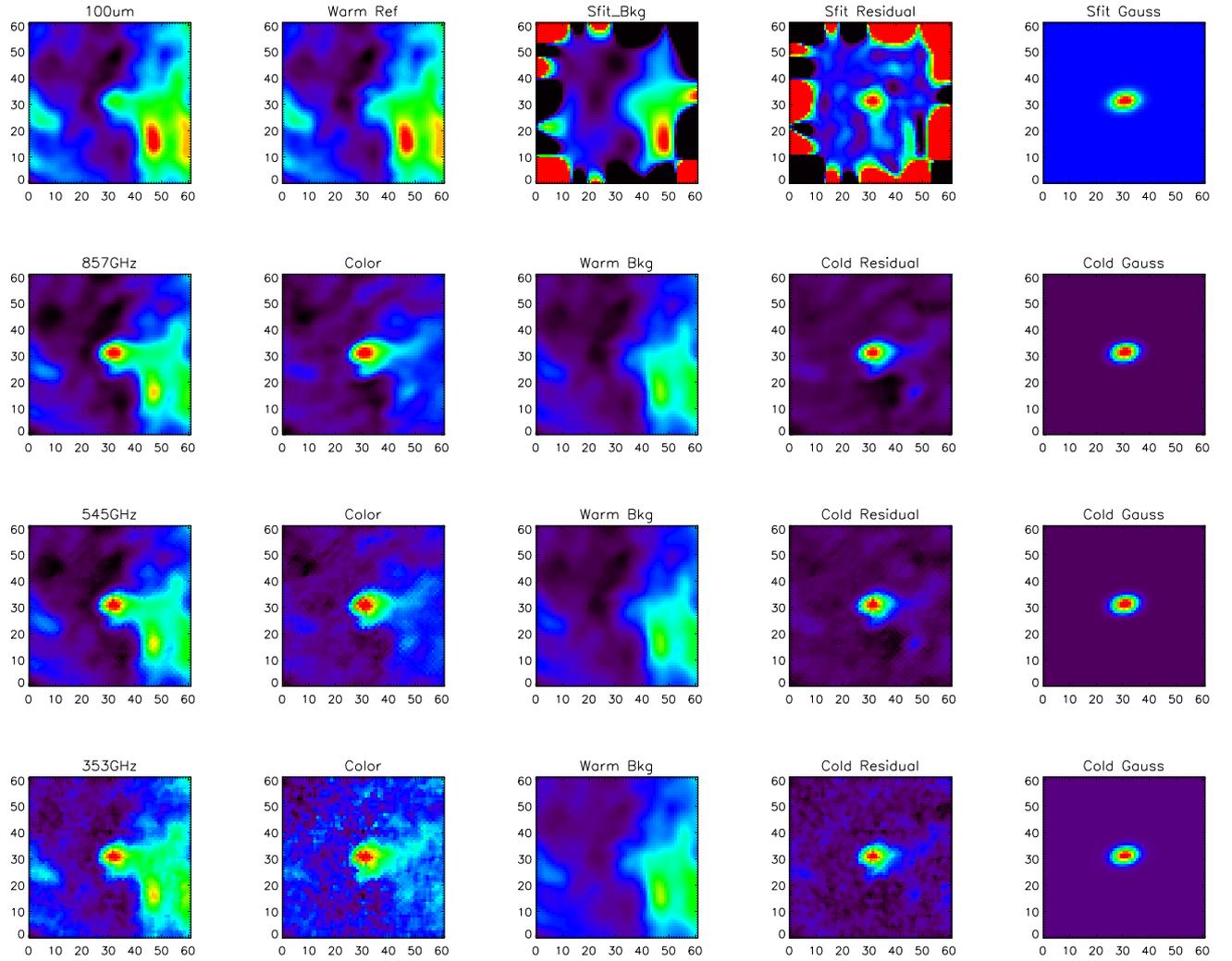}

\caption{Illustration of the steps used in the detection and core extraction
methods, as applied to source S1. 
The first column shows the initial surface brightness emission maps in the IRIS
and HFI bands.  The rows show the process applied to the 100\,$\mu$m, 857, 545,
and 353\,GHz data, respectively.  
As described in Paper~I, the method follows the following main steps: 
a ``Warm Ref'' is built from the IRIS 100\,$\mu$m map 
and extrapolated to the \Planck\ bands using the local colour estimated
around each pixel of the \Planck\ map (these colour maps are shown in the
second column); the `Warm Bkg' map obtained at a given frequency is then
removed from the \Planck\ map, revealing the ``Cold Residual'';
and finally an elliptical Gaussian fit is performed on the colour maps to
derive core the size parameters, while aperture photometry is performed on
the HFI bands, integrating the signal within the elliptical Gaussian. 
}
\label{fig:fig_detect}
\end{figure*}

\begin{figure}
\center
\newdimen\digitwidth
\setbox0=\hbox{\rm 0}
\digitwidth=\wd0
\catcode`*=\active
\def*{\kern\digitwidth}
\begin{tabular}{cc}
 ***S1 & ***S2 \\
\includegraphics[width=4cm]{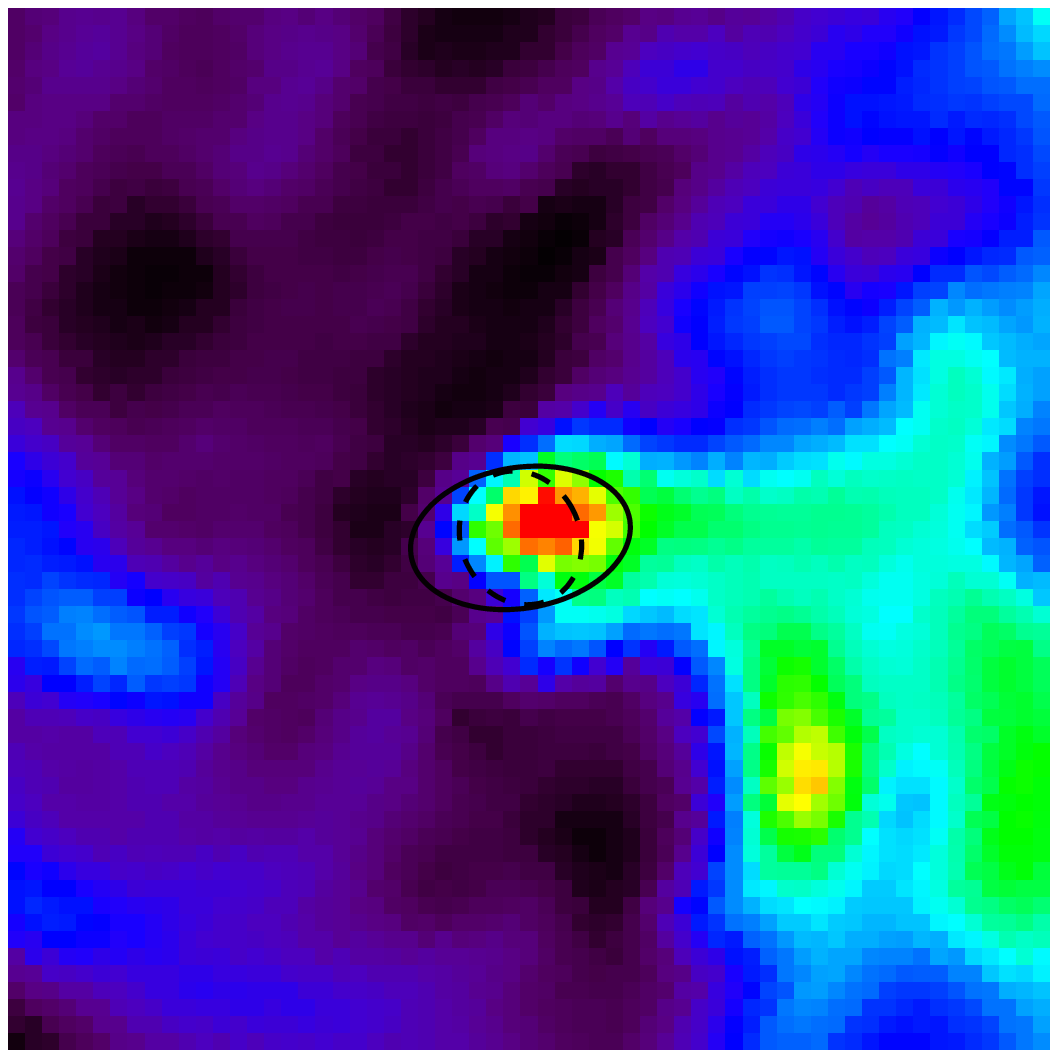}  &
\includegraphics[width=4cm]{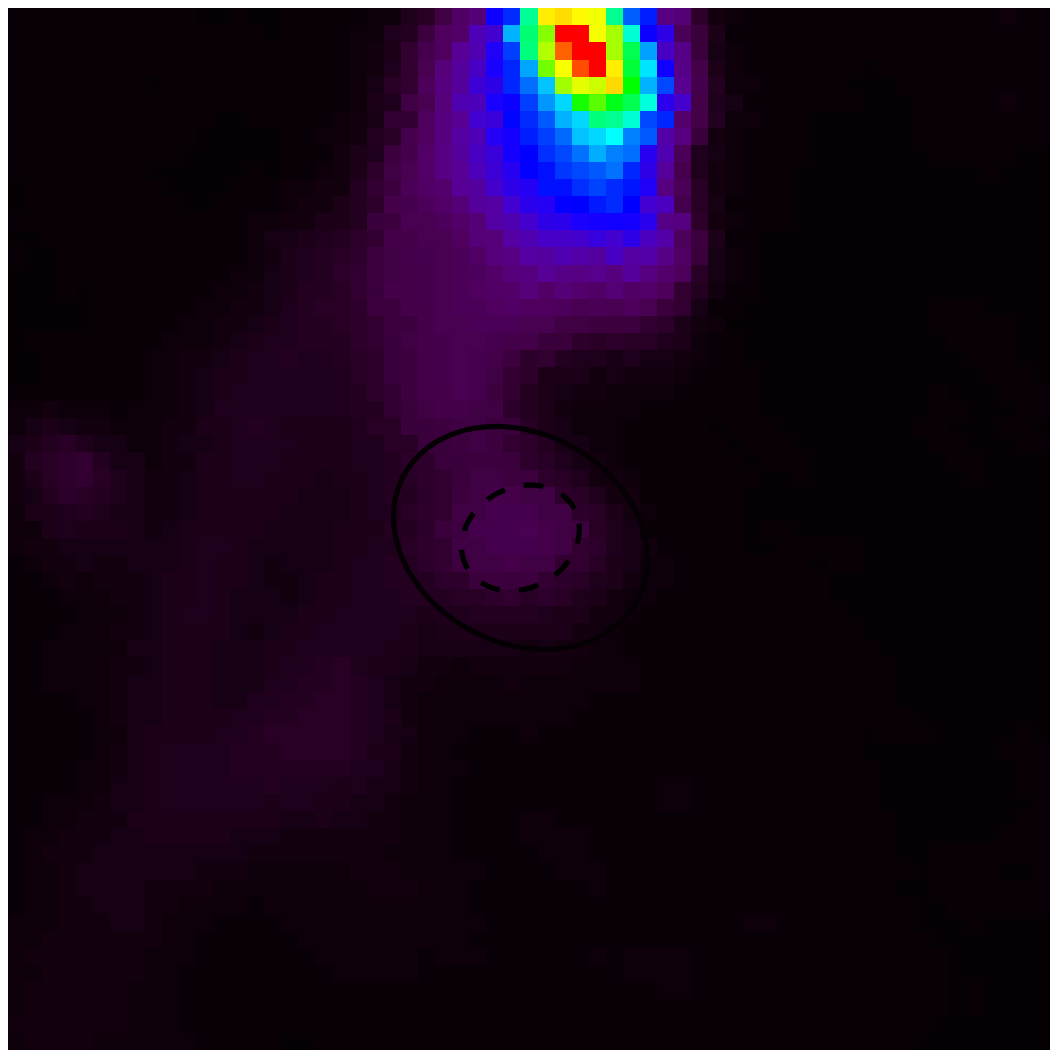} \\
 
 ***S3 & ***S4 \\
\includegraphics[width=4cm]{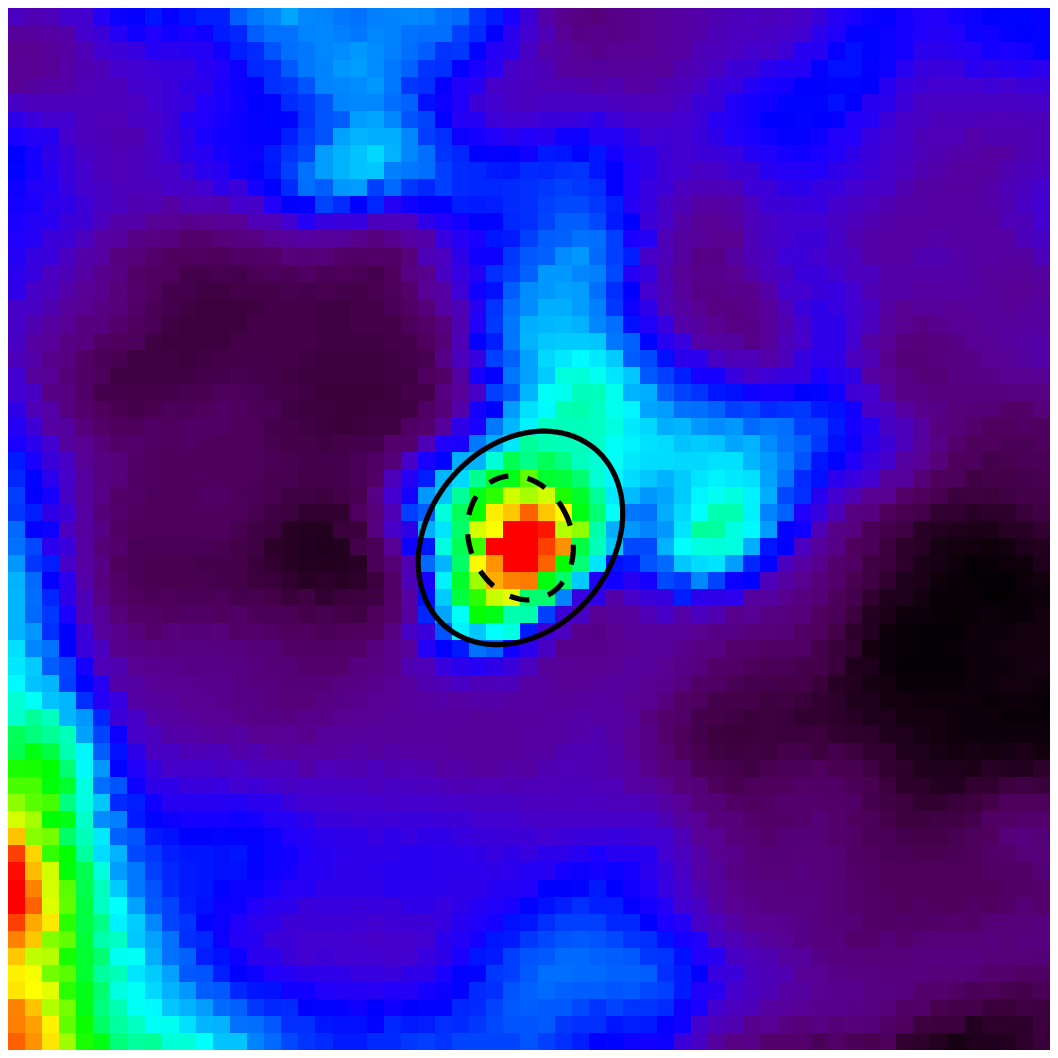}  &
\includegraphics[width=4cm]{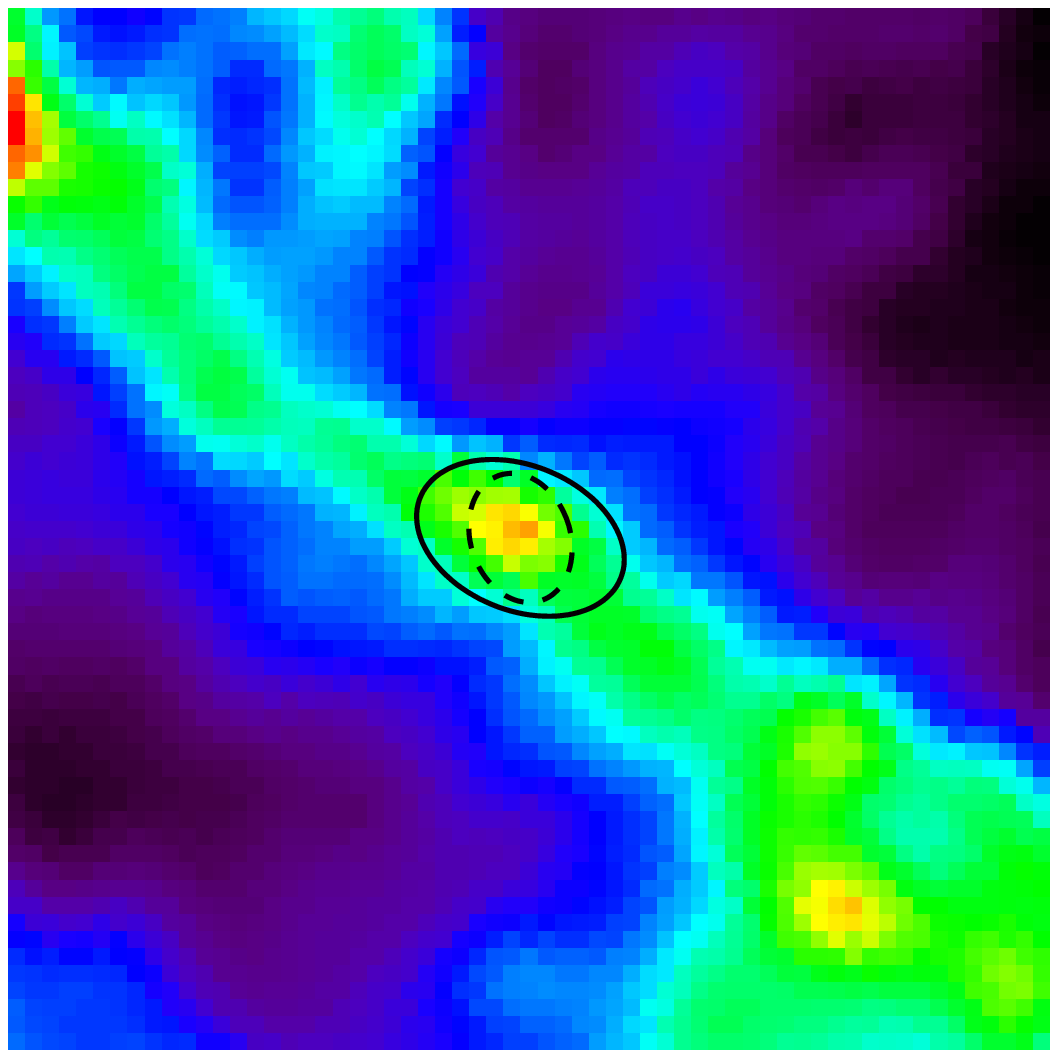} \\

 \\

 ***S5 & ***S6 \\
\includegraphics[width=4cm]{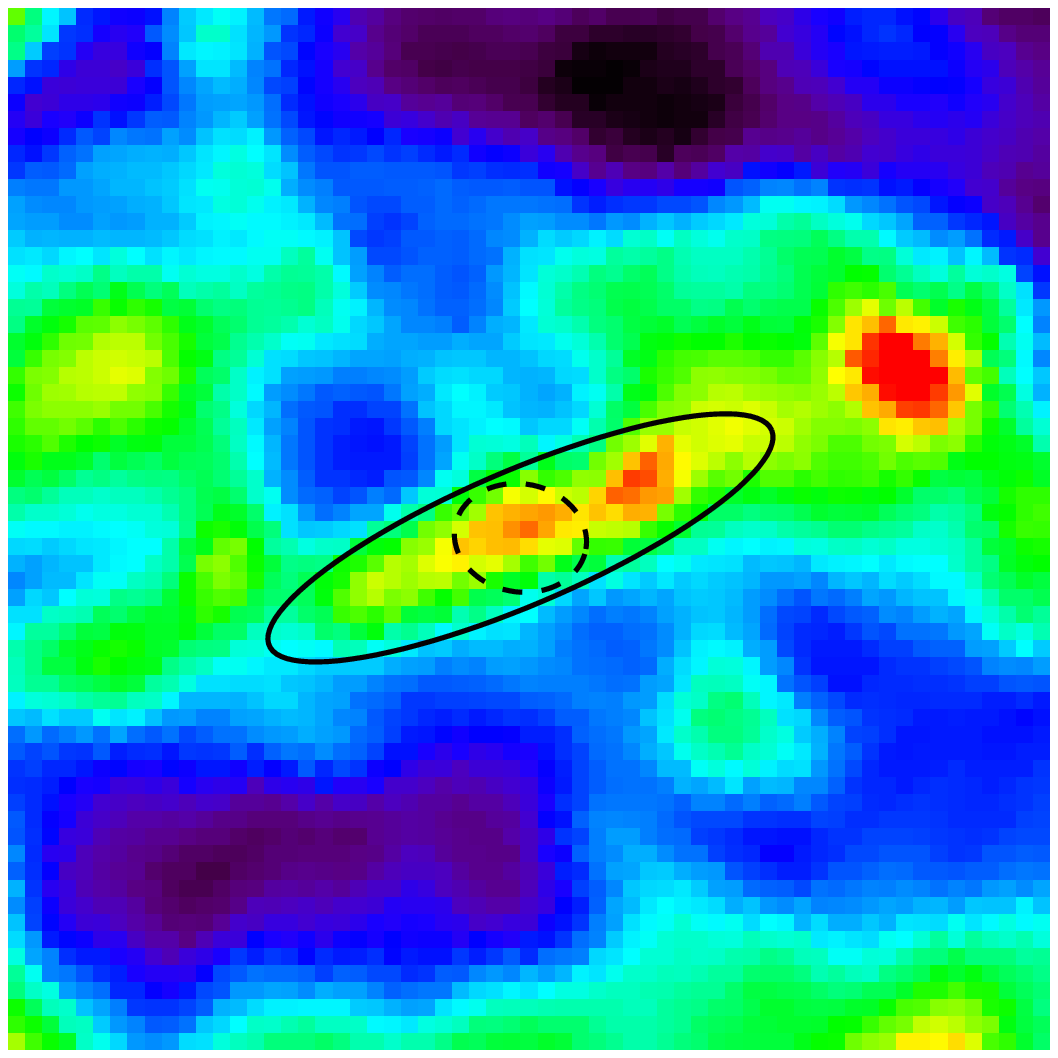}  &
\includegraphics[width=4cm]{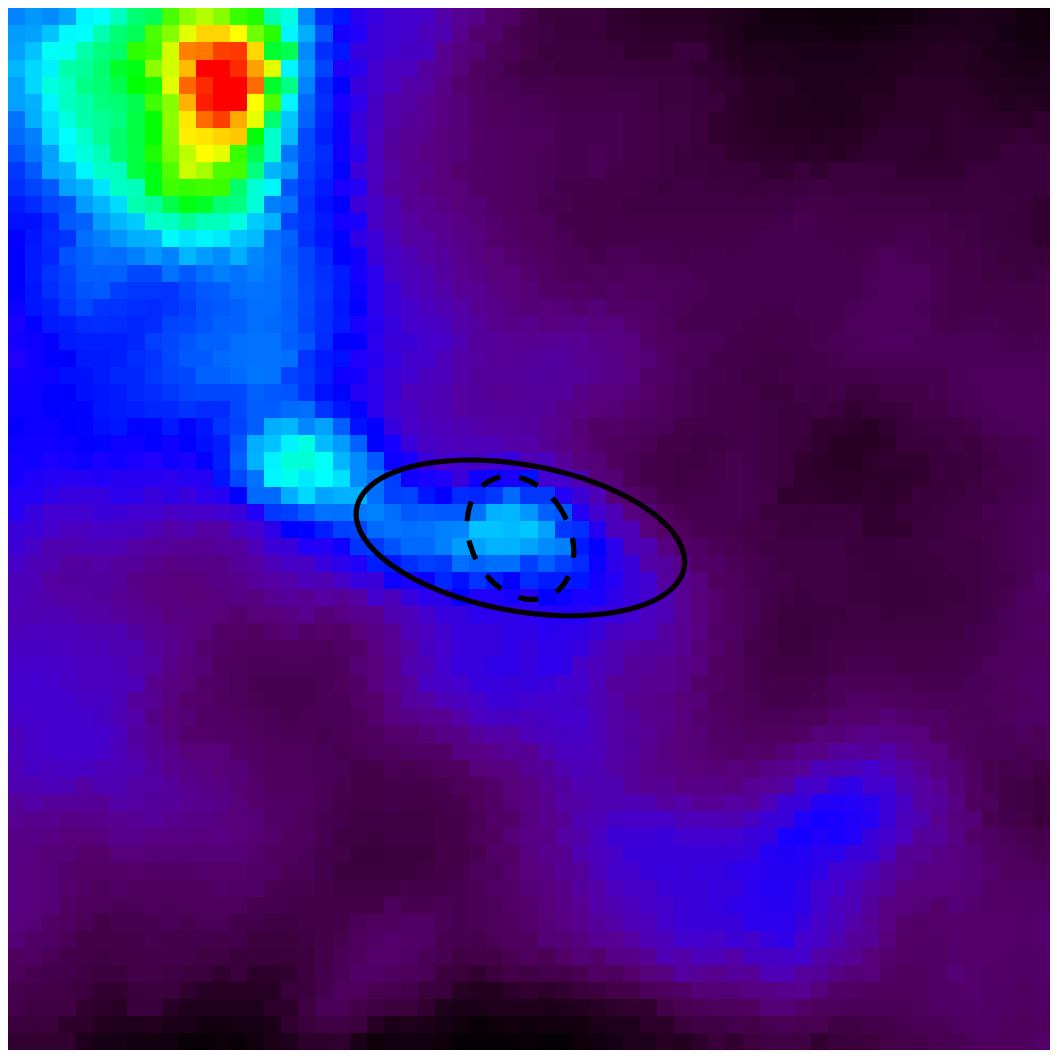} \\

 ***S7 & ***S8 \\
\includegraphics[width=4cm]{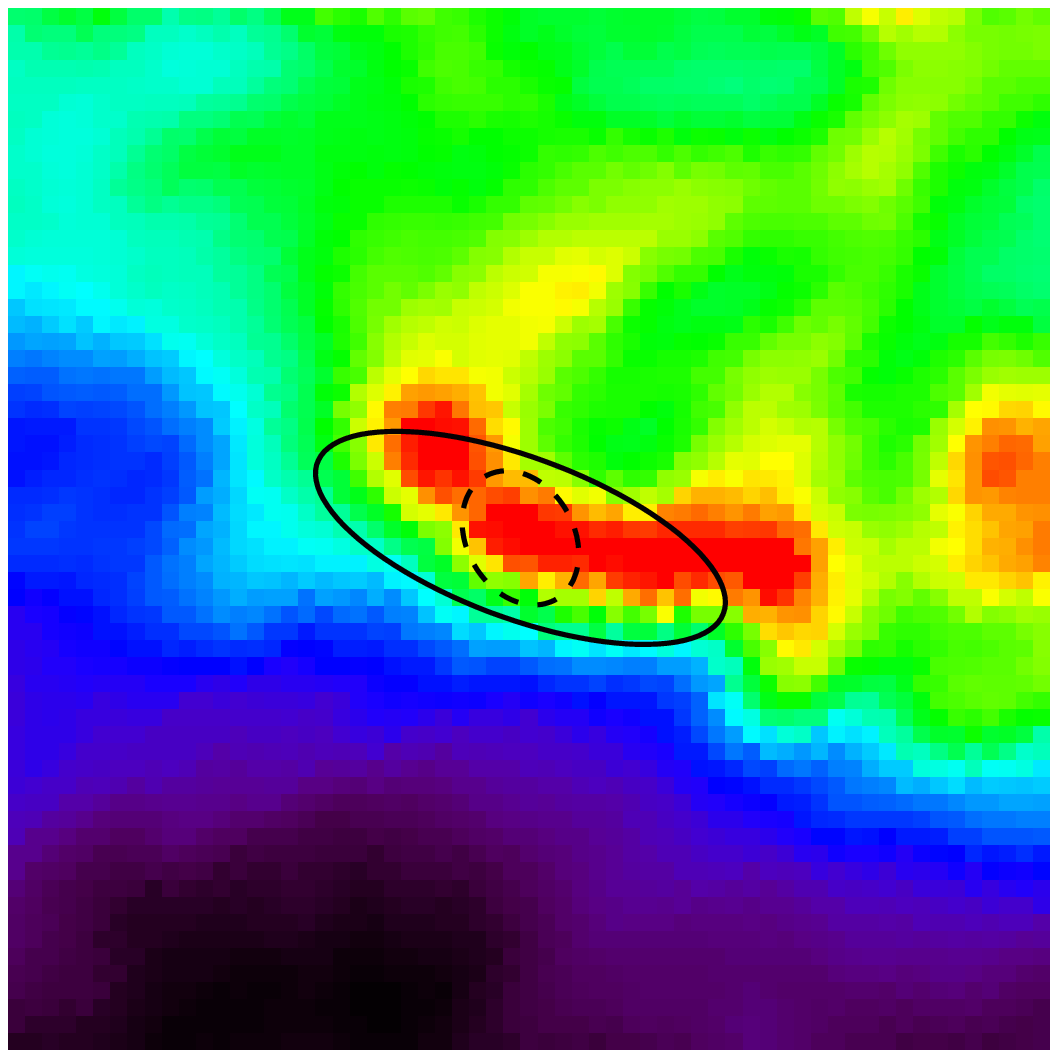}  &
\includegraphics[width=4cm]{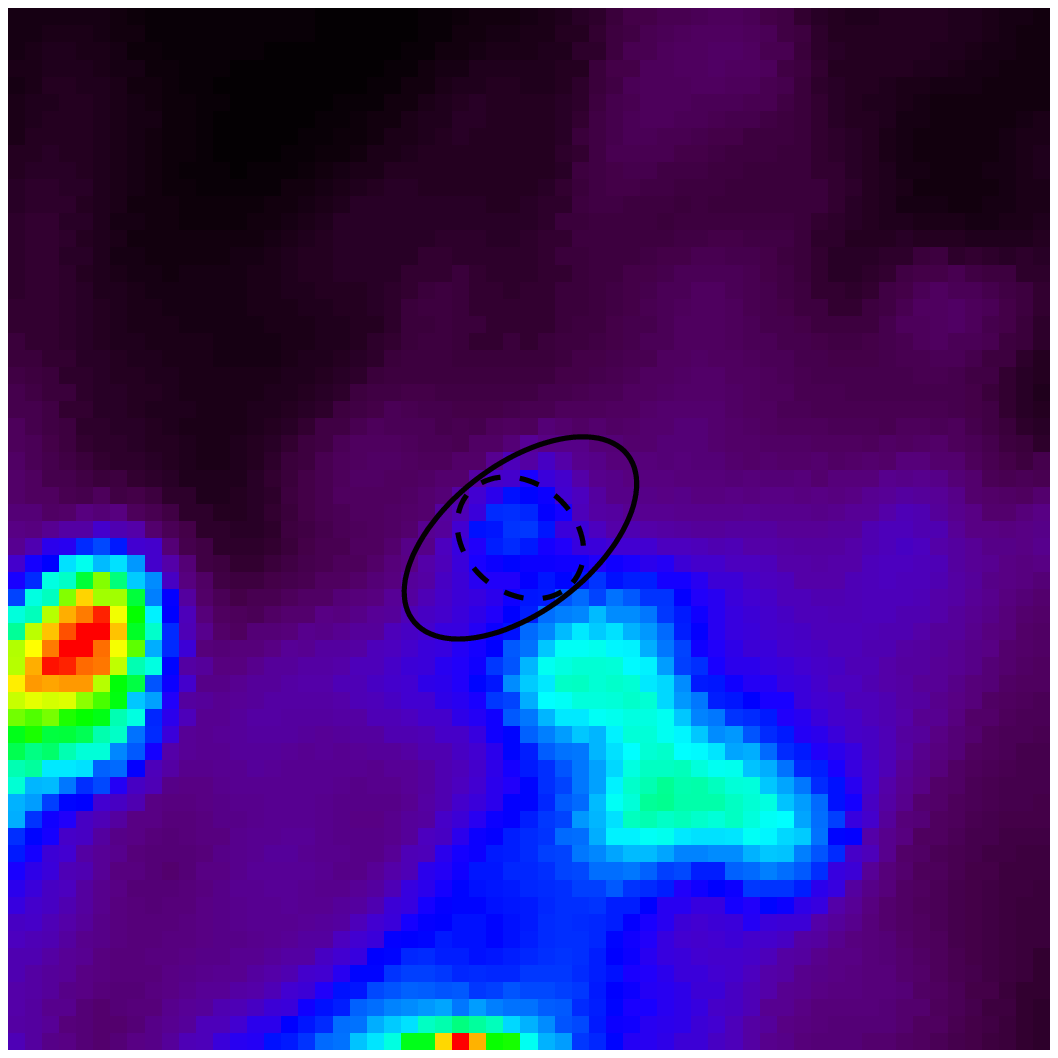} \\

 ***S9 & **S10 \\
\includegraphics[width=4cm]{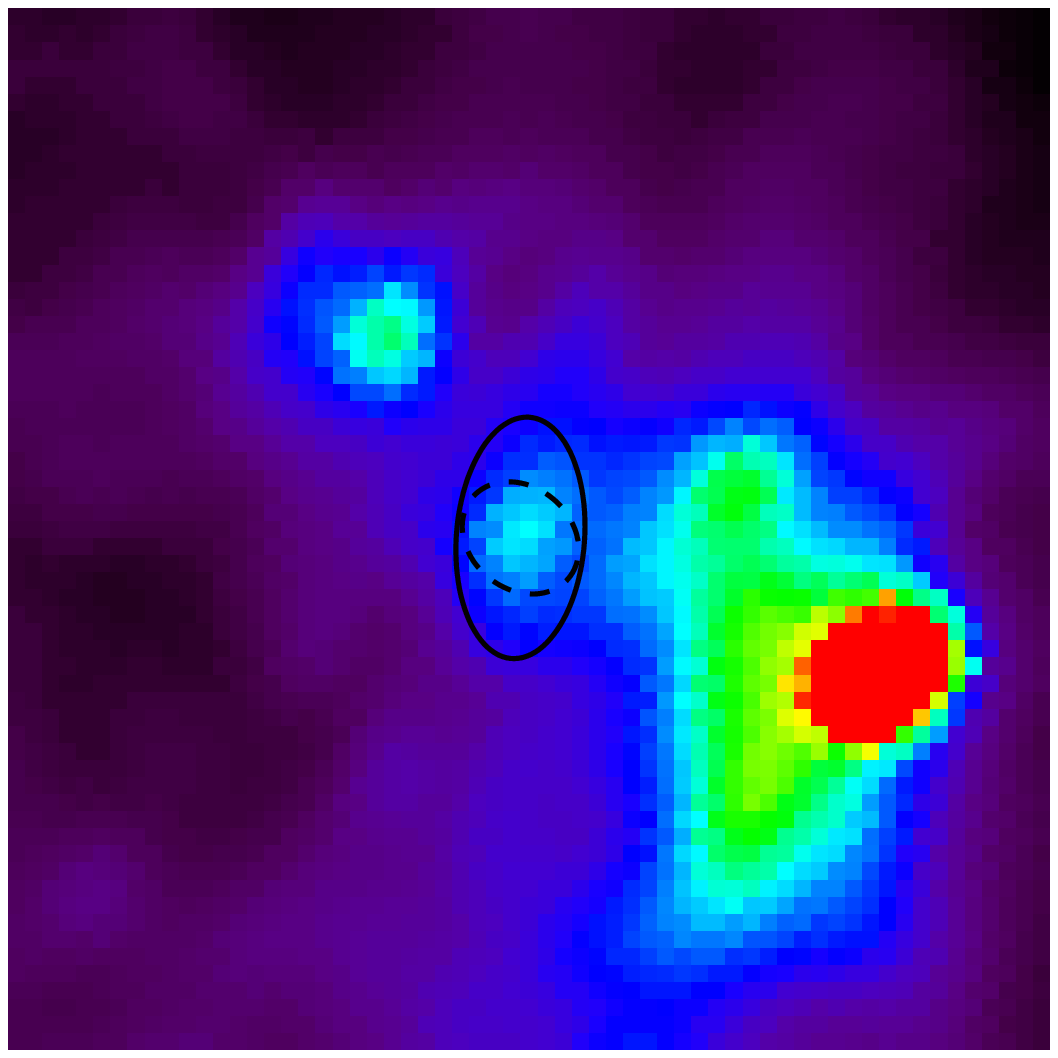}  &
\includegraphics[width=4cm]{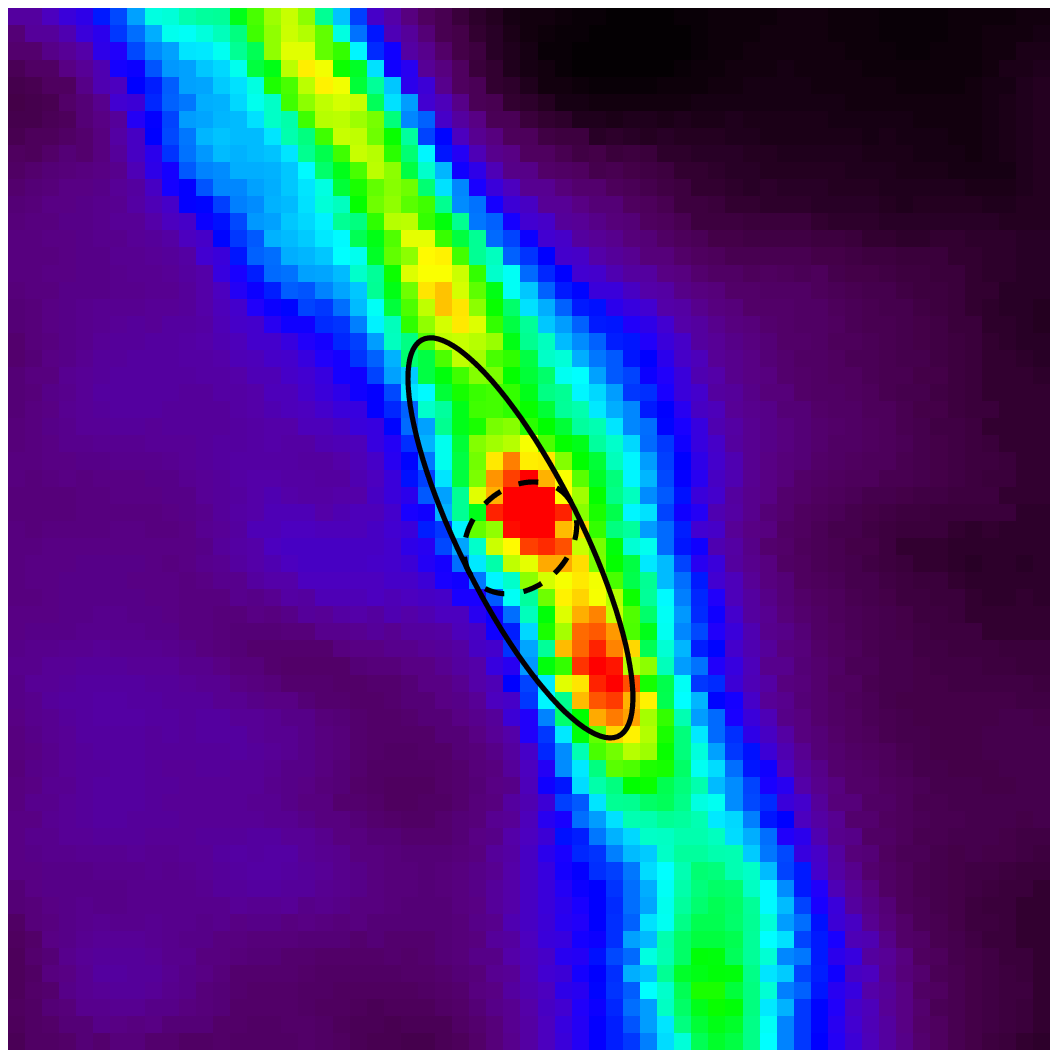} \\

\end{tabular}
\caption{Comparison of the source elongation with the beam shape determined
with the ``FEBeCoP'' tool \citep{Mitra2010}. The ellipses correspond to the estimated size
of the \Planck\ cold clumps, and the dashed circles trace the local beam shape.}
\label{fig_FEBeCoP}
\end{figure}

%\section{Telescope settings and sensitivity of IRAM observations}
\label{sect:HFI_maps}

\begin{table}
\caption{Observational parameters and sensitivity of IRAM observations}
\label{30m}
\nointerlineskip
\setbox\tablebox=\vbox{
\newdimen\digitwidth
\setbox0=\hbox{\rm 0}
\digitwidth=\wd0
\catcode`*=\active
\def*{\kern\digitwidth}
\newdimen\signwidth
\setbox0=\hbox{+}
\signwidth=\wd0
\catcode`!=\active
\def!{\kern\signwidth}
\newdimen\smallwidth
\setbox0=\hbox{\footnotesize{a}}
\smallwidth=\wd0
\catcode`@=\active
\def@{\kern\smallwidth}

\halign{#\hfil\tabskip=0.5em&
\hfil#\hfil&
\hfil#\hfil&
\hfil#\hfil&
\hfil#\hfil&
\hfil#\hfil\tabskip=0pt\cr
\noalign{\doubleline}
Molecule   & Transition & Frequency & Beam size & $\delta v$ & rms \cr
           &            & (MHz) & (arcsec) & ${\rm m}\,{\rm s}^{-1}$) & (K) \cr
\noalign{\vskip 4pt\hrule\vskip 3pt}
1st seting &&&&&\cr
\noalign{\vskip 4pt\hrule\vskip 3pt}
$^{12}$CO  & $J$=1$\rightarrow$0 & 115271.197@ & 23 & 200 &0.24@\cr
$^{12}$CO  & $J$=2$\rightarrow$1 & 230537.990@ & 12 & 100 & 0.15@\cr
\noalign{\vskip 4pt\hrule\vskip 3pt}
2nd seting &&&&&\cr
\noalign{\vskip 4pt\hrule\vskip 3pt}
$^{13}$CO  & $J$=1$\rightarrow$0 & 110201.370@ & 24 & *27 & 0.15@\cr
$^{13}$CO  & $J$=2$\rightarrow$1 & 220398.686@ & 12 & *53 & 0.15@\cr
$^{13}$CO  & $J$=1$\rightarrow$0 & 110201.370@ & 24 & *27 & 0.90$^{\rm a}$ \cr
$^{13}$CO  & $J$=2$\rightarrow$1 & 220398.686@ & 12 & *53 & 0.90$^{\rm a}$ \cr
C$^{18}$O  & $J$=1$\rightarrow$0 & 109782.182@ & 24 & *27 & 0.15@\cr
C$^{18}$O  & $J$=1$\rightarrow$0 & 109782.182@ & 24 & *27 & 0.90$^{\rm a}$ \cr
\noalign{\vskip 4pt\hrule\vskip 3pt}
3rd seting &&&&&\cr
\noalign{\vskip 4pt\hrule\vskip 3pt}
N$_2$H$^+$ & $J$=1$\rightarrow$0 & *93173.764$^{\rm b}$ & 27 & *31 & 0.04@\cr
C$^{18}$O  & $J$=2$\rightarrow$1  & 219560.358@ & 12 & *53 & 0.05@\cr
\noalign{\vskip 4pt\hrule\vskip 3pt}
} }
\endPlancktable
\tablenote a Raw value for on-the-fly mapping.\par
\tablenote b Main hyperfine components, frequency from
\cite{Pagani2009}.\par
\end{table}

\begin{acknowledgements}
A description of the Planck Collaboration and a list of its members can
be found at http://www.rssd.esa.int/index.php?project=\ PLANCK\&page=Planck\_Collaboration.
This publication makes use of data products from the Two Micron All
Sky Survey, which is a joint project of the University of
Massachusetts and the Infrared Processing and Analysis
Center/California Institute of Technology, funded by the National
Aeronautics and Space Administration and the National Science
Foundation.
This research has made use of the SIMBAD database, operated at CDS,
Strasbourg, France.
This research is based on observations with {\it AKARI}, a JAXA project with
the participation of ESA.

\end{acknowledgements}

\allearlypapers 

\raggedright
\end{document}